\documentclass[twocolumn]{aastex631}
\usepackage{hyperref}
\usepackage{multirow}

\newcommand{\s}[1]{\ensuremath{_\mathrm{#1}}}
\newcommand{\kms}{km s$^{-1}$}

\shorttitle{The SVS 13 Protobinary System}
\shortauthors{Diaz-Rodriguez et al.}

\begin{document}

\title{The Physical Properties of the SVS 13 Protobinary System: Two Circumstellar Disks and a Spiraling Circumbinary Disk in the Making}

\correspondingauthor{Guillem Anglada}
\email{guillem@iaa.es}

\author[0000-0001-9112-6474]{Ana K. Diaz-Rodriguez}
\affiliation{Instituto de Astrof{\'i}sica de Andaluc{\'i}a, CSIC,
Glorieta de la Astronom{\'i}a s/n, 18008 Granada, Spain}
\affiliation{UK ALMA Regional Centre Node, Jodrell Bank Centre for Astrophysics, Department of Physics and Astronomy, \\ The University of Manchester, Oxford Road, Manchester M13 9PL, UK}

\author[0000-0002-7506-5429]{Guillem Anglada}
\affiliation{Instituto de Astrof{\'i}sica de Andaluc{\'i}a, CSIC,
Glorieta de la Astronom{\'i}a s/n, 18008 Granada, Spain}

\author[0000-0002-7078-2373]{Guillermo Bl{\'a}zquez-Calero}
\affiliation{Instituto de Astrof{\'i}sica de Andaluc{\'i}a, CSIC,
Glorieta de la Astronom{\'i}a s/n, 18008 Granada, Spain}

\author[0000-0002-6737-5267]{Mayra Osorio}
\affiliation{Instituto de Astrof{\'i}sica de Andaluc{\'i}a, CSIC,
Glorieta de la Astronom{\'i}a s/n, 18008 Granada, Spain}

\author[0000-0002-7065-542X]{Jos{\'e} F. G{\'o}mez}
\affiliation{Instituto de Astrof{\'i}sica de Andaluc{\'i}a, CSIC,
Glorieta de la Astronom{\'i}a s/n, 18008 Granada, Spain}

\author[0000-0001-8509-1818]{Gary A. Fuller}
\affiliation{UK ALMA Regional Centre Node, Jodrell Bank Centre for Astrophysics, Department of Physics and Astronomy, \\ The University of Manchester, Oxford Road, Manchester M13 9PL, UK}
\affiliation{Instituto de Astrof{\'i}sica de Andaluc{\'i}a, CSIC,
Glorieta de la Astronom{\'i}a s/n, 18008 Granada, Spain}
\affiliation{I. Physikalisches Institut, University of Cologne, Z\"ulpicher Str. 77, 50937 K\"oln, Germany}

\author[0000-0001-7341-8641]{Robert Estalella}
\affiliation{Departament de F{\'i}sica Qu\`antica i Astrof{\'i}sica, Institut de Ci{\`e}ncies del Cosmos, Universitat de Barcelona, IEEC-UB, Mart{\'i} i Franqu{\`e}s, 1, 08028 Barcelona, Spain}

\author[0000-0002-6896-6085]{Jos\'e M. Torrelles}
\affiliation{Institut de Ci{\`e}ncies de l'Espai (ICE, CSIC), Can Magrans s/n, E-08193 Cerdanyola del Vall{\`e}s, Catalonia, Spain}
\affiliation{Institut d'Estudis Espacials de Catalunya (IEEC), E-08034 Barcelona, Catalonia, Spain}

\author[0000-0002-1593-3693]{Sylvie Cabrit}
\affiliation{Univ. Grenoble Alpes, CNRS, IPAG, 38000 Grenoble, France}
\affiliation{Observatoire de Paris, PSL University, Sorbonne Universit{\'e}, CNRS, LERMA, F-75014, Paris, France}

\author[0000-0003-2737-5681]{Luis F. Rodr{\'i}guez}
\affiliation{Instituto de Radioastronom{\'i}a y Astrof{\'i}sica, Universidad Nacional Aut{\'o}noma de M{\'e}xico, P.O. Box 3-72, 58090, Morelia, Michoac{\'a}n, M{\'e}xico}

\author[0000-0001-7349-6113]{Charl{\`e}ne Lef{\`e}vre}
\affiliation{Institut de RadioAstronomie Millim{\'e}trique (IRAM), 300 rue de la Piscine, 38406 Saint Martin d'H{\`e}res, France}

\author[0000-0003-1283-6262]{Enrique Mac{\'i}as}
\affiliation{Joint ALMA Observatory, Alonso de C{\'o}rdova 3107, Vitacura, Santiago 763-0355, Chile}
\affiliation{European Southern Observatory (ESO), Alonso de C{\'o}rdova 3107, Vitacura, Santiago 763-0355, Chile}

\author[0000-0003-2862-5363]{Carlos Carrasco-Gonz\'alez}
\affiliation{Instituto de Radioastronom{\'i}a y Astrof{\'i}sica, Universidad Nacional Aut{\'o}noma de M{\'e}xico, P.O. Box 3-72, 58090, Morelia, Michoac{\'a}n, M{\'e}xico}

\author[0000-0003-2343-7937]{Luis A. Zapata}
\affiliation{Instituto de Radioastronom{\'i}a y Astrof{\'i}sica, Universidad Nacional Aut{\'o}noma de M{\'e}xico, P.O. Box 3-72, 58090, Morelia, Michoac{\'a}n, M{\'e}xico}

\author[0000-0003-4518-407X]{Itziar de Gregorio-Monsalvo}
\affiliation{European Southern Observatory (ESO), Alonso de C{\'o}rdova 3107, Vitacura, Santiago 763-0355, Chile}

\author[0000-0002-3412-4306]{Paul T. P. Ho}
\affiliation{Academia Sinica Institute of Astronomy and Astrophysics, P.O. Box 23-141, Taipei 106, Taiwan}
\affiliation{East Asian Observatory, Hilo 96720, HI, USA}




\begin{abstract}

We present VLA and ALMA observations of the close ($0\farcs3$ = 90 au separation) protobinary system SVS 13. We detect two small circumstellar disks (radii $\sim$12 and $\sim$9 au in dust, and $\sim$30 au in gas) with masses of $\sim$0.004-0.009 $M_{\sun}$ for VLA 4A (the western component) and $\sim$0.009-0.030 $M_{\sun}$ for VLA 4B (the eastern component). A circumbinary disk with prominent spiral arms extending $\sim$500 au and a mass of $\sim$0.052 $M_{\sun}$ appears to be in the earliest stages of formation. The dust emission is more compact and with a very high optical depth toward VLA 4B, while toward VLA 4A the dust column density is lower, allowing the detection of stronger molecular transitions. We infer rotational temperatures of $\sim$140 K, on scales of $\sim$30 au, across the whole source, and a rich chemistry. Molecular transitions typical of hot corinos are detected toward both protostars, being stronger toward VLA 4A, with several ethylene glycol transitions detected only toward this source. There are clear velocity gradients, that we interpret in terms of infall plus rotation of the circumbinary disk, and purely rotation of the circumstellar disk of VLA 4A. We measured orbital proper motions and determined a total stellar mass of 1 $M_{\sun}$. From the molecular kinematics we infer the geometry and orientation of the system, and stellar masses of $\sim$0.26 $M_{\sun}$ for VLA 4A and $\sim$0.60 $M_{\sun}$ for VLA 4B.

\end{abstract}




\section{Introduction} \label{sec:intro}

Although most stars belong to binary systems \citep{Leinert1993,Raghavan2010,Duchene2013}, the way in which such systems form still remains poorly known, especially in the first stages of the process. Two main different mechanisms have been suggested for binary formation (e.g., \citealt{Bate2015}), namely disk fragmentation (e.g., \citealt{Bonnell1994,Bonnell1994a,Bonnell1994b,Kratter2010}) and the large-scale fragmentation of cores and filaments, either without turbulence \citep{Matsumoto2003} or through turbulent fragmentation (e.g., \citealt{Padoan1999,Offner2010}).  Although several observational studies have investigated which mechanism is dominant (e.g., \citealt{Tobin2018}), the debate is not yet settled. On the other hand, dissipative dynamical interactions within a cluster are also a major factor to be considered, especially for understanding the formation of close binary systems \citep{Bate2012,Bate2015}.

At the early stages, a binary protostellar system will usually be embedded in a more massive infalling envelope from which it accretes to its final mass. Accretion of this material changes the masses of the components of the binary, the binary's orbit, and it is also largely responsible for forming the final configuration of disks in the system \citep{Bate2015}. Thus, besides the investigation of the fragmentation mechanism, a fundamental issue is to study the early evolution of the binary during the formation process, as this process determines not only the final stellar masses, but also the formation of disks and, consequently, the formation of planets (e.g., \citealt{Bate2015,Bate2018}).

In a binary system, three disks are possible: a circumprimary disk, a circumsecondary disk, and a circumbinary disk. According to numerical simulations, the specific angular momentum of the accreted gas relative to that of the binary is the most important parameter to determine the evolution of the binary properties and, in particular, the formation of a circumbinary disk (e.g., \citealt{Bate2015}).

Since circumbinary disks are not geometrically thin, high resolution 3D simulations are required to study these structures in young binary systems. The recent 3D simulations of \citet{Matsumoto2019} provide fine details of the formation and evolution of circumstellar disks, spiral arms, and circumbinary structures in protobinary systems. These simulations show the formation of spirals, a bridge structure between the circumstellar disks, and the development of a slowly rotating asymmetric pattern in the circumbinary disk. Some of these features have been observationally detected (e.g., in the systems L1551 NE; \citealt{Takakuwa2014,Takakuwa2017}, and L1551 IRS 5; \citealt{CruzSaenz2019,Takakuwa2020}), but this kind of work is still scarce and more observational studies providing good estimates of physical parameters suitable for making detailed comparisons between observations and 3D simulations are needed.

Therefore, it is imperative to conduct in-depth observational studies of the earliest stages of the binary star formation process, i.e., to study protobinary systems. 
Binaries with separations $\sim$1000 au are called wide binaries and are attributed to core fragmentation, while those with separations of $\sim$100 au are in the boundary between wide and close binaries. In the latter, the interaction between the two stars significantly affects their evolution. This is particularly relevant when the two stars have different initial properties. Thus, protobinaries with separations of the order of 100 au or less, whose observation is currently achievable with the resolution of the large radio interferometers, are excellent targets to study their properties and to compare them with those predicted by numerical simulations (e.g., \citealt{Bate2018,Matsumoto2019}). 

So far, only for a limited number of these systems there is available (but partial) observational information about their physical properties, such as temperatures, kinematics (from radial velocities and proper motions), presence of circumbinary disks, stellar and disk masses (from dust and molecular emission and from dynamics), chemical properties, interaction with the surrounding medium (e.g., outflows and interaction with the ambient molecular cloud, flyby encounters and tidal interaction with nearby young stars). Some examples where these studies have been performed are the binary systems L1551 IRS 5 \citep{Rodriguez1998,CruzSaenz2019,Takakuwa2020}, RW Aur \citep{Cabrit2006,Rodriguez2018}, L1551 NE \citep{Takakuwa2014,Takakuwa2017}, XZ Tau \citep{Carrasco2009,Zapata2015,Osorio2016,Ichikawa2021}, GG Tau \citep{Phuong2020}, BHB2007 \citep{Alves2019}, UX Tau \citep{Zapata2020}, and IRAS 16293$-$2422 \citep{Maureira2020}.

A particularly remarkable case is SVS 13, the proposed exciting source of the HH 7-11 system \citep{Strom1976}. SVS 13 is a nearby ($d$ = 300 pc; \citealt{Ortiz2018,GaiaEDR3}, see also \citealt{Hirota2008}) optically visible star (e.g., \citealt{Eisloeffel1991}) but also a strong mm source (e.g., \citealt{Looney2000}). VLA 3.6 cm observations \citep{Anglada2000} revealed that SVS 13 is a close binary system, with the two components (VLA 4A and VLA 4B) separated in projection by $0\farcs3$ (90 au). Further observations \citep{Anglada2004} revealed that only one of the components, VLA 4B, presents a strong 7 mm emission, that was attributed to dust. These results were interpreted as implying that the development of a protoplanetary disk occurs preferentially in one of the components, with the disk being much weaker in the other component. Strong maser emission has been detected in the proximity of SVS 13 and toward other nearby sources in the field \citep{Rodriguez2002}. More recently, the close environment of the SVS 13 system has been explored in the near-IR using laser adaptive-optics at the Keck telescope \citep{Hodapp2014}, revealing a $0\farcs2$ long microjet traced by [Fe II] emission and a sequence of H$_2$ bubbles extending a few arcsec away from the position of VLA 4B. The system has also been observed at mm wavelengths with the 30 m and PdBI of IRAM (e.g., \citealt{Lefevre2017,Bianchi2019,Belloche2020}) tracing the dust and several molecular transitions with angular resolutions ranging from $0\farcs3$ to $\sim$30$\arcsec$. The source has also been observed as part of the VANDAM survey at cm wavelengths with the VLA \citep{Tychoniec2018} and at 1.3 mm continuum and several molecular line transitions with a resolution of $\sim0\farcs2$-$0\farcs3$ with ALMA \citep{Tobin2018}.

 In this paper, we present new VLA results in the continuum from 3 cm to 7 mm, as well as ALMA 0.9 mm continuum and line results with improved angular resolution and sensitivity. These results, combined with recent Gaia astrometry, shed new light on this protobinary system probing scales as small as $\sim0\farcs05$ ($\sim$15 au).

Our paper is focused on characterizing as well as possible the physical and chemical properties of the SVS 13 protobinary system. We defer to a subsequent paper a detailed comparison with both model simulations and observations from the literature of other sources reaching a similar degree of detail. The rest of this paper is organized as follows. In Section \ref{sec:obs} we describe the observations and data reduction procedures. In Section \ref{sec:results} we report the sources detected and discuss in detail our main results on the SVS 13 protobinary: In Section \ref{sec:vla1} we describe our main VLA results on the ionized emission and proper motions of the protostars; in Section \ref{sec:dust1} we describe the circumstellar and circumbinary dust observed with ALMA; in Section \ref{sec:sed} we analyze the overall spectral energy distribution; in Section \ref{sec:alma} we present our main ALMA results regarding the distribution and kinematics of the molecular gas, as well as the molecular transitions identified; in Section \ref{sec:phys} we estimate the physical parameters of the system (temperature, disk masses, and stellar masses), and we discuss about their implications; in Section \ref{sec:chem} we discuss the chemical properties and in particular the hot corino nature of the protostellar sources. Finally, Section \ref{sec:conclusions} summarizes our main results and conclusions.

\section{Observations and Data Reduction} \label{sec:obs}

We present continuum observations in the 0.7-3 cm (10-43 GHz) range carried out with the Karl G. Jansky Very Large Array (VLA) of the National Radio Astronomy Observatory (NRAO)\footnote{The National Radio Astronomy Observatory is a facility of the National Science Foundation operated under cooperative agreement by Associated Universities, Inc.}, as well as continuum and line observations at $\sim$0.9 mm ($\sim$333 GHz) taken with the Atacama Large Millimeter/submillimeter Array (ALMA). Except when noted otherwise, the images presented in this work have been corrected by the primary beam response.

\subsection{VLA Continuum and H$_2$O Maser Observations}\label{sec:vlaobs}

The VLA observations were carried out as part of our Project 15A-260 (PI: Ana K. Diaz-Rodriguez) in the B configuration at Ka (9 mm) and Q (7 mm) bands on 2015 February 14 and March 7, respectively, and in the A configuration at X (3 cm), K (1.3 cm), Ka, and Q bands on 2015 June 28, June 27, July 7, and June 20, respectively. Data at X band had poor quality, and we repeated the observations at this band in the A configuration on 2016 November 22 as part of DDT 16B-417 project (PI: Ana K. Diaz-Rodriguez). The results presented in this paper at X band are obtained using only the 2016 data.

The bandwidth was 4 GHz at X band, and 8 GHz at K, Ka, and Q bands. We used wideband mode setup for X, Ka, and Q bands, and spectral line mode setup for the K band to observe the H$_2$O maser emission at 22.23508 GHz. For this, we included a narrow subband of 4 MHz covering the maser line, with 128 channels of 31.25 kHz (0.42 km~s$^{-1}$) each; the other 63 subbands had 128 MHz bandwidth each, and were used for the continuum imaging. The phase center was at coordinates $\alpha$(J2000) = 03$^{\rm h}$29$^{\rm m}$3$\fs$75, $\delta$(J2000)=+31\degr16\arcmin3\farcs9 in all VLA observations. Absolute flux calibrators were J0137+3309 (3C48) at X, K, and Ka bands, and J0542+4951 (3C147) at Q band. Bandpass calibration was performed using J0137+3309 (3C48) at X band, and J0319+4130 (3C84) for the rest of the bands. J0336+3218 was used for complex gain calibration at all bands. The time on source in A configuration was 16.5 min at Q band, 23.6 min at Ka band, 39.7 min at K band, and 39.2 min at X band, while in B configuration was 20.4 min at Q band and 21.7 min at Ka band.

\begin{figure*}[htb]
\gridline{\fig{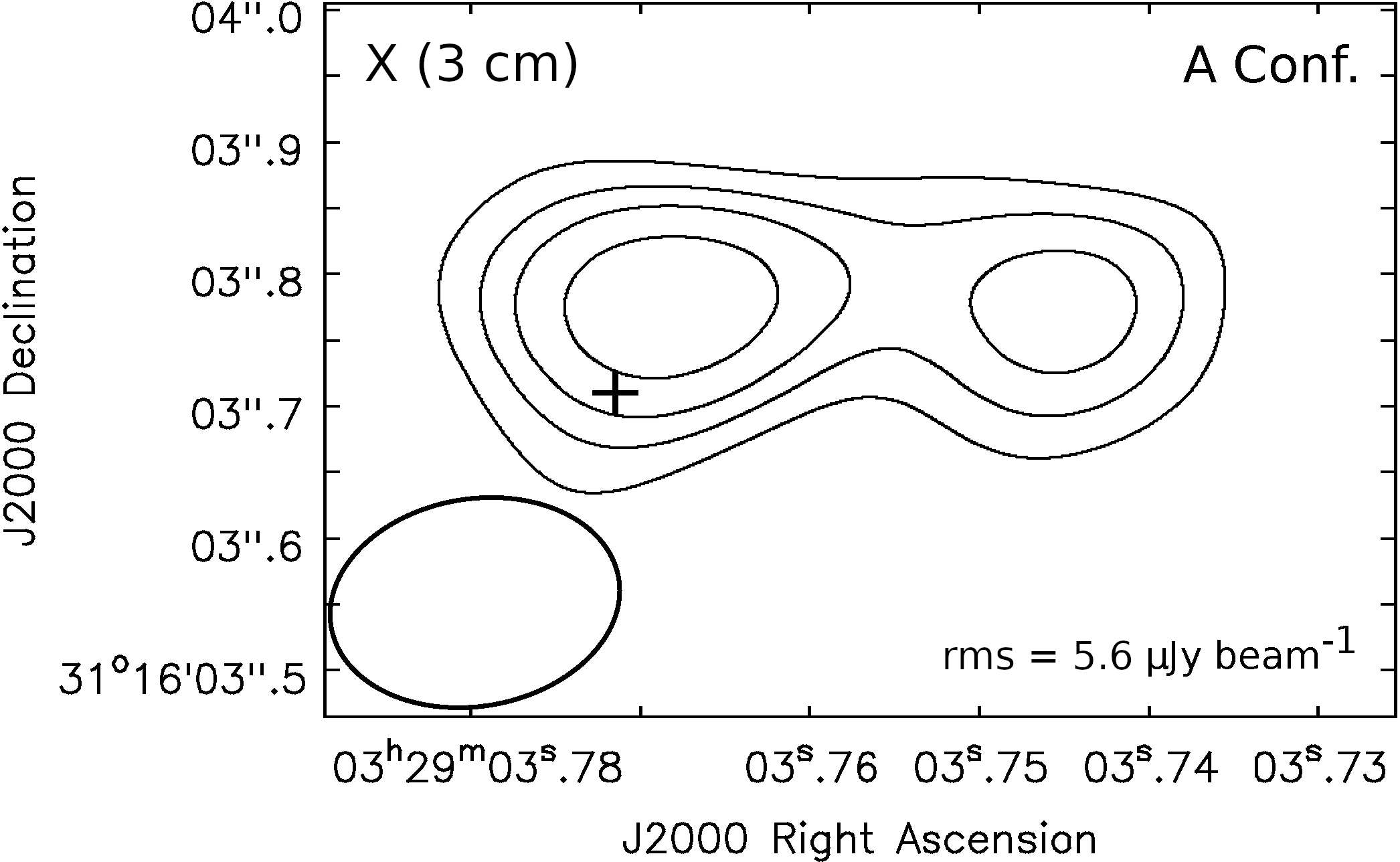}{0.46\textwidth}{(a)}
\hfill\fig{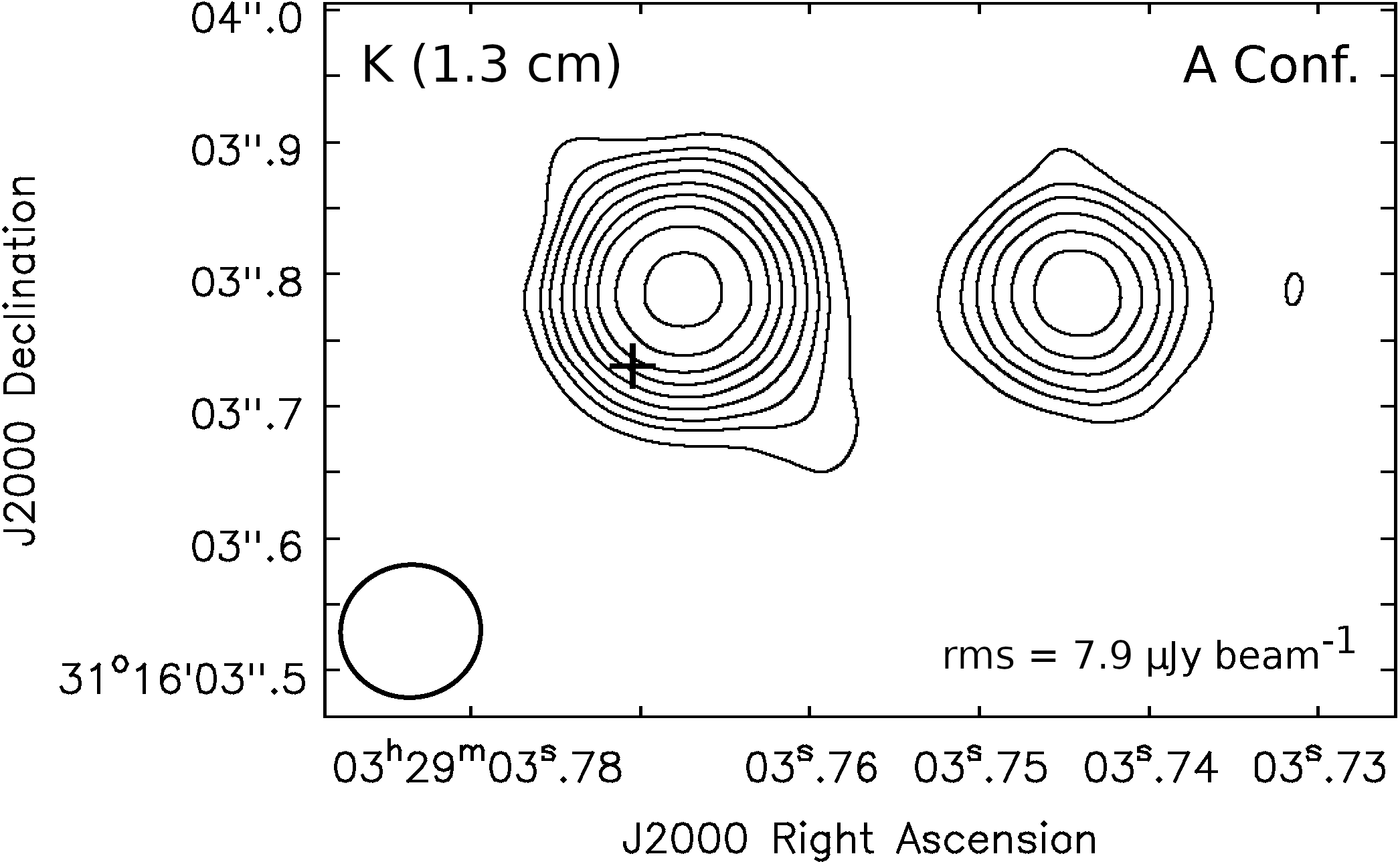}{0.46\textwidth}{(b)}
          }
\gridline{\fig{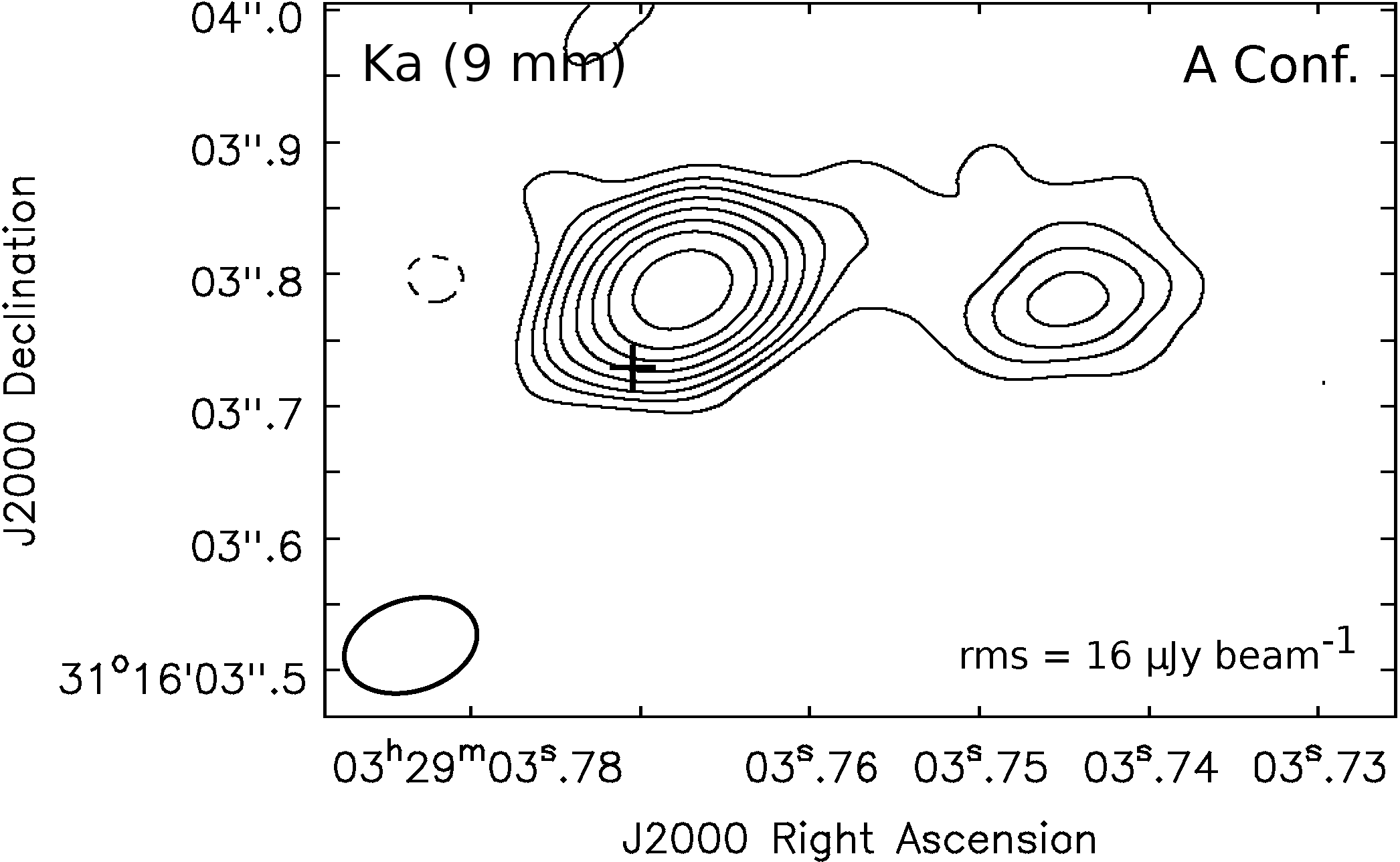}{0.46\textwidth}{(c)}
\hfill\fig{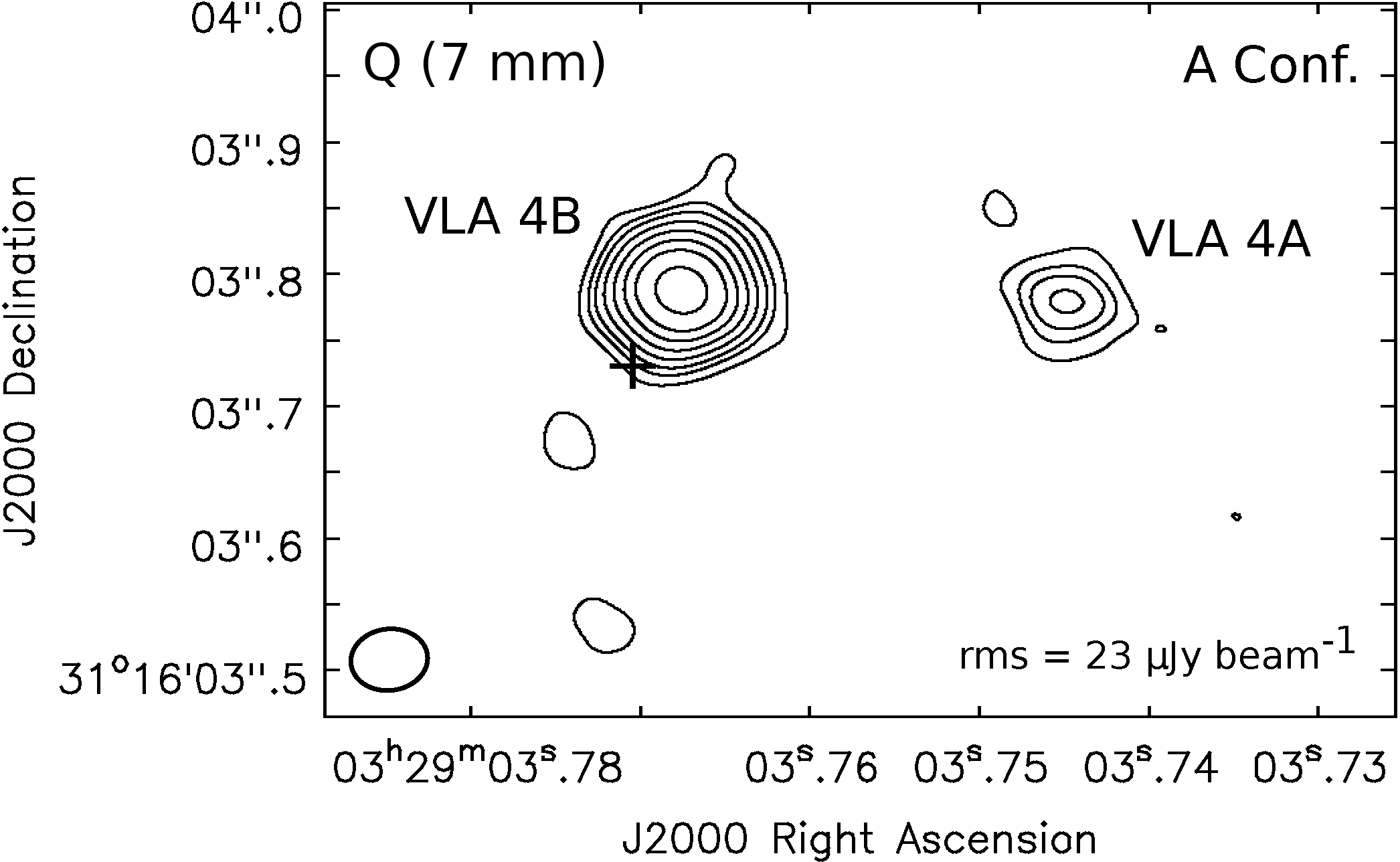}{0.46\textwidth}{(d)}
          }
\caption{VLA images of SVS 13 at X (3 cm), K (1.3 cm), Ka (9 mm), and Q (7 mm) bands obtained with the A configuration. The plus sign marks the optical position of the star SVS 13 from Gaia EDR3 \citep{GaiaEDR3} extrapolated to the epoch of the VLA observations (a correction $<0\farcs014$; see text); after this correction the Gaia position is accurate to within a fraction of mas, well below the size of the plus sign. Contour levels are $-$3, 3, 5, 7, 10, 14, 19, 25, 35, and 50 times the rms of each image. The synthesized beams and the rms are shown in the bottom left and right corners, respectively, of the images. We used natural weighting for images at K, Ka, and Q bands, and Briggs weighting with the robust parameter set to 0 for the image at X band. The synthesized beams are $0\farcs22\times0\farcs16$, PA = $-80.3\degr$ at X band; $0\farcs11\times0\farcs10$, PA = $-80.7\degr$ at K band; $0\farcs10\times0\farcs07$, PA = $-72.9\degr$ at Ka band; and $0\farcs06\times0\farcs05$, PA = $-83.4\degr$ at Q band.
\label{fig:vla}}
\end{figure*}

The data were first pipeline calibrated and then manually edited and calibrated again using the Common Astronomy Software Applications package (CASA) version 4.3.1. Self-calibration was attempted but no improvement of the signal-to-noise ratio was obtained. Images were made with the task CLEAN of CASA version 4.6.0. The synthesized beams of the continuum images range from $\sim0\farcs20$ (X band) to $\sim0\farcs06$ (Q band) and the rms noise from 6 $\mu$Jy beam$^{-1}$ to 23 $\mu$Jy beam$^{-1}$, respectively (see Fig.~\ref{fig:vla}).
 In the K-band cube we detect strong H$_2$O maser emission at a few Jy level in the proximity of SVS 13, indicating that the previously reported masers \citep{Rodriguez2002} are still active; these results will be presented in a forthcoming paper.

\subsection{ALMA Continuum and Line Observations}\label{sec:almaobs}

The ALMA observations were carried out in Band 7 (0.9 mm) as part of our projects 2015.1.01229.S (Cycle 3; PI: Guillem Anglada) and 2016.1.01305.S (Cycle 4; PI: Guillem Anglada), with angular resolutions of $\sim0\farcs1$ and $\sim0\farcs3$, respectively, and a maximum recoverable scale of $2\farcs6$ in both projects. Data from Cycle 3 were obtained during two identical runs on 2016 September 9 and 10, with 36 antennas operating with baselines ranging from 15 m to 3247 m. The total time spent on source was 1.6 h. The precipitable water vapor was 0.44 mm and 0.41 mm, during the first and second day of observations, respectively. Data from Cycle 4 were obtained on 2016 November 24, with 40 antennas operating with baselines ranging from 15 m to 704 m. The time on source was 4.15 min and the precipitable water vapor was 0.63 mm.  In all cases, the calibrators J0237+2848, J0238+1636, and J0336+3218 were used for bandpass, absolute flux, and complex gain calibration, respectively.

In the Cycle 3 observations, the correlator was configured with one baseband set to low spectral resolution continuum mode, and three basebands set to high spectral resolution line mode. Each line baseband was subdivided into two equal spectral windows of 0.234 GHz bandwidth. The continuum baseband was centered at 333 GHz and had 1.875 GHz of bandwidth. The line spectral windows were centered on the sky frequencies of the $^{13}$CO (3-2), SO$_2$ (11-12), CS (7-6),  HC$^{15}$N (4-3),  H$^{13}$CN (4-3), and $^{12}$CO (3-2), whose rest frequencies are 330.587965 GHz, 331.580244 GHz, 342.882857 GHz, 344.200109 GHz, 345.339756 GHz, and 345.795990 GHz, respectively. The channel width was 0.1058 km s$^{-1}$ (0.122 MHz) for the CO spectral windows, 0.212 km s$^{-1}$ (0.244 MHz) for H$^{13}$CN, CS, and SO$_2$, and 0.426 km s$^{-1}$ (0.488 MHz) for the HC$^{15}$N spectral window. The line free-channels of the line spectral windows (totaling $\sim$ 0.3 GHz) and of the continuum baseband were used to form the continuum image, covering a total bandwidth of $\sim$2.175 GHz. 

In the Cycle 4 observations, the correlator was configured with three basebands set to low spectral resolution continuum mode and one set to high spectral resolution line mode. Continuum basebands had 1.875 GHz of bandwidth each and were centered at 331.5, 333.5, and 344.0 GHz. The high spectral resolution baseband was centered on the sky frequency of $^{12}$CO (3-2) (rest frequency 345.795990 GHz), with a channel width of 0.426 km s$^{-1}$ (0.488 MHz) and 0.9375 GHz bandwidth.

After the standard pipeline calibration, the continuum data from Cycle 3 were phase-only self-calibrated to increase the signal to noise ratio. We used two rounds in which the solutions were found in intervals that spanned the entire scan length of 6.07 minutes (first round), and 6.05 s, which is the length of a single integration (second round). The self-calibration solutions were applied to the spectral line data and the continuum emission was subtracted. Briggs-weighted (robust = 0.5) images obtained in this way have a synthesized beam of $0\farcs170\times0\farcs084$, PA = $-2.3\degr$, and an rms of 0.075 mJy beam$^{-1}$ for the continuum, and 2-3.5 mJy beam$^{-1}$ per channel (depending on the spectral window) for the line images (cubes). Self-calibrated Cycle 3 data were concatenated with standard-calibrated Cycle 4 data, and we followed the same steps to self-calibrate the whole measurement continuum data set. The Briggs-weighted (robust = 0.5) image obtained with the concatenated data has a synthesized beam of $0\farcs176\times0\farcs088$, PA = $-2.0\degr$ and an rms of 0.10 mJy beam$^{-1}$. In our discussion, we use the image obtained with only the Cycle 3 data which has slightly better angular resolution and rms.

The ALMA observations were pipeline calibrated with CASA versions 4.7.01 (Cycle 3) and 4.7.2 (Cycle 4). For both self-calibration and imaging, CASA version 4.5.2 was used. The images were made using the task CLEAN of CASA.

\section{Results and Discussion} \label{sec:results}

\subsection{VLA Results}\label{sec:vla1}

\subsubsection{Radio Continuum Sources}\label{sec:vlafield}

Figure \ref{fig:vla} shows VLA images of the continuum emission toward SVS 13 at 3, 1.3, 0.9, and 0.7 cm. 
The two compact sources previously identified as VLA 4A (the western one) and VLA 4B (the eastern one) by \citet{Anglada2000,Anglada2004} and by \citet{Tobin2016mult} are detected at all the observed wavelengths. Table \ref{tab:posSVS13} lists the positions of the two sources and Table \ref{tab:fluxSVS13} their flux densities obtained from our images and from the literature.

 Additional sources detected in the observed field are VLA 2, VLA 3, VLA 17, VLA 20, and VLA 22. All these sources were previously reported at cm wavelengths by \citet{Rodriguez1999}. 
 Our VLA images are shown in Figures \ref{fig:vla3} to \ref{fig:vla22}, and their positions and flux densities are given in Table \ref{tab:posSources}.
Source VLA 22 splits into two components, VLA 22N and VLA 22S, separated 40 au in projection (Fig.~\ref{fig:vla22}), and both present a high degree of circular polarization. The high degree of circular polarization of VLA 22 was already noted by \citet{Rodriguez1999}, but the binary was not resolved in those observations. Interestingly, \citet{Rodriguez1999} reported a left-handed circularly polarized source, while in our observations right-handed polarization is dominant.
 In the following sections we discuss in more detail the properties of the sources VLA 4A and VLA 4B, which are associated with SVS 13. Further study of the remaining sources detected in the field is deferred to a forthcoming paper (A. K. Diaz-Rodriguez, in prep.).

\begin{deluxetable*}{lcll@{\extracolsep{12pt}}l@{\extracolsep{8pt}}lc}[htb]
\tabletypesize{\small}
\tablewidth{0pt}
\tablecaption{Positions of the VLA 4A and VLA 4B Radio Sources\tablenotemark{a} \label{tab:posSVS13}}
\tablehead{
Epoch & Wavelength &\multicolumn2c{VLA 4A} & \multicolumn2c{VLA 4B} \\
\cline{3-4} \cline{5-6}
(yr) & (mm) & \colhead{RA (J2000) } & \colhead{DEC (J2000)} & \colhead{RA (J2000)} & \colhead{DEC (J2000)} & Ref.
}
\startdata
1989.04 & 36 & 03 29 03.730$\pm$0.002 & 31 16 04.02$\pm$0.02 & 03 29 03.751$\pm$0.002 & 31 16 04.08$\pm$0.02 & 1 \\
1994.31 & 36 & 03 29 03.730$\pm$0.002 & 31 16 03.95$\pm$0.02 & 03 29 03.758$\pm$0.004 & 31 16 04.01$\pm$0.02 & 1 \\
1996.98 & 36 & 03 29 03.731$\pm$0.002 & 31 16 03.94$\pm$0.02 & 03 29 03.757$\pm$0.002 & 31 16 04.01$\pm$0.02 & 1 \\
1998.32\tablenotemark{b} & 36 & 03 29 03.7376$\pm$0.0017 & 31 16 03.957$\pm$0.026 & 03 29 03.7575$\pm$0.0015 & 31 16 03.968$\pm$0.023 & 1, 6 \\
1999.50 & 36 & 03 29 03.734$\pm$0.002 & 31 16 03.96$\pm$0.02 & 03 29 03.757$\pm$0.001 & 31 16 03.91$\pm$0.02 & 1 \\
2001.34 & 7  & 03 29 03.7405$\pm$0.0020 & 31 16 03.946$\pm$0.030 & 03 29 03.7594$\pm$0.0008 & 31 16 03.937$\pm$0.012 &  2\tablenotemark{c} \\
2002.29 & 13 & 03 29 03.7357$\pm$0.0015 & 31 16 03.935$\pm$0.022 & 03 29 03.7596$\pm$0.0011 & 31 16 03.914$\pm$0.016 & 3 \\
2003.40 & 13 & 03 29 03.7386$\pm$0.0011 & 31 16 03.876$\pm$0.017 & 03 29 03.7609$\pm$0.0007 & 31 16 03.908$\pm$0.011 &  4 \\
2014.56 & 9  & 03 29 03.7430$\pm$0.0007 & 31 16 03.790$\pm$0.010 & 03 29 03.7660$\pm$0.0007 & 31 16 03.810$\pm$0.010 &  5 \\
2015.47 & 7  & 03 29 03.7448$\pm$0.0007 & 31 16 03.780$\pm$0.010 & 03 29 03.7676$\pm$0.0007 & 31 16 03.788$\pm$0.010 &  6 \\
2015.49 & 13 & 03 29 03.7442$\pm$0.0007 & 31 16 03.785$\pm$0.010 & 03 29 03.7674$\pm$0.0007 & 31 16 03.788$\pm$0.010 &  6 \\
2015.52 & 9  & 03 29 03.7455$\pm$0.0008 & 31 16 03.794$\pm$0.012 & 03 29 03.7676$\pm$0.0007 & 31 16 03.789$\pm$0.011 &  6 \\
2016.69 & 0.9 & 03 29 03.7460$\pm$0.0003 & 31 16 03.769$\pm$0.005 & 03 29 03.7691$\pm$0.0003 & 31 16 03.761$\pm$0.005 & 6\tablenotemark{d} \\
2016.89 & 30 & 03 29 03.7457$\pm$0.0010 & 31 16 03.771$\pm$0.015 & 03 29 03.7689$\pm$0.0008 & 31 16 03.775$\pm$0.012 & 6 \\
\enddata
\tablenotetext{a}{Positions derived from elliptical Gaussian fits. Uncertainties correspond to those of absolute positions, and are calculated by adding in quadrature a systematic error of 0\farcs01 (for VLA data, \citealt{Dzib2017}) or 0\farcs005 (for ALMA data, \citealt{Remijan2020}) to the formal error of the fit, $0.5\,\theta / {\rm SNR}$ \citep{Reid1988}. The systematic error accounts for uncertainties introduced by the phase calibration process, and the error of the fit for those due to the convolved source size ($\theta$) and signal-to-noise ratio (SNR).}
\tablenotetext{b}{From an image obtained concatenating data from epochs J1998.24 and J1998.40 reported in \citet{Anglada2000} and \citet{Carrasco2008}.}
\tablenotetext{c}{After recalibration of the originally published data.}
\tablenotetext{d}{From the ALMA image obtained using baselines $>$750 k$\lambda$, where the extended emission has been filtered out (Figure \ref{fig:cont750kl}).}
\tablerefs{(1) \citet{Carrasco2008}; (2) \citet{Anglada2004}; (3) G. Anglada, private communication; (4) C. Carrasco-Gonz\'alez, private communication; (5) \citet{Tobin2016mult}; (6) This work.}
\end{deluxetable*}

\begin{deluxetable*}{ccccccl}[htb]
\tabletypesize{\small}
\tablecaption{Flux Densities of the VLA 4A and VLA 4B Radio Sources\label{tab:fluxSVS13}}
\tablewidth{0pt}
\tablehead{Frequency & Wavelength & \multicolumn{2}{c}{Flux density (mJy)\tablenotemark{a}} & Epoch\tablenotemark{b} & Resolution & \\
\cline{3-4}
(GHz)  &  (mm)  & VLA 4A & VLA 4B & (yr) & (\arcsec) & Ref. }
\startdata
4.68  & 64  &    0.065$\pm$0.008 & 0.059$\pm$0.006 & 2014.2-2014.3   & $\sim$0.4  & 1\\
7.31  & 41  &  0.109$\pm$0.008 & 0.093$\pm$0.007 & 2014.2-2014.3     & $\sim$0.4  & 1\\
8.33  & 36  &  0.17$\pm$0.03   & 0.12$\pm$0.03 & 1994.3-1999.5       & $\sim$0.2  & 2 \\
10.0  & 30  &  0.045$\pm$0.009 & 0.084$\pm$0.013 & 2016.9            & $\sim$0.2  & 3 \\
22.1  & 13  & 0.24$\pm$0.03   & 0.61$\pm$0.06 & 2015.5               & $\sim$0.1  & 3 \\
30.0  & 10  & 0.38$\pm$0.05 & 1.12$\pm$0.11 & 2013.7-2015.9      & $\sim$0.1  & 4 \\
33.0  &  9  & 0.46$\pm$0.05    & 1.36$\pm$0.14   & 2015.1-2015.5       & $\sim$0.2  & 3 \\
33.0  &  9     &   0.41$\pm$0.06   & 1.22$\pm$0.13 & 2015.5          & $\sim$0.1  & 3 \\
37.5  &  8  &  0.51$\pm$0.06   & 1.76$\pm$0.18 & 2013.7-2015.9     & $\sim$0.1  & 4 \\
42.8  &  7 & 0.34$\pm$0.12     & 1.29$\pm$0.19 & 2001.3              & $\sim$0.1  & 5\tablenotemark{c}\\
44.0  &  7  & 0.61$\pm$0.07  & 1.96$\pm$0.20 & 2015.2-2015.5        & $\sim$0.1  & 3\\
44.0  &  7      &  0.37$\pm$0.06   & 2.08$\pm$0.22 & 2015.5          & $\sim$0.05 & 3  \\
86.8  & 3.5 & \multicolumn{2}{c}{26$\pm$3\tablenotemark{d}} & 1998.2              & $\sim$3.5  & 6 \\
94    & 3.2  & \multicolumn{2}{c}{21$\pm$2\tablenotemark{d}}  & 2011.9                & $\sim$1.2  & 7 \\
107.7 & 2.7  & \multicolumn{2}{c}{38$\pm$10\tablenotemark{d}} & 1996.3-1998.2      & $\sim$0.6  & 8 \\
230.5 & 1.3   & \multicolumn{2}{c}{320$\pm$34\tablenotemark{d}}& 1998.2              & $\sim$1.5  & 6  \\
231   & 1.3  & \multicolumn{2}{c}{120$\pm$14\tablenotemark{d}} & 2011.1                  & $\sim$0.4   & 7 \\
232.5 & 1.3  & \multicolumn{2}{c}{400$\pm$44\tablenotemark{d}} & 2015.7 & $\sim$0.2  & 9 \\
338.2 & 0.9  &  83$\pm$9\tablenotemark{e} & 110$\pm$11\tablenotemark{e} & 2016.7  & $\sim$0.08 & 3 \\
338.2 & 0.9  & \multicolumn{2}{c}{1102$\pm$110\tablenotemark{f}} & 2016.7-2016.9   & $\sim$0.1 & 3 \\
\enddata 
\tablenotetext{a}{Uncertainties account for the rms noise of the images and 10\% of absolute flux calibration uncertainty, except for the data reported by (1), the 231 GHz data point reported by (7), and the data reported by (8), whose calibration uncertainties are 5\%, 15\%, and 20\%, respectively.}
\tablenotetext{b}{When a range of epochs is indicated, flux densities have been measured in an image made by concatenating data obtained at different epochs.}
\tablenotetext{c}{After recalibration of the originally published data.}
\tablenotetext{d}{Total observed flux density (the binary is unresolved). Assumed to be dominated by the emission of the circumbinary disk.}
\tablenotetext{e}{Measured in the image made only with the Cycle 3 ALMA data and baselines $>$750 k$\lambda$, where the extended emission has been filtered out (Fig.~\ref{fig:cont750kl}). Assumed to be dominated by the emission of the circumstellar disk.}
\tablenotetext{f}{Measured in an image made using both the Cycle 3 and 4 ALMA data, to better recover the extended emission. Assumed to be dominated by the emission of the circumbinary disk.}
\tablerefs{(1) \citet{Tychoniec2018}; (2) \citet{Anglada2000}, (3) This work, (4) \citet{Tobin2016mult}, (5) \citet{Anglada2004}, (6) \citet{Bachiller1998}, (7) \citet{Maury2019}, (8) \citet{Looney2000}, (9) \citet{Tobin2018}}
\end{deluxetable*}

\begin{deluxetable*}{lcc@{\extracolsep{8pt}}c@{\extracolsep{6pt}}c@{\extracolsep{6pt}}c@{\extracolsep{6pt}}c@{\extracolsep{6pt}}c@{\extracolsep{6pt}}c}[htb]
\tabletypesize{\small}
\tablewidth{0pt}
\tablecaption{Positions and Flux Densities of Other Radio Sources in the SVS 13 Field\tablenotemark{a}\label{tab:posSources}}
\tablehead{
 & \multicolumn2c{Position} & \multicolumn6c{Flux Density (mJy)} \\
\cline{2-3} \cline{4-9}
\colhead{Source} & \colhead{RA (J2000) } & \colhead{DEC (J2000)} & \colhead{3 cm} & \colhead{1.3 cm} & \colhead{9 mm} & \colhead{7 mm} & \colhead{1.3 mm\tablenotemark{b}} & \colhead{0.9 mm}
}
\startdata
VLA 2   & 03 29 01.9719$\pm$0.0007 & 31 15 38.044$\pm$0.010 & 0.64$\pm$0.07            & 1.59$\pm$0.17   & 1.76$\pm$0.18   & 2.11$\pm$0.25 &  & \nodata\tablenotemark{c} \\
VLA 17  & 03 29 03.0813$\pm$0.0007 & 31 15 51.701$\pm$0.010 & 0.023$\pm$0.006  & 0.31$\pm$0.04   & 1.18$\pm$0.12   & 2.12$\pm$0.22 & 222.2$\pm$22.3 & 400$\pm$40   \\
VLA 3   & 03 29 03.3888$\pm$0.0007 & 31 16 01.571$\pm$0.010 & 0.067$\pm$0.013          & 0.183$\pm$0.024 & 0.34$\pm$0.04   & 0.28$\pm$0.04 & 13.9$\pm$1.7 & 27$\pm$3 \\
VLA 20  & 03 29 04.2545$\pm$0.0007 & 31 16 09.067$\pm$0.010 & 0.087$\pm$0.016          & 0.127$\pm$0.018 & 0.16$\pm$0.03 & 0.13$\pm$0.04 & 4.4$\pm$1.1 & 8.4$\pm$0.9  \\
VLA 22N & 03 29 05.7621$\pm$0.0008 & 31 16 39.617$\pm$0.012 & 0.009$\pm$0.007\tablenotemark{d} & 0.090$\pm$0.017 & 0.25$\pm$0.05\tablenotemark{d} & 0.39$\pm$0.15\tablenotemark{d} & & \nodata\tablenotemark{c} \\
VLA 22S & 03 29 05.7649$\pm$0.0008 & 31 16 39.503$\pm$0.012 & 0.029$\pm$0.007\tablenotemark{d} & 0.099$\pm$0.018 & 0.05$\pm$0.05\tablenotemark{d} & 0.23$\pm$0.16\tablenotemark{d} &  & \nodata\tablenotemark{c} \\
\enddata
\tablenotetext{a}{Positions derived from elliptical Gaussian fits to the ALMA image at 0.9 mm, except for sources VLA 2 and VLA 22, for which the VLA image at K band (1.3 cm) has been used. Positional uncertainties are calculated as in Table \ref{tab:posSVS13}. VLA flux densities are measured in A-configuration (3 and 1.3 cm) and combined A+B-configuration (9 and 7 mm) total intensity images, except for VLA 22 where polarized A-configuration (3 cm and 7 mm) and B-configuration (9 mm) images are used. Flux density uncertainties include a 10\% absolute flux calibration uncertainty.}
\tablenotetext{b}{From the ALMA observations by \citet{Tobin2018}.}
\tablenotetext{c}{Source outside the imaged field of view.}
\tablenotetext{d}{The two sources appear blended in the total intensity ($I$) image at this wavelength. Flux densities are obtained from two-Gaussian fits to the LCP and RCP images shown in Figure~\ref{fig:vla22}, assuming unresolved sources at the positions (taken as fixed parameters) of VLA 22N and VLA 22S. The total intensity is calculated as $I=(LCP+RCP)/2$.}
\end{deluxetable*}

\subsubsection{Binary Protostars Traced by the Ionized Gas}\label{sec:vla2}

 As shown in Figure \ref{fig:vla}, the two sources detected toward SVS 13 are separated by $\sim0\farcs30$, corresponding to 90 au in projection at a distance of 300 pc. The sources appear compact, although there is a hint of weak emission, which is more evident at X and Ka bands, connecting the two sources. As noted previously by \citet{Anglada2004}, source VLA 4B is significantly stronger than VLA 4A at mm wavelengths (intensity ratios of 3.8 at 7 mm and 3.7 at 9 mm), while at longer wavelengths the difference in the observed intensities of the two sources is less significant (intensity ratios of 2.4 at 1.3 cm and 1.6 at 3 cm, see further discussion in Section \ref{sec:sed}).

In the figures, we plot the optical position of the star SVS 13 as obtained from Gaia EDR3 Catalog (\citealt{GaiaEDR3}; given at epoch J2016.0), corrected by the reported Gaia proper motions (see Table~\ref{tab:3D}) to the epoch of the VLA observations (a small correction of not more than $0\farcs014$) and converted to equinox J2000 coordinates with the task ICRSToJ2000 of the CASA Extensions \href{https://safe.nrao.edu/wiki/bin/view/Main/CasaExtensions}{AnalysisUtils} Python package. As can be seen in the figure, the optical position falls closer to VLA 4B than to VLA 4A. This association is in disagreement with the initial identification of the optical star with VLA 4A made by \citet{Anglada2000} from the limited astrometric precision ($\sim$0\farcs3) optical data available at that time. The identification of VLA 4B with the optically visible star was already proposed by \citet{Hodapp2014}, who noted the positional agreement of the near-IR position measured by 2MASS (with an astrometric accuracy of $0\farcs1$, better than that of the previous optical data) and the radio source VLA 4B, assuming that the 2MASS source and the optical star were the same object. The identification was corroborated by \citet{Lefevre2017} by comparing more precise absolute astrometry ($\sim$0\farcs02) PdBI mm data of the base of the SiO/CO jet, which was found to be associated with the near-IR jet imaged by \citet{Hodapp2014}.

Source VLA 4A is closer to the densest part of the ambient cloud, as traced by the moderately high angular resolution ($\sim 4''$) VLA ammonia images \citep{Rudolph2001,Diaz2021}. These images show that the ammonia emission extends preferentially toward the west and south-west, suggesting that the VLA 4A component of the binary could be more extincted by foreground material than VLA 4B, consistent with the association of VLA 4B with the optical star.

Nevertheless, we note that the coincidence between the optical and the radio position is not perfect (see Fig. \ref{fig:vla} and Table \ref{tab:posSVS13}). The discrepancy is $\sim 0\farcs07$, larger than the estimated astrometric uncertainty, which is $\sim 0\farcs01$ (see discussion in Section \ref{sec:pm}). 

\begin{figure}[h]
\centering
\includegraphics[width=0.46\textwidth]{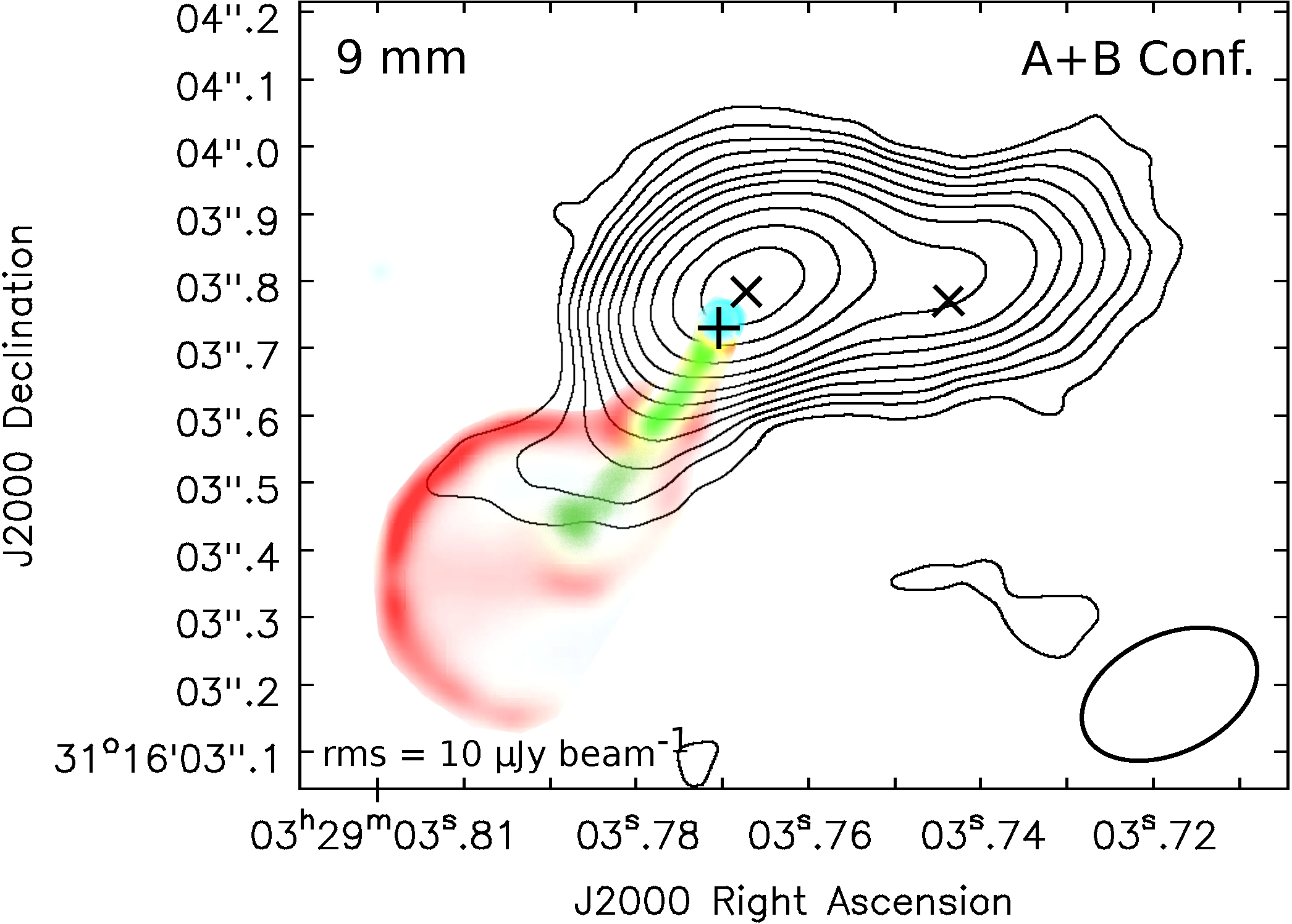}\hfill
\caption{VLA image of SVS 13 at Ka band (9 mm) obtained from the concatenated data of the A and B configurations using natural weighting (beam = 0\farcs28 $\times$ 0\farcs18, PA = $-64.3\degr$), with an overlay of the bottom panel of Figure 3 of \citealt{Hodapp2014}, showing the IR continuum emission of the SVS 13 star (light blue), the bubble traced by H$_2$ S(1) (red) and the jet traced by [Fe II] (green). The plus sign marks the optical position of the star SVS 13 from Gaia EDR3 \citep{GaiaEDR3} extrapolated to the epoch of the Ka-band observations (see text). The crosses mark the positions of the two radio sources. Note that the extension to the SE of the (free-free) radio emission traces the direction of the [Fe II] jet, and possibly part of the NE wall of the H$_2$ S(1) bubble. Contour levels are $-$3, 3, 5, 7, 10, 14, 19, 25, 35, 50, 70, and 95 times the rms of the image. The synthesized beam and the rms are shown in the bottom right and left corners, respectively, of the image.
 \label{fig:VLA4B}}
 \end{figure}

The VLA Ka-band image in Figure \ref{fig:vla} also shows a hint of a SE elongation of the source VLA 4B, which is difficult to confirm since the beam is elongated in the same direction. In Figure \ref{fig:VLA4B} we show the Ka-band image obtained from the concatenated A+B configuration data. Although with a lower angular resolution, this image shows quite clearly an extension of the emission from the position of VLA 4B toward the SE. Interestingly, this SE protuberance coincides very well with the axis, at PA = 145$^\circ$, of the $0\farcs2$ long [Fe II] jet imaged in the near-IR by \citet{Hodapp2014}; this microjet pierces inside the first of a series of bubbles and fragments of bubbles traced in the H$_2$ 1--0 S(1) line to the SE of SVS 13. A false-color image of the microjet and its associated bubble is overlaid on the contour image of the 9 mm emission shown in Figure \ref{fig:VLA4B}. The IR and radio images were aligned by matching the IR emission of the SVS 13 star (in light blue in the image) with the Gaia EDR3 optical position of SVS 13 corrected to the epoch of our VLA observations (plus sign). We note that \citet{Hodapp2014} found evidence of expanding motion of the bubble, with its front end moving at a projected velocity of 20 mas yr$^{-1}$ (28.5 km s$^{-1}$); since their image was obtained $\sim$ 1.5 yr earlier than our VLA image, the actual size of the bubble at the epoch of our observations could be $\sim0\farcs05$ larger than what is shown in the figure, which does not affect the accuracy of our comparison. 

Therefore, it seems that the 9 mm continuum emission of our observations traces the free-free emission of the ionized radio jet associated with the high excitation shock-excited [Fe II] jet. Additionally, in Figure \ref{fig:VLA4B}, there is a hint that the 9 mm emission extending to the SE of VLA 4B, further bends toward the east, partially tracing the NE edge of the bubble observed by \citet{Hodapp2014} in the lower excitation emission of shock-excited H$_2$. Unfortunately, our A-configuration data alone are not sensitive enough to trace further details of this jet-bubble system, but additional, more sensitive VLA A-configuration observations could reveal with a high degree of detail the emission of the ionized material associated with this shock interaction, and trace the jet back to its origin to clarify whether the optical/IR star SVS 13 is the same object as the radio source VLA 4B.

\begin{figure}[htb]
\centering
\includegraphics[width=0.46\textwidth]{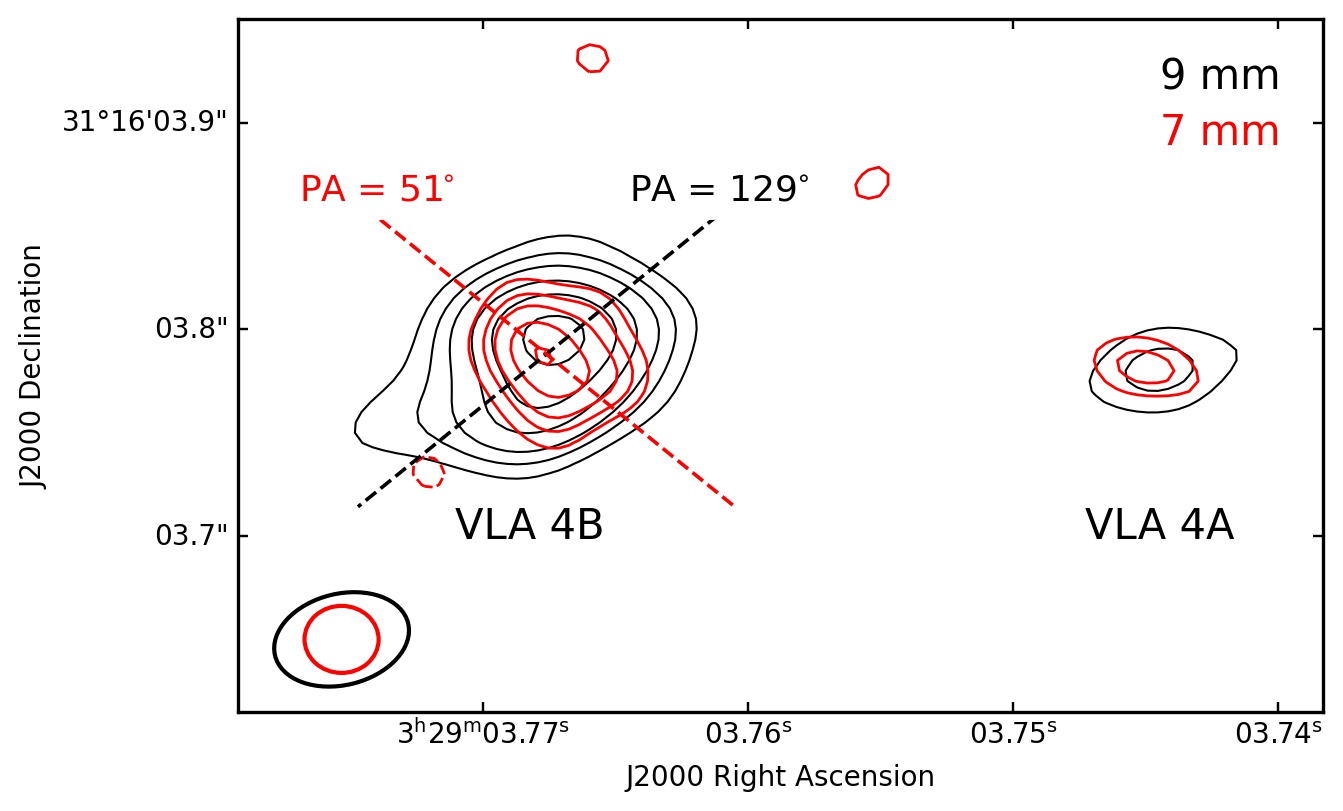}\hfill
\caption{Overlay of the enhanced angular resolution images of SVS 13 at Q band (7 mm; red contours) and Ka band (9 mm; black contours) obtained from the A configuration VLA data using robust = 0 weighting. Only baselines $>$1000 k$\lambda$ (beam = 0\farcs036 $\times$ 0\farcs032, PA = $89.2\degr$; rms = 34 $\mu$Jy beam$^{-1}$) and $>$600 k$\lambda$ (beam = 0\farcs066 $\times$ 0\farcs044, PA = $-76.3\degr$; rms = 26 $\mu$Jy beam$^{-1}$) have been used for the Q-band and Ka-band images, respectively. The beam-deconvolved sizes and position angles obtained from Gaussian fits to the source VLA 4B in these images are 60.3$\pm$3.5 mas $\times$ 49.0$\pm$3.1 mas, PA = 51$^\circ\pm25^\circ$ (7 mm) and 71.3$\pm$8.8 mas $\times$ 53.4$\pm$7.5 mas, PA = 129$^\circ\pm14^\circ$ (9 mm).
 Contour levels are $-$3, 3, 5, 7, 10, 13, and 17 times the rms of the images. The synthesized beams are shown in the bottom left corner of the image.
 \label{fig:Ka+Q}}
 \end{figure}

Figure~\ref{fig:Ka+Q} shows an overlay of the Q (7 mm) and Ka (9 mm) band images obtained with robust=0 weighting. This improved angular resolution image illustrates the change in the orientation of the major axis of the emission associated with VLA 4B, from PA = 51$^\circ\pm25^\circ$ (7 mm) to PA = 129$^\circ\pm14^\circ$ (9 mm). Thus, the morphology at $<0\farcs1$ scales further suggests that the 9 mm image of VLA 4B is dominated primarily by free-free emission arising from the jet, observed in [Fe II] at PA = 145$^\circ$ \citep{Hodapp2014}, while the 7 mm image is dominated by the dust emission of its associated, roughly perpendicular, circumstellar disk. VLA 4A remains unresolved at all bands in Figures \ref{fig:vla} and \ref{fig:Ka+Q}, and a similar morphological analysis cannot be performed.

\subsubsection{Proper motions}\label{sec:pm}

\begin{figure}[htb]
\centering
\includegraphics[height=0.33\textheight]{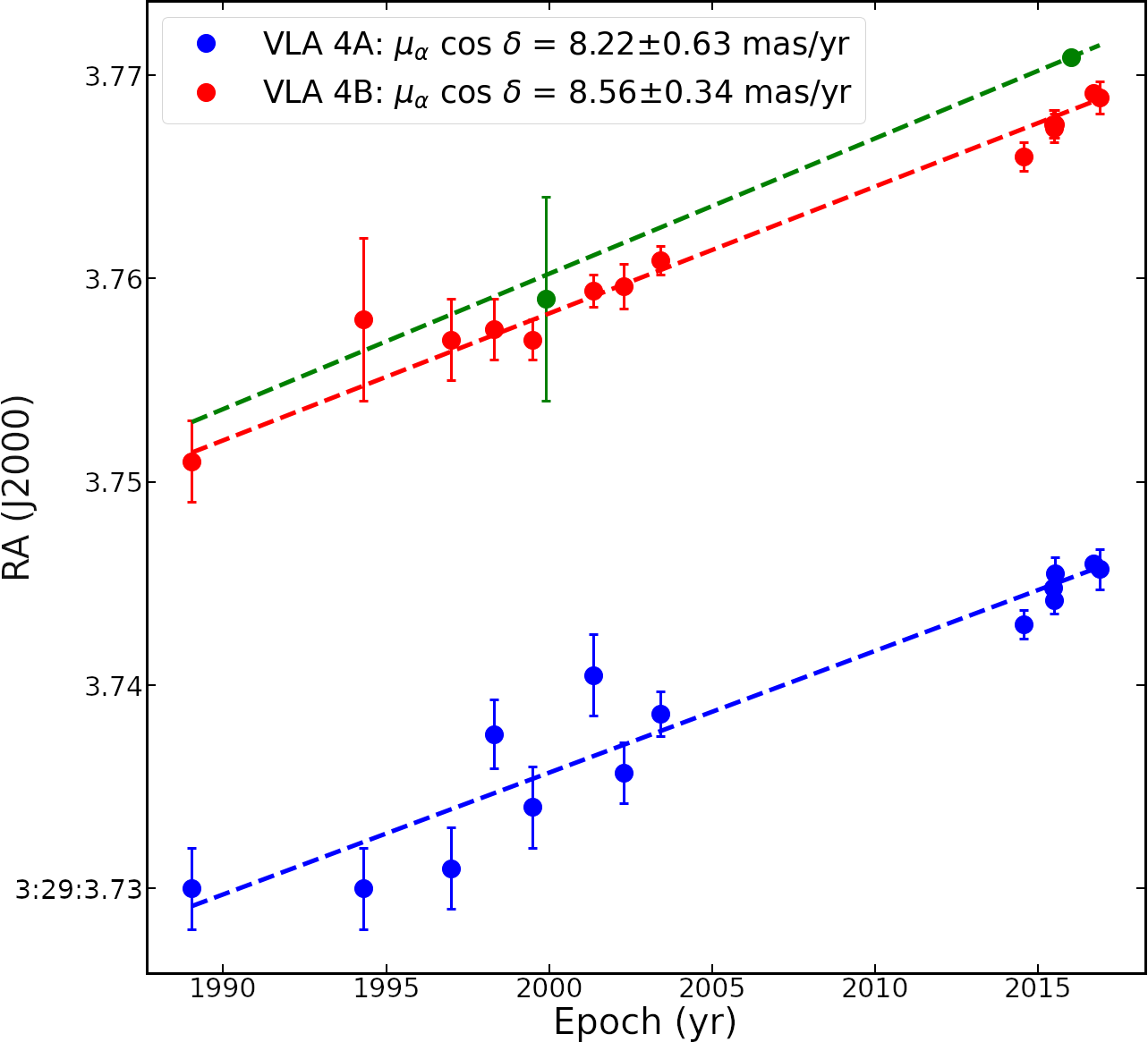}\hfill
\includegraphics[height=0.33\textheight]{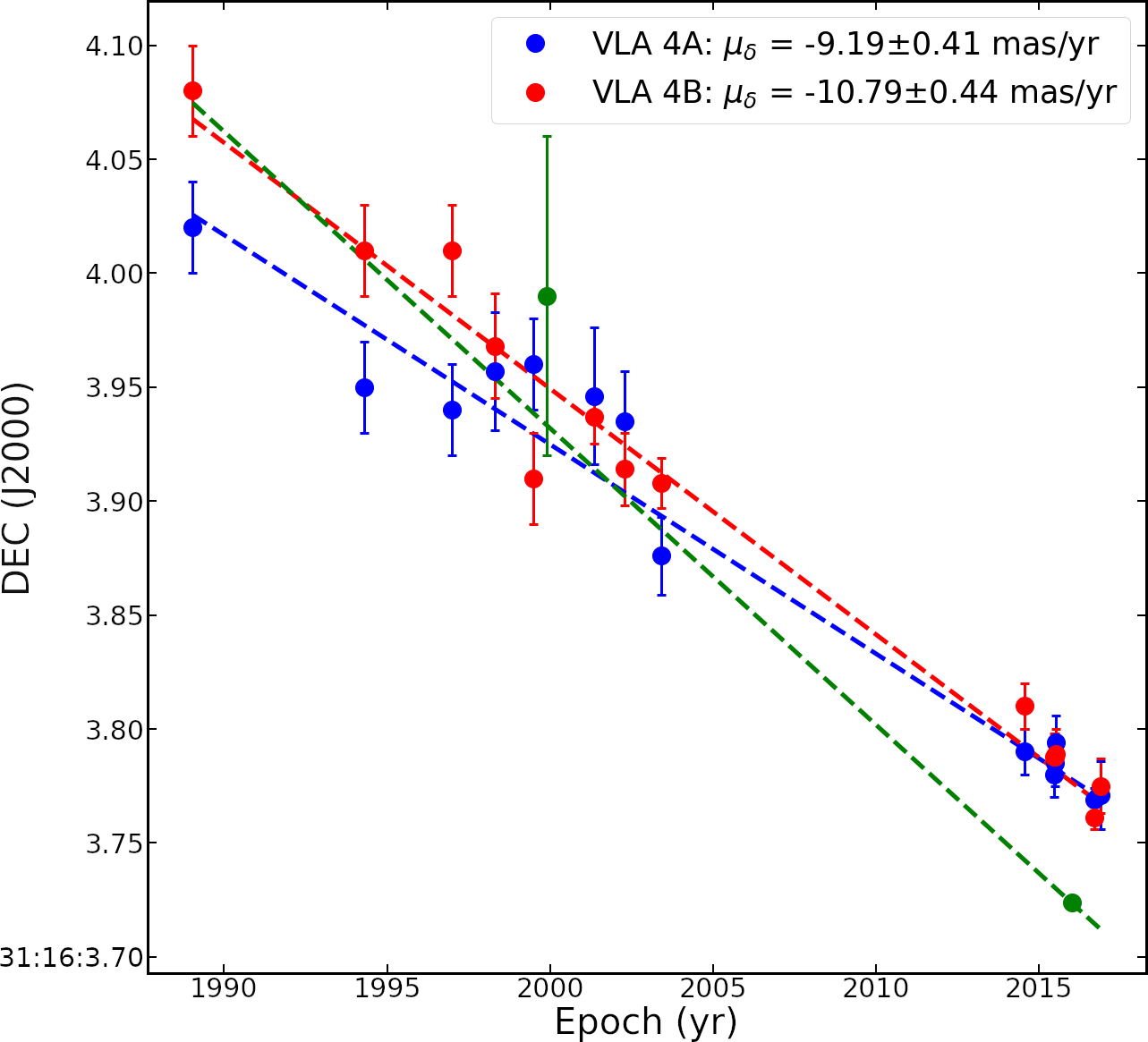}
\caption{J2000 right ascension and declination of the two radio components of SVS 13, VLA 4A (blue) and VLA 4B (red), as a function of the epoch of observation. The positions used in this plot are listed in Table~\ref{tab:posSVS13}. The proper motions are obtained through linear least-squares fits (dashed lines) to the radio positions. The near-IR 2MASS position of SVS 13, $\alpha$(J2000) = 03$^h$29$^m$03.759$^s \pm 0.005^s$, $\delta$(J2000) = 31$^\circ$16$'$03$\farcs99 \pm 0\farcs07$, at epoch J1999.90 \citep{Skrutskie2006} and the optical Gaia EDR3 position, $\alpha$(J2000) = 03$^h$29$^m$03.77088$^s \pm 0.00002^s$, $\delta$(J2000) =  31$^\circ$16$'$03$\farcs7237 \pm 0\farcs0002$, at epoch J2016.0 \citep{GaiaEDR3} are also plotted (green) as a reference. The Gaia EDR3 proper motions (Table~\ref{tab:3D}) have been plotted as green dashed lines. For consistency, the Gaia EDR3 coordinates, originally referred to ICRS have been transformed to J2000 using the task ICRSToJ2000 of the \href{https://safe.nrao.edu/wiki/bin/view/Main/CasaExtensions}{AnalysisUtils} Python package. 
 \label{fig:pmotions}}
 \end{figure}

Proper motions of SVS 13 and nearby radio sources have been reported by \citet{Carrasco2008} and by \citet{Lefevre2017}. VLA data are currently available for SVS 13 spanning nearly 30 years, and with a high enough angular resolution to separate the two components, VLA 4A and VLA 4B, of the binary. In Table~\ref{tab:posSVS13} we list the positions of the two radio sources obtained from the literature and from the data presented in this paper. From these data we can obtain more accurate proper motions and try to measure differences in the relative positions of the two sources to trace, or at least to constrain, their orbital motion. In Figure \ref{fig:pmotions} we plot the right ascension and declination coordinates of VLA 4A and VLA 4B as a function of the epoch of observation, from 1989 to 2017. In addition to the cm VLA data we use the ALMA positions at 0.9 mm obtained from an image made using baselines $>$750 k$\lambda$ in order to minimize the contribution of the extended emission (see Fig.~\ref{fig:cont750kl} in Section \ref{sec:dust2}). Table~\ref{tab:3D} gives the measured proper motions as derived from the linear least-square fit to the data shown in Figure \ref{fig:pmotions}. The absolute proper motions are 12.33$\pm$0.52 mas yr$^{-1}$ at PA = $138.2 \pm 2.4^\circ$ for source VLA 4A, and 13.77$\pm$0.38 mas yr$^{-1}$ at PA = $141.6 \pm 1.5^\circ$ for source VLA 4B. These values are in agreement with previous estimates \citep{Carrasco2008,Lefevre2017} within the uncertainties, but our values are more accurate and allow us to search for evidence of possible orbital motions.

\begin{deluxetable*}{lcc@{\extracolsep{12pt}}c@{\extracolsep{10pt}}c@{\extracolsep{10pt}}c}[htb]
\tablewidth{0pt}
\tablecaption{Proper Motions and 3D Velocities \label{tab:3D}}
\tablehead{
 & \multicolumn2c{Proper Motions\tablenotemark{a}} & \multicolumn3c{3D Velocity\tablenotemark{b}} \\
 \cline{2-3} \cline{4-6}
  & $\mu_\alpha \cos \delta$ & $\mu_\delta$ & $V_\alpha$ & $V_\delta$ & $V_{\rm LSR}$ \\
  & (mas yr$^{-1}$) & (mas yr$^{-1}$) & (km s$^{-1}$) & (km s$^{-1}$) & (km s$^{-1}$)}
\startdata
 VLA 4A  & 8.22$\pm$0.63 & $-$9.19$\pm$0.41  & 11.69$\pm$0.90 & $-$13.07$\pm$0.58 & 7.36$\pm$0.21 \\
 VLA 4B  & 8.56$\pm$0.34 & $-$10.79$\pm$0.44 & 12.17$\pm$0.48 & $-$15.34$\pm$0.58 & 9.33$\pm$0.18 \\
Gaia EDR3 & 9.98$\pm$0.26 & $-$13.02$\pm$0.20 & 14.19$\pm$0.37 & $-$18.52$\pm$0.28 & \nodata \\
 VLA (4B$-$4A)\tablenotemark{c} & 0.34$\pm$0.72 & $-$1.60$\pm$0.58 & 0.48$\pm$1.02 & $-$2.28$\pm$0.82 & 1.97$\pm$0.28 \\
\enddata
\tablenotetext{a}{Proper motions in the plane of the sky, derived from the linear least-square fit to the data shown in Table \ref{tab:posSVS13} and Figure \ref{fig:pmotions}.}
\tablenotetext{b}{Full 3D velocity. $V_\alpha$ and $V_\delta$ are the velocity components in the direction of the right ascension and declination axes, respectively, obtained from the proper motions in the plane of the sky, assuming a distance of 300 pc; $V_{\rm LSR}$ is the line-of-sight velocity component, obtained from our ALMA molecular line emission data toward each component (Section \ref{sec:mol}).}
\tablenotetext{c}{Motion of VLA 4B relative to VLA 4A.}
\end{deluxetable*}

\begin{figure}[htb]
\centering
\includegraphics[height=0.33\textheight]{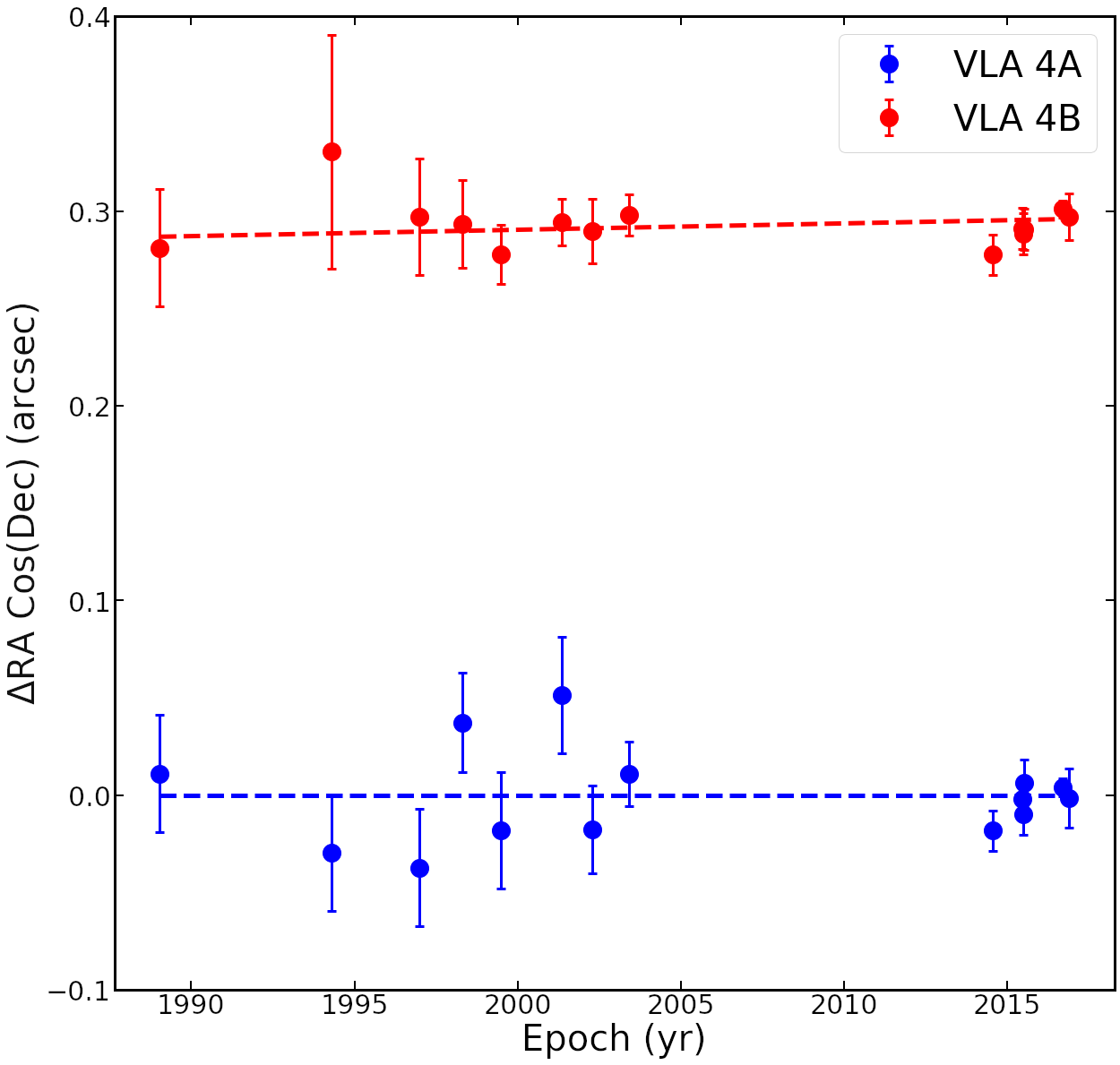}\hfill
\includegraphics[height=0.33\textheight]{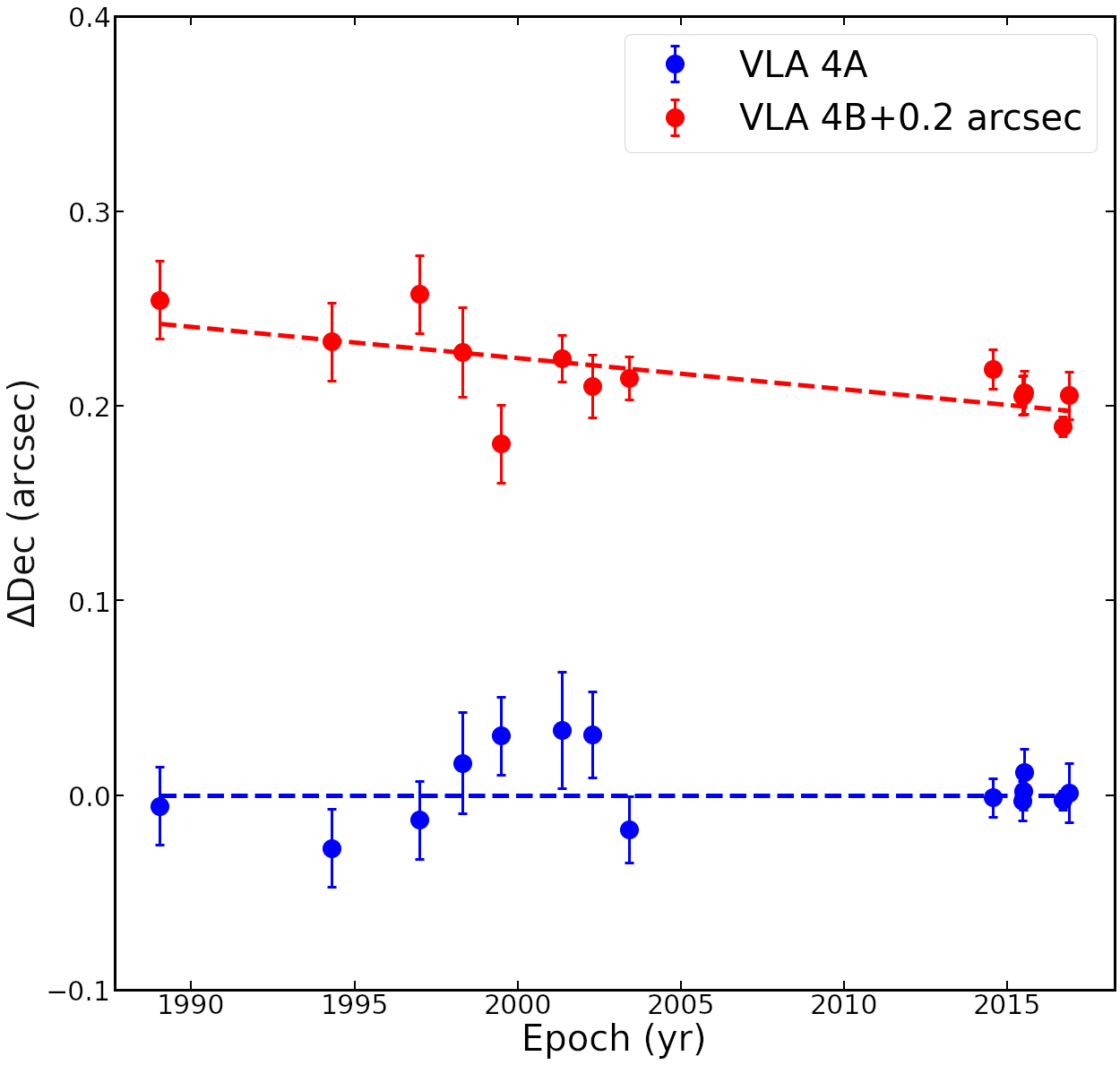}
\caption{
Residual offsets in right ascension and declination as a function of time for each radio component of SVS 13, VLA 4A (blue) and VLA 4B (red), relative to the fitted linear proper motion of VLA 4A (blue dashed lines in Fig.~\ref{fig:pmotions}). 
An offset of $0\farcs2$ has been added to the difference in declination of VLA 4B for improved visualization. The dashed lines correspond to the fits obtained in Fig.~\ref{fig:pmotions}. The figure is intended to better illustrate the relative motion between the two sources using a co-moving reference system that is reasonably independent of sudden changes in the position of one of them (e.g., produced by ejection events, as it possibly occurred in source VLA 4A around epoch J2000). The center of mass would be a natural reference for relative motions, but is not known here. The use of VLA 4A instead of VLA 4B as a reference is arbitrary.
 \label{fig:difpmotions}}
 \end{figure}

Figure \ref{fig:pmotions} and Table~\ref{tab:3D} show indications that the declination of VLA 4B relative to VLA 4A decreases with time (VLA 4B moving to the south relative to VLA 4A) at a rate of 1.60$\pm$0.58 mas yr$^{-1}$ (2.28$\pm$0.82 km s$^{-1}$) with a total displacement of 48.0$\pm$17.4 mas (14.40$\pm$5.22 au) in 30 yr, while the rate and the total displacement in right ascension is considerably smaller, 0.34$\pm$0.72 mas yr$^{-1}$ (0.48$\pm$1.02 km s$^{-1}$) with a total displacement of 10.2$\pm$21.5 mas (3.06$\pm$6.44 au). Relative positions should not be affected by errors in absolute astrometry from one epoch to another, and a high precision can be reached, in principle. However, small differences in the protostellar positions are difficult to measure from their cm radio emission because these sources usually trace free-free jets that can suffer apparent changes in their position because of morphological changes due to the ejection of new ionized knots (see, e.g., the multiepoch study of the radio sources in NGC 2071 by \citealt{Carrasco2012}). In this respect, positions obtained at Q band and shorter wavelengths (tracing the dust contribution of the disk) are more stable and provide more reliable proper motion measurements. This has been noticed, for example, in the well-studied A1-A2 Class 0 close binary of the IRAS 16293$-$2422 triple system, whose proper motions are affected by the positional changes due to the ejection of ionized knots at cm wavelengths, while positions from data at shorter wavelengths appear more reliable (see discussion in \citealt{Loinard2007,Hernandez2019,Maureira2020}). In the case of SVS 13, the consistent departure of the position of VLA 4A in declination from the overall motion (Fig. \ref{fig:pmotions}) suggests that such an ejection event may have taken place around epoch J2000, resulting in a reversal of the relative positions of the two sources in the declination axis. For a better visualization, we plot in Figure \ref{fig:difpmotions} the positions of the two sources after subtraction of the proper motion derived from the linear least-squares fit to the VLA 4A positions (blue dashed lines in Fig.~\ref{fig:pmotions}), as a function of time.
As this figure better illustrates, despite the anomaly around epoch J2000, the currently available data show a systematic trend in the relative proper motions, with VLA 4B moving to the south relative to VLA 4A. This result deserves further investigation and can be easily confirmed by future Q-band VLA and mm/sub-mm ALMA observations at high angular resolution.

The observed trend in the proper motions, with VLA 4B moving to the south relative to VLA 4A, corresponds to a counter-clockwise orbital motion as seen (projected on the plane of the sky) from the Earth. Since the observed blue-shifted jet and molecular outflow lobe point southeast, and assuming that its associated disk in VLA 4B is roughly co-planar with the binary orbit, this implies that the angular momentum vector points away from the molecular cloud, toward the observer. That is, we are observing the orbital motion from the head of the angular momentum vector. These results are in agreement with what is inferred from the ALMA data (see Sects.~\ref{sec:spirals}, \ref{sec:gas}, and \ref{sec:mol}).

Absolute proper motions can, in principle, inform us about the relative masses of the two components in the binary, since the center of mass is closer to the more massive component, and therefore this component is expected to be the one that shows smaller absolute proper motions. However, when the system has a significant overall proper motion, which is not independently known, as in our case, disentangling both types of motions becomes extremely challenging unless the data cover a significant fraction of the orbital period (e.g., in \citealt{Carrasco2009,Ichikawa2021}). We note that measuring the difference in the absolute proper motions of the two components of the A1-A2 IRAS 16293$-$2422 binary is insufficient to derive the relative masses and that resolved gas kinematics is required for this purpose (see \citealt{Maureira2020} for details).

In Figure \ref{fig:pmotions} we also plot, as a reference, the proper motions of the optical position of the SVS 13 star reported by Gaia (green dashed line), together with the Gaia position (epoch J2016.0) and the 2MASS IR position (epoch J1999.90) plotted as green points; the 2MASS position is consistent with the Gaia proper motions but the 2MASS positional errors are large and the agreement is not significant enough. We note that there is a significant difference between the radio proper motions of VLA 4B and the Gaia EDR3 proper motions of SVS 13 (Table~\ref{tab:3D}). This difference is  
1.42$\pm$0.43 mas yr$^{-1}$ in right ascension and $-$2.23$\pm$0.46 mas yr$^{-1}$ in declination, which is larger than the dispersion in proper motions in the whole NGC 1333 region, 0.92$\pm$0.05 mas yr$^{-1}$ in right ascension and 0.60$\pm$0.03 mas yr$^{-1}$ in declination, as estimated by \citet{Ortiz2018}. Thus, the observed differences in position and proper motions between VLA 4B and the optical star SVS 13 (Tables \ref{tab:posSVS13} and \ref{tab:3D}, and Fig. \ref{fig:pmotions}) would be compatible with the two being independent objects belonging to the same molecular cloud. However, the uncertainties of the Gaia EDR3 catalog may possibly be underestimated in the case of SVS 13 (see below). Indeed, the position, proper motions and parallax given in Gaia EDR3 for SVS 13 fall outside the uncertainty range of the previous values in the DR2 \citep{Gaia2018}. 

A reasonable explanation for this discrepancy is that what is seen at optical wavelengths is not the direct light of the SVS 13 star but scattered light from its circumstellar environment, namely, from the walls of a cavity created by the jet imaged by \citet{Hodapp2014}. It has been known for a long time that jets are frequently driven by embedded stars, whose actual position is better traced by the cm emission peak, while the observed optical intensity peak appears displaced by a fraction of arcsec (e.g., HL Tau and HH 30: \citealt{Weintraub1995,Ray1996}). The observed difference of $0\farcs07$ between the optical and radio positions of SVS 13 can be easily accounted by this fact. Indeed, HST images of SVS 13 suggest that what is seen at optical wavelengths is not a point-like star but an elongated source consistent with scattered light from a cavity carved by the jet (G. Bl\'azquez-Calero et al., in prep.). Although Gaia provides extremely accurate positions for point-like stellar sources, it does not work well when substructure is present at scales below $0\farcs7$, and works very poorly below $0\farcs4$ \citep{Fabricius2021}. Therefore, Gaia cannot identify the presence of extended emission below these scales, which could evidence the non-stellar nature of some sources. For extended sources this would also result in increased errors in the Gaia-determined proper motions and parallax, as may be occurring in SVS 13.

We also note that the relative proper motion of the Gaia star with respect to the radio source VLA 4B is oriented along PA = 147$^\circ$, which is very close to the [Fe II] jet PA of 145$^\circ$. This PA agreement makes it even less likely that the Gaia position traces a different object, since optical proper motion along jet axis could be easily caused by a tiny change in jet/cavity brightness distribution (e.g., because of the ejection of new knot, as it is also frequently observed in free-free radio jets; see discussion above). 

In summary, the Gaia results show that the optical position of SVS 13 is very close to VLA 4B and different from that of VLA 4A, definitely favoring VLA 4B, and not VLA 4A, as the radio counterpart of the optically identified star SVS 13. However, the optical and radio positions differ by an amount larger than the formal uncertainties, and the possibility that they trace different objects (and, therefore, a triple system is present) cannot be completely discarded. We favor the identification of the optical star and the radio source VLA 4B as a single object, and we attribute the small difference in the observed optical/radio positions to the fact that SVS 13 is an embedded object (likely a Class I source) whose direct light is not optically visible but is scattered in its circumstellar environment. As a consequence, its true position is better traced by the radio emission, as it has already been observed to occur in other cases, such as the examples mentioned in a previous paragraph.

\subsection{ALMA Continuum Results}\label{sec:dust1}

\subsubsection{Millimeter Continuum Sources}\label{sec:almafield}

\begin{figure}[htb]
\centering
\includegraphics[width=0.46\textwidth]{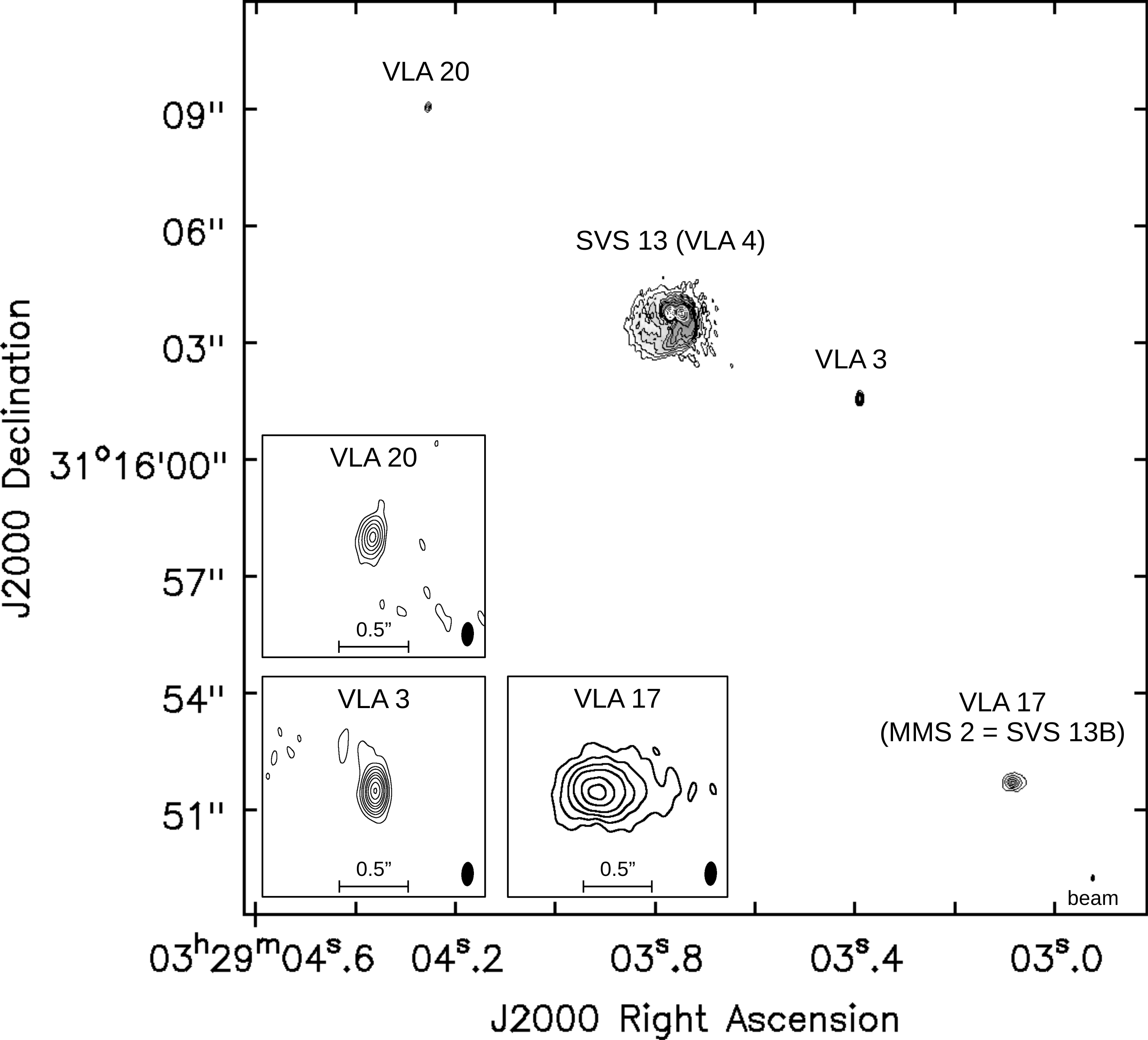}\hfill
\caption{ALMA image of the continuum emission at 0.9 mm of a region $\sim$18$''$$\times$20$''$ in size in the vicinity of SVS 13. The region shown corresponds essentially to the ALMA field of view (the FWHM of the primary beam is $18''$, centered on SVS 13), and includes the continuum dust emission from the nearby YSOs VLA 20, VLA 3, and VLA 17 (MMS 2 = SVS 13B), whose positions are given in Table \ref{tab:posSources}. A close-up of these nearby sources is shown as an inset in the bottom-left corner. For better visualization, the wide-field image has not been corrected for the primary beam response, but the values given below for the contour levels have been corrected. Contours are $-$12, 12, 24, 40, 56, 72, 88, 120, 200, 400, 800, and 1200 times 0.075 mJy beam$^{-1}$, the rms at the center of the image (SVS 13); $-$6, 6, 12, 20, 30, 50, 70, 100, 150, and 200 times 0.094 mJy beam$^{-1}$ (VLA 3); $-$3, 3, 6, 12, 20, 30, and 40 times 0.135 mJy beam$^{-1}$ (VLA 20); and $-$6, 6, 12, 20, 30, 50, and 70 times 0.75 mJy beam$^{-1}$ (VLA 17). The synthesized beam ($0\farcs170\times0\farcs084$, PA = $-2.3\degr$) is shown in the bottom right corner of the images.
\label{fig:field}}
\end{figure}

Figure \ref{fig:field} illustrates the 0.9 mm dust continuum emission detected in the ALMA field of view (the FWHM of the ALMA primary beam is $\sim$18$''$, centered on SVS 13). The image is quite similar to the 1.3 mm ALMA image obtained by \citet{Tobin2018} with a larger synthesized beam. Our field of view encloses the SVS 13 close binary, which stands out prominently as the strongest and more extended continuum source near the center of the field, but also dust emission from the nearby YSOs VLA 20, VLA 3, and VLA 17 (MMS 2 = SVS 13B) is detected. All these sources are detected in our VLA observations (Sect.~\ref{sec:vlafield}). Sources VLA 3, VLA 17, and VLA 20 were also detected in the 1.3 mm ALMA images of \citet{Tobin2018}. 

We note that, although the Class 0 object MMS 2 (= VLA 17) has been considered in the past a (wide) binary companion of SVS 13 (e.g., \citealt{Bachiller2000}; hence its alternative name, SVS 13B, and the use by some authors of SVS 13A to refer to the star SVS 13), at first glance, there is no direct connection between these two objects in the continuum emission images, nor with the rest of the sources which, indeed, are closer in projection to SVS 13 than VLA 17 (MMS 2 = SVS 13B). For the sake of clarity in the identification of the true (close) binary SVS 13, we will avoid, as far as possible, this terminology in terms of SVS 13A/B.

The positions and flux densities of the ALMA sources detected in the field are given in Tables \ref{tab:posSVS13}, \ref{tab:fluxSVS13}, and \ref{tab:posSources}. As noted in Section~\ref{sec:vlafield}, in this paper we focus our discussion on SVS 13. A study of the proper motions and SEDs of the remaining sources will be presented in a forthcoming paper (A. K. Diaz-Rodriguez, in prep.).

\subsubsection{Circumstellar and Circumbinary Dust}\label{sec:dust2}

\begin{figure}[htb]
\centering
\includegraphics[width=0.46\textwidth]{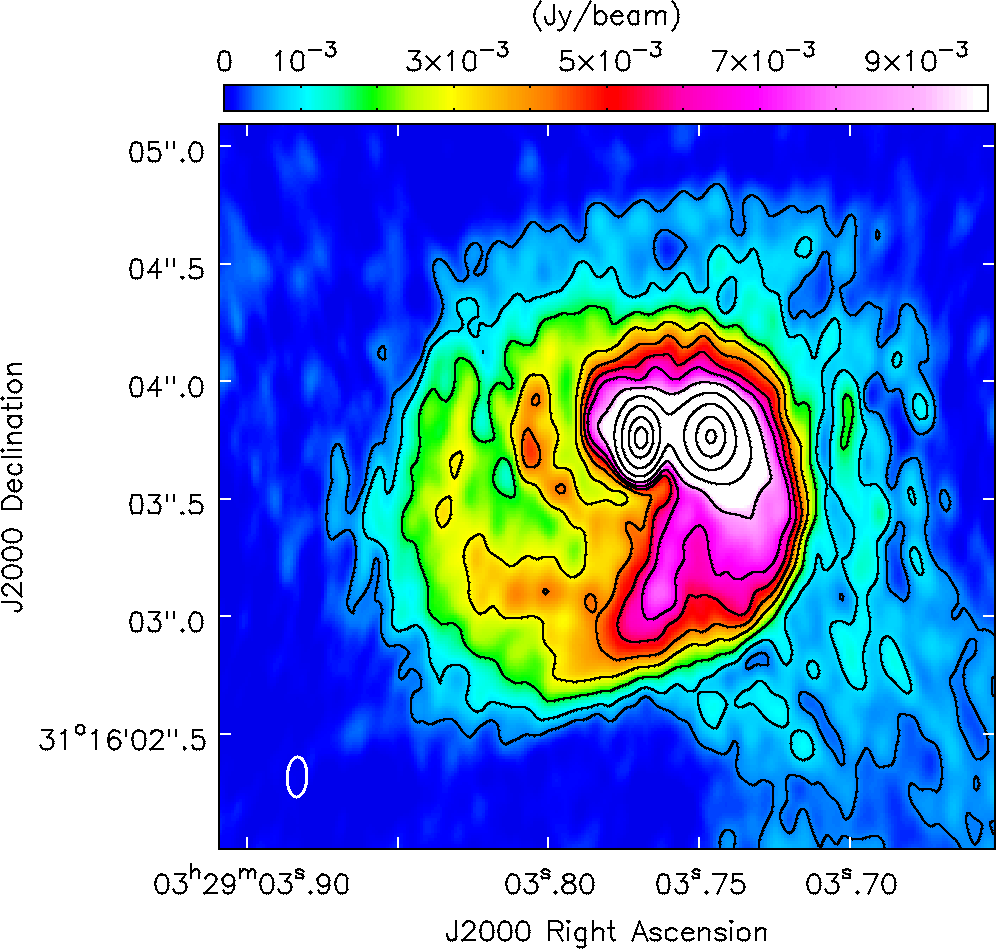}
\caption{ALMA image at 0.9 mm toward SVS 13. The two emission peaks are associated with the previously identified cm sources, VLA 4A (the western one) and VLA 4B (the eastern one). The image has been made from the highest angular resolution data (Cycle 3) using all the baselines and Briggs weighting with robust parameter of 0.5. Contours are $-6$, 6, 12, 24, 40, 56, 72, 88, 120, 200, 400, 800, and 1200 times 0.075 mJy beam$^{-1}$, the rms of the image. Contour levels in units of beam-averaged brightness temperature are $-$0.3, 0.3, 0.6, 1.3, 2.2, 3.0, 3.9, 4.8, 6.5, 10.8, 21.6, 43.2, and 64.9 K. The synthesized beam ($0\farcs170\times0\farcs084$, PA = $-2.3\degr$) is shown in the bottom left corner of the image. The image 
has been corrected by the primary beam response.
\label{fig:continuum}}
\end{figure}

Figure \ref{fig:continuum} shows a close-up toward SVS 13 of the 0.9 mm continuum emission imaged with ALMA. The image shows two strong dust peaks, coincident with the sources VLA 4A and VLA 4B, surrounded by extended emission in the form of spirals. A similar morphology can be seen in the ALMA image at 1.3 mm obtained by \citet{Tobin2018} (see their Fig.~1). We interpret the two compact emission peaks as tracing circumstellar disks of dust associated with the two protostars, VLA 4A and VLA 4B, at scales $\lesssim 0\farcs2$. A large fraction of the 0.9 mm continuum emission is distributed at larger scales (within $\lesssim 1\farcs5$ from the protostars), probably tracing dust from a circumbinary disk that is still in the process of forming. The dust emission shows an inner and an outer spiral arms, which are better differentiated to the east of the binary. Most of this dust emission seems to converge toward the position of source VLA 4A through the outer arm. For the inner arm it is unclear whether it is connected to VLA 4A or to an intermediate position in between VLA 4A and VLA B. Overlapped with the emission from these two spiral features, a slight emission enhancement is observed extending to the southeast of VLA 4A, which appears to follow a straight path along a PA=160$^\circ$ angle, similar to the PA of the CO/H$_2$ jet that appears to originate in VLA 4B (\citealt{Hodapp2014,Lefevre2017}; G. Bl\'azquez-Calero et al. in prep.). 
With the present data, it is unclear whether this is just the result of the superposition of the emission from the spiraling features observed south of the protostars, or whether it could potentially be related to the southern lobe of the outflow. Further analysis is needed to clarify the nature of this feature. There is also some weaker fluffy emission extending further out to the southwest, which seems to be associated with the ambient high density molecular core that continues to the west and southwest of SVS 13, as is seen in VLA ammonia images \citep{Rudolph2001,Diaz2021}.

\begin{figure}[htb]
\begin{center}
\includegraphics[width=0.46\textwidth]{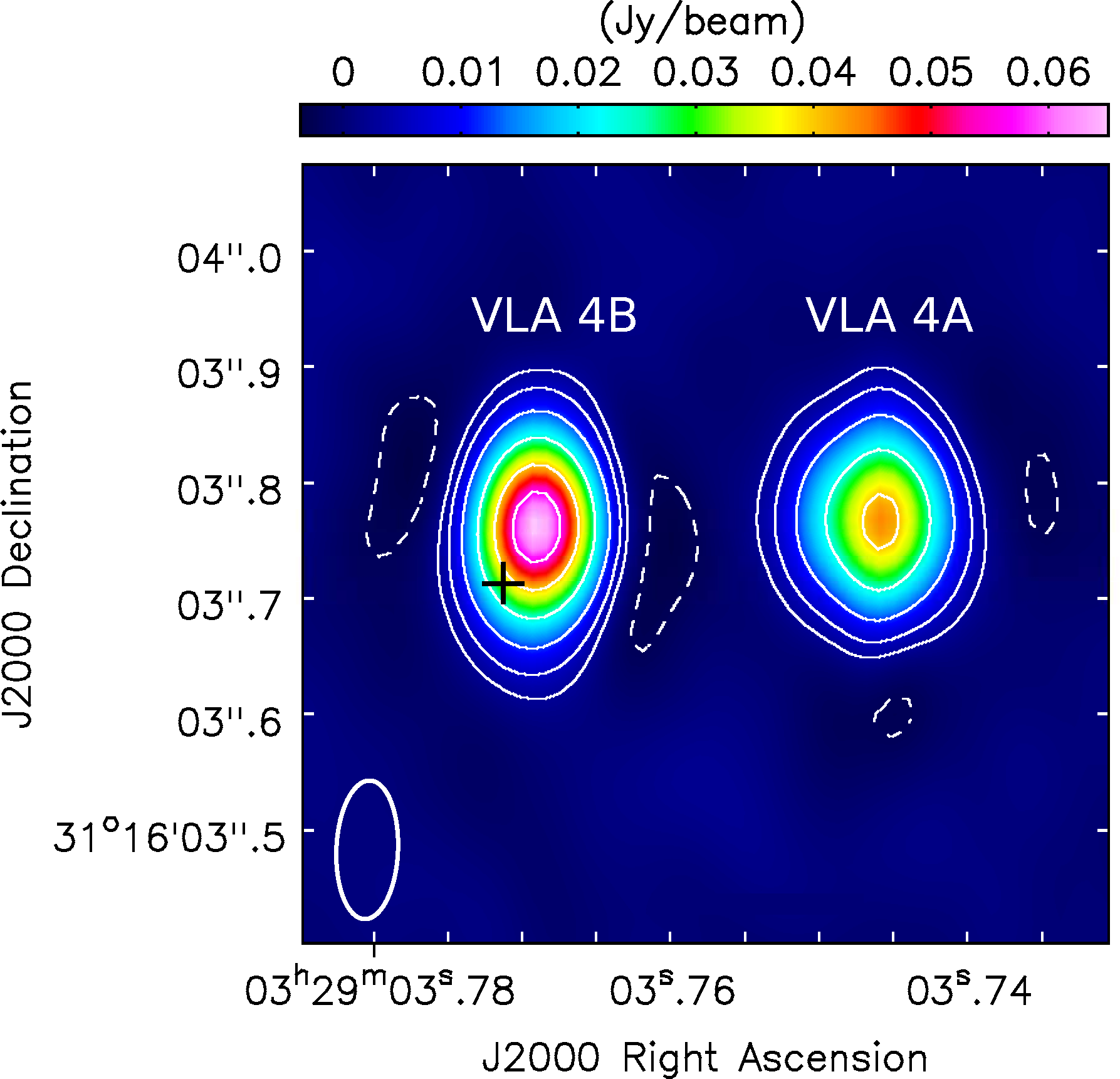}
\caption{Close-up of the 0.9 mm continuum emission near the two components of the SVS 13 protobinary. The image has been obtained from the same data and with the same weighting as in Fig. \ref{fig:continuum}, but using only baselines $>$750 k$\lambda$ to filter out the extended emission. The black plus sign marks the J2000 position of the Gaia EDR3 \citep{GaiaEDR3} optical source at the epoch of our observations (see Section \ref{sec:pm}, Fig.~\ref{fig:pmotions}).  Contours are $-$6, 6, 14, 30, 60, 100, and 140 times 0.40 mJy beam$^{-1}$, the rms of the image. The synthesized beam ($0\farcs120\times0\farcs053$, PA = $-2.4\degr$) is shown in the bottom left corner of the image.
\label{fig:cont750kl}}
\end{center}
\end{figure}

If the extended emission is suppressed by restricting the uv-range of the data ($>$750 k$\lambda$), we obtain the image shown in Figure \ref{fig:cont750kl}. This image is expected to preferentially trace the compact circumstellar dust associated with each protostar. From this image, accurate positions for the two sources are obtained, coincident within $\sim$10 mas with the positions inferred from the cm data (see Table \ref{tab:posSVS13}). It is interesting to note that, although in the image using all the baselines (Fig.~\ref{fig:continuum}) the 0.9 mm emission concentrates toward the western side, apparently mainly associated with VLA 4A, when excluding the extended emission the peak toward VLA 4B is brighter than the peak toward VLA 4A. This suggests that there is a larger amount of emitting dust surrounding source VLA 4A, extending over a larger region than in VLA 4B. The emitting dust associated with VLA 4B appears to be distributed on a more compact scale and, at this scale, the column density is higher and/or the dust is hotter than near VLA 4A.

Specifically, in the 0.9 mm image obtained using only baselines $>$750 k$\lambda$ (Fig. \ref{fig:cont750kl}), a Gaussian fit gives a flux density toward VLA 4A of 82.9$\pm$3.0 mJy (Table \ref{tab:disks}), arising from a compact region with a deconvolved size (FWHM of the major axis) of 80.6$\pm$3.8 mas (or 24$\pm$1 au), at PA=83$\degr$$\pm$8$\degr$. Given the small size obtained, we interpret this emission as dust emission from an inclined circumstellar disk. The total flux density that is actually associated with VLA 4A is difficult to estimate, as there is not a clearly defined boundary in the image obtained using all the baselines (Fig.~\ref{fig:continuum}), but we estimate it as $\sim$220$\pm$25 mJy (within a region of $\sim$100 au in size), from this image.

\begin{deluxetable*}{lccc@{\extracolsep{8pt}}c@{\extracolsep{6pt}}c@{\extracolsep{6pt}}c}[htb]
\tabletypesize{\small}
\tablewidth{0pt}
\tablecaption{Parameters of the 0.9 mm Circumstellar Dust Emission\tablenotemark{a} \label{tab:disks}}
\tablehead{
 & \multicolumn3c{$>$750 k$\lambda$} & \multicolumn3c{$>$500 k$\lambda$} \\
 \cline{2-4} \cline{5-7}
 Parameter & VLA 4A & VLA 4B & Beam & VLA 4A & VLA 4B & Beam}
\startdata
 Size (mas\,$\times$\,mas)  & 80.6$\pm$3.8$\times$34.2$\pm$20.2 &  61.3$\pm$2.6$\times$45.7$\pm$8.5 & 120$\times$53 & 99.7$\pm$4.0$\times$53.5$\pm$13.3 & 65.8$\pm$3.5$\times$45.4$\pm$15.1 & 150$\times$60 \\
  Size (au\,$\times$\,au)  & 24.2$\pm$1.1$\times$10.3$\pm$6.1  &  18.4$\pm$0.8$\times$13.7$\pm$2.5 &  36$\times$16 & 29.9$\pm$1.2$\times$16.1$\pm$4.0 & 19.7$\pm$1.1$\times$13.6$\pm$4.5 & 45$\times$18 \\
  PA ($\degr$)           & 83$\pm$8     & 93$\pm$17      & $-$2   &    80$\pm$7    &    93$\pm$20    & $-$2 \\
  $i$ ($\degr$)\tablenotemark{b}            & 64$\pm$16    & 40$\pm$12      &        &    57$\pm$9    &    43$\pm$18    & \\
  $S_{\nu}$ (mJy)         & 82.9$\pm$3.0 & 109.7$\pm$2.8  &        & 121.6$\pm$3.4  & 122.0$\pm$3.5   & \\
  $T_{\rm dust}$ (K)\tablenotemark{c}              & 430$\pm$190  & 430$\pm$80     &        &   260$\pm$60   &   480$\pm$145   & \\
\enddata
\tablenotetext{a}{Sizes, position angles and flux densities are obtained from elliptical Gaussian fits. Uncertainties in these parameters only account for the formal error of the fit. Sizes correspond to the deconvolved FWHM of the major and minor axes. Since the sources are elongated roughly along the direction of the minor axis of the beam, where the angular resolution is higher, we consider that the values of the size of the major axis are reliable but the uncertainty of the minor axis can be quite underestimated.}
\tablenotetext{b}{Inclination angle of the disk plane with respect to the plane of the sky, inferred from the aspect ratio. Uncertainties are derived following the procedures explained in Appendix \ref{sec:errors}. Only the formal errors of the Gaussian fits have been taken into account in the error calculation. Therefore, the uncertainties of the inclination angles can be notably underestimated.}
\tablenotetext{c}{Optically thick lower limit to the dust temperature obtained from $S_\nu/\Omega_S = B_\nu(T_{\rm dust})$, where $B_\nu$ is the Planck function and $S_{\nu}$ the flux density of the source, assumed to be a Gaussian of solid angle $\Omega_S = \pi\,a\,b\,/(4 \ln 2)$, where $a$ and $b$ are its FWHM dimensions. Uncertainties in the $S_\nu/\Omega_S$ ratio are estimated as explained in Appendix~\ref{sec:errors}.}
\end{deluxetable*}

On the other hand, in the image obtained using only baselines $>$750 k$\lambda$, the flux density toward VLA 4B is 109.7$\pm$2.8 mJy, coming from a region with a deconvolved size (FWHM of the major axis) of 61.3$\pm$2.6 mas (or 18$\pm$1 au), at PA=93$\degr$$\pm$17$\degr$ (Table \ref{tab:disks}). We interpret this 0.9 mm emission as dust emission from a  circumstellar disk associated with VLA 4B. From the image obtained using all the baselines, we estimate that the total flux density of the dust associated with VLA 4B (within a region of $\sim$80 au in size) is $\sim$160$\pm$16 mJy.

 In summary, the flux density of the whole dust emission associated with VLA 4B appears to be $\sim$ 30\% lower than in VLA 4A, but the emission at very small scales (within a radius of $\sim$9 au) toward the position of VLA 4B is $\sim$30-50\% higher than toward VLA 4A. This suggests that, while VLA 4A appears to be associated with a (total) larger mass of dust in its surroundings, the dust column density (and therefore the dust opacity) at very small scales from the protostar is significantly higher toward VLA 4B.

We note that the synthesized beam of the $>$750 k$\lambda$ image is elongated in the north-south direction, with a minor axis of 50 mas in the east-west direction, very close to the direction of the major axes of the two disks (Table \ref{tab:disks}). Therefore, we consider that, although the sizes of the minor axes are rather uncertain, the deconvolved sizes of the major axes of the disks are quite well determined and reliable. Another indication of the significance of the disk sizes is the fact that we obtain comparable values of the sizes and flux densities by fitting an image obtained with a less restrictive uv-range, using baselines $>$500 k$\lambda$, and a somewhat larger beam (Table \ref{tab:disks}). In this last case, we obtain a major axis size of 99.7$\pm$4.0 mas (30$\pm$1 au), at PA=80$\degr$$\pm$7$\degr$, for VLA 4A; and 65.8$\pm$3.5 mas (20$\pm$1 au), at PA=93$\degr$$\pm$20$\degr$, for VLA 4B (Table \ref{tab:disks}). This suggests that there is a fairly well-defined separation between the boundary of the circumstellar disks and the surrounding envelope, and that the angular sizes of $\sim$80 and $\sim$60 mas (radii of 12 au and 9 au, respectively) for the circumstellar disks of dust obtained from the $>$750 k$\lambda$ images are therefore significant. The position angles of the major axes are also consistently determined to be $\sim$90$^\circ$, further supporting the significance of the fits.

However, the sizes of the disk minor axes are poorly constrained, as the beam is highly elongated in the north-south direction. From the source aspect ratios, and assuming they trace circular disks, the inclination angle of the disk plane with respect to the plane of the sky can be calculated (see results in Table \ref{tab:disks}, and Appendix \ref{sec:errors} for the procedures followed in the error calculations). But we want to caution that the values of the disk inclination we have obtained are very uncertain, probably considerably more than indicated by the uncertainties given in the table, as these uncertainties only take into account the formal errors of the Gaussian fits.

Thus, from the results described above, we conclude that the radii of the circumstellar disks of dust associated with VLA 4A and VLA 4B are likely of the order of $\sim$12 au and $\sim$9 au, respectively. These values of the radii are smaller than those typically found for disks around single stars. In binary systems, tidal truncation to disk radii of $\sim$1/3 of the binary separation is expected \citep{Papaloizou1977,Artymowicz1994}, resulting  in smaller disks. Since the projected distance between the two components of the binary is 90 au, disks with radii of $\sim$30 au are expected to result from tidal truncation assuming circular orbits. Thus, the circumstellar disks in SVS 13 seem to be somewhat smaller than expected, if only tidal truncation is at work and unless the orbit of one (or both) of the binary components has a high eccentricity. In any case, we note that we are comparing the values of the radii at half the maximum intensity, but the dust emission certainly extends beyond this radius; it should also be noted that, as has been observed in other disks, the gas is usually distributed up to larger radii than the dust. Thus, in the case of the circumstellar disks of SVS 13, the deviation from expectations is not severe and is roughly consistent with a tidal truncation origin. An extreme case of a still smaller disk was reported by \citet{Osorio2016} in the XZ Tau A/B young system (separation between components of 42 au), where a disk of dust with a radius of only 3 au was imaged.

The small sizes and large flux densities measured for the 0.9 mm emission toward VLA 4A and VLA 4B  imply relatively high brightness temperatures, and therefore warm dust. From the flux densities and sizes measured in the highest resolution images described above (Table \ref{tab:disks}), and assuming optically thick emission (which gives a lower limit for the dust temperature), we obtain dust temperatures of the order of $\sim$300 K toward VLA 4A and $\sim$450 K toward VLA 4B. Since implausibly higher temperatures would be obtained if the emission were optically thin, and because there is independent evidence that the dust emission is optically thick toward the protostars (Sect.~\ref{sec:gas}), we conclude that the dust emission is very optically thick in the SVS 13 circumstellar disks, and that the dust reaches kinetic temperatures of the order of $\sim$300-450 K at radii of $\sim$10 au from the protostars.

Under certain conditions, and at these small radii, these relatively high dust temperature values are not unrealistic. Indeed, similar values have been inferred from the modeling of the SEDs and mm images of other low-mass protobinaries such as L1551 IRS 5 \citep{Osorio2003}.
In general, flared disks are warmer at the disk surface and cooler near the equatorial plane. But in sources with relatively high mass accretion rates, such as the circumstellar disks of L1551 IRS 5, models predict temperatures on the order of 300-400 K near the equatorial plane, at radii of 5-10 au \citep{Osorio2003}.
The model predictions on the configuration and physical properties of the L1551 IRS 5 system by \citet{Osorio2003} have been nicely confirmed through recent high-resolution ($\sim$$0\farcs1$) ALMA observations at 0.9 mm by \citet{Takakuwa2020}. In particular, these authors measured brightness temperatures reaching $\sim$260 K, implying high dust temperatures in the L1551 IRS 5 circumstellar disks, similar to the values we infer in the circumstellar disks of SVS 13. However, the current observational uncertainties in SVS 13 remain large (e.g., the disk minor axes and, thus, the solid angle of the sources are poorly constrained with the current angular resolution) and, also, more specific model predictions, tailored to the properties of the SVS 13 disks, are needed to test whether their properties are compatible with a high accretion regime and which scales above the disk midplane the optically thick 0.9 mm emission traces.

As one moves away from the protostars, the brightness temperature decreases sharply, down to values of $\sim$80 K at scales of the order of the beam size of the robust = 0.5 image shown in Figure~\ref{fig:continuum}. Since the temperatures inferred from the molecular emission at these scales are $\sim$140 K (see Sect.~\ref{sec:temp}), we attribute the decrease in brightness temperature to the combined effect of a true decrease of the kinetic temperature together with a decrease in the optical depth and/or the beam filling factor.

Toward other positions, in the circumbinary material and in the spirals of SVS 13, we obtain dust brightness temperature values in the range 5-12 K. Since the temperatures for the circumbinary disk inferred from the molecular emission (see Sect.~\ref{sec:temp}) are of the order of 140 K, we conclude that the 0.9 mm dust emission is optically thin (optical depth $\sim$0.05-0.1) across the circumbinary structure and the spiral arms. This is in contrast with the warmer temperatures and high optical depths inferred for the close environment of the protostars.

\subsubsection{Spirals of Dust}\label{sec:spirals}

An outstanding feature of the ALMA continuum emission (Fig.~\ref{fig:continuum}) are the prominent spiral arms associated with the SVS 13 binary. Trailing spiral arms should point away from the direction of rotation, therefore implying an SVS 13 disk counter-clockwise rotation, as seen from the Earth. This is in agreement with our conclusion about the sense of the orbital motion inferred from the observed proper motions (Section \ref{sec:pm}) and with the analysis of radial velocities from the molecular line emission (Section \ref{sec:gas} and Fig.~\ref{fig:cartoon}).

\begin{figure}[htb]
\begin{center}
\includegraphics[width=0.46\textwidth]{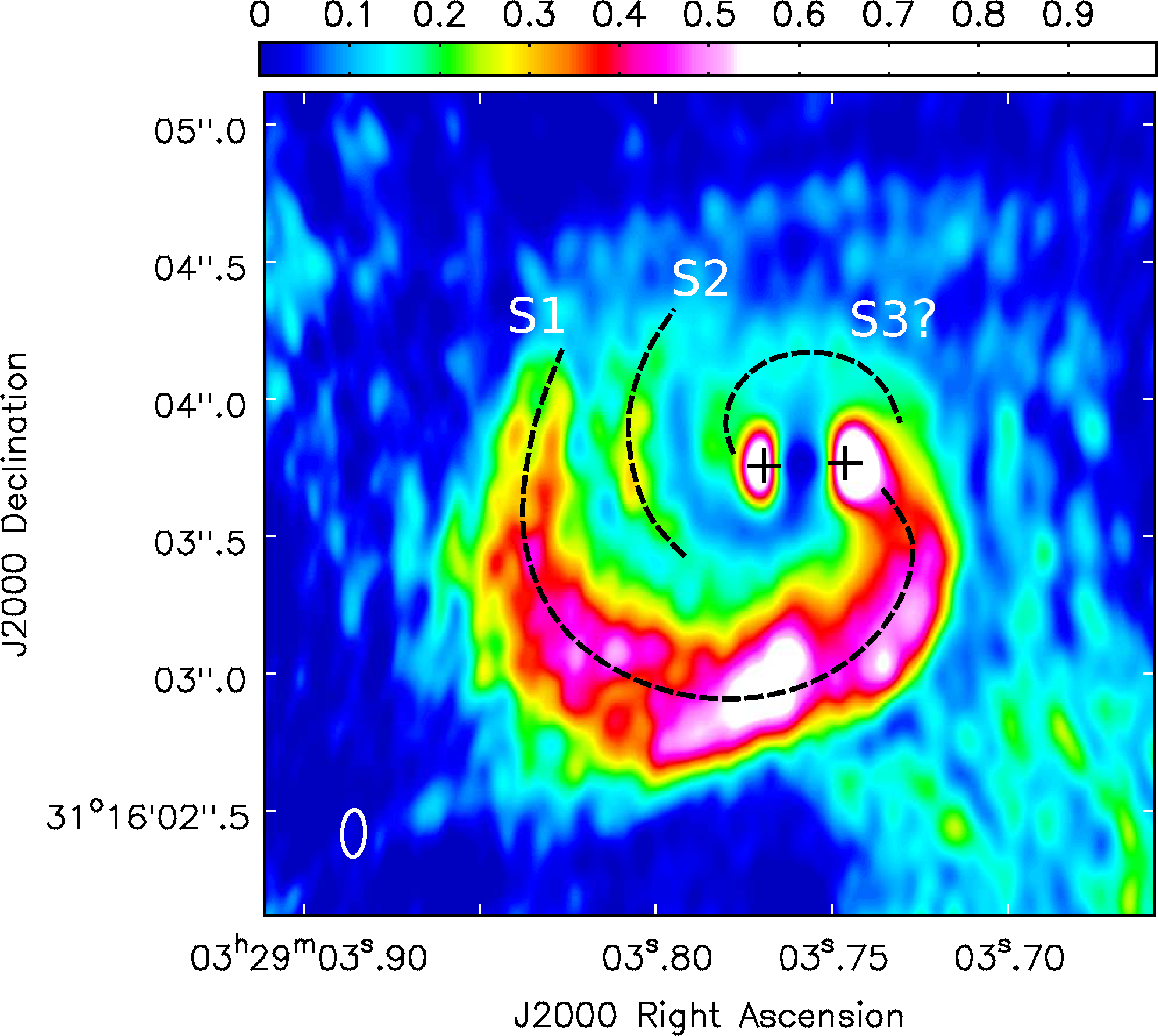}
 \caption{Same as in Fig. \ref{fig:continuum} but with the intensity scaled by a factor $\propto r^{1.5}$, where $r$ is the projected distance to the midpoint between VLA 4A and VLA 4B in arbitrary units, to compensate for the decreasing intensity and to enhance the visibility of the spiral features at large distances. The resulting intensity has been normalized to the maximum. No correction has been applied for projection effects. Three spirals (S1, S2, and S3) are tentatively identified in the image and marked with dashed lines. The plus signs mark the positions of VLA 4A (western) and VLA 4B (eastern) sources.
\label{fig:spirals}}
\end{center}
 \end{figure}

To better emphasize the spiral structures, in Figure \ref{fig:spirals} we present the same 0.9 mm ALMA image as in Figure \ref{fig:continuum} but normalized by a factor $\propto r^{1.5}$, where $r$ is the distance to the midpoint between VLA 4A and VLA 4B, to compensate for the natural decrease of intensity with distance to the source. We identify three features, S1, S2, and S3, in order of apparent intensity (see Fig. \ref{fig:spirals}). S1 seems to originate near the location of VLA 4A. For S2, it is unclear whether it originates near VLA 4A or at an intermediate position between VLA 4A and VLA 4B. S3, the weakest feature, seems to originate near the location of VLA 4B. S1 and S2 are quite clearly defined spirals, but for S3 it is unclear whether it is a spiral arm or a feature of the northern part of the circumbinary disk.

Spirals have been observed in a number of protoplanetary disks and various mechanisms, which can possibly act together, have been suggested to explain their origin, including planet-disk interactions (e.g., \citealt{Ogilvie2002,Muto2012,Dong2015,Phuong2020}), gravitational instabilities (e.g., \citealt{Lodato2005}) or shadows \citep{Montesinos2016}. In the hydrodynamical simulations of the formation of binary systems, spirals also appear as a natural outcome (e.g., \citealt{Bate1997,Bate2018,Matsumoto2019}).

 In particular, the overall evolution of the fiducial model in the \citet{Matsumoto2019} simulations shows two spiral arms on the sides of the primary and secondary stars, where the secondary arm is connected directly to the circumsecondary disk, while the primary spiral arm runs around the circumprimary disk and is connected to the bridge between the two circumstellar disks. This result from the numerical simulations appears to be broadly consistent with our observations of SVS 13 if the secondary were VLA 4A (the component with the more prominent spiral arm, S1) and the primary were VLA 4B (with the S2 spiral surrounding this protostar and being connected to some undetermined point between VLA 4A and VLA 4B). This scenario is indeed consistent with the estimates of the individual protostellar masses obtained in Sect.~\ref{sec:dyn} and Appendices~\ref{sec:keplerian} and \ref{sec:infall} since we obtain a VLA 4B mass higher than that of VLA 4A. Unfortunately, our present results only provide rough mass estimates.
 Future observations of accurate proper motions at mm wavelengths, such as those outlined in Section~\ref{sec:pm}, combined with small-scale kinematics obtained from new molecular data with even improved angular resolution, should be able to shed more light on the masses of the SVS 13 binary and to reliably identify the primary and secondary components. This will make it possible to test the models of the evolution of a binary system which is undergoing the development of a circumbinary disk and spiral arms, as seems to be the case for SVS 13.

\subsection{The Spectral Energy Distribution} \label{sec:sed}

\begin{figure}[htb]
\begin{center}
\includegraphics[width=0.46\textwidth]{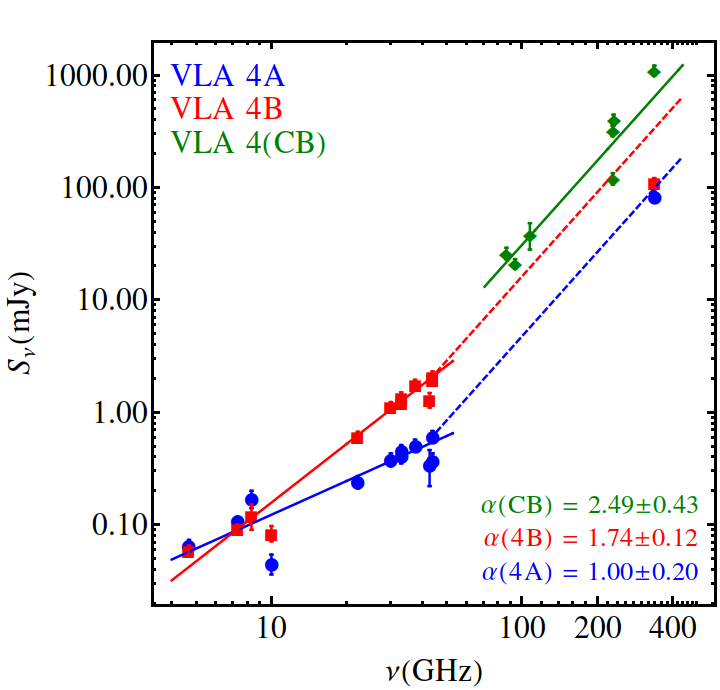}
 \caption{Spectral energy distribution (SED) of SVS 13 from cm to mm wavelengths. Data used are reported in Table~\ref{tab:fluxSVS13}; blue and red data points represent the flux density toward the individual components (VLA 4A and VLA 4B), while green data points represent extended emission toward the whole system (identified with footnotes $d$ and $f$ in Table 2, and labeled as ``CB" in this figure), tracing mainly the circumbinary dust. The spectral indices ($\alpha$, where $S_\nu\propto\nu^\alpha$) are obtained from least-squares fits (in log-log scale) that are shown as solid lines. For the individual components, VLA 4A and VLA 4B, the free-free spectral indices are obtained using only data with a frequency below 40 GHz ($\lambda > 7.5$ mm). The dashed lines are extrapolations from our 7 mm measurements, assuming optically thin dust emission with the same spectral index obtained in the circumbinary dust. Note that, in the case of VLA 4B, the observed 0.9 mm flux density is well below the optically thin extrapolation, indicating a high dust optical depth at this wavelength (see Sect. \ref{sec:dust2}).\label{fig:SED}}
\end{center}
 \end{figure}

Our VLA and ALMA observations angularly resolve the SVS 13 system over a range of wavelengths from 6 cm to 0.9 mm. There are not many observational data on close ($<$100 au) protobinary systems that angularly separate the emission of the two components. Our data, combined with previous data from the literature allow us to characterize in depth the nature of the emission and the physical properties of this system. In Table \ref{tab:fluxSVS13} we list the measured flux densities and in Figure \ref{fig:SED} we plot the resulting SED.

Regarding the VLA data, we note that in 2016.9 there was an important decrease of the X-band (3 cm) flux density of VLA 4A, by a factor close to 4 with respect to previous dates ($\sim$1995-2014), which is atypical for thermal free-free emission, where the flux density variability usually does not exceed $\sim$20\% (see \citealt{Anglada2018}). This variability is usually attributed to sudden variations in the mass accretion rate, which trigger subsequent variations in the outflow mass loss rate. But, in general, they imply sudden increases in flux density, sometimes associated with the ejection of new ionized knots, rather than sudden decreases. Further monitoring of the sources with high sensitivity and angular resolution are necessary to improve the previous data and to clarify the origin of the X-band variability. At shorter wavelengths, $\sim$7 mm, a smaller increase in flux density, by a factor of $\sim$1.5, took place in 2015 in both sources, but only in VLA 4B persisted in the subsequent observations.

\citet{Anglada2004} noticed that the flux density of VLA 4B was much stronger than that of VLA 4A at 7 mm, while the fluxes of the two sources were comparable at longer wavelengths. We confirm this observational result, which can also be seen in the data of \citet{Tychoniec2018} and \citet{Tobin2016mult}. Since the observed 7 mm flux of VLA 4B coincided with the extrapolation of the mm data available at that time (which did not separate the binary), \citet{Anglada2004} attributed the excess of 7 mm emission to dust and proposed that, while VLA 4B was associated with a bright circumstellar disk of dust, VLA 4A was not. This observational result was interpreted by \citet{Anglada2004}, in terms of the \citet{Bate1997} theoretical simulations, as suggestive that the formation of the SVS 13 binary system occurred via the accretion of material with low specific angular momentum. In this case, the primary may have a large circumstellar disk, while the secondary is essentially naked; thus, VLA 4B was proposed as the primary, more massive, component and VLA 4A as the secondary. Also,
according to \citet{Bate1997} the systems formed by this mechanism should not develop a significant circumbinary disk.

Our present results clarify the previous conclusions by \citet{Anglada2004} and shed new light on the interpretation of the cm and mm continuum data. Despite the difficulties in distinguishing between circumstellar and circumbinary contributions at such early stages of the formation process of a binary system, as is the case of SVS 13, we identify three main dust contributions. We see that both VLA 4A and VLA 4B protostars are associated with circumstellar dust, although with different properties regarding its distribution, as discussed in Section \ref{sec:dust2}. We also clearly see that there is dust (and gas; see Sect. \ref{sec:alma}) in between and surrounding the two protostars, and with prominent spiral arms. We interpret this dust as material being assembled into a proto-circumbinary disk, which we will refer to simply as ``the circumbinary disk''.

In Figure \ref{fig:SED} we plot the SED in the cm/mm regimes, where we identify several contributions, namely: (i) The free-free emission associated with VLA 4A (with a spectral index of 1.0 at wavelengths longer than $\sim$7 mm), likely tracing a thermal radio jet; (ii) The free-free emission associated with VLA 4B (that dominates the emission of this object at long wavelengths), also consistent with the presence of a thermal radio jet; (iii) The dust emission of the circumstellar disk associated with VLA 4A (traced by the 0.9 mm blue data point) that implies a change to a steeper spectral index, $\sim$2.5, at short wavelengths; (iv) The circumstellar disk around VLA 4B (traced by the 0.9 mm red data point), with a flux density consistent with the extrapolation of the long-wavelength data with a single slope of 1.7. Since a spectral index $<$2 cannot be produced by dust emission alone, this value suggests a free-free contribution at long wavelenghts and optically thick dust emission at short wavelengths; (v) The dust emission of the circumbinary disk and its associated spiral arms (green diamonds with a spectral index of 2.5) which dominates the total emission observed at short wavelengths.

Based on the previous discussion on the morphology of the sources at Q and Ka bands (Sect.~\ref{sec:vla2}) and on the dust brightness temperatures (Sect.~\ref{sec:dust2}), it is reasonable to assume that the 9 mm emission is dominated by free-free emission, while the 7 mm and 0.9 mm emissions are dominated by dust. The 0.9 mm emission in the circumbinary disk is optically thin dust, while the dust emission is optically thick in the VLA 4A circumstellar disk, and very optically thick in the disk of VLA 4B. The 7 mm emission of both circumstellar disks is expected to be optically thin, but it can contain some residual free-free contamination.

\subsection{ALMA Line Emission Results}\label{sec:alma}

\subsubsection{Distribution and Overall Kinematics of the Molecular Gas}\label{sec:gas}

In this Section, we are focusing on the molecular line emission that appears to be more directly associated with the SVS 13 binary and associated disks, both spatially (within a few arcsec) and in velocity (within a few km s$^{-1}$ of the adopted systemic velocity of the cloud, +8.5 km s$^{-1}$; \citealt{Rudolph2001,Diaz2021}), and that is most useful to infer the physical properties of the system and its close environment. The ALMA observations also reveal complex molecular structures at large scales, $\gtrsim$$10''$ ($\gtrsim$3000 au), and high-velocity features likely associated with the blue outflow lobe that will be discussed in forthcoming papers (A. K. Diaz-Rodriguez et al., in prep., and G. Bl\'azquez-Calero et al. in prep.).

\begin{figure*}[htb]
\includegraphics[height=0.29\textheight]{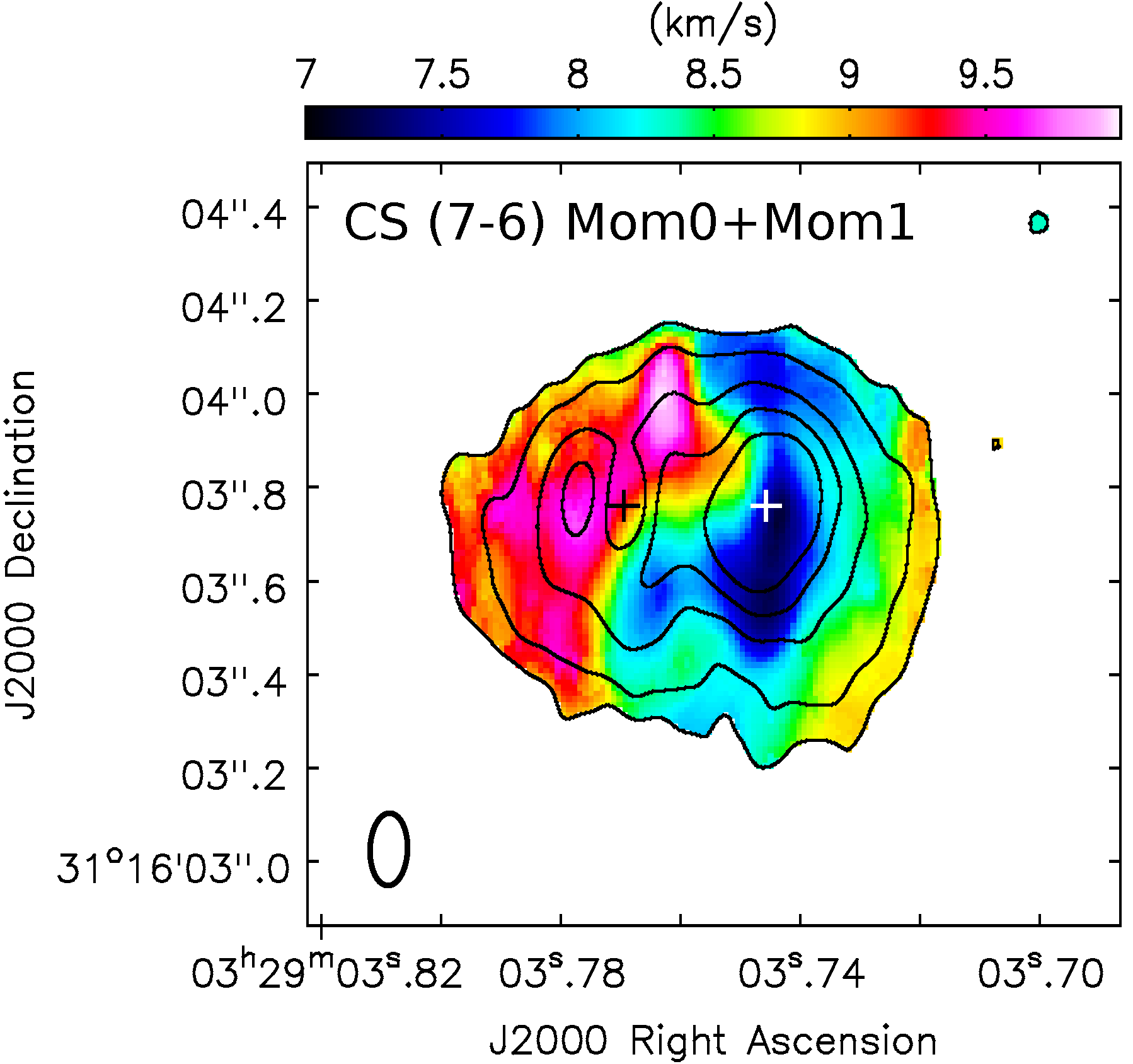}\hfill\includegraphics[height=0.29\textheight]{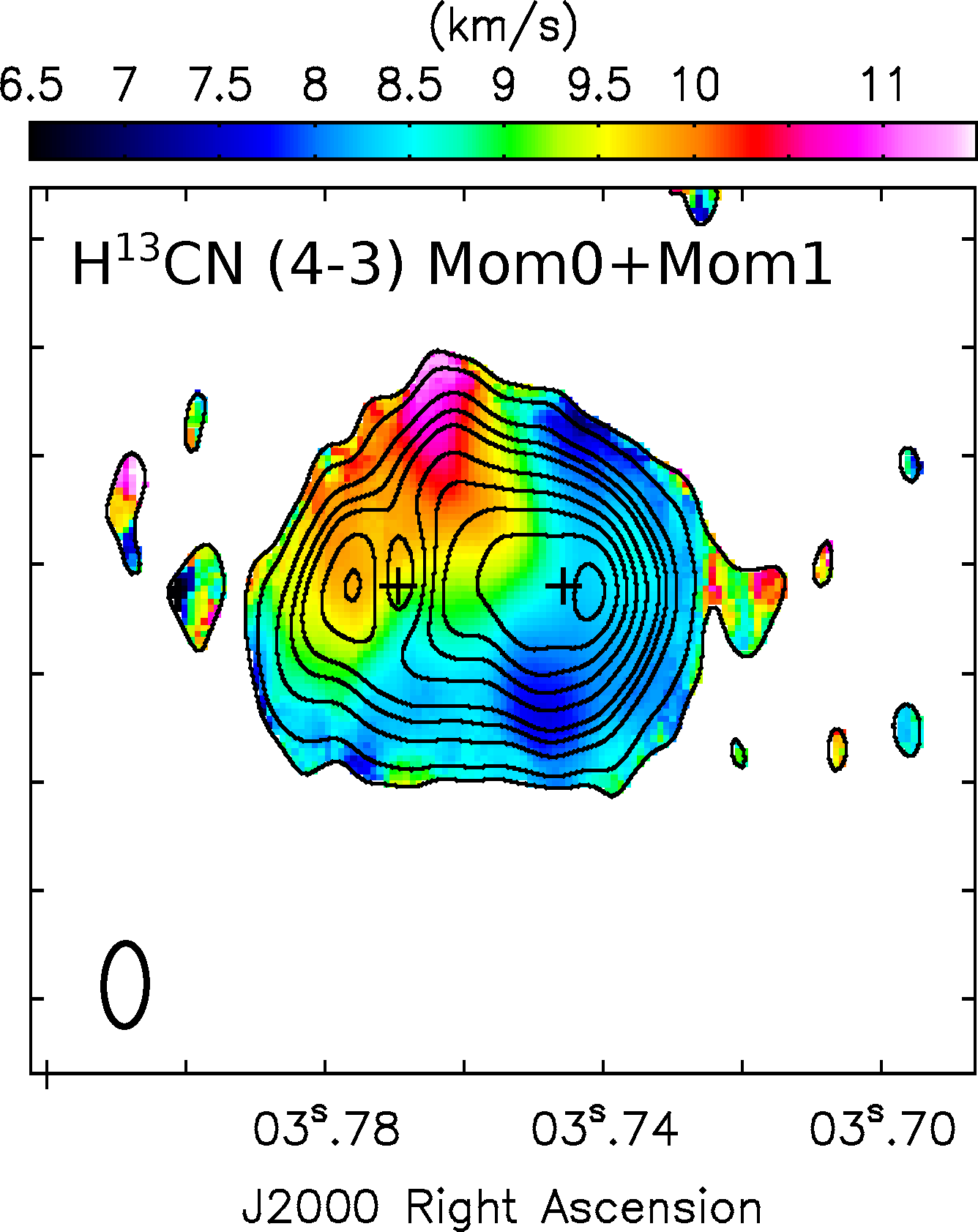}\hfill\includegraphics[height=0.29\textheight]{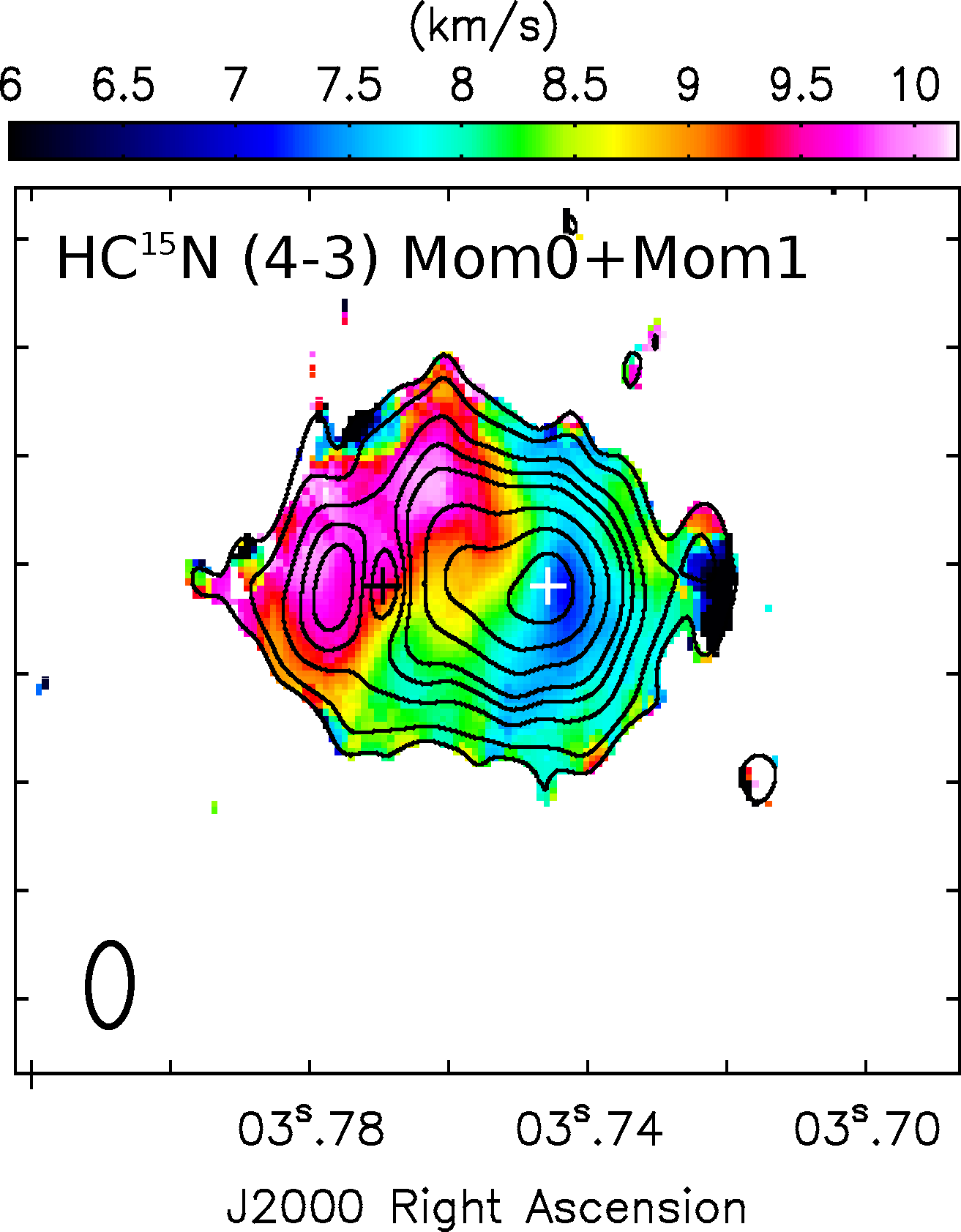}\\ 
\\
\includegraphics[height=0.29\textheight]{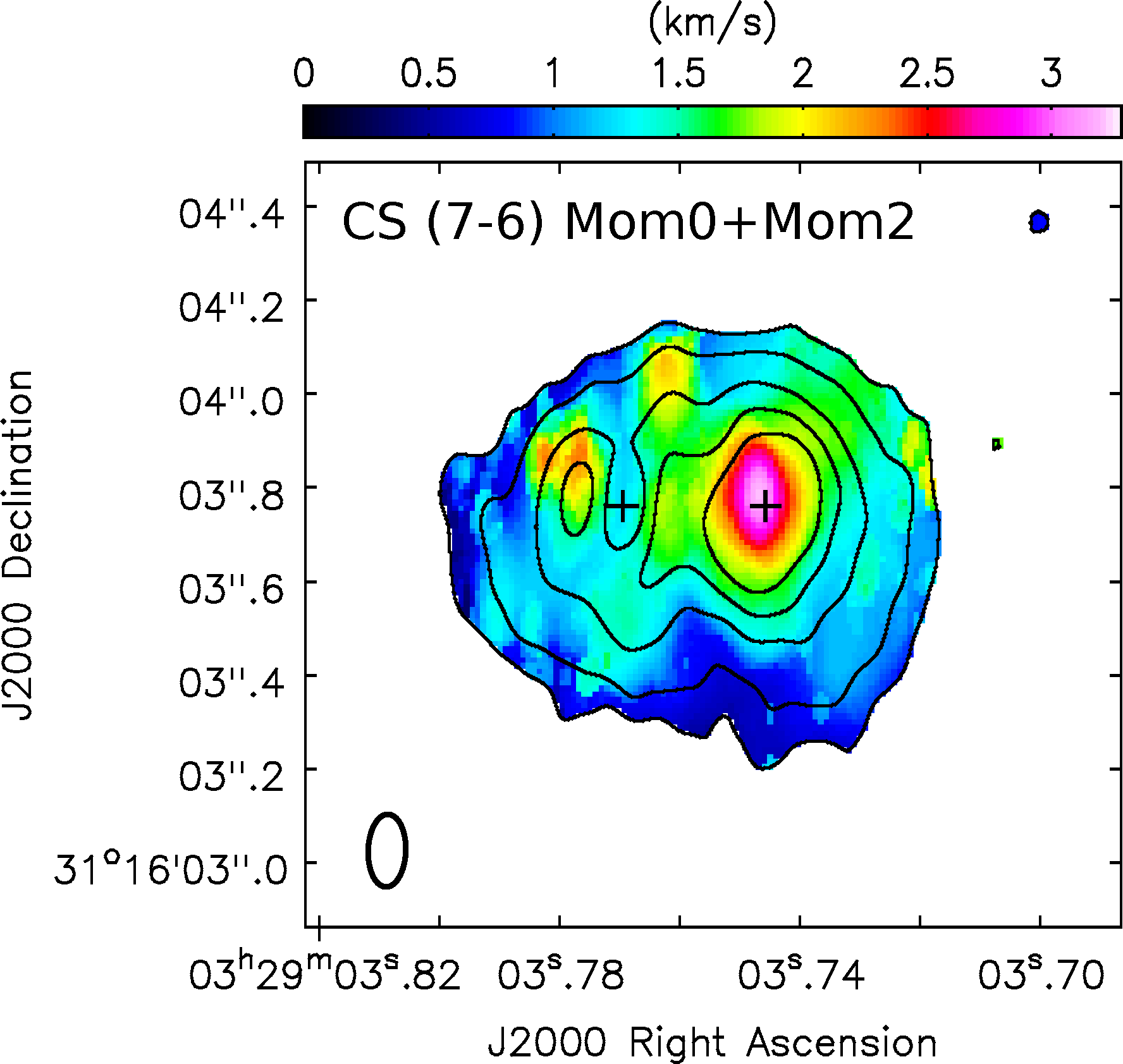}\hfill\includegraphics[height=0.29\textheight]{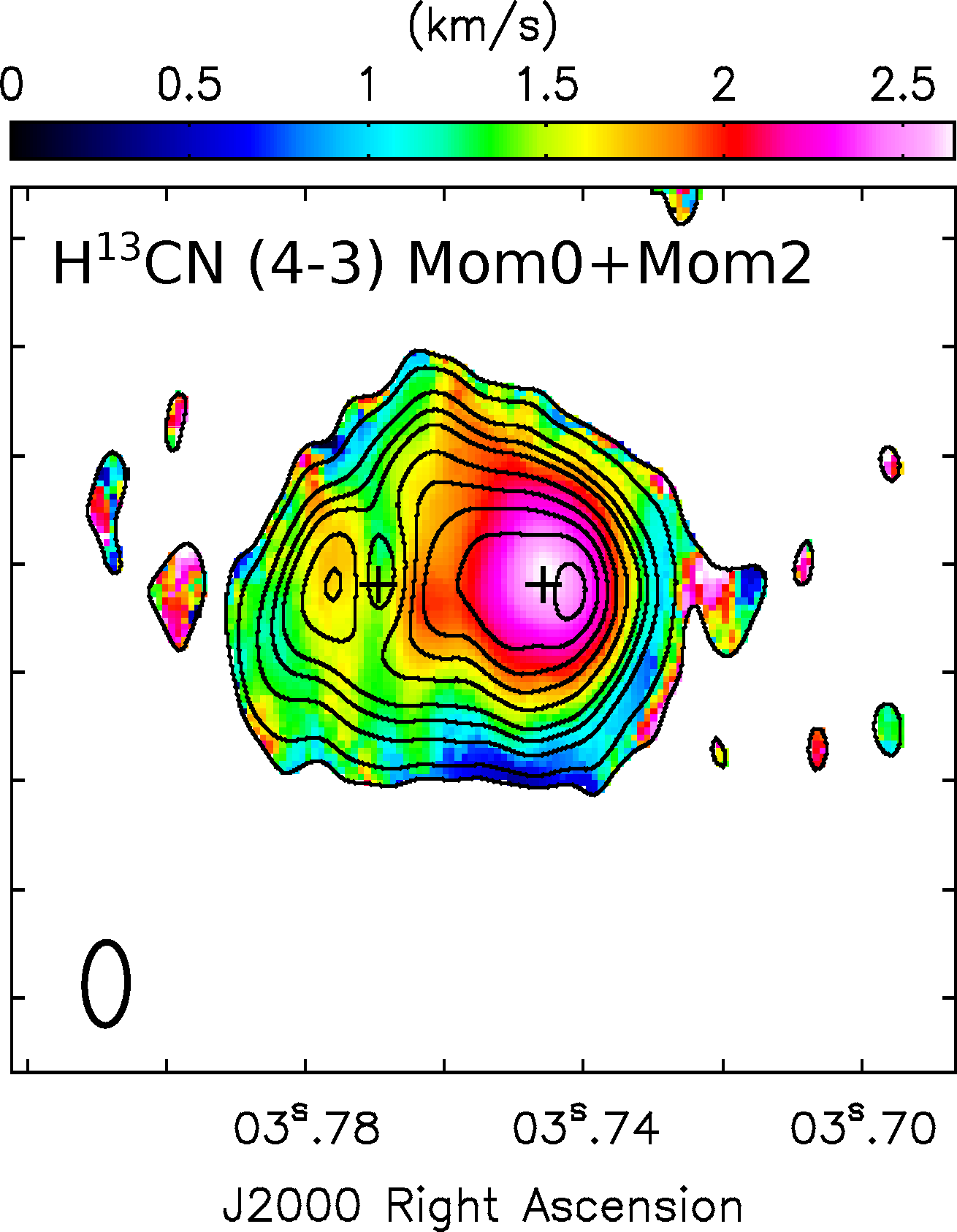}\hfill\includegraphics[height=0.29\textheight]{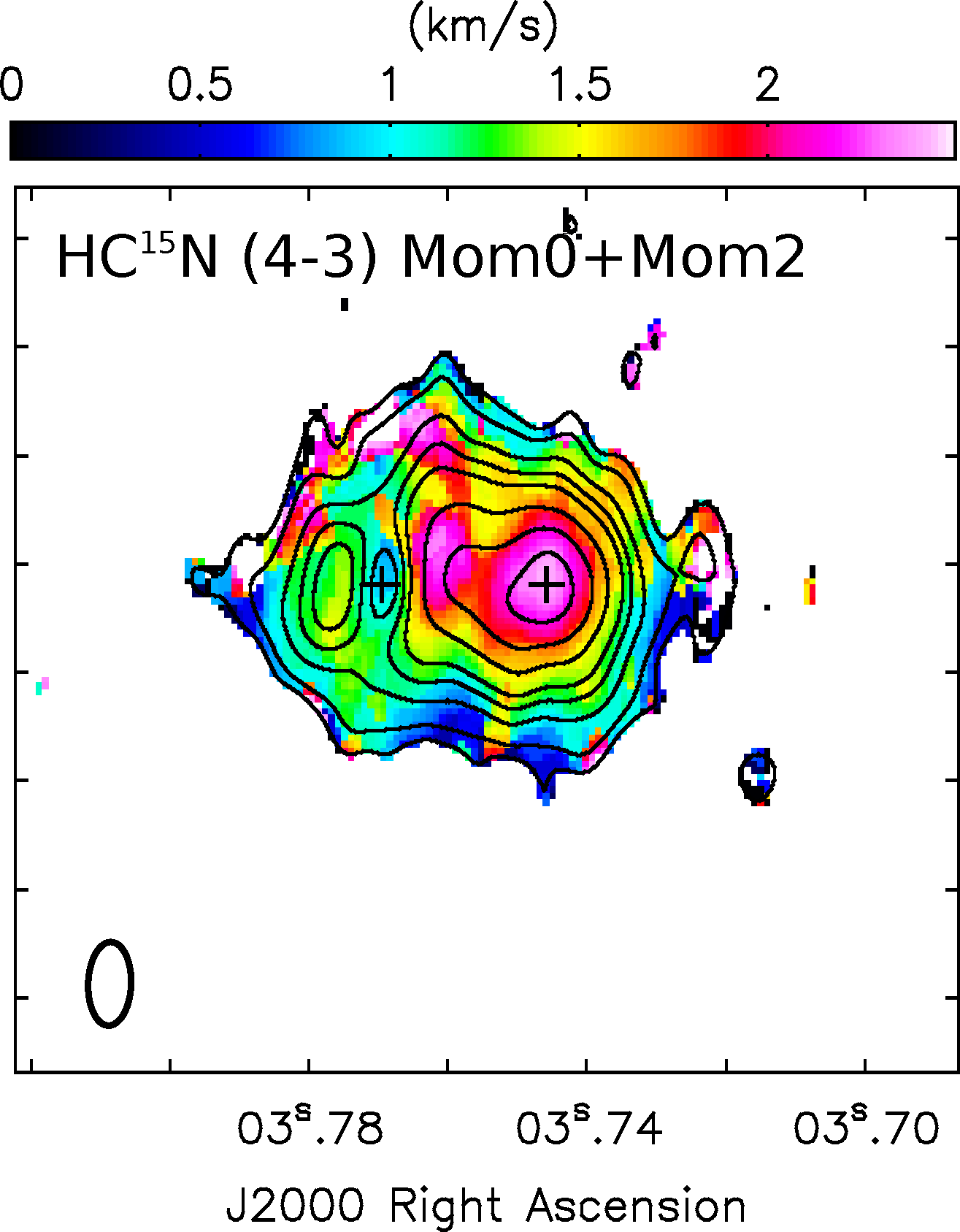}
\caption{Continuum-subtracted velocity-integrated emission (zeroth-order moment; contours), intensity-weighted mean velocity (first-order moment; color scale in top panels) and velocity dispersion (second-order moment; color scale in bottom panels) of the CS (7-6), H$^{13}$CN (4-3), and HC$^{15}$N (4-3) molecular lines observed with ALMA toward SVS 13. Contours are $-$3, 3, 5, 10, 15, 20, 30, 40, 50, 60, and 70 times the rms of each map (25 mJy beam$^{-1}$ km s$^{-1}$ for CS, and 10 mJy beam$^{-1}$ km s$^{-1}$ for H$^{13}$CN and HC$^{15}$N). The synthesized beam (indicated in the bottom left corner of the maps) is $0\farcs15\times0\farcs08$, PA = $-$2$^\circ$. The positions of VLA 4A (western) and VLA 4B (eastern) are marked with plus signs.
\label{fig:CS}}
\end{figure*}

Figure \ref{fig:CS} shows the continuum-subtracted velocity-integrated emission (ze\-roth-order moment), the intensity-weighted mean velocity (first-order moment) and the velocity dispersion (second-order moment) of the CS (7-6), H$^{13}$CN (4-3), and HC$^{15}$N (4-3) molecular lines. These lines are the centering lines of three of the six spectral windows in our observations and have been selected as representative of the emission most directly associated with the binary system. As can be seen in the figures, the emission appears distributed surrounding the binary, and we interpret it as mostly tracing circumbinary gas from which the stars are being assembled. The integrated emission clearly peaks toward source VLA 4A, while there is an apparent decrease of emission toward source VLA 4B that we think is probably not real but the result of an imperfect continuum subtraction, because of the non-additivity of optically thick intensities. As discussed in Section \ref{sec:dust2}, the dust continuum emission toward VLA 4B is very compact and intense, so it probably has a very high optical depth and is saturated, making it difficult for line emission to emerge above the continuum.

The integrated intensity maps presented in Figure \ref{fig:CS} (in contours) show a roughly elliptical molecular emission ($\sim$$1\farcs0 \pm 0\farcs1 \times 0\farcs8 \pm 0\farcs1$, or $\sim$300$\pm$30 au $\times$ 240$\pm$30 au in size for CS, and slightly smaller, $\sim$$0\farcs8 \pm 0\farcs1 \times 0\farcs6 \pm 0\farcs1$, or $\sim$240$\pm$30 au $\times$ 180$\pm$30 au
 for the HCN isotopologues), with the major axis roughly E-W. If the real structure is assumed to be roughly circular, the observed ellipticity implies an inclination angle of the disk (angle of the disk rotation axis and the line-of-sight, where 0$^\circ$ corresponds to face-on) of $i \simeq38^\circ\pm14^\circ$ (see Appendix~\ref{sec:errors} for uncertainty calculations).

\begin{figure}[h]
\begin{center}
\epsscale{1.15}
\plotone{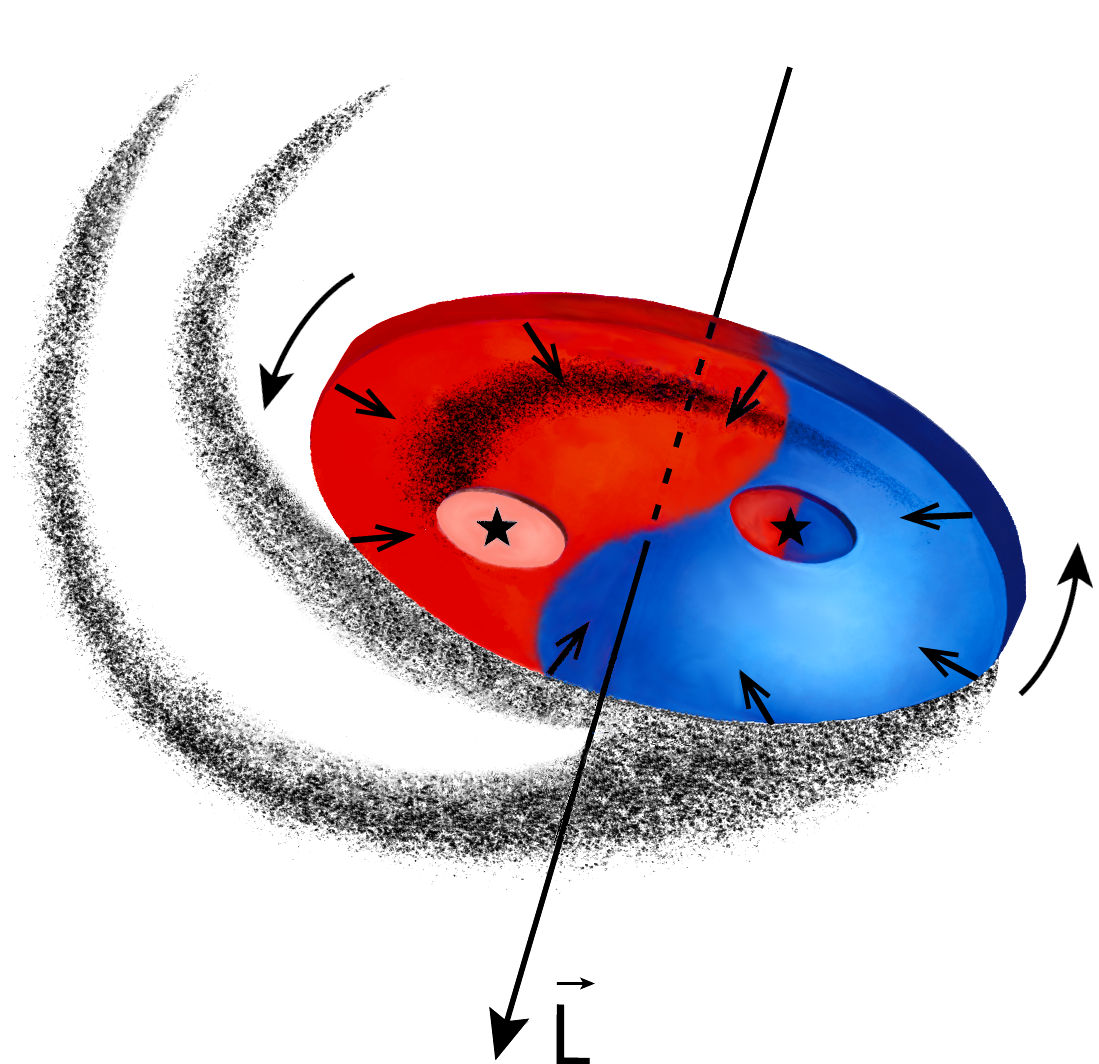}
 \caption{Sketch of the proposed geometrical and kinematical configuration of the SVS 13 protobinary system and its circumbinary material. The red and blue colors represent the appropriate velocity shift of the gas with respect to the systemic velocity with the inferred angular momentum axis also shown.
 \label{fig:cartoon}}
\end{center}
 \end{figure}

The first-order moment maps (in color in the top panels of Fig.~\ref{fig:CS}) show a clear velocity gradient roughly in the SW-NE direction, with blue-shifted velocities toward the SW side. The ALMA maps obtained by \citet{Tobin2018} for other molecules and lower frequency transitions show differences from one molecule to another, but are roughly consistent with ours in the shape and in the direction of the overall velocity gradient. However, our higher angular resolution and better sensitivity allow us to perform a more detailed analysis. Actually, our maps show a kind of ``yin-yang''-shaped velocity distribution that we attribute to the simultaneous presence of both infall and rotation with comparable velocities (see Fig.~\ref{fig:cartoon}). Given that the southern lobe of the outflow associated with SVS 13 is blue-shifted (i.e., pointing toward the observer; \citealt{Bachiller2000,Raga2013,Hodapp2014}), and assuming that the circumbinary emission is mainly distributed in a disk-like structure close to the equatorial plane of the circumstellar disk associated with (and thus perpendicular to) the outflow, this implies that we are observing these disks from ``below''. In other words, we are observing the circumbinary disk in such a way that its northern half is the one closest to the observer. Thus, the observed velocity gradient, with the eastern side red-shifted, implies a counter-clockwise rotation as seen from the observer. This direction of the circumbinary disk rotation is in agreement with that inferred from the observed orbital proper motions (Sect.~\ref{sec:pm}) and the dust spirals in the 0.9 mm continuum emission (see Sect.~\ref{sec:spirals} and Figs.~\ref{fig:continuum}, \ref{fig:spirals}). Given the orientation of the disk-like structure, the presence of infall motions would produce additional red-shifted emission in its northern part and blue-shifted emission in its southern part, resulting in the observed ``yin-yang''-shaped velocity distribution (see Fig. \ref{fig:cartoon}). However, given that the northern lobe of the bipolar outflow is red-shifted and the southern one is blue-shifted (e.g., \citealt{Bachiller2000,Raga2013,Hodapp2014}), a contribution to the observed velocity field from the outflow cannot be discarded.

The second-order moment maps (in color in Fig.~\ref{fig:CS} bottom panels) show a large increase in velocity dispersion toward VLA 4A.
 The second-order moment increases from a typical value of 1.2 km s$^{-1}$ in the circumbinary structure up to $\sim$3.2 km s$^{-1}$ toward source VLA 4A. This local increase in the velocity dispersion could be due to infalling motions, increased turbulence
associated with accretion shocks caused by the impact of infalling material onto the disk, or unresolved rotation in the disk.
The origin of the main spiral arm (S1), near the position of VLA 4A, also strongly suggests that this object is a major accretion focus (see Sect.~\ref{sec:spirals}). In the proximity of VLA 4B there is only a modest increase in the velocity dispersion, with a local decrease toward the position of the star. However, it is unclear whether this apparent decrease in the second-order moment toward VLA 4B is real, with the turbulence and/or accretion being smaller toward this object, or is an artifact due to the suppression of line emission in the presence of very optically thick continuum emission.

\subsubsection{Detection and Identification of Molecular Transitions}\label{sec:mol}

Figure \ref{fig:spectra} shows the full 2.8 GHz spectral band, obtained over a region with the size of the beam ($\sim$$0\farcs16\times0\farcs08$, PA = 0$\degr$) toward source VLA 4A (blue) and toward source VLA 4B (red), as well as toward the circumbinary material (CB) in a position centered between the two protostars (black). It is evident from the figure that there is a shift in the velocity of the centroids of the molecular lines, with the lines toward VLA 4A being blue-shifted and the lines toward VLA 4B being red-shifted with respect to the LSR velocity of the lines observed toward the central position CB. This velocity shift is fully consistent with the overall velocity gradient observed in the first-order moment shown in the top panels of Figure \ref{fig:CS}. By measuring the LSR velocity centroid of 15 lines toward VLA 4A and VLA 4B, and 20 lines toward CB, we estimate mean values of $V_{\rm LSR}$(VLA\,4A) = 7.36$\pm$0.21  km s$^{-1}$, $V_{\rm LSR}$(VLA\,4B) = 9.33$\pm$0.18  km s$^{-1}$, and $V_{\rm LSR}$(CB) = 8.77$\pm$0.16 km s$^{-1}$, where the quoted uncertainties correspond to the standard deviation of these centroids among the measured lines (see also Table~\ref{tab:3D}).

In general, the intensity of a given line is stronger toward the central (CB) position than in the rest of the molecular structure, and lines observed toward VLA 4A are stronger than toward VLA 4B. On the other hand, lines are broader toward VLA 4A than toward other positions of the molecular structure, as was already inferred from the second-order moment images shown in Figure \ref{fig:CS} (bottom panels). As can also be seen in Figure~\ref{fig:CS}, the lines are broader and more intense toward the western part of the circumbinary structure than toward the eastern one, and are narrower toward the southern part than the northern one. Additionally, the intensity of the lines in general decreases as we move farther from the center of the circumbinary disk. The strongest lines often have a double peaked profile, characteristic of absorption, which is particularly evident toward VLA 4A (see Fig.~\ref{fig:spectra}). This suggests that these lines are optically thick, but it could also be indicative of two velocity components. Lines are weaker toward VLA 4B, but the emission at the stellar position could be saturated by the intense, and likely very optically thick continuum emission, making the lines hardly detectable above the continuum.

To identify the detected transitions we compared the observed spectra with synthetic ones modeled in \href{http://cassis.irap.omp.eu}{CASSIS} \citep{Vastel2015} applying the LTE radiative transfer equations to a homogeneous column of gas along the line-of-sight, where the distribution of the population of all the rotational energy levels for a given molecule is described by a single excitation temperature, also known as the rotational temperature. To obtain an initial guess of the expected intensities of the different transitions, in our model we considered T$_{\rm ex}$ = 100 K based on previous works \citep{Belloche2020}, a source size of $0\farcs16$ (the region where the spectrum was taken), and an initial FWHM of 2 km s$^{-1}$, and changed manually the species column density and the FWHM to match the observed spectrum. We note that our goal during this procedure was the identification of the lines, not to derive parameters from the model. In a first round, we used the species in the ``Lowmass hot corino" and ``ISM" filters of CASSIS. We only included in our model transitions with upper level energies ($E_{\rm up}$) below 500 K. To identify the remaining unknown lines after this first stage, we included in the model any species with transitions close to the unidentified lines and with upper level energies up to 700 K. 

Our proposed identifications are listed in Table \ref{tab:lines}. We report 180 transitions of 29 species, counting the different isotopologues, as well as 28 unidentified lines. Among the identified species, 13 are complex organic molecules (COMs), seven of which are reported here for the first time in the SVS 13 system (see further details in Section \ref{sec:COM}).

We note that in some cases (marked as ``nLTE'' in Table \ref{tab:lines}) the intensity of the lines is not well reproduced by the model (differences $\gtrsim 50\%$), which could be due to departures from LTE or other effects such as optical depth or gradients.

\subsubsection{Distribution and Kinematics of the Circumstellar Molecular Gas}\label{sec:CSD}

We interpreted the presence of compact 0.9 mm continuum emission directly associated with VLA 4A and VLA 4B as tracing circumstellar disks of dust with sizes of $\sim$ 20 au associated with these objects (see Fig.~\ref{fig:cont750kl} and discussion in Section \ref{sec:dust2}). On the other hand, the maps presented in Figure~\ref{fig:CS} and discussed in Section \ref{sec:gas} show that both the integrated intensity (zero-order moment) and the velocity dispersion (second-order moment) of the observed molecular emission clearly peak toward VLA 4A and (less clearly, likely because of the high optical depth of the dust) tend to increase in the proximity of VLA 4B. This suggests that the two protostars are also associated with circumstellar molecular emission.

\begin{figure*}[htb]
\includegraphics[height=0.25\textheight]{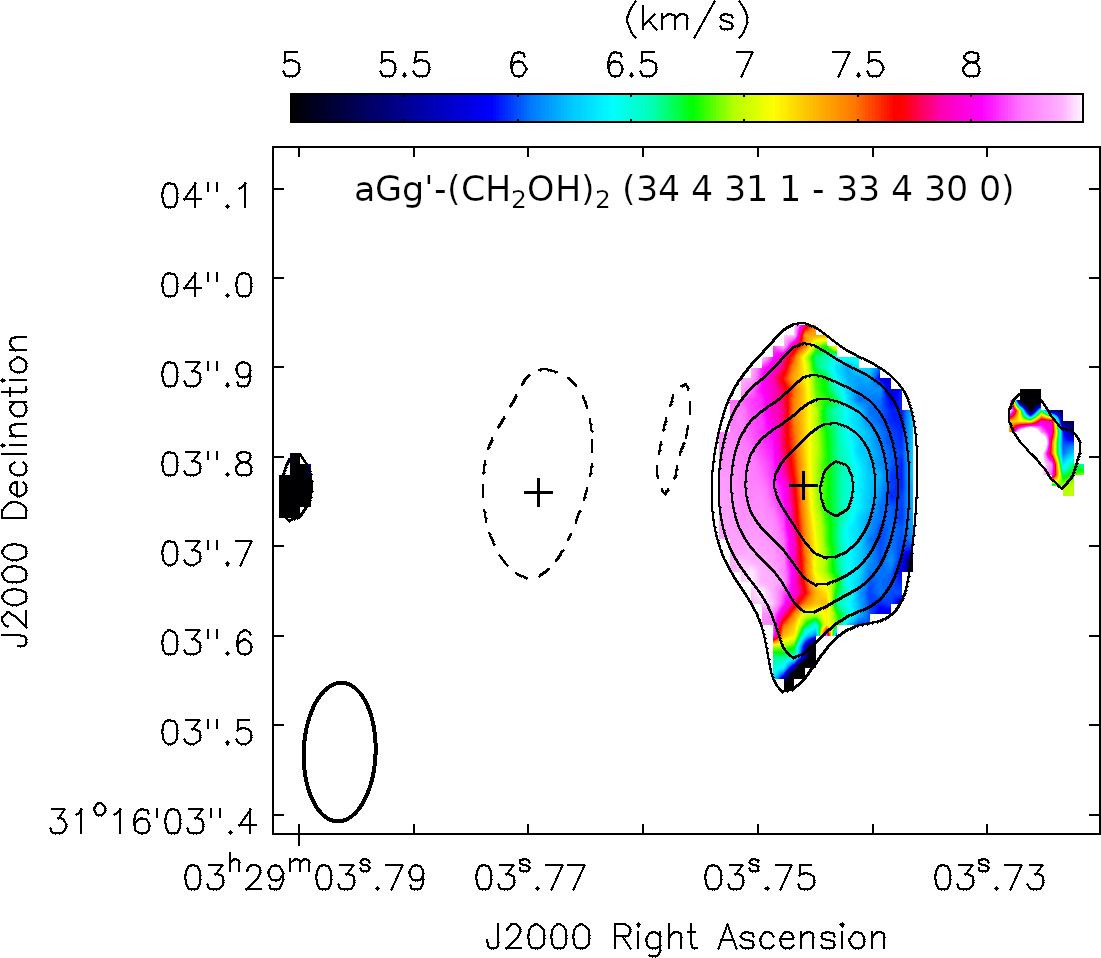}\hfill\includegraphics[height=0.25\textheight]{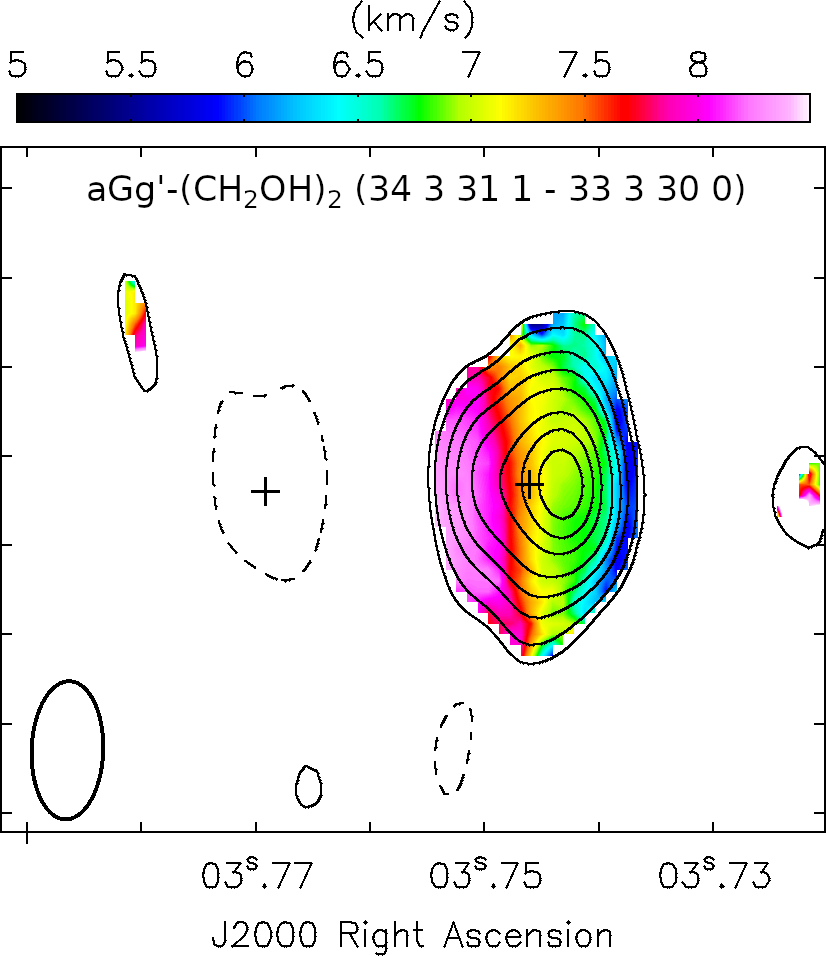}\hfill\includegraphics[height=0.25\textheight]{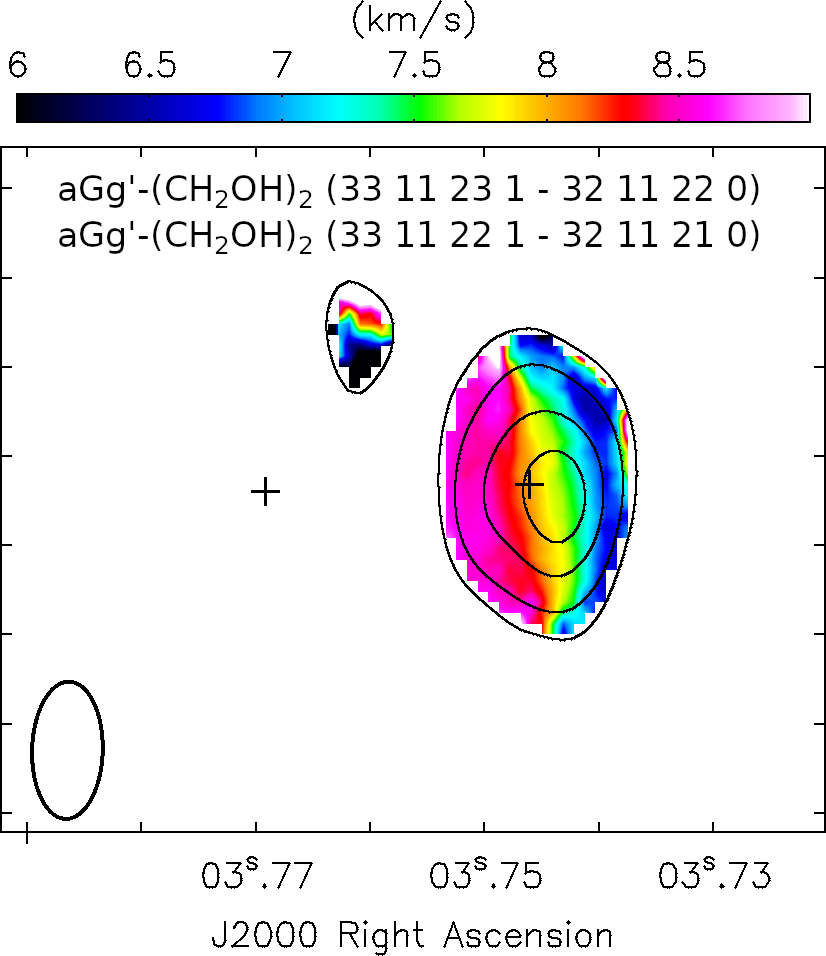}\\
\\
\hspace*{3.2em}\includegraphics[width=0.31\textwidth]{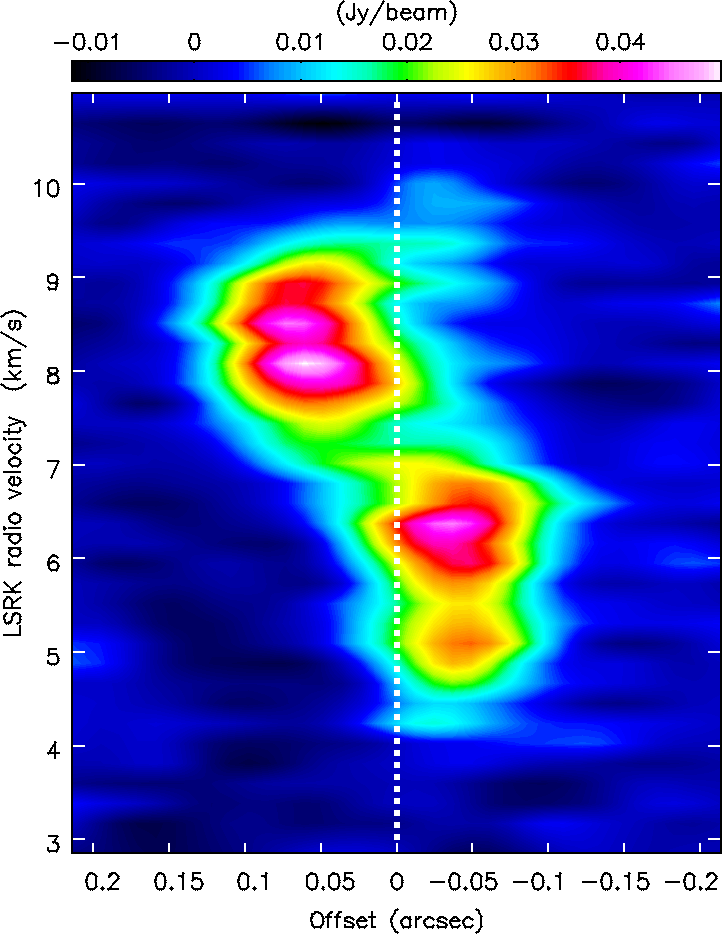}\hspace{0.5em}\includegraphics[width=0.31\textwidth]{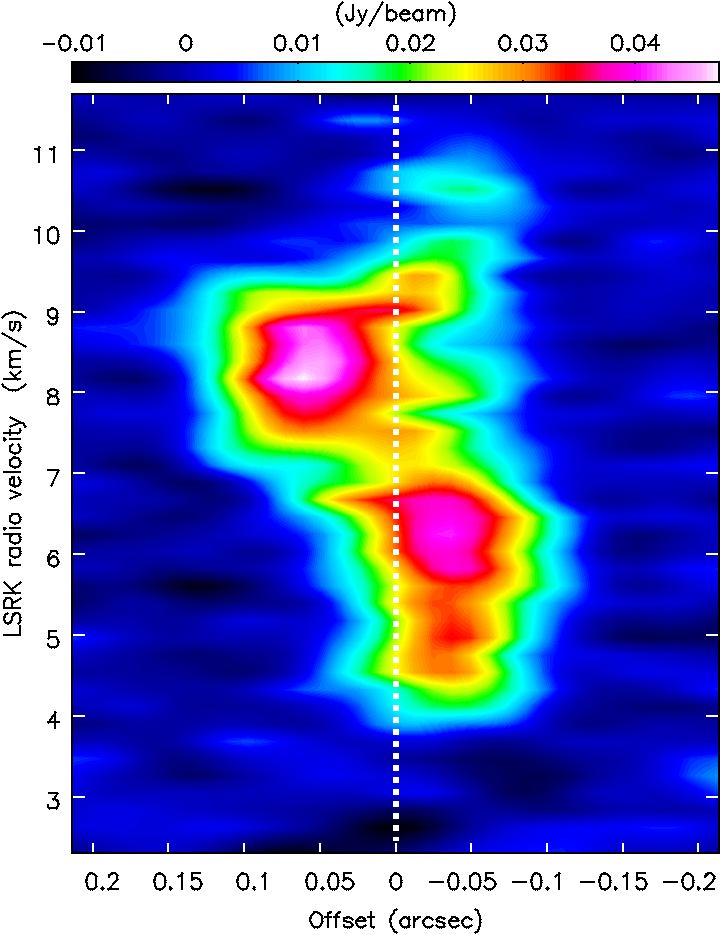}\hspace{0.5em}\includegraphics[width=0.31\textwidth]{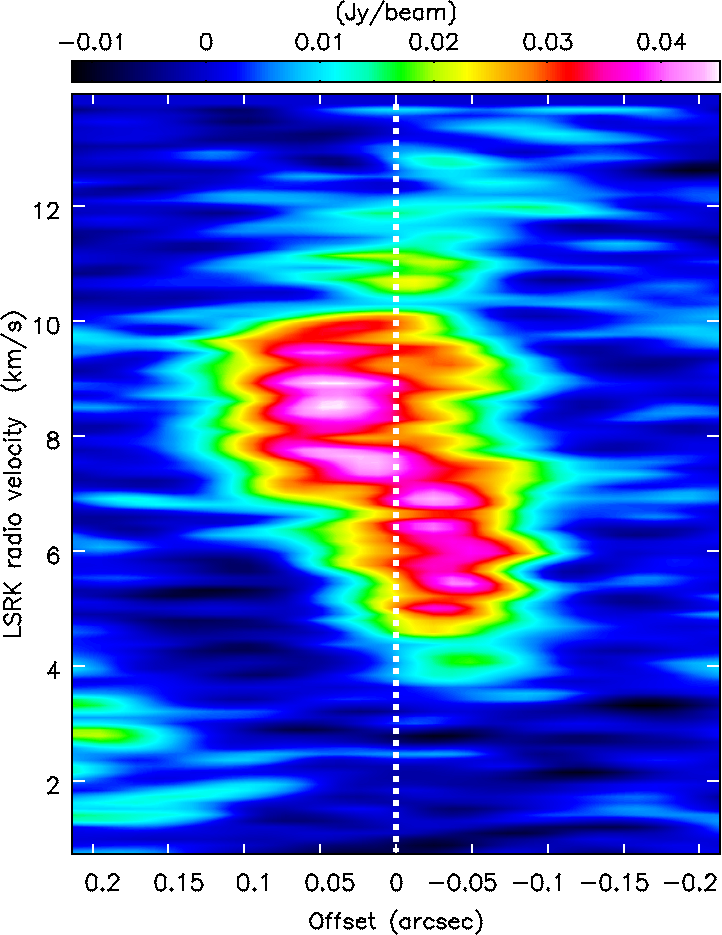}
\caption{Continuum-subtracted velocity-integrated emission (zeroth-order moment; contours) and intensity-weighted mean velocity (first-order moment; color scale in top panels) of the ethylene glycol (aGg'-(CH$_2$OH)$_2$ conformer) lines directly associated with component VLA 4A of SVS 13 that have been detected with ALMA. Contours are $-$3, 3, 5, 9, 13, 18, 25, 30, and 35 times the rms of each map (4.5 mJy beam$^{-1}$ km s$^{-1}$ for the left and middle top panels, and 15 mJy beam$^{-1}$ km s$^{-1}$ for the right top panel). Synthesized beam (indicated in the bottom left corner of the maps) is $0\farcs15\times0\farcs08$, PA = $-$2$^\circ$. The positions of VLA 4A (western) and VLA 4B (eastern) are marked with plus signs. The E-W velocity gradients are clearly observed. 
We note that there is a 3-$\sigma$ negative contour toward VLA 4B in the left and central top panels (see also Fig.~\ref{fig:A}); it is unclear whether it is related to a weak absorption signature or just the result of a somewhat imperfect continuum subtraction.
The lower panels show the corresponding position-velocity diagrams along the E-W direction centered on the position of VLA 4A. Offsets are in right ascension coordinates relative to VLA 4A. Transitions are indicated in the upper panels. Note that the rightmost panels include the emission of two blended transitions.
\label{fig:EGly}}
\end{figure*}

\begin{figure}[htb]
\begin{center}
\epsscale{01.15}
\plotone{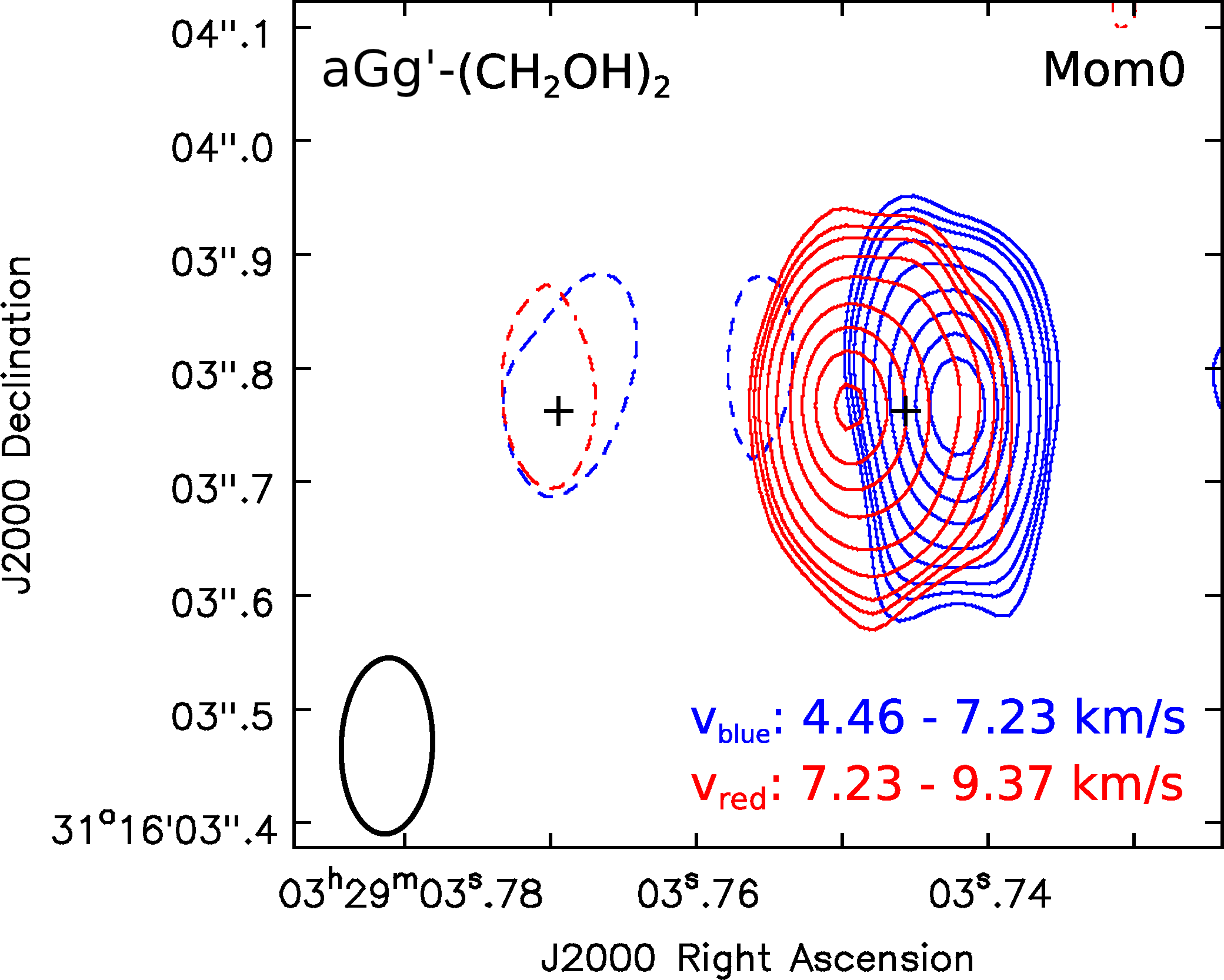}
\caption{Continuum-subtracted emission of the ethylene glycol aGg'-(CH$_2$OH)$_2$ conformer (34 4 31 1 - 33 4 30 0) transition (left panels in Fig.~\ref{fig:EGly}), 
 associated with the component VLA 4A of SVS 13. The emission has been integrated over two velocity ranges, from 4.46 to 7.23 km s$^{-1}$ (blue contours) and from 7.23 to 9.37 km s$^{-1}$ (red contours), to illustrate the two velocity components observed toward VLA 4A. Contours are $-3$, 3, 4, 5, 7, 10, 15, 20, 25, and 30 times 2.5 mJy beam$^{-1}$ km s$^{-1}$. The positions of VLA 4A (western) and VLA 4B (eastern) are marked with plus signs. The synthesized beam is indicated in the bottom left corner.
\label{fig:A}}
\end{center}
\end{figure}

This association of molecular gas at circumstellar scales is nicely illustrated by some transitions of the ethylene glycol molecule which are only detected toward VLA 4A, as Figure~\ref{fig:EGly} shows. The top panels in Figure~\ref{fig:EGly} show the integrated emission of different transitions of ethylene glycol which is clearly associated with VLA 4A at circumstellar scales of $\sim$30 au. These panels also show the presence of a clear velocity gradient in the E-W direction which is consistently detected in all these transitions (note that the rightmost panels include two blended transitions). The bottom panels in Figure~\ref{fig:EGly} show position-velocity cuts along a PA = $90^\circ$ to better unveil the structure of this gradient, which is suggestive of rotation motions associated with VLA 4A. Figure~\ref{fig:A} further illustrates the distribution of the blue- and red-shifted emission, integrated over two separate velocity ranges, of the transition displayed in the left panel of Figure~\ref{fig:EGly}.

\subsection{Physical Parameters of the SVS 13 System}\label{sec:phys}

\subsubsection{Temperature}\label{sec:temp}

We can use the detected molecular transitions to estimate the kinetic temperature by inferring the excitation temperature of the rotational levels, also known as the rotational temperature, $T_{\rm rot}$. For this, we used CASSIS to build rotational diagrams to derive the temperature and the column density, assuming optically thin lines and LTE. In this case, the upper-level column density

\begin{equation}
N_u = \frac{8\pi k\nu^2}{h c^3 A_{ul}}\int{T_{\rm mb}\,dV},
\end{equation}
and the rotational temperature $T_{\rm rot}$ are related as
\begin{equation} \label{eq:rotational}
\ln{\frac{N_u}{g_u}}= \ln{N_{\rm mol}}-\ln{Q\left(T_{\rm rot}\right)}-\frac{E_u}{k T_{\rm rot}},
\end{equation}
 where $k$ is Boltzmann's constant, $h$ is Planck's constant, $c$ is the speed of light, $\nu$ is the frequency of the transition and $A_{ul}$ its Einstein coefficient for spontaneous emission, $g_u$ is the statistical weight of the upper level and $E_u$ its energy, $N_{\rm mol}$ is the total column density of the molecule, $Q(T_{\rm rot})$ is the partition function for $T_{\rm rot}$, and $\int{T_{\rm mb}\,dV}$ is the integrated main beam brightness temperature of the line, which is derived from a Gaussian fit to the observed molecular line.

We use only molecular lines that have been positively identified and well reproduced by the LTE model, with energies that span over a relatively large range, and that are non-blended, or, in case they are blended, that can be effectively isolated by a multi-component Gaussian fitting taking into account all the blended lines involved. These selection criteria allow us to populate the rotational diagrams only for a few molecules (CH$_3$OCHO, CH$_2$DOH, and aGg'-(CH$_2$OH)$_2$); in the case of aGg'-(CH$_2$OH)$_2$ this is possible only toward VLA 4A.

\begin{figure*}[htb]
\gridline{\fig{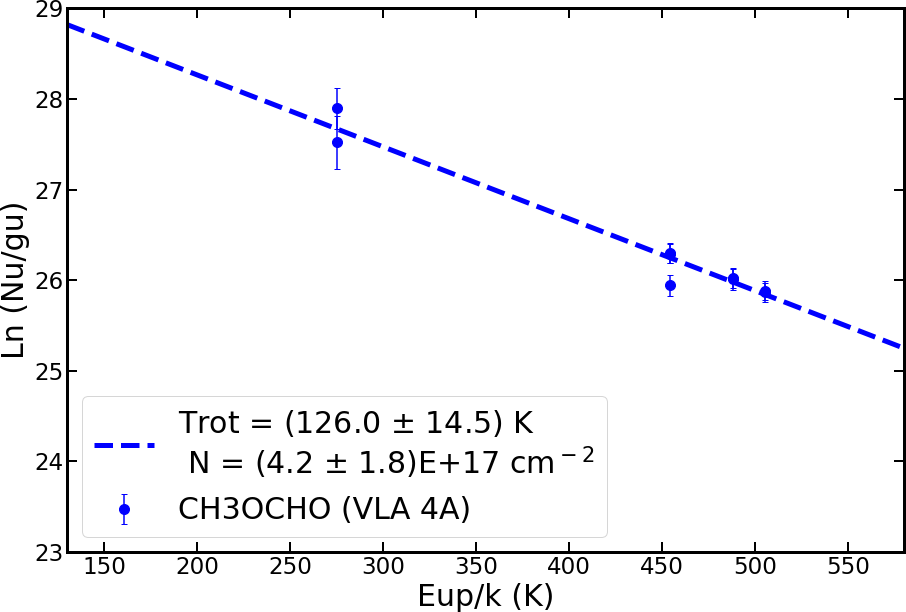}{0.32\textwidth}{(a)}
\hfill\fig{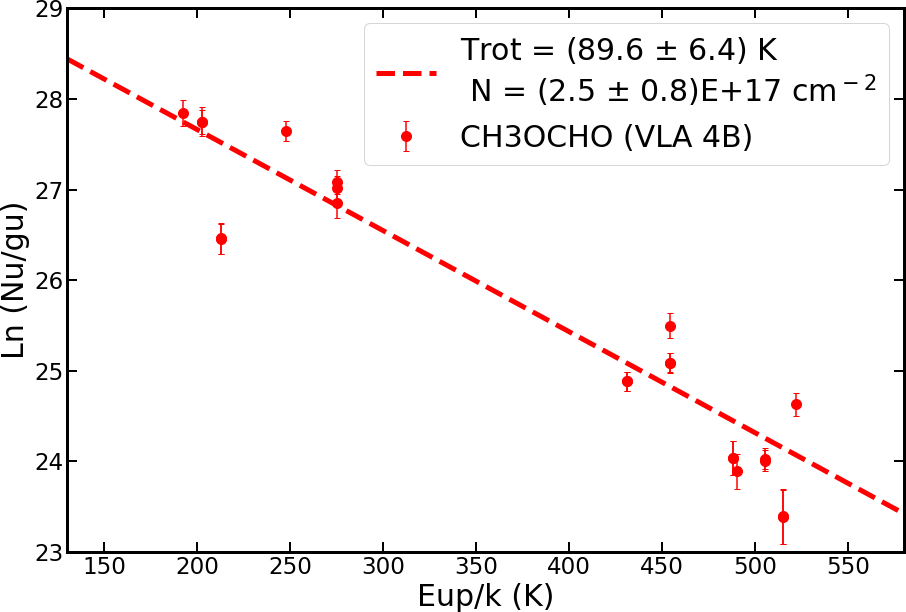}{0.32\textwidth}{(b)}
\hfill\fig{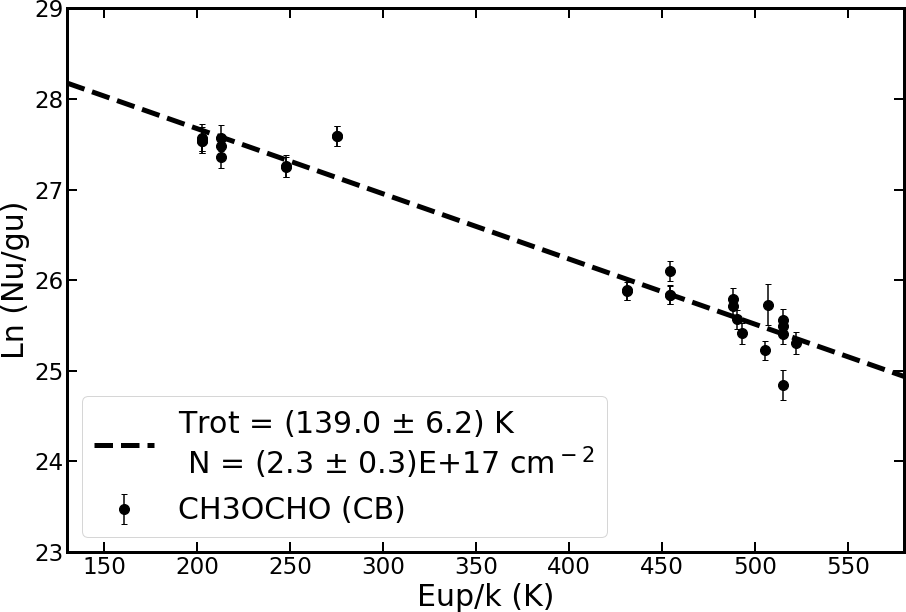}{0.32\textwidth}{(c)}}
\gridline{\fig{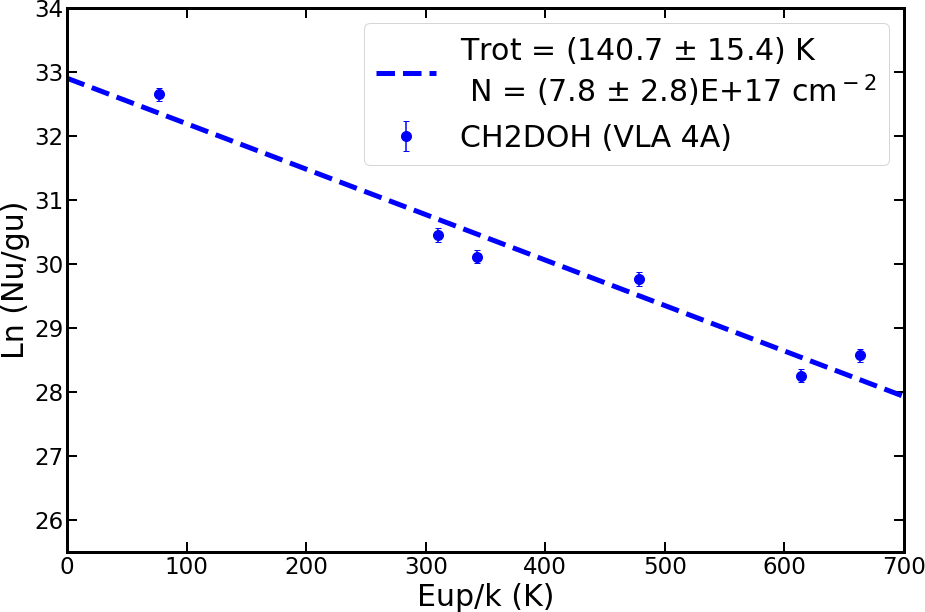}{0.32\textwidth}{(d)}
\hfill\fig{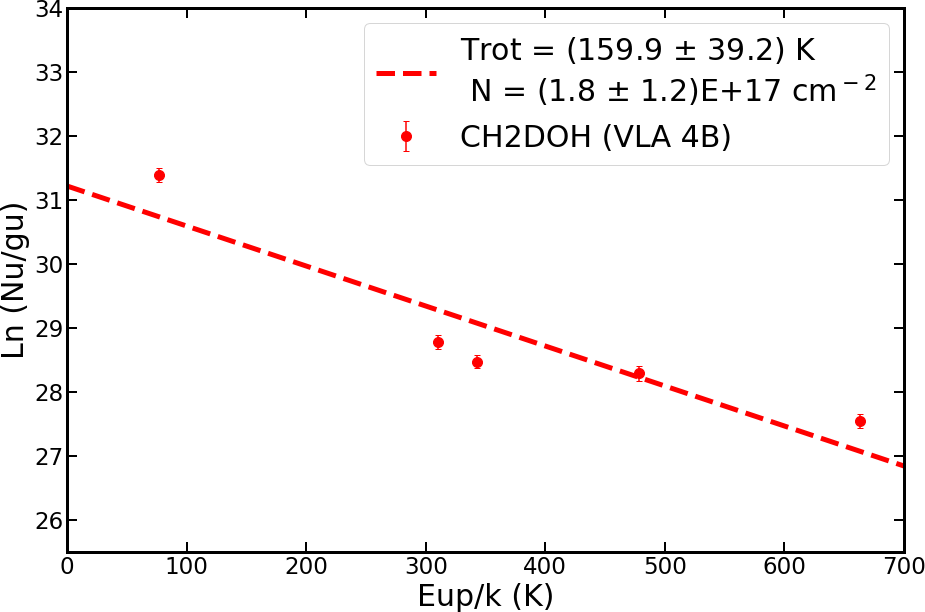}{0.32\textwidth}{(e)}
\hfill\fig{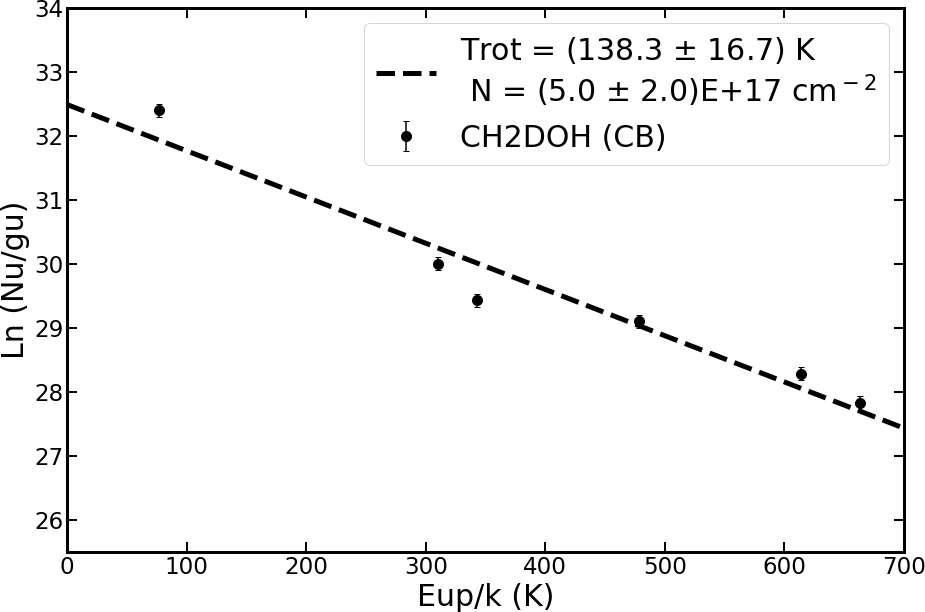}{0.32\textwidth}{(f)}}
\gridline{\fig{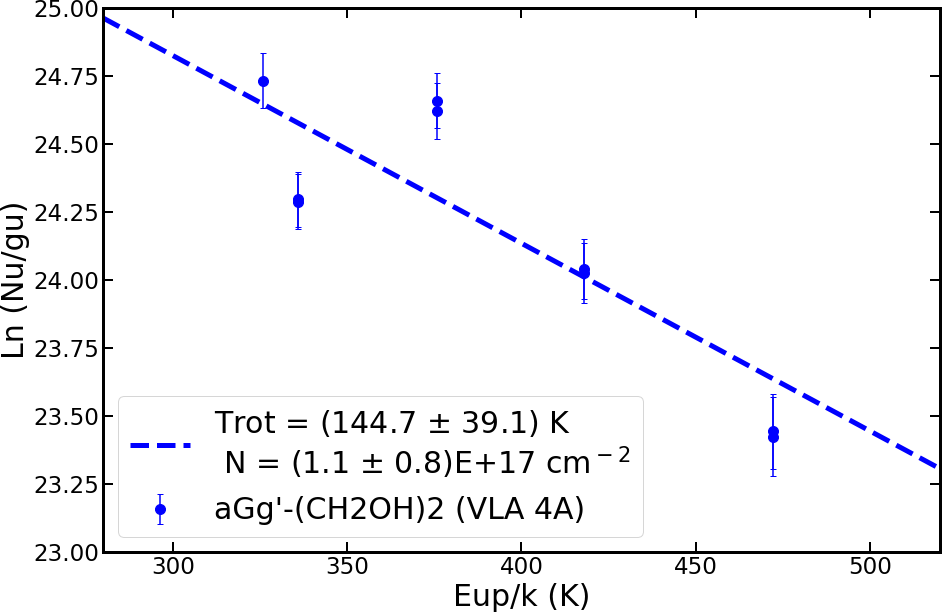}{0.32\textwidth}{(g)}\hspace{122mm}}
\caption{Rotational diagrams for methyl formate (CH$_3$OCHO), deuterated methanol (CH$_2$DOH) and ethylene glycol (aGg'-(CH$_2$OH)$_2$), for VLA 4A (blue), VLA 4B (red), and the circumbinary disk (black), using spectra taken in beam-size regions ($0\farcs16\times0\farcs08$, PA = 0$\degr$) toward each component and toward the geometric center of the system (shown in Fig. \ref{fig:spectra}). We only include in the plot lines that have been positively identified and well reproduced by the LTE model, that are non-blended, or, in case they are blended, that we were able to isolate by a multi-component Gaussian fit. In the case of aGg'-(CH$_2$OH)$_2$ this is possible only toward VLA 4A.
\label{fig:trot}}
\end{figure*}

\begin{deluxetable*}{lcc@{\extracolsep{6pt}}cc@{\extracolsep{6pt}}cccccc}[htb]
\tabletypesize{\small}
\tablewidth{0pt}
\tablecaption{Rotational Temperatures, Molecular Column Densities and Abundances \label{tab:trot}}
\setlength{\tabcolsep}{1.6pt}
\tablehead{
 & \multicolumn3c{VLA 4A} & \multicolumn3c{VLA 4B} & \multicolumn4c{CB\tablenotemark{a}}\\
\cline{2-4} \cline{5-7} \cline{8-11}
 & $T_{\rm rot}$\tablenotemark{b} & $N_{\rm mol}$\tablenotemark{b} & 
 & $T_{\rm rot}$\tablenotemark{b} & $N_{\rm mol}$\tablenotemark{b} & 
 & $T_{\rm rot}$\tablenotemark{b} & $N_{\rm mol}$\tablenotemark{b} &  & $X_{\rm H_2}$\tablenotemark{d} \\
Molecule 
& (K) & ($10^{17}$\,cm$^{-2}$) & $X_{\rm CH_3OH}$\tablenotemark{c}
& (K) & ($10^{17}$\,cm$^{-2}$) & $X_{\rm CH_3OH}$\tablenotemark{c}
& (K) & ($10^{17}$\,cm$^{-2}$) & $X_{\rm CH_3OH}$\tablenotemark{c} & ($\times$$10^{-7}$)}
\startdata
CH$_3$OCHO          & 127$\pm$15 & 4.2$\pm$1.8 & 2.6$\times$$10^{-2}$ &
                       90$\pm$6  & 2.5$\pm$0.8 & 6.9$\times$$10^{-2}$ &
                      139$\pm$6  & 2.3$\pm$0.3 & 2.3$\times$$10^{-2}$ & 1.3\\
aGg'-(CH$_2$OH)$_2$ & 145$\pm$39 & 1.1$\pm$0.8 & 7.1$\times$$10^{-3}$ &
                                      \nodata & \nodata & \nodata &
                                      \nodata & \nodata & \nodata \\
CH$_2$DOH           & 141$\pm$15 & 7.8$\pm$2.8 & \nodata &
                      160$\pm$39 & 1.8$\pm$1.2 & \nodata &
                      138$\pm$17 & 5.0$\pm$2.0\tablenotemark{e} & 5$\times$$10^{-2}$ & 2.7 \\
$^{13}$CH$_3$OH &  \nodata & \nodata & \nodata &
                   \nodata & \nodata & \nodata &
                   \nodata & 0.8\tablenotemark{f} & 8$\times$$10^{-3}$ & 0.44 \\
CH$_3$$^{18}$OH &  \nodata & \nodata & \nodata &
                   \nodata & \nodata & \nodata &
                   \nodata & 0.3\tablenotemark{f} & 3$\times$$10^{-3}$ & 0.17 \\
CH$_3$OH        &  \nodata & \nodata & \nodata &
                   \nodata & \nodata & \nodata &
                   \nodata & 100\tablenotemark{e,g} & 1 & 56 \\
\enddata
\tablenotetext{a}{Central position, between VLA 4A and VLA 4B, taken as representative of the circumbinary material.}
\tablenotetext{b}{Rotational temperature and molecular column density obtained from the rotational diagram fits shown in Fig.~\ref{fig:trot}, except when indicated.}
\tablenotetext{c}{Molecular abundance relative to CH$_3$OH obtained from the CH$_2$DOH column density, assuming a deuteration fraction D/H = 0.05, estimated as the ratio of the CH$_2$DOH and CH$_3$OH column densities toward CB (see notes e, g).}
\tablenotetext{d}{Molecular abundance relative to H$_2$. The adopted H$_2$ column density is 1.8$\times$10$^{24}$ cm$^{-2}$, calculated from the dust emission, assuming optically thin emission, a dust temperature of 140 K, and a gas-to-dust ratio of 100. See text for further details.}
\tablenotetext{e}{From the CH$_2$DOH and CH$_3$OH column density ratio a deuteration fraction D/H = 0.05 is estimated, which is used for the calculation of the molecular abundances relative to methanol.}
\tablenotetext{f}{Obtained from the observed spectra assuming $T_{\rm rot}$ = 140 K.} 
\tablenotetext{g}{The CH$_3$OH lines are optically thick and saturated, and therefore unsuitable for deriving the column density. The value listed in the table is obtained as the geometrical mean of the CH$_3$OH column densities of 0.6$\times$10$^{19}$ cm$^{-2}$ and 1.7$\times$10$^{19}$ cm$^{-2}$, derived from the $^{13}$CH$_3$OH and CH$_3$$^{18}$OH isotopologues, assuming isotopic ratios of $^{12}$C/$^{13}$C = 77 and $^{16}$O/$^{18}$O = 560 \citep{Wilson1994}, respectively.}
\end{deluxetable*}

In Figure \ref{fig:trot} we show the rotational diagrams toward the positions of VLA 4A, VLA 4B, and toward the geometric center of the system (CB, that we adopt as representative of the circumbinary disk). In Table \ref{tab:trot} we report the rotational temperatures and column densities derived from these diagrams. From CH$_3$OCHO and CH$_2$DOH we obtain similar temperatures of $\sim$140 K toward both VLA 4A and the central position CB. From ethylene glycol, aGg'-(CH$_2$OH)$_2$, a molecule emitting in a small region of $\sim$$0\farcs2$ ($\sim$60 au) in size around VLA 4A, we find a slightly higher value, $T_{\rm rot} \simeq 145$ K, toward this position. Therefore, we conclude that there are no substantial differences in the rotational temperatures toward the CB and the VLA 4A positions, measured at scales of the order of the beam size ($\sim$48 au $\times$ 24 au), which is the size of the region where the spectra used in the $T_{\rm rot}$ analysis have been obtained. 

On the other hand, our two temperature measurements toward VLA 4B give values of 90$\pm$6 K and 160$\pm$39 K. Since the observed molecular emission toward this position could be more affected by dust opacity (see Sects.~\ref{sec:gas} and \ref{sec:mol}), we consider these values of $T_{\rm rot}$ toward VLA 4B as less reliable.  As noted in Section \ref{sec:dust2}, the brightness temperature of VLA 4B
inferred from our highest angular resolution continuum image (Fig.~\ref{fig:cont750kl} and Table \ref{tab:disks}) is $\sim$450 K, suggesting a possible temperature increase at distances much closer to the star ($r$ $\la$ 10 au), compared to the scales tested in the $T_{\rm rot}$ analysis. In fact, we expect molecular emission to be unobservable above the continuum toward the compact region of high optical depth in VLA 4B, identified in the highest angular resolution continuum images (baselines $>$750 k$\lambda$; Fig.~\ref{fig:cont750kl}), so we expect $T_{\rm rot}$ to be insensitive to the region with the brightest and hottest circumstellar dust. We will adopt 140 K as a representative value for the kinetic temperature, at scales of the order of the beam size ($\sim$25-50 au), throughout the source but with the caveat that in the circumstellar disks, at very small radii, dust can reach higher temperatures.

\citet{Belloche2020}, from a large number of molecular transitions, but with data at lower frequencies and poorer angular resolution ($\sim0\farcs33-1\farcs75$) that does not separate the binary well, obtained a wide range of rotational temperatures, from $\sim$30 K to $\sim$350 K. However, when considering only the values with signal-to-noise ratio higher than 3.5$\sigma$ their temperatures range from $\sim$100 to $\sim$200 K, which are in better agreement with ours. For the two molecules we have in common, CH$_3$OCHO and aGg'-(CH$_2$OH)$_2$, the temperatures obtained by \citet{Belloche2020} are $\sim$100 K, which are somewhat lower than our values.

The molecular column density of CH$_3$OCHO obtained by these authors is $2.1\times 10^{17}$ cm$^{-2}$, which is a factor $\sim$2 lower than the value we find toward VLA 4A, but similar to the column density we find in the rest of the source (Table \ref{tab:trot}). For aGg'-(CH$_2$OH)$_2$, which we only detect toward the position of VLA 4A, \citet{Belloche2020} found a column density of $1.3\times 10^{16}$ cm$^{-2}$, which is about one order of magnitude lower than the value we obtain. We attribute these discrepancies to the non-uniform distribution of this molecular emission, as revealed by our higher angular resolution data, which is smeared within the larger beam of the \citet{Belloche2020} observations, resulting in lower averaged values. We point out that the rotational diagrams presented in \citet{Belloche2020} have an error in the labeling of the y-axis units (see \citealt{Belloche2020-err}). Our diagrams are consistent with those after the proper correction is performed. The rotational temperatures obtained from these diagrams, and the column densities reported by these authors, which were obtained from radiative transfer models, are not affected by the error.

\subsubsection{Disk Masses from Dust}\label{sec:masses}

Once we have an estimate of the temperature, we can derive the masses of the circumstellar and circumbinary disks of SVS 13 from the observed 0.9 mm continuum emission, which traces thermal dust emission (since the expected free-free contribution at this short wavelength is negligible). Given the uncertainties involved and the fact that we do not observe large variations of rotational temperature from one position to another across the circumbinary structure (Sect.~\ref{sec:temp}), we will assume a single value of 140 K for the dust temperature in the circumbinary disk.

For the circumstellar disks, the highest angular resolution 0.9 mm continuum results discussed in Section~\ref{sec:dust2} suggest dust temperatures of $\sim$300 K for VLA 4A and of $\sim$450 K for VLA 4B, at scales $\la$10 au (Table~\ref{tab:disks}). The molecular data are consistent with rotational temperatures of $\sim$140 K toward the positions of both VLA 4A and VLA 4B, at beam scales of 25-50 au. So, in the calculation of dust masses we will consider a range of dust temperatures from 140 K to 300 K (VLA 4A) or from 140 K to 450 K (VLA 4B).

In addition, as noted in Section \ref{sec:dust2}, the observed 0.9 mm dust brightness temperatures indicate that this dust emission is optically thin in the circumbinary disk, but not in the circumstellar disks, particularly in VLA 4B where the dust emission can be very optically thick.

For optically thin emission from an isothermal source, the observed flux density is directly related to the total mass of emitting material as:
 \begin{equation} \label{eq:mass-dust}
M = \frac{S_\nu D^2}{\kappa_\nu B_\nu(T_{\rm dust})},
 \end{equation}
 where $D$ is the distance to the source, $S_\nu$ is the flux density, $\kappa_\nu$ is the dust opacity at the observed frequency $\nu$, $B_\nu(T_{\rm dust})$ is the Planck function, and $T_{\rm dust}$ is the dust temperature. Therefore, we can use eq.~(\ref{eq:mass-dust}) to estimate the mass of circumbinary material from the observed flux density at 0.9 mm.

If the dust emission is not optically thin, as we suspect is the case for the circumstellar disks, then eq.~(\ref{eq:mass-dust}) gives a lower limit of the mass. A suitable upper limit to the circumstellar masses can be obtained from the 7 mm emission, which is expected to be optically thin and dominated by the dust emission (see Sect.~\ref{sec:vla2}). We take this as an upper limit because of a possible residual free-free contamination. Therefore, using the 0.9 mm and 7 mm data we can constrain the masses of the circumstellar disks.

We assume a gas-to-dust ratio of 100 \citep{Bohlin1978} and adopt a dust opacity $\kappa_\nu$(0.9 mm) = \makebox{1.65 cm$^2$ g$^{-1}$} (per unit mass of dust). This value of $\kappa_\nu$(0.9 mm) was obtained by interpolating the values at 0.7 mm and 1 mm from \citet{Ossenkopf1994} and should be adequate for protostellar objects where the dust grain growing is still incipient. These dust opacities have been obtained by including thin ice mantles in an MRN model \citep{Mathis1977}, consisting of a mixture of spherical silicate and graphite grains with separated power-law particle size distributions.

Since the longest wavelength in the \citet{Ossenkopf1994} opacities is 1.3 mm, we need to extrapolate the opacities up to 7 mm. For this, we assume a power-law dependence of the dust opacity with frequency ($\kappa_\nu\propto\nu^\beta$, with 0$<$$\beta$$<$2), where $\beta$ is related to the dust grain-size distribution and composition. In the case of optically thin dust emission, $\beta$ can be inferred from the observed SED as $\beta = \alpha - 2$, where $\alpha$ is the observed spectral index ($S_\nu\propto\nu^\alpha$). Assuming that the overall emission of the circumbinary dust of SVS 13 (green diamonds in Fig.~\ref{fig:SED}) is optically thin, we can use the spectral index obtained from this SED ($\alpha\simeq2.5$) to infer a value of $\beta$ = 0.5 for the circumbinary dust. By further assuming that $\beta$ has the same value in the circumstellar and circumbinary material, we can use $\beta$ = 0.5 to extrapolate the \citet{Ossenkopf1994} opacity $\kappa_\nu$(1.3 mm) = 0.899 cm$^2$ g$^{-1}$ to \makebox{$\kappa_\nu$(7 mm) = 0.387 cm$^2$ g$^{-1}$} (per unit mass of dust).  

Using eq.~(\ref{eq:mass-dust}), a dust temperature of $\sim$140 K (from the rotational diagrams presented in Section \ref{sec:temp}), and a flux density at 0.9 mm of 909 mJy, obtained by subtracting the estimated flux densities of VLA 4A and VLA 4B from the total flux density given in Table~\ref{tab:fluxSVS13}, we obtain a total (dust+gas) mass for the circumbinary disk $M_{\rm disk}$(CB) $\simeq$ 0.052 $M_\odot$.

For the circumstellar disks, assuming a range of dust temperatures 140-300 K (VLA 4A) and 140-450 K (VLA 4B), we obtain strict lower limits of the disk masses of $M_{\rm disk}$(VLA\,4A) $\simeq$ 0.002-0.005 $M_\odot$ and $M_{\rm disk}$(VLA\,4B) $\simeq$ 0.002-0.006 $M_\odot$, from the likely optically thick 0.9 mm emission.  From the optically thin 7 mm emission, we obtain strict upper limits of $M_{\rm disk}$(VLA\,4A) $\simeq$ 0.004-0.009 $M_\odot$ and $M_{\rm disk}$(VLA\,4B) $\simeq$ 0.009-0.030 $M_\odot$. The masses obtained from the 7 mm emission are higher by a factor of $\sim$2 for VLA 4A (partially overlapping with the 0.9 mm mass range) and $\sim$5 for VLA 4B, strongly suggesting very optically thick emission in this last case. Given that the very high optical depth of the 0.9 mm emission would likely affect the mass calculations more than the possible free-free contamination at 7 mm (which is expected to be small, as discussed in Sect.~\ref{sec:vla2}), in the following we will adopt the range of circumstellar disk masses obtained from the 7 mm data, $\sim$0.004-0.009 $M_\odot$ for VLA 4A and $\sim$0.009-0.030 $M_\odot$ for VLA 4B, as representative of the true values.

\subsubsection{Dynamical Masses of the Stars}\label{sec:dyn}

A powerful tool to obtain the stellar masses is from the measured 3D velocities. For a gravitationally bound system of two objects in orbital motion, the orbital period is given by:
 \begin{equation} \label{eq:period}
\left(\frac{P}{\rm yr}\right) = \left(\frac{a}{\rm au}\right)^{3/2} \left(\frac{M_{\rm tot}}{M_\odot}\right)^{-1/2},
 \end{equation}
 where $M_{\rm tot}$ is the total mass of the system and $a$ is the separation between the two components.

Equivalently,
 \begin{equation} \label{eq:mass-stars}
\left(\frac{M_{\rm tot}}{M_\odot}\right) = 1.12\times10^{-3} \left(\frac{a}{\rm au}\right) \left(\frac{V_{\rm rel}}{\rm km\:s^{-1}}\right)^2,
 \end{equation}
 where $V_{\rm rel}$ is the relative velocity between the two objects.

 From the observed proper motions in right ascension and declination (Sect.~\ref{sec:pm}) and the line-of-sight velocities inferred from the molecular lines (Sect.~\ref{sec:mol}), we can obtain the 3D relative velocity between the VLA 4A and VLA 4B protostars, which is given in Table~\ref{tab:3D}. This velocity is $\vec{V}_{\rm rel}$ = (+0.48$\pm$1.02 km\,s$^{-1}$, $-$2.28$\pm$0.82 km\,s$^{-1}$, +1.97$\pm$0.28 km\,s$^{-1}$).

By adding quadratically the three velocity components we obtain $V_{\rm rel}$ = 3.06$\pm$0.65 km s$^{-1}$. Then, taking into account that the separation is $a$ = 90 au (actually, we measure the projected separation but, since the two stars are aligned in a direction close to the observed major axis of the circumbinary structure, we do not expect an important difference between the real and the projected separation), from eq.~(\ref{eq:mass-stars}) we obtain a total mass of the system of 1.0$\pm$0.4 $M_\odot$, and, from eq.~(\ref{eq:period}), an orbital period of 850 yr. Also, the inclination angle of the binary orbital plane with respect to the plane of the sky can be calculated as the arctan of the ratio between the line-of-sight (LSR) and the plane-of-the-sky (proper motion) velocity components, resulting $i$ = $41.7^\circ\pm9.5^\circ$ (see Appendix~\ref{sec:errors} for the uncertainty calculation).

We note that the total mass of the stars is much higher than those of the circumstellar and circumbinary disks estimated from the dust emission in the previous Section~\ref{sec:masses} (the total mass of the disks is $\sim$6-9\% of the total stellar mass). Therefore, it is reasonable to neglect these disks in the calculation of the masses of the stars from the orbital parameters. 

In principle, one could obtain a dynamical estimate of the total mass (stars+disks) from the observed gradients in the first-order moment maps (Fig. \ref{fig:CS}) of the circumbinary molecular emission. However, the likely high optical depth of the brighter molecular transitions, the complex overall structure and kinematics of the source, with two central objects, spiral arms, and other asymmetries prevent us from attempting this calculation. Nevertheless, we still can obtain interesting results for the individual objects, if we restrict our calculation to some specific molecular lines and the regions that are less perturbed by asymmetries.

First, we discuss the implications of the kinematics of the four ethylene glycol transitions that are only detected toward VLA 4A (Figs.~\ref{fig:EGly} and \ref{fig:A}). The first-order moments and position-velocity diagrams in Figure \ref{fig:EGly} show the presence of a clear velocity gradient in the E-W direction and centered on VLA 4A, which is consistently detected in all these transitions (note that the rightmost panels include two blended transitions). This gradient is suggestive of rotation motions associated with VLA 4A, with a rotation axis projected on the plane of the sky at PA $\simeq$ 0$^\circ$. This is consistent with all rotation axes having a similar position angle, PA $\simeq$ 0$^\circ$, as suggested by the fact that the major axes of dust and molecular disks (observed in projection on the plane of the sky) all have a PA $\simeq$ 90$^\circ$, as inferred in Sections \ref{sec:dust2}, \ref{sec:gas}, and \ref{sec:CSD}. As can be seen in Figures \ref{fig:EGly} and \ref{fig:A}, toward VLA 4A there are two sharply differentiated velocity components that we interpret may be tracing the receding (eastern) and approaching (western) sides of a circumstellar rotating disk around VLA 4A. 

Thus, assuming a rotating disk around the VLA 4A protostar, from Figures~\ref{fig:EGly} and \ref{fig:A} we infer a line-of-sight projected rotation velocity of $\sim$1.5 km s$^{-1}$ at a radius of $\sim$15 au, resulting in a mass $M_*$ = 0.04 $\sin^{-2} i~M_\odot$ for VLA 4A. Unfortunately, the value of this mass depends strongly on the inclination angle, which is not known. Assuming coplanarity between the circumstellar disk and the binary orbit ($i$ $\simeq$ 40$^\circ$) we obtain $M_*$ = 0.09 $M_\odot$, but this value of the mass increases significantly for smaller inclination angles (e.g., $M_*$ = 0.2 $M_\odot$ for $i$ = 25$^\circ$).

Alternatively, since the characteristic shape of a Keplerian velocity profile is not clearly seen with the angular resolution of our observations, we considered the possibility that the two velocity components of the ethylene glycol transitions observed toward VLA 4A (Figs.~\ref{fig:EGly} and \ref{fig:A}) are tracing two different objects orbiting one around each other.
In this case, VLA 4A itself would be a binary with a difference of $\sim$3 km s$^{-1}$ in the line-of-sight velocity of the two subcomponents and a projected separation $a\simeq30$ au (Fig. \ref{fig:EGly}). Using eq.~(\ref{eq:mass-stars}) and assuming an orbit coplanar with that of the main SVS 13 binary ($i=40^\circ$), we obtain a total mass of $\sim$0.73 $M_\odot$ for VLA 4A. Furthermore, the presence of dust toward VLA 4A and the relatively wide velocity range ($\sim$2.5 km s$^{-1}$) of each subcomponent in the position-velocity diagrams suggest that each of them could be associated, in its turn, with an angularly unresolved ($\la 0\farcs1$ in size) disk, implying an individual mass of each subcomponent of $\la$0.064 $M_\odot$, if all the orbits were coplanar. Thus, the total mass of VLA 4A obtained from the dynamics of the disk of each subcomponent separately ($\la$0.13 $M_\odot$) would be significantly smaller than the mass inferred from their overall orbital motion ($\sim$0.73 $M_\odot$). A misalignment of $\sim$25$^\circ$ between this orbital plane and the circumstellar disks of the potential VLA 4A binary would be required to reach an agreement in the total mass, making this binary scenario for VLA 4A somewhat speculative.

In order to obtain more accurate values of the mass and an estimate of additional parameters such as the inclination angle, in Appendix \ref{sec:keplerian} we perform a fit of the observed channel maps of the ethylene glycol transitions to a Keplerian thin disk model (see \citealt{Zapata2019} for further details). The best-fitted physical and geometrical parameters of the Keplerian disk surrounding VLA 4A are presented in Tables \ref{tab:4A_fit} and \ref{tab:4A_deriv}. The best fit is obtained for a disk model with an inner radius of $\sim$2~au, an outer radius of 33$\pm$2 au, an inclination of 22$^\circ$$\pm$7$^\circ$, and a Keplerian rotation velocity of 2.7 km s$^{-1}$ at a reference radius of 30 au, implying a central mass of 0.25$\pm$0.15 $M_\odot$. It is remarkable that the three fits to  different ethylene glycol transitions (one of them consisting of two blended lines) consistently give very similar results (see Tables \ref{tab:4A_fit} and \ref{tab:4A_deriv}).

We note, however, that the disk inclination inferred from the gas rotation in VLA 4A is significantly smaller than the inclination inferred from the aspect ratio of dust continuum sizes obtained from a Gaussian fit (Sect.~\ref{sec:dust2}). We consider the former to be more reliable, as it is obtained from a model fitting over several spectral channels, but it is still very uncertain. Since the beams are elongated in the direction of the minor axes of the disks, the actual uncertainties on these axes, and on the inclinations derived from them, are larger than the formal fitting errors for both the dust and the line images.

\begin{figure*}[hbt]
\begin{center}
\epsscale{01.15}
\plotone{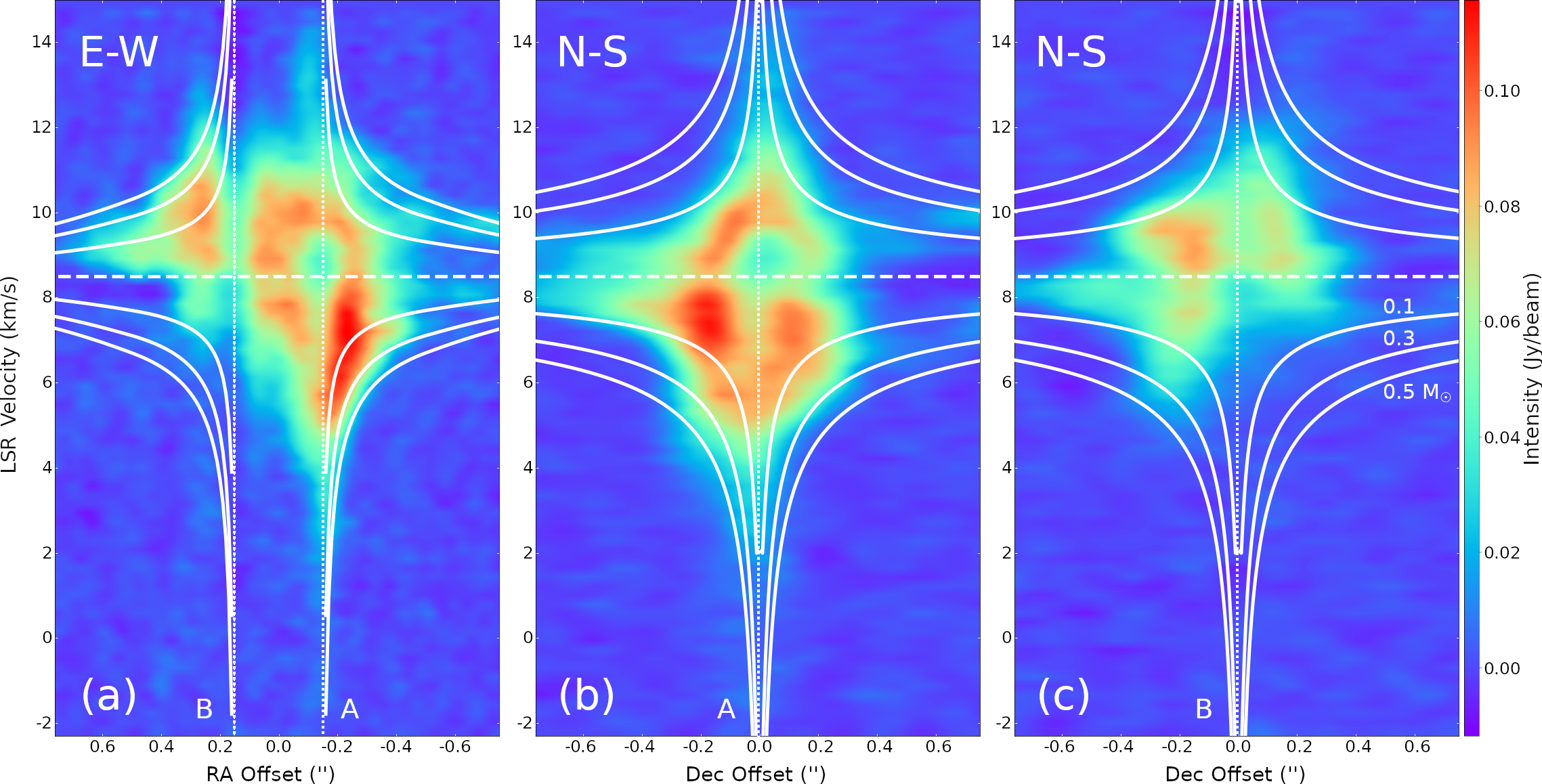}
\caption{Observed position-velocity diagrams of the continuum-subtracted CS(7-6) line emission. (a) Along the E-W direction (PA = 90$^\circ$) toward the central position (CB) midway between VLA 4A and VLA 4B. Offsets are in increasing right ascension relative to this central position. (b) Along a N-S direction (PA = 0$^\circ$), toward VLA 4A, with offsets in increasing declination relative to this position. (c) Same as (b) but for VLA 4B. The dashed line indicates the systemic velocity of the cloud (8.5 km s$^{-1}$). The solid curves, which are plotted as a reference, correspond to the line-of-sight velocity of a particle in free-fall toward a point-like central mass of 0.1, 0.3, and 0.5 $M_\odot$, as a function of the projected separation.
\label{fig:CSPV}}
\end{center}
\end{figure*}

Unfortunately a similar analysis in terms of disk rotation cannot be performed in VLA 4B since there are no molecular transitions that trace its potential circumstellar disk, likely because of the high dust opacity at these very small scales ($\sim$10-20 au). However, there are signs of gravitational acceleration at scales somewhat larger ($\sim$100-200 au), traced by the molecular emission of the infalling material onto the disk. The signature of this gravitational acceleration, with the highest observed velocity along a given line-of-sight increasing as the (projected) distance to the central protostar decreases, can be identified in the CS(7-6) position-velocity plots shown in Figure~\ref{fig:CSPV} (the free-fall velocity of a particle for different values of the central mass is plotted for reference). This behavior can be more clearly identified on the western side of VLA 4A and on the eastern side of VLA 4B, which appear less perturbed by the presence of the respective companion. 
In addition, the presence of a jet originating in VLA 4B and pointing to the southeast (see Fig. \ref{fig:VLA4B}) has apparently swept away the material in this direction, which appears devoid of molecular emission; therefore, for the study of VLA 4B we have considered only the northeastern part of the PV diagram.

In order to obtain an estimate of the masses of the protostars VLA 4A and VLA 4B, we calculated the CS (7-6) line intensity resulting from a simple model consisting of a spherical envelope in gravitational collapse onto a central compact object with velocity and temperature power-law dependences with the radius with indices of $-0.5$. We are aware that such a simple model will provide only a crude estimate of the physical parameters, as the lack of spherical symmetry is evident. However since we restrict the fitting to the less perturbed parts of the source and to a relatively narrow range of radii, we consider that this local approximation can give satisfactory results as a local measurement of the velocity and gravitational acceleration is enough to measure the enclosed mass. The parameter space was explored by comparing the results of the model, after convolution with the observing beam, with the observations. A range of values of the free parameters  (central mass, inner and outer radii of the envelope) for which the model agrees with the observed position-velocity diagrams were obtained for VLA 4A and VLA 4B (see details in Appendix \ref{sec:infall}). 

The parameters of the best-fit models for VLA 4A and VLA 4B are given in Tables \ref{tab:env_fit} and \ref{tab:env4b_fit}, respectively. In Figures \ref{fig:cs_pvcuts}, \ref{fig:cs_pvcut_model}, and  \ref{fig:cs4b_pvcuts} we show a comparison of the model and observed PV plots. From the best-fit models we obtain for VLA 4A an infalling envelope with an inner radius of $\sim$15 au, an outer radius of $\sim$105 au, and a central mass of 0.27$\pm$0.10 $M_\odot$, where the quoted uncertainty is only the formal fitting error and does not include the effect of possible non-sphericity and optical depth. For VLA 4B we obtain an envelope inner radius of $\sim$12 au, an outer radius of $\sim$150 au, and a central mass of 0.6$\pm$0.2 $M_\odot$. Interestingly, these results are fully consistent with the dynamical masses obtained above from the ethylene glycol transitions and from the orbital motion of the SVS 13 system. Also, we note that the inner radii of the envelopes nicely fit the outer radii of the circumstellar disks, so we consider these results as satisfactory.

Finally, we can assess the importance of self-gravity in the SVS 13 disks based on the analysis of the Toomre Q parameter. Theoretically, $Q>1$ indicates that the self-gravity of the disk is negligible compared with the gravity of the central object. Such a disk would be stable against local gravitational instabilities and would not be expected to undergo fragmentation. For Keplerian disks $Q\simeq 2M_* H/(M_{\rm disk} R)$ \citep{Kratter2016,Tobin2016trip,Tobin2020}, where $M_*$ and $M_{\rm disk}$ are the masses of the central protostar/s and the disk, respectively, and $H$ is the scale height of the disk at radius $R$. For a Keplerian disk, $H$ = $c_s/\Omega_K$, where $c_s$ is the sound speed in the gas (in this case H$_2$) and $\Omega_K$ is the Keplerian angular velocity at radius $R$. Using the values derived from our observations (see also Sections \ref{sec:dust2} and \ref{sec:masses}), we obtain for the circumbinary disk, $c_s$ = 0.9 km s$^{-1}$ (for $T$ = 140 K), $H$ = 40 au (at the disk outer radius of 120 au), and $Q$ = 13. For the VLA 4A circumstellar disk we obtain $c_s$ = 0.9-1.3 km s$^{-1}$ (for $T$ = 140-300 K), $H$ = 2.4-3.5 au (at the disk outer radius of 12 au), and $Q$ = 22-81. Finally, for the VLA 4B circumstellar disk we obtain $c_s$ = 0.9-1.6 km s$^{-1}$ (for $T$ = 140-450 K), $H$ = 1.1-1.9 au (at the disk outer radius of 9 au), and $Q$ = 5-25. Therefore, all the disks are expected to be stable and self-gravity is not likely important in the disks of SVS 13 at the present time. This does not exclude, however, that the binary could have originated through the fragmentation of a circumstellar disk, which in the past was more massive and gravitationally unstable.

\subsection{Chemistry of the Circumstellar and Circumbinary Environment}\label{sec:chem}

\subsubsection{Hot Corinos in the SVS 13 Protobinary System}\label{sec:COM}

Complex Organic Molecules (COMs, molecules with six or more atoms with at least one carbon atom; \citealt{Herbst2009}) are usually detected in association with the so-called hot molecular cores (HMC), which are hot and dense regions that characterize one of the first stages in the formation of massive stars \citep{Osorio1999}. However, in a few cases ($\sim$20 currently known), these complex molecules have been also detected in association with low-mass protostars (e.g., IRAS 16293$-$2422: \citealt{Dishoeck1995,Schoier2002,Jorgensen2016,Jorgensen2018}, NGC1333 IRAS 4A: \citealt{Sahu2019,Desimone2020}), and these regions are called ``hot corinos''.

Hot corinos are small, dense regions very close to their host protostar with temperatures higher than the water ice desorption temperature of $\sim$100 K, and are considered the low-mass counterparts of HMCs. COMs are the trademark of hot corinos. COMs have been detected previously toward SVS 13 with poorer angular resolution (from $0\farcs3$ to few arcseconds) observations that do not resolve well the emission of the two components \citep{LopSep2015,Lefevre2017,DeSimone2017,Bianchi2019,Belloche2020}. Since in these observations the peak emission of the detected COMs was found to be closer to the compact source VLA 4A, \citet{Lefevre2017} proposed that this component of the binary was the hot corino. With our higher angular resolution observations ($0\farcs08\times0\farcs$16), we are able to detect COMs associated with both components of the binary, VLA 4A and VLA 4B, as well as in the circumbinary material (Table \ref{tab:lines}).

We confirm the previous detection of acetaldehyde, ethanol \citep{Bianchi2019,Belloche2020}, formamide \citep{LopSep2015,Bianchi2019}, ethylene glycol, methanol \citep{Lefevre2017,Belloche2020}, and methyl formate \citep{Lefevre2017,Bianchi2019,Belloche2020} toward SVS 13. We additionally report the detection, for the first time in this source, of acetone (CH$_3$COCH$_3$), ethyl cyanide (C$_2$H$_5$CN), ethylene oxide (c-C$_2$H$_4$O), methyl cyanide (CH$_3$$^{13}$CN), deuterated acetaldehyde (CH$_3$CDO), deuterated methanol (CH$_2$DOH)\footnote{The detection of the CH$_3$OD isomer was reported by \citet{Lefevre2017}.}, and deuterated methyl cyanide (CH$_2$DCN).

In general, these COMs emit in a region that appears to be associated with circumbinary gas. In a few transitions (marked 4A in Table \ref{tab:lines}), the emitting region is restricted to a smaller region surrounding VLA 4A (e.g., Fig. \ref{fig:EGly}). Interestingly, deuterated acetaldehyde is only detected toward the position between the two protostars (marked as CB in Table \ref{tab:lines}).

We find that, in general, the COMs detected toward VLA 4A are also present toward VLA 4B. The weaker transitions are not detected toward VLA 4B, and a few transitions are only detected toward VLA 4A. This suggests that the chemistry in VLA 4A could be much richer than in VLA 4B, as previous studies with lower angular resolution also suggest (e.g., \citealt{Lefevre2017}). As noted in Section \ref{sec:gas}, the velocity dispersion toward VLA 4A is much larger than toward VLA 4B; also, VLA 4A appears to be the origin of the main spiral arm. All these facts together suggest that a higher kinematical and heating activity is taking place in VLA 4A. Alternatively, the lack of detection of several molecular transitions toward VLA 4B could reflect the fact that the continuum emission peak toward VLA 4B has a very high optical depth that makes difficult the observation of line emission at this position. Recently, \citet{Desimone2020} have suggested that some reported differences in the COM detections toward binary protostars that were attributed to differences in the chemical activity of the two binary components, could, in fact, be due to a high dust opacity that would hide the molecular lines. This effect could be taking place in the SVS 13 binary leading to a reduction, or even a suppression, of the detectable continuum-subtracted molecular line emission toward VLA 4B. However, the high sensitivity of our observations allow us to confirm that COMs are present in both components of the SVS 13 binary and, therefore, that both components are ``bona fide'' hot corinos.

We note that there is a 3-$\sigma$ negative contour toward VLA 4B in the left and middle top panels of Figure~\ref{fig:EGly}, and in Figure~\ref{fig:A}. The present analysis does not allow us to distinguish whether this is a weak absorption feature or just the result of a somewhat imperfect continuum subtraction. Detailed analysis of these possible absorption features will be investigated in a future work.

\subsubsection{Chemical Abundances}

We detected four transitions of the main isotopologue of methanol (CH$_3$OH). However, these transitions appear to be optically thick, so we cannot derive reliable values of the molecular column density directly from these data. Fortunately, we detected the isotopologues $^{13}$CH$_3$OH and CH$_3$$^{18}$OH, which are optically thinner, and are therefore more suitable for deriving column densities. However, the number of detected transitions of these isotopologues is insufficient to construct rotational diagrams for a good and reliable temperature and column density fitting. So, we have adopted a temperature of 140 K, derived from other molecular species, to obtain the column densities that best reproduce the observed spectra, assuming optically thin emission and LTE. Finally, the methanol column density can be obtained assuming isotopic ratios ($^{12}$C/$^{13}$C = 77 and $^{16}$O/$^{18}$O = 560; \citealt{Wilson1994}). 

In Table~\ref{tab:trot} we report the values of the $^{13}$CH$_3$OH and CH$_3$$^{18}$OH column densities toward the center of the circumbinary disk (CB), at a position midway between VLA 4A and VLA 4B, as well as the adopted column density of CH$_3$OH. The CH$_3$OH methanol column density derived from the $^{13}$CH$_3$OH isotopologue is 0.6$\times$$10^{19}$ cm$^{-2}$, while the value derived from CH$_3$$^{18}$OH is 1.7$\times$$10^{19}$ cm$^{-2}$. If the isotopic ratios were as assumed \citep{Wilson1994}, and the emission of both isotopologues were optically thin, we would expect the observed  $^{13}$CH$_3$OH line intensity (as well as the estimated column density) to be a factor 560/77 $\simeq$ 7 higher than that of the CH$_3$$^{18}$OH, resulting in the same derived CH$_3$OH column density in both cases. However, the ratio of the observed isotopologue line intensities is $\sim$3, a value smaller than expected, indicating that either the $^{13}$CH$_3$OH emission is optically thick or the isotopic ratios are somewhat different than assumed.

To clarify this issue, we obtained additional spectra of the two isotopologues toward positions slightly offset from CB, where the column density has decreased and the emission is expected to become optically thinner. We found that the intensities drop by a factor of $\sim$6-8 and $\sim$3-4 toward positions offset $0\farcs2$ north (close to the S3 feature) and south (close to spiral S1) of CB, respectively. However, the isotopologue line intensity ratio toward these positions is $\sim$2 and $\sim$4, respectively, not much different from the ratio of $\sim$3 observed toward CB. Therefore, we conclude that the discrepancy in the inferred methanol column densities does not seem to be due to opacity effects, and that the $^{13}$C isotope is probably somewhat less abundant and/or the $^{18}$O more abundant than we have initially assumed; that is, $^{12}$C/$^{13}$C $\ga$ 77 and/or $^{16}$O/$^{18}$O $\la$ 560. Thus, we adopt in Table~\ref{tab:trot} an intermediate value of $\sim$1$\times10^{19}$ cm$^{-2}$ for the methanol column density toward CB.

\citet{Belloche2020} found a lower value of $2\times10^{18}$ cm$^{-2}$ for the methanol column density in SVS 13. As discussed above (e.g., Sect.~\ref{sec:temp}), we attribute this discrepancy in SVS 13 to the larger beam of those observations, probing scales of $\sim$100 au, as compared with our data that trace smaller scales of $\sim$30 au. On the other hand, methanol column densities of $\sim$$10^{19}$ cm$^{-2}$, similar to the value we obtain for SVS 13, have been reported in other hot corinos, even at larger scales of $\sim$60 au in IRAS 16293$-$2422B \citep{Jorgensen2016,Jorgensen2018} and $\sim$120 au in L483 \citep{Jacobsen2019}.

One of the molecules for which we derived a rotational temperature (Fig.~\ref{fig:trot}) is methyl formate (CH$_3$OCHO).
Since the CH$_3$OCHO column density toward CB is $2.3\times10^{17}$ cm$^{-2}$ (Table~\ref{tab:trot}), we infer an abundance relative to methanol [CH$_3$OCHO/CH$_3$OH] $\simeq$ 0.023 in SVS~13. Our value is lower than the value of 0.10 reported by \citet{Belloche2020} in SVS 13, likely because the column density of methanol obtained by these authors is lower (see above). The CH$_3$OCHO abundances relative to methanol reported for other hot corinos span two orders of magnitude (see Fig.~7 of \citealt{Bergner2019} or Fig.~17 of \citealt{Yang2021} for a comparison). Our value, similar to the 0.026 value found in IRAS 16293$-$2422B by \citet{Jorgensen2016,Jorgensen2018}, lies in the middle of the range. Additionally, from the CH$_3$OH and CH$_2$DOH column densities toward CB (Table~\ref{tab:trot}) we obtain a deuteration fraction D/H = 0.05, similar to values found in other low-mass protostars \citep{Jorgensen2018,Taquet2019}.

The analysis toward the positions of the individual protostars is more challenging since the $^{13}$CH$_3$OH and CH$_3$$^{18}$OH lines are wider and difficult to separate from other lines there. To obtain abundances relative to methanol toward these positions we infer the methanol column density from the deuterated methanol, CH$_2$DOH, which is detected toward all the positions (Table~\ref{tab:trot}), using the deuteration fraction. In this way, for aGg'-(CH$_2$OH)$_2$, which we only detect toward the position of VLA 4A, we obtain an abundance relative to methanol of $7.1\times10^{-3}$, which is quite similar to the value of $6.4\times10^{-3}$ reported by \citet{Belloche2020}.

Abundance ratios and their variation across the source can be obtained for other detected molecules and a more complete analysis can be performed from the data collected with ALMA. This further chemical analysis is postponed for a separate paper.

Nevertheless, a rough estimate of the absolute abundances, i.e., relative to H$_2$, can be obtained for methanol and the other molecules listed in Table~\ref{tab:trot}, toward the central position (CB) using the beam-averaged column density obtained from the dust emission. As discussed in Section~\ref{sec:dust2}, the dust emission is optically thin toward this position, and we can further assume that there the dust temperature coincides with the rotational temperature of 140 K obtained from molecular observations. We cannot make a similar assumption toward the positions of VLA 4A and VLA 4B, because it is unclear whether toward these positions the dust originates from the same region as the observed molecular emission since, as discussed in Section~\ref{sec:dust2}, at least part of the dust emission appears to come from a different, more compact and hotter region. Thus, assuming a gas-to-dust ratio of 100, and a dust opacity \makebox{$\kappa_\nu$(0.9 mm) = 1.65 cm$^2$ g$^{-1}$} (per unit mass of dust), as in Section \ref{sec:masses}, we estimate an H$_2$ column density of 1.8$\times$10$^{24}$ cm$^{-2}$ toward CB from which we infer the abundances relative to H$_2$ of the molecules listed in Table \ref{tab:trot}. 

We obtain an abundance of methanol relative to H$_2$ of 5.6$\times$10$^{-6}$ toward the central position (CB). This abundance of methanol is consistent with the results of \citet{Bianchi2019}, who report a value of the $^{13}$CH$_3$OH abundance of 3$\times$10$^{-8}$ in SVS 13, similar to the value of 4.4$\times$10$^{-8}$ listed in our Table~\ref{tab:trot} for the same isotopologue. The methanol abundance we have estimated in SVS 13 is of the same order than the values obtained in other protobinaries (e.g., $<$3$\times$10$^{-6}$ in IRAS 16193$-$2422; \citealt{Jorgensen2016}). We note that the presence of temperature gradients, uncertainties in the dust opacity and in the gas-to-dust ratio can be important and difficult to estimate. Therefore the values of the abundances relative to H$_2$ should be taken with caution. 

Nevertheless, we can still measure variations from one position to another as they are less affected by the uncertainties in the absolute values. We find evidence of a decrease of the methanol abundance toward the north of the central position CB, since when moving 0\farcs2 (60 au) to the north of this central position, the estimated column densities of the $^{13}$CH$_3$OH and CH$_3$$^{18}$OH (and therefore, the methanol abundance) decrease by a larger factor than the H$_2$ column density. Assuming that there are no significant variations in the dust properties at these scales, we infer that the methanol abundance relative to H$_2$ toward this northern position decreases by a factor 2.5-3.2 with respect to the CB position. We do not see a similar behavior, and the abundance remains the same, when moving the same distance to the south of the central position. We speculate on whether the presence of more prominent spiral arms connected to the southern part of the circumbinary disk could be related with an overabundance of methanol relative to the northern part.

\section{Summary and Conclusions}\label{sec:conclusions}

We carried out a comprehensive study of the protobinary system associated with the SVS 13 star in the star-forming region of NGC 1333. We have clarified the geometry of the system and the physical properties, even at very small scales of $\sim$10-20 au, thanks to sensitive, high-angular resolution VLA and ALMA observations. These observations provided us with proper motions measurements in the plane of the sky and radial velocities along the line-of-sight, allowing us to infer the kinematical and dynamical properties of the system. Overall, temperatures, masses and other physical parameters have been inferred to characterize the SVS 13 system. 

Despite the uncertainties involved, we favor the scenario of a single component for VLA 4A and for VLA 4B. The SVS 13 system would, then, consist of two protostars, with masses of $\sim$0.26 $M_\odot$ and $\sim$0.6 $M_\odot$ for VLA 4A and VLA 4B respectively, each of them surrounded by a circumstellar protoplanetary disk, and encompassed by a larger molecular and dust structure associated with at least two spiral arms that we interpret as an incipient circumbinary disk (see Fig.~\ref{fig:cartoon}). The mass of the whole system is dominated by the protostars, with a total mass of 1 $M_\odot$, as suggested both by the orbital motions and the dynamical calculations of the rotating and/or infalling gas for each individual protostar. The mass in the disks is much smaller ($<10\%$). In this configuration, VLA 4B is likely the primary and VLA 4A the secondary, with a ratio between the secondary and primary masses, q$\simeq$0.26/0.6=0.4. However, uncertainties are still large, and a configuration with two objects of similar mass (or even with VLA 4A being the more massive) cannot be discarded with the current data. This result can be substantially improved with sensitive new observations of higher angular resolution that can better resolve the PV diagrams and potentially reveal a Keplerian rotation profile. Obtaining molecular data of high angular resolution but at lower frequencies, where the dust contribution should be smaller, could help to minimize the difficulties in detecting molecular emission from the circumstellar disk of VLA 4B, which has been hindered in our observation by the very high opacity of the continuum. These observations will be easily performed with the next generation Very Large Array, that will provide unprecedented sensitivity and angular resolution in the 40 to 120 GHz regime.


\startlongtable
\begin{deluxetable*}{lcccc}
\tablecaption{Summary of Physical Parameters of the SVS 13 Protobinary System\label{tab:summary}}
\tablewidth{0pt}
\tablehead{ Parameter & VLA 4A & VLA 4B & Circumbinary & Notes}
\startdata
Disk Size (dust) & 24$\pm$1 au $\times$ 10$\pm$6 au &18$\pm$1 au\,$\times$ 14$\pm$3 au &750$\pm$75 au $\times$ 600$\pm$75 au & 1\\
Disk PA (dust) & 83$\degr$$\pm$8$\degr$ & 93$\degr$$\pm$17$\degr$ & $\sim$90$\degr$ & 1\\
$R_{\rm disk}$, $i$ (dust) & 12$\pm$1 au, 64$\degr$$\pm$16$\degr$ & 9$\pm$1\,au, 40$\degr$$\pm$12$\degr$ & 375$\pm$38 au, 34$\degr$$\pm$12$\degr$ & 2 \\
$T_{\rm dust}$ & 260$\pm$60 K & 430$\pm$80 K & \nodata & 3 \\
Disk Size (gas) &  67$\pm$3 au $\times$ 62$\pm$3 au & \nodata & 240$\pm$30 au $\times$ 180$\pm$30 au & 4 \\
Disk PA (gas) & $\sim$90$\degr$ & \nodata & $\sim$90$\degr$ & 4 \\
$R_{\rm disk}$, $i$ (gas) & 33$\pm$2 au, 22$\degr$$\pm$7$\degr$ & \nodata & 120$\pm$15 au, 38$\degr$$\pm$14$\degr$ & 5 \\
$M_{*}$ (EthyGly) & 0.25$\pm$0.15 $M_{\sun}$ & \nodata & \nodata & 6 \\
$M_{*}$ (CS) & 0.27$\pm$0.10 $M_{\sun}$ & 0.60$\pm$0.20 $M_{\sun}$ & \nodata & 7 \\
$T_{\rm rot}$ & 126-145 K & 90-160 K & 138-139 K & 8 \\
$M_{\rm disk}$ & 0.004-0.009 $M_{\sun}$ & 0.009-0.030 $M_{\sun}$ & 0.052 $M_{\sun}$ & 9 \\
$\mu_\alpha \cos \delta$ & 8.22$\pm$0.63 mas yr$^{-1}$ & 8.56$\pm$0.34 mas yr$^{-1}$ &  & 10 \\
$\mu_\delta$ &  $-$9.19$\pm$0.41 mas yr$^{-1}$ & $-$10.79$\pm$0.44 mas yr$^{-1}$ &  & 10 \\
$V_{\rm LSR}$ & 7.36$\pm$0.21 km s$^{-1}$ & 9.33$\pm$0.18 km s$^{-1}$ & 8.77$\pm$0.16 km s$^{-1}$ & 11 \\
$\vec{V}_{\rm rel}$(B$-$A) & \multicolumn2c{(0.48$\pm$1.02 km s$^{-1}$, $-$2.28$\pm$0.82 km s$^{-1}$, 1.97$\pm$0.28 km s$^{-1}$) } & & 12 \\
$i$ (orbit) & \multicolumn2c{41.7$\degr$$\pm$9.5$\degr$} & & 13 \\
$a$ (orbit) & \multicolumn2c{90 au} & & 14 \\
$M_{*}$ (orbit) & \multicolumn2c{1.0$\pm$0.4 $M_{\sun}$} & & 15 \\
$P$ (orbit) & \multicolumn2c{850 yr} & & 16 \\
$c_s$ & 0.90-1.3 km s$^{-1}$  & 0.9-1.6 km s$^{-1}$ & 0.90 km s$^{-1}$ & 17 \\
$H(R_{\rm disk})$ & 2.4-3.5 au & 1.1-1.9 au & 40 au & 18 \\
$Q(R_{\rm disk})$ & 22-81 & 5-25 & 13 & 19 \\
\enddata
\tablecomments{Assuming $d$ = 300 pc.}
\tablenotetext{\scriptsize 1}{ VLA 4A and VLA 4B: Deconvolved FWHM size obtained from a Gaussian fit to the continuum image made using baselines $>$750~k$\lambda$, where the extended emission has been filtered out (Fig.~\ref{fig:cont750kl}). The errors given are the formal errors of the fits. Similar sizes of 30$\pm$1\,au\,$\times$\,16$\pm$4\,au, PA=80$\degr$$\pm$7$\degr$ (VLA 4A) and 20$\pm$1\,au\,$\times$\,14$\pm$5\,au, PA=93$\degr$$\pm$20$\degr$ (VLA 4B) are obtained from a fit to an image made using baselines $>$500~k$\lambda$. 
Circumbinary emission: Size and PA of the emission above a 6-$\sigma$ level, measured on the image made using all the visibilities (Fig.~\ref{fig:continuum} in Sect. \ref{sec:dust2}), including the emission of the spirals.}
\tablenotetext{2}{ Disk radius and inclination angle. The disk radius is obtained as half the major axis; since the sources are elongated roughly along the direction of the minor axis of the beam, where the angular resolution is higher, the disk radii are well constrained. The disk inclination angle ($i$=0$\degr$ means face-on) is obtained from the aspect ratio of the emission (see note~1);  uncertainties are derived as explained in Appendix~\ref{sec:errors}, but they may be significantly underestimated for the circumstellar disks since their minor axes are poorly resolved because of the elongated beam.}
\tablenotetext{3}{ Dust temperature in the circumstellar disks, taken as the brightness temperature derived from  Gaussian fits to the highest angular resolution 0.9 mm continuum images (Table~\ref{tab:disks} in Section~\ref{sec:dust2}). The values listed are those with the smallest uncertainties, obtained from the image made using baselines $>$500 k$\lambda$ for VLA 4A and $>$750 k$\lambda$ for VLA 4B. The dust emission in the circumbinary disk appears to be optically thin and the brightness temperature is not useful as a constraint on the dust temperature.}
\tablenotetext{4}{ VLA 4A: Derived from a model fitting of a disk to ethylene glycol transitions (Fig.~\ref{fig:EGly} in Sect.~\ref{sec:CSD}; see Appendix~\ref{sec:keplerian}). VLA 4B: No suitable molecular emission imaged. Circumbinary: From a Gaussian fit to the zero-order moments of the H$^{13}$CN and HC$^{15}$N isotopologues; a slightly larger size of $\sim$300 $\times$ 240 au is obtained from the CS(7-6) transition, which is probably optically thicker (Fig.~\ref{fig:CS} in Sect.~\ref{sec:gas}).}
\tablenotetext{5}{ VLA 4A: Disk radius and inclinaton angle derived from the model fitting (Appendix~\ref{sec:keplerian}). Circumbinary: See notes 2 and 4.}
\tablenotetext{6}{ Stellar mass derived from the model fitting of a thin Keplerian disk to the ethylene glycol transitions associated with VLA 4A (Appendix~\ref{sec:keplerian} and Sect.~\ref{sec:dyn}).}
\tablenotetext{7}{ Stellar mass derived from the model fitting of a spherical infalling envelope to the CS(7-6) emission associated with VLA 4A and VLA 4B (Appendix~\ref{sec:infall} and Sect.~\ref{sec:dyn}).}
\tablenotetext{8}{ Rotational temperature derived from a LTE fit to several molecular transitions (see Sect. \ref{sec:temp}).} 
\tablenotetext{9}{ Total mass (gas+dust) of the circumstellar/circumbinary material, obtained from the dust emission assuming a gas-to-dust ratio of 100 and optically thin emission at 0.9 mm (circumbinary) and 7 mm (circumstellar). For the circumbinary mass we assume a dust temperature of 140 K. The lower limits for the VLA 4A and VLA 4B circumstellar disk masses have been obtained for temperatures of 300 K and 450 K, respectively, while the upper limits have been obtained using 140 K (Sect.~\ref{sec:masses}).}
\tablenotetext{10}{  Absolute proper motions, in right ascension and declination, of the two protostars (Sect. \ref{sec:pm}).}
\tablenotetext{11}{ LSR line-of-sight velocity derived from molecular spectra toward the positions of VLA 4A, VLA 4B, and toward the geometric center of the system (CB) that we adopt as representative of the circumbinary disk (Fig.~\ref{fig:spectra}). We give the mean velocity and standard deviation of 20 lines toward CB and 15 lines toward VLA 4A and VLA 4B. The assumed systemic velocity of the ambient cloud is $V_{\rm sys}=8.5$ km s$^{-1}$.}
\tablenotetext{12}{  Components of the 3D relative velocity of VLA 4B with respect to VLA 4A, along the right ascension, declination and line-of-sight axes, obtained from the observed differences in the proper motions and in the LSR velocity (Table~\ref{tab:3D} and Sect.~\ref{sec:dyn}).}
\tablenotetext{13}{  Inclination of the orbital plane with respect to the plane of the sky, derived from the 3D relative velocity of the two protostars. Uncertainties are estimated as explained in Appendix~\ref{sec:errors}. The orbital parameters are such that the observed orbital motion is counter-clockwise and the angular momentum points toward the Earth (Fig. \ref{fig:cartoon}).}
\tablenotetext{14}{  Projected distance between VLA 4A and VLA 4B, taken as the separation between the two components of the binary. The small relative proper motion in right ascension as compared with the proper motion in declination (Fig.~\ref{fig:difpmotions} in Sect.~\ref{sec:pm}), and the alignment of the sources along the major axes of the disks (PA $\simeq$ 90$\degr$; this table) suggests that this is a good approximation.}
\tablenotetext{15}{  Total mass of the stellar system, derived from the orbital motions (Sect.~\ref{sec:dyn}).}
\tablenotetext{\small 16}{  Orbital period of the binary, derived from the orbital motions (Sect.~\ref{sec:dyn}).}
\tablenotetext{17}{  Sound speed in the disks. See note 8 for the adopted temperatures.}
\tablenotetext{18}{  Scale height (Sect.~\ref{sec:dyn}) at the disk radius (12 au for VLA 4A, 9 au for VLA 4B, and 120 au for the circumbinary). See note 8 for the adopted temperatures.}
\tablenotetext{19}{  Toomre Q parameter (see Sect.~\ref{sec:dyn}). All these disks are found to be stable ($Q>1$).}
\end{deluxetable*}

In Table~\ref{tab:summary} we provide a characterization of this protobinary system in terms of its physical properties. In a forthcoming paper we will compare these properties, and other parameters inferred from them with the predictions of existing numerical simulations. This will provide a deeper understanding of the binary formation process and will guide future theoretical and observational works.



Here we summarize our main conclusions:

\begin{enumerate}

\item We confirm that SVS 13 is a multiple protostellar system with at least two embedded protostars, separated by 90 au (for an adopted distance of 300 pc), known as VLA 4A and VLA 4B. We cannot rule out a higher order multiple system, as each of the two protostars could have a companion at scales below 20-30 au. However, we favor the interpretation in terms of a binary system. 

\item Both protostars are associated with free-free emission (likely from jets), circumstellar dust and molecular gas (likely from disks), and complex organic molecules characteristic of hot corinos. The radii (half the FWHM obtained from a Gaussian fit) of the VLA 4A and VLA 4B circumstellar disks of dust are $\sim$12 au and $\sim$9 au, respectively. In both disks, the PA of the major axis is $\sim$90$^\circ$. However, we cannot reliably obtain the inclination angle from the aspect ratio because of insufficient angular resolution in the direction of the minor axes due to a very elongated synthesized beam.

\item In addition to the circumstellar disks, this protostellar system has an associated circumbinary structure that appears to be a circumbinary disk in the early stages of formation. Its radius, as inferred from the molecular emission, is $\sim$120 au. The circumbinary disk image appears elongated along PA$\sim$90$^\circ$, similar to the circumstellar disks, and from the aspect ratio we infer an inclination angle of the molecular disk plane with respect to the plane of the sky of $\sim$40$\degr$. The circumbinary dust emission presents a larger radius of $\sim$375 au, with prominent spiral arms extending to distances of $\sim$500 au.

\item The dust emission appears more intense and compact toward VLA 4B (the eastern component), suggesting a very high column density and optical depth, at very small scales, that makes difficult the detection of molecular lines toward the stellar position. The dust column density toward VLA 4A is lower than toward VLA 4B, resulting in a lower dust opacity and stronger observed molecular lines. However, the dust emission associated with VLA 4A is more extended, resulting in a larger total amount of dust in its surroundings. 

\item Molecular transitions typical of hot corinos are detected toward both VLA 4A and VLA 4B, as well as in the circumbinary disk. We are able to estimate rotational temperatures and molecular column densities, indicating warm temperatures of $\sim$140 K for the molecular gas across the various components of the source, and a rich chemistry. The dust emission suggests still higher temperatures of $\sim$400 K at very small radii of $\sim$10 au.

\item We detect line-of-sight velocity gradients in both the circumstellar and circumbinary disks. These, together with the spiral arms and the stellar proper motions, all indicate a rotation in a counter-clockwise direction as observed from the Earth, with the northern half of the disk closer to the observer. The angular momentum vector therefore points toward the observer. 

\item The orbital proper motions have been inferred from high-resolution continuum data taken over 30 years. From these data and the line-of-sight velocities, the 3D relative orbital velocity is obtained. From this, a total mass (dominated by the stellar component) of 1 $M_\sun$, an orbital period of 850 yr, and an inclination angle of the orbit with respect to the plane of the sky of 42$\degr$ are derived. Most of the positions used in the proper motion analysis were obtained from free-free emission that can be affected by episodic ejections of blobs of ionized material that reduce the accuracy of the measurements. The emission at 7 mm and shorter wavelengths, which traces circumstellar dust and is not affected by this issue, is strong enough in SVS 13 that future observations at these wavelengths
 will give more accurate positions to more tightly constrain the orbital parameters and mass of this young system. These will provide important tests of models for the formation of binary star systems.

\item We identified several ethylene glycol transitions, which are associated only with VLA 4A, showing a clear east-west velocity gradient. By fitting a rotating disk, we consistently get a VLA 4A stellar mass of 0.25$\pm$0.15 $M_\sun$, a disk radius of 33$\pm$2 au, and an inclination angle of $22\degr\pm7\degr$, from independent fits to the different observed transitions. The molecular disk radius appears to be larger than the dust disk radius, as is often observed in other protoplanetary disks.

\item The CS (7-6) position-velocity plots show indications of gravitational acceleration. From a simple model of spherical infall onto each of the protostars, we obtain rough estimates of their stellar masses of 0.27$\pm$0.10 $M_\sun$ for VLA 4A and 0.60$\pm$0.20 $M_\sun$ for VLA 4B, which are consistent with the total mass inferred from the orbital motion and the mass obtained from the ethylene glycol fitting of the VLA 4A circumstellar disk. These calculations suggest that VLA 4B is the primary, but the uncertainties are still large.

\item From the dust emission we estimate the masses of the circumstellar and the circumbinary disks, obtaining 0.004-0.009 $M_\sun$ for the VLA 4A circumstellar disk, 0.009-0.030 $M_\sun$ for the VLA 4B circumstellar disk, and 0.052 $M_\sun$ for the circumbinary disk. We find that the disks have masses of the order of 1.5-5\% the mass of the stars they surround, and the Toomre parameter is Q$>1$ for all the disks, indicating that they are stable at the present time.
 However, this does not exclude the possibility that the binary could have originated through the past fragmentation of a circumstellar disk around one of the stars.

\item We report the detection, to the best of our knowledge for the first time in SVS 13, of acetone (CH$_3$COCH$_3$), ethyl cyanide (C$_2$H$_5$CN), ethylene oxide (c-C$_2$H$_4$O), methyl cyanide (CH$_3$$^{13}$CN), deuterated acetaldehyde (CH$_3$CDO), deuterated methanol (CH$_2$DOH), and deuterated methyl cyanide (CH$_2$DCN). 

\item We estimate a methanol column density of $\sim$1$\times$10$^{19}$ cm$^{-2}$ and an abundance relative to H$_2$ of $\sim$6$\times$10$^{-6}$ toward the central position, in between VLA 4A and VLA 4B, with signs of detectable variations in abundance at scales of $\sim$60 au. Similar values of the methanol column density and abundance have been reported in the literature in other protobinary sources. However, lower values have been inferred from previous observations toward SVS 13 performed with poorer angular resolution. We also derive a deuteration fraction D/H = 0.05, similar to values found in other low-mass protostars. Finally, we infer an abundance of methyl formate relative to methanol of $\sim$0.023 in SVS~13, similar to the value reported in the literature toward the IRAS 16293$-$2422B protostar, which also belongs to a multiple system.

\end{enumerate}

In conclusion, SVS 13 appears to be an excellent testbed for testing numerical simulations of the earliest stages in the formation of binary and multiple stellar systems.


\begin{acknowledgments}

 We thank Jim Moran, who tackled some years ago the ``divide problem'', for his helpful comments and inspiration. 
We also thank the referee for a constructive and helpful report.
G.A., G.B.-C., I.dG.-M., A.K.D.-R., R.E., G.A.F., J.F.G., M.O. and J.M.T. acknowledge support from the Spanish MINECO/AEI through the AYA2017-84390-C2 grant (co-funded by FEDER) and MCIN/AEI/10.13039/501100011033 through the PID2020-114461GB-I00 and PID2020-117710G-I00 grants. G.A., G.B.-C., A.K.D.-R., G.A.F., J.F.G., and M.O. acknowledge financial support from the State Agency for Research of the Spanish MCIU through the ``Center of Excellence Severo Ochoa'' award for the Instituto de Astrof{\'i}sica de Andaluc{\'i}a (SEV-2017-0709). G.B.-C. also acknowledges support from the Spanish MICINN/AEI through the PRE2018-086111 (SEV-2017-0709-18-1) grant (co-funded by ESF). 
 G.A.F also acknowledges support from the Collaborative Research Centre 956, funded by the Deutsche Forschungsgemeinschaft (DFG) project ID 184018867. L.A.Z. acknowledges financial support from CONACyT-280775 and UNAM-PAPIIT IN110618 grants, M{\'e}xico. This paper makes use of the following ALMA data: ADS/JAO.ALMA\#2015.1.01229.S, ADS/JAO.ALMA\#2016.1.01305.S. ALMA is a partnership of ESO (representing its member states), NSF (USA) and NINS (Japan), together with NRC (Canada) and NSC and ASIAA (Taiwan) and KASI (Republic of Korea), in cooperation with the Republic of Chile. The Joint ALMA Observatory is operated by ESO, AUI/NRAO and NAOJ. Based on analysis carried out with the \href{http://cassis.irap.omp.eu}{CASSIS} software \citep{Vastel2015} and JPL, CDMS and VASTEL databases. CASSIS has been developed by IRAP-UPS/CNRS. This research made use of \href{http://www.astropy.org}{Astropy} a community-developed core Python package for Astronomy \citep{astropy2013, astropy2018}.

\end{acknowledgments}

%

\vspace{5mm}
\facilities{ALMA, VLA}


\software{Astropy (\href{http://www.astropy.org}{http://www.astropy.org}; \citealt{astropy2013, astropy2018}),
          CASA (\href{http://casa.nrao.edu}{http://casa.nrao.edu}; \citealt{McMullin2007}),
          CASSIS (\href{http://cassis.irap.omp.eu}{http://cassis.irap.omp.eu}; \citealt{Vastel2015})
          }



\newpage

\appendix

\onecolumngrid

\section{VLA Images of Other Radio Sources in the Field}\label{ap:vla}

\restartappendixnumbering

\begin{figure}[h]
\begin{center}
\includegraphics[height=0.3\textheight]{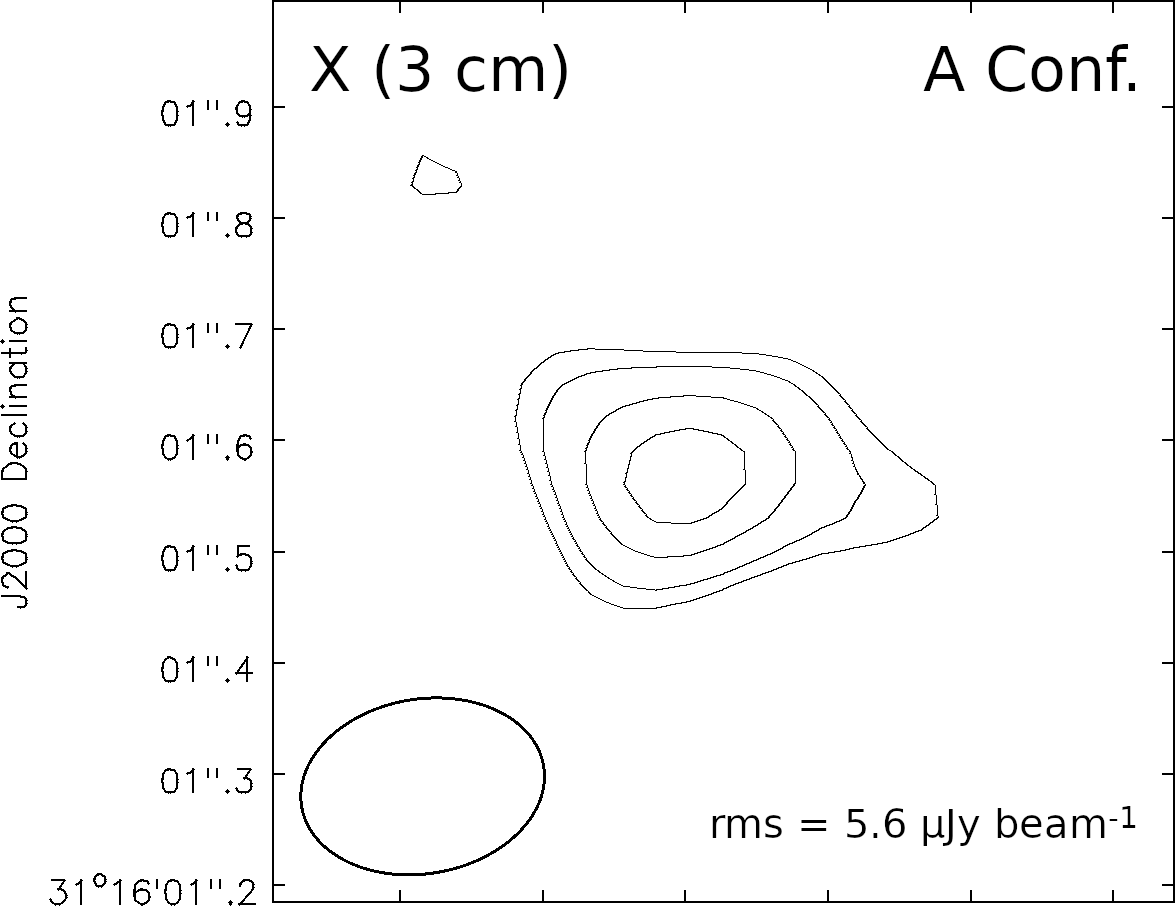}\hspace*{1.5em}\includegraphics[height=0.3\textheight]{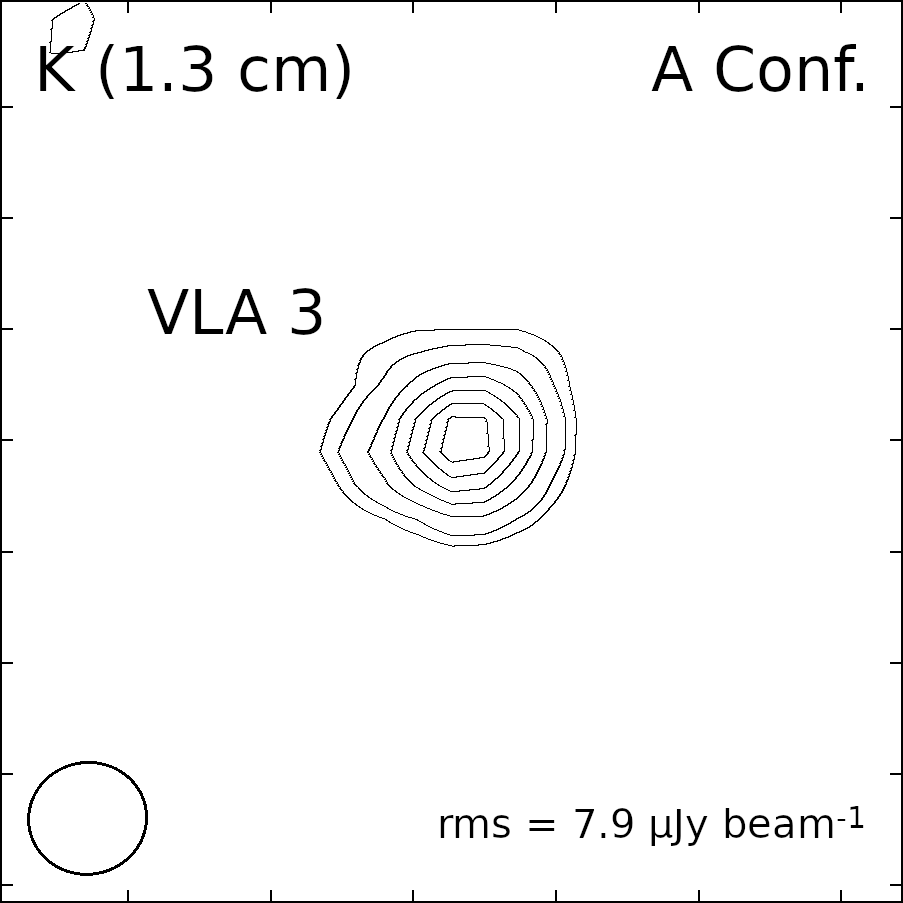}\hfill\\
\vspace*{1.5em}
\includegraphics[height=0.34\textheight]{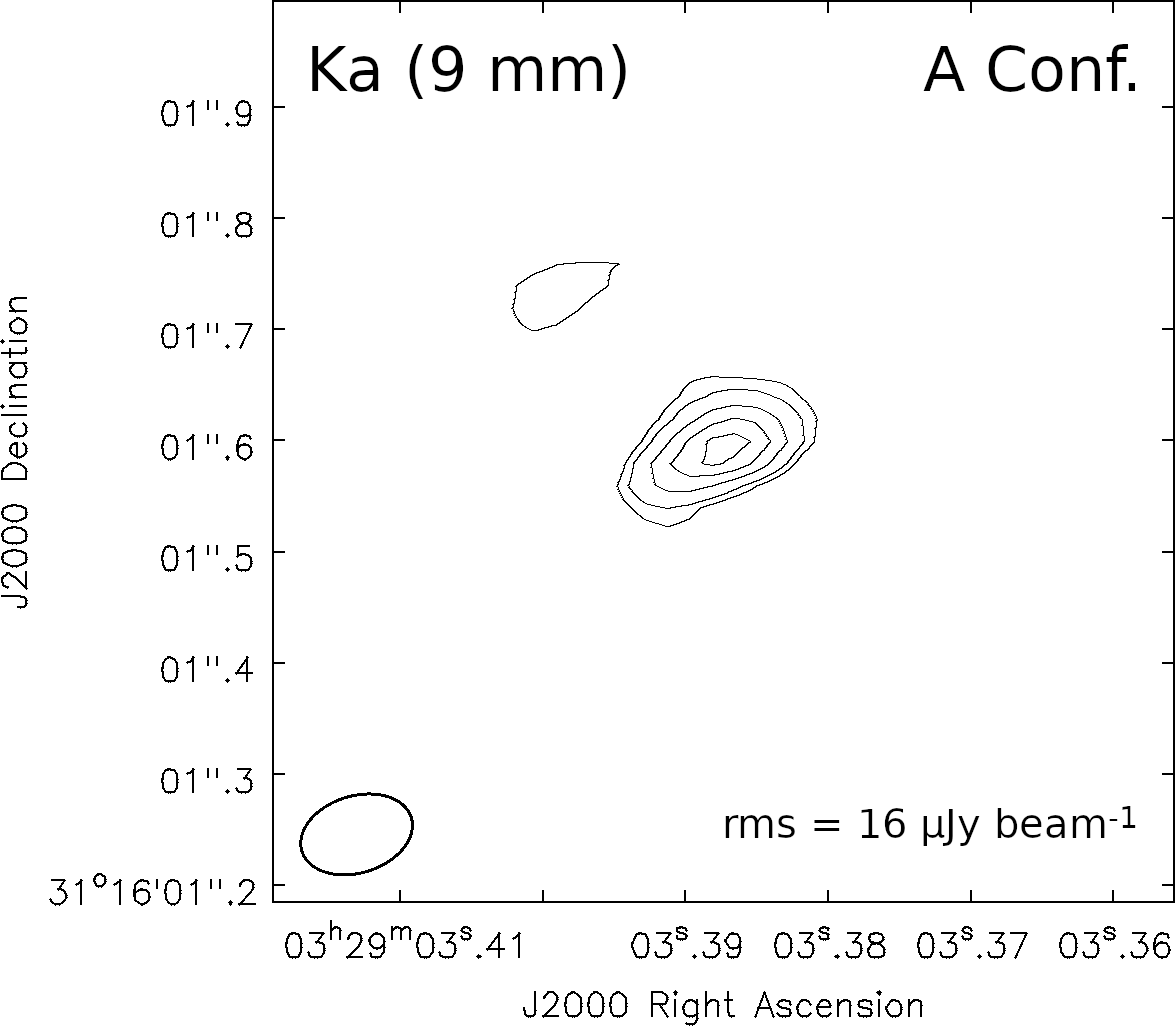}\hspace*{1.5em}\includegraphics[height=0.34\textheight]{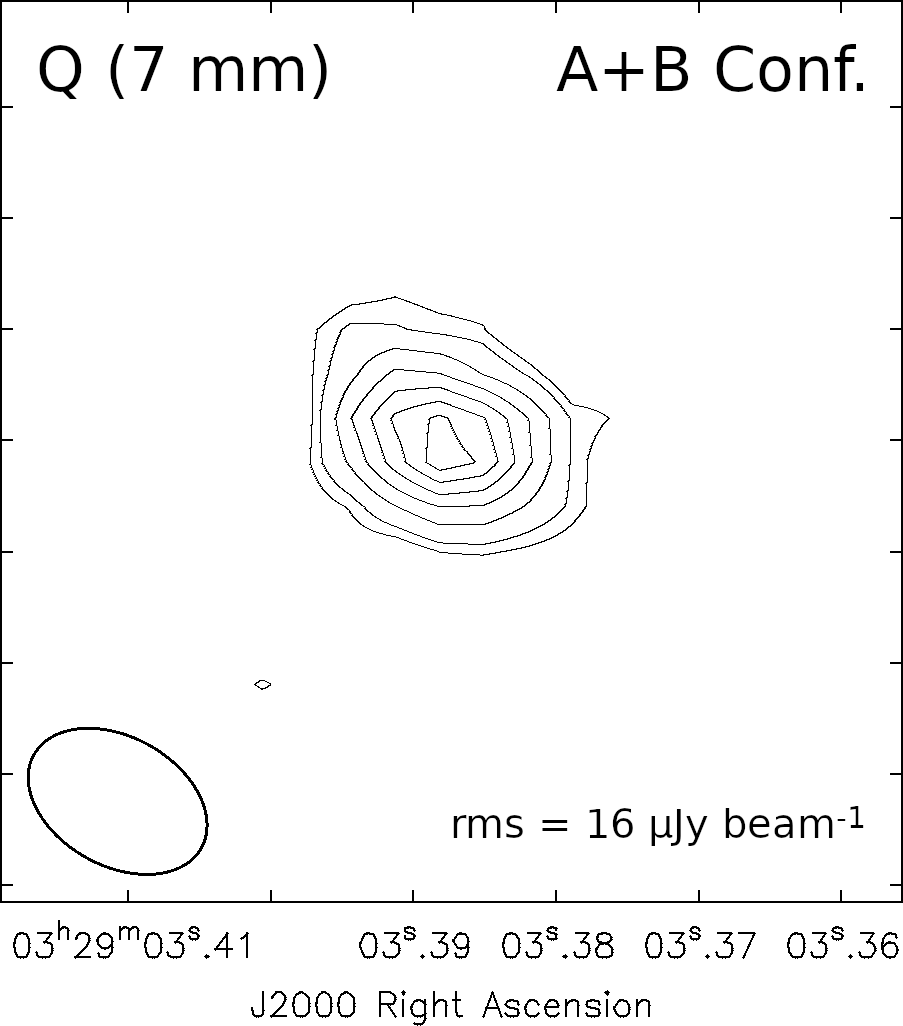}\hfill
\caption{VLA images of source VLA 3 at X (3 cm), K (1.3 cm), Ka (9 mm), and Q (7 mm) bands. Contour levels are $-$3, 3, 4, 6, 8, 10, 12, and 14 times the rms of each image. The synthesized beams and the rms are shown in the bottom left and right corners, respectively, of the images.
We used Briggs weighting with the robust parameter set to 0 for the image at X band, and natural weighting for the remaining images.
 The synthesized beams are $0\farcs22\times0\farcs16$, PA = $-80.3\degr$ at X band; $0\farcs11\times0\farcs10$, PA = $-80.7\degr$ at K band; $0\farcs10\times0\farcs07$, PA = $-72.9\degr$ at Ka band; and $0\farcs17\times0\farcs12$, PA = $60.9\degr$ at Q band.
\label{fig:vla3}}
\end{center}
\end{figure}

\begin{figure}[h]
\begin{center}
\includegraphics[height=0.3\textheight]{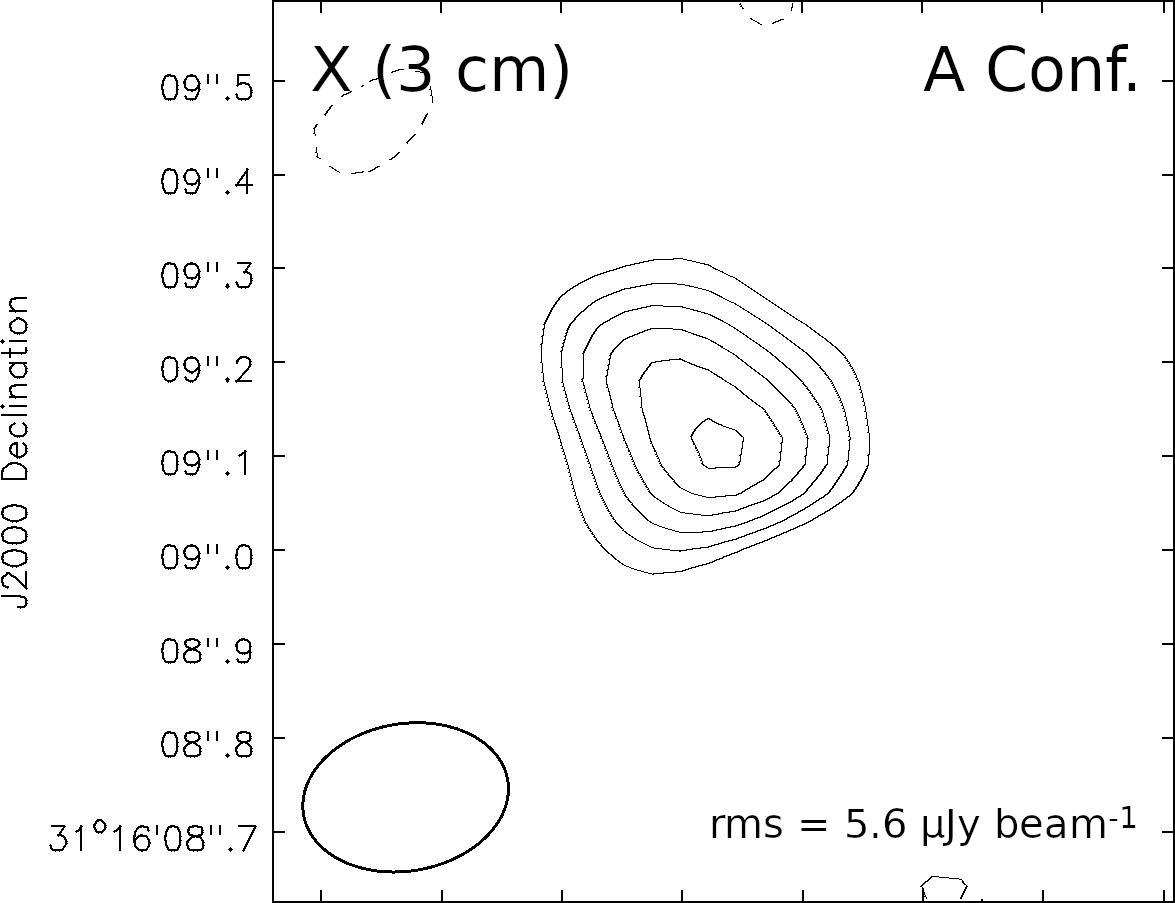}\hspace*{1.5em}\includegraphics[height=0.3\textheight]{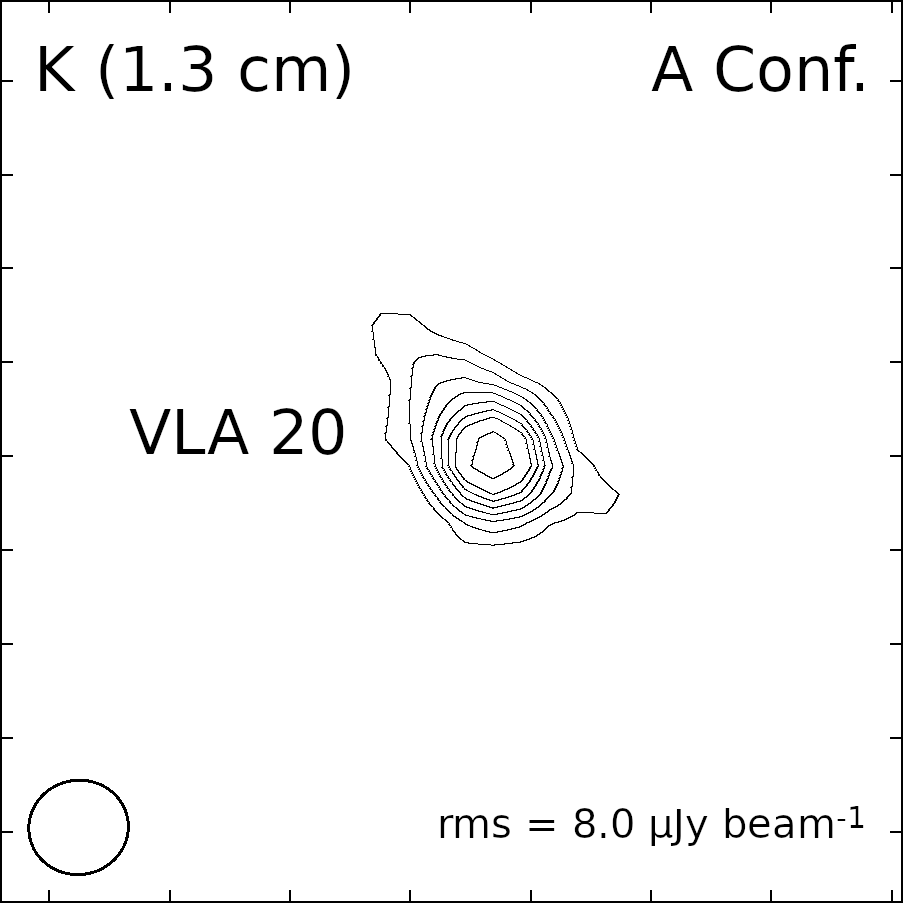}\hfill\\
\vspace*{1.5em}
\includegraphics[height=0.34\textheight]{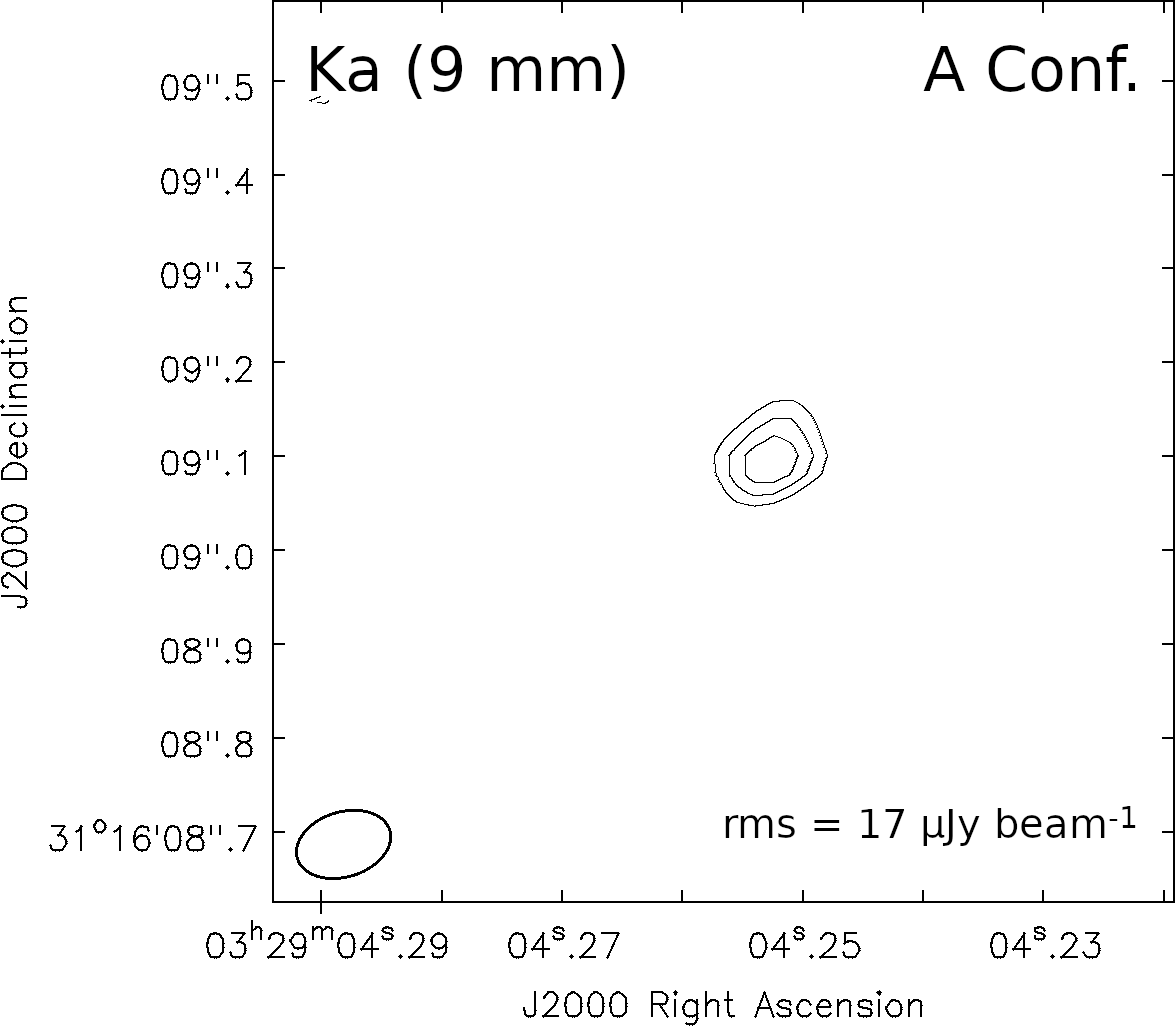}\hspace*{1.5em}\includegraphics[height=0.34\textheight]{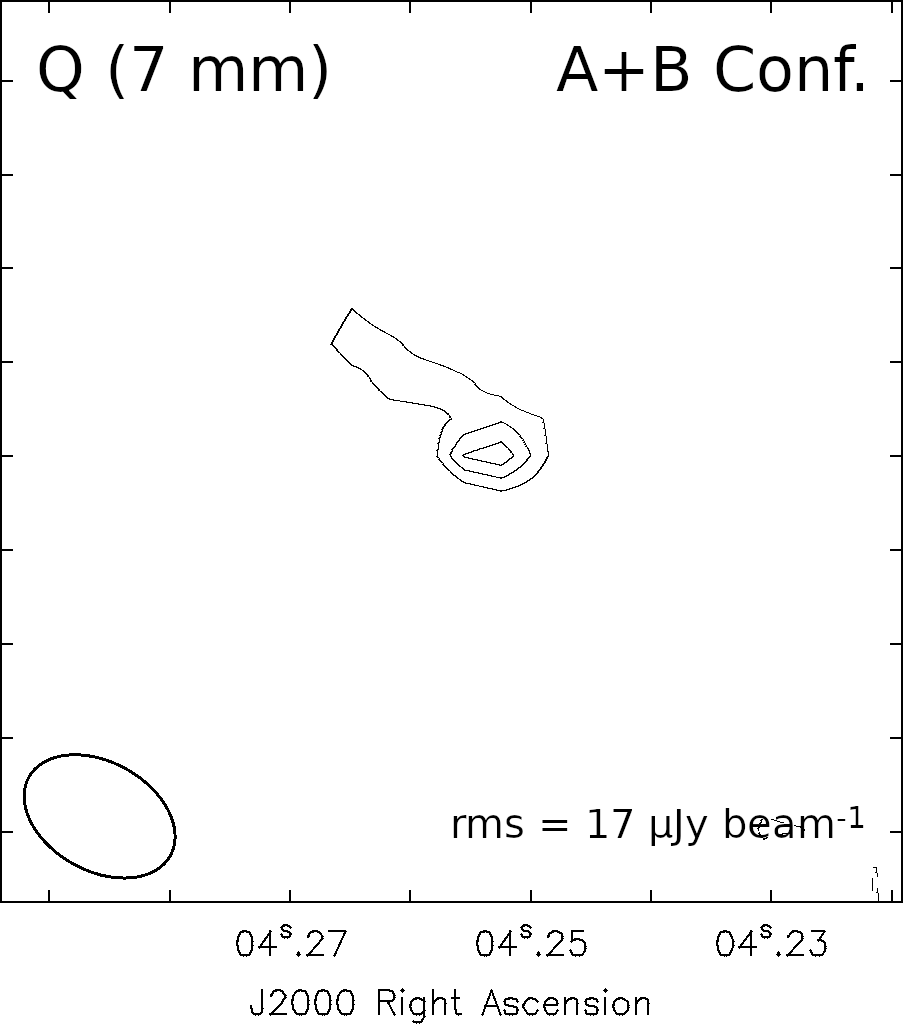}\hfill
\caption{Same as Fig.~\ref{fig:vla3} but for source VLA 20. Contour levels are $-$3, 3, 4, 5, 6, 7, 8, 9, and 11 times the rms of each image. 
\label{fig:vla20}}
\end{center}
\end{figure}

\begin{figure}[h]
\begin{center}
\includegraphics[height=0.3\textheight]{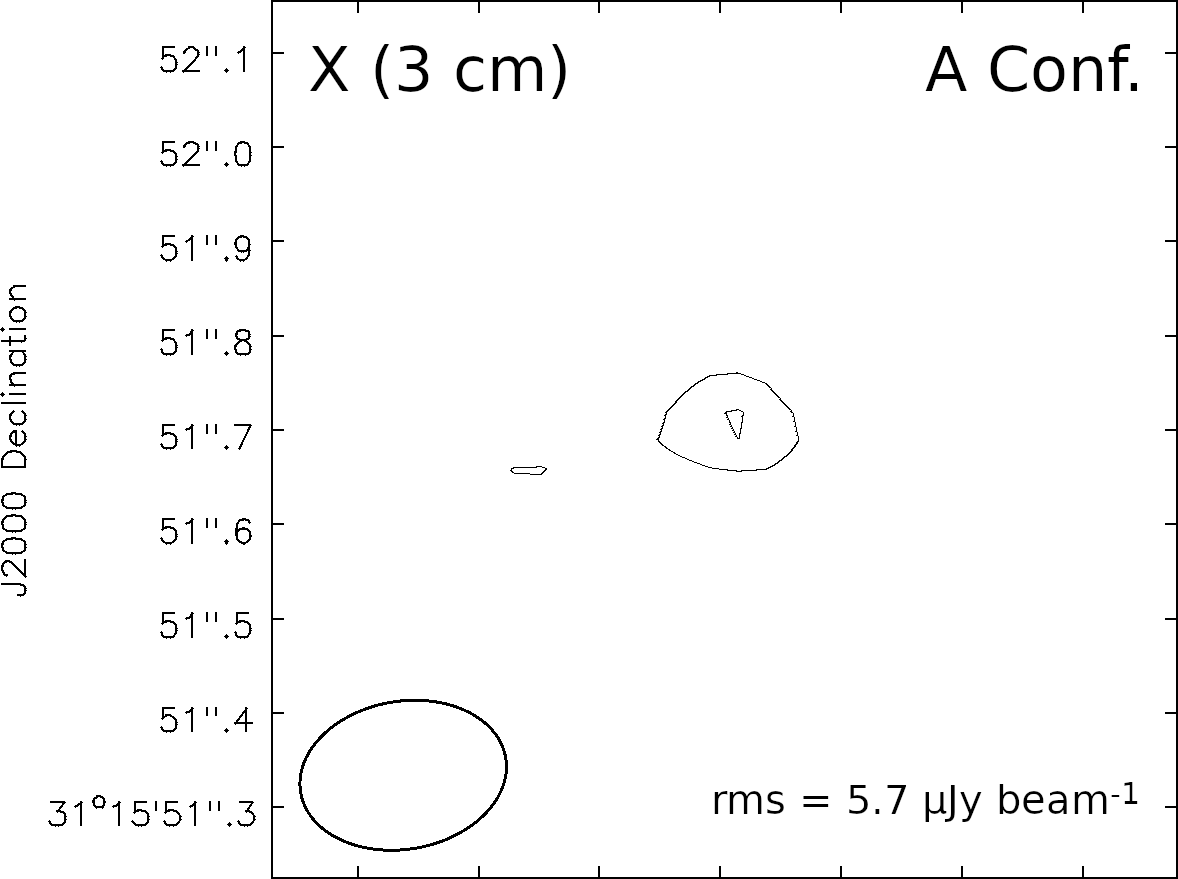}\hspace*{1.5em}\includegraphics[height=0.3\textheight]{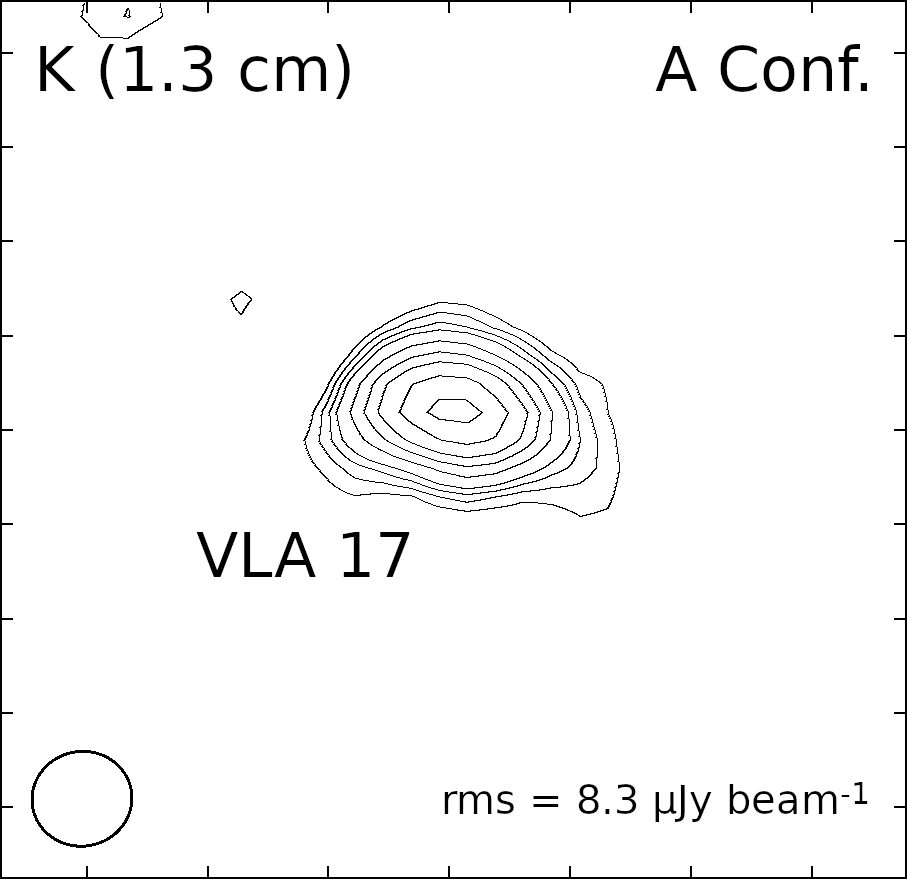}\hfill\\
\vspace*{1.5em}
\includegraphics[height=0.34\textheight]{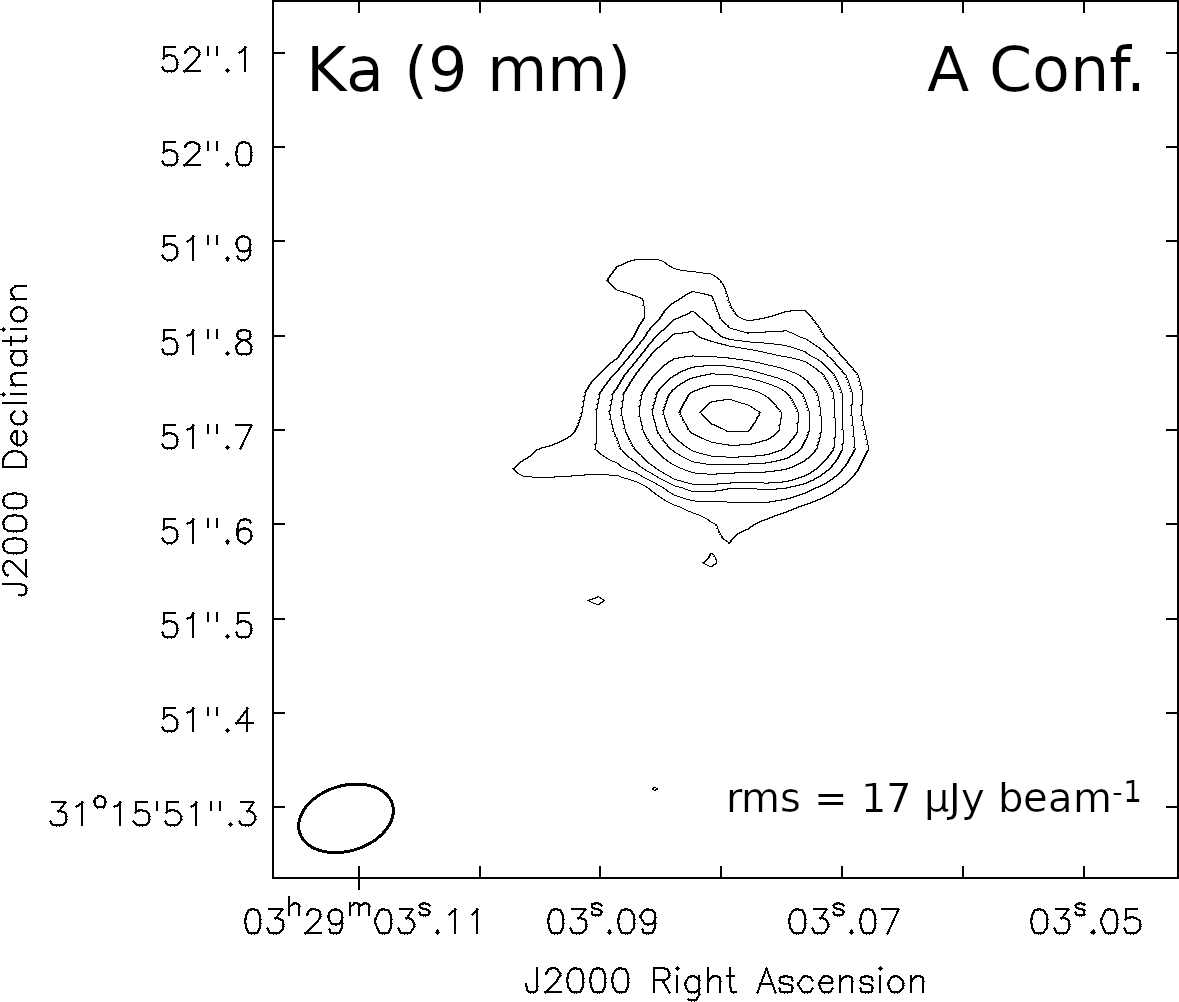}\hspace*{1.5em}\includegraphics[height=0.34\textheight]{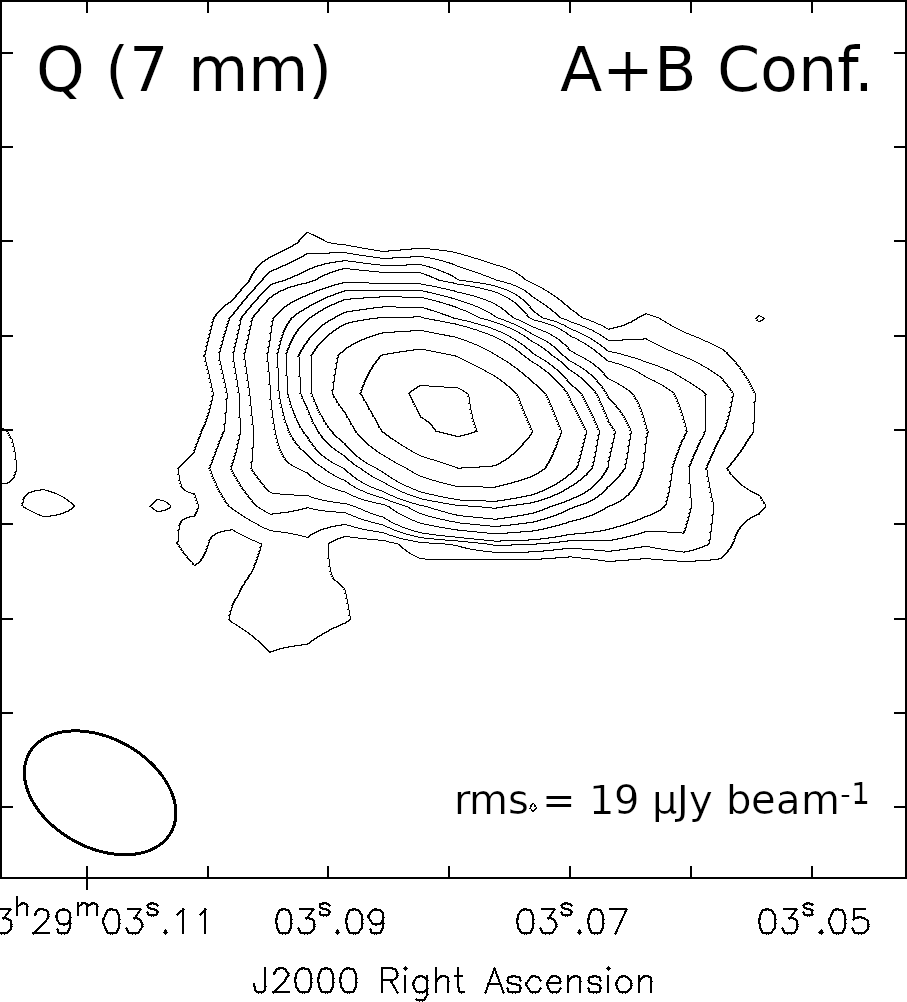}\hfill
\caption{Same as Fig.~\ref{fig:vla3} but for source VLA 17. Contour levels are $-$3, 3, 4, 5, 6, 8, 10, 12, 15, 18, 25, 35, and 50 times the rms of each image.
\label{fig:vla17}}
\end{center}
\end{figure}

\begin{figure}[h]
\begin{center}
\includegraphics[height=0.3\textheight]{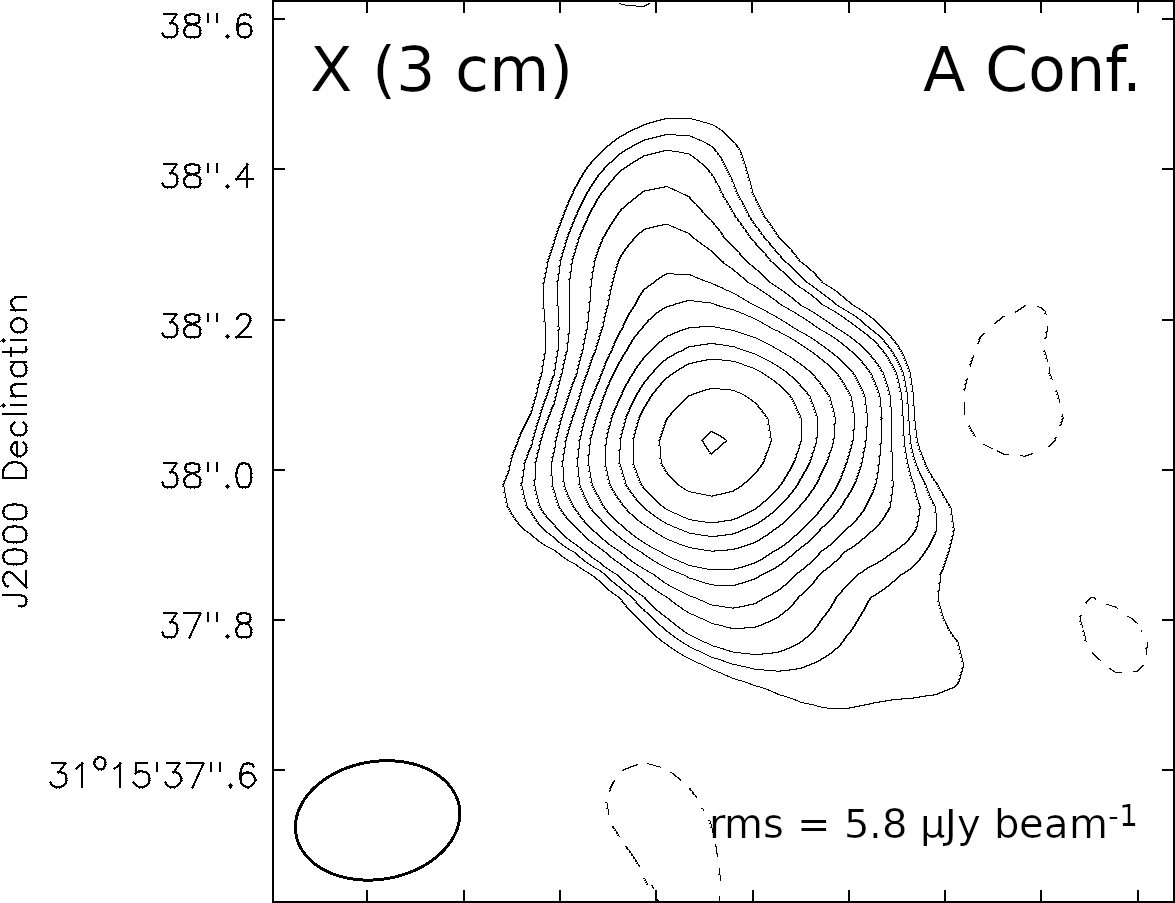}\hspace*{1.5em}\includegraphics[height=0.3\textheight]{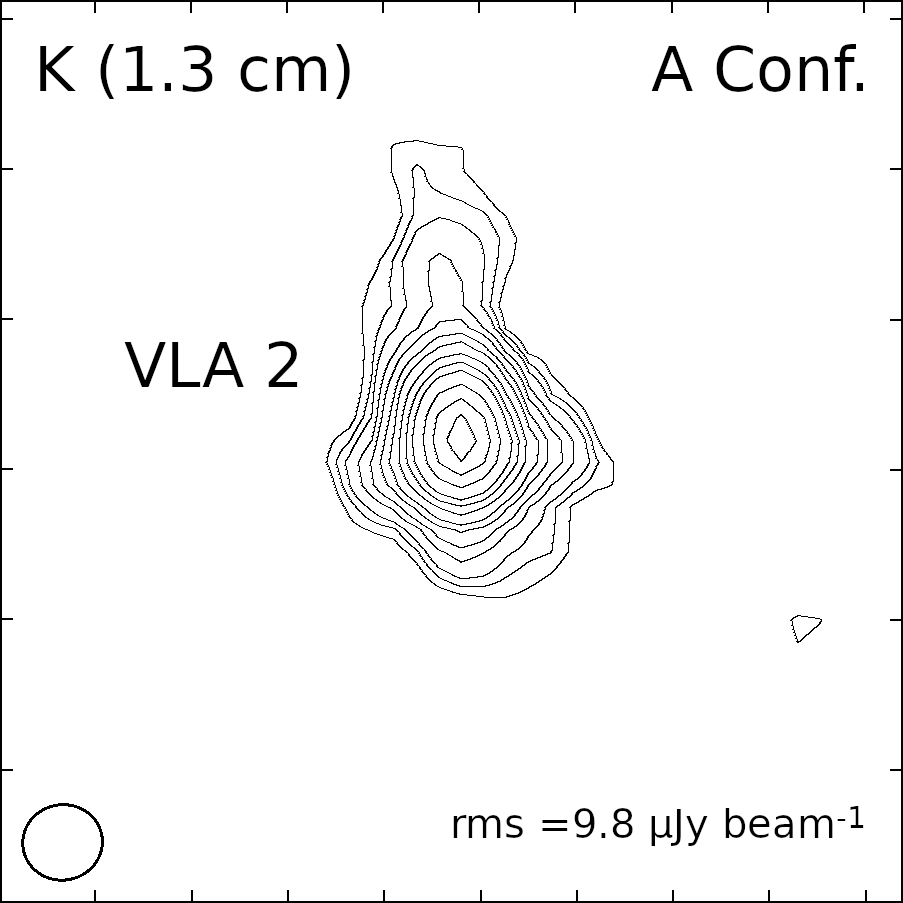}\hfill\\
\vspace*{1.5em}
\includegraphics[height=0.34\textheight]{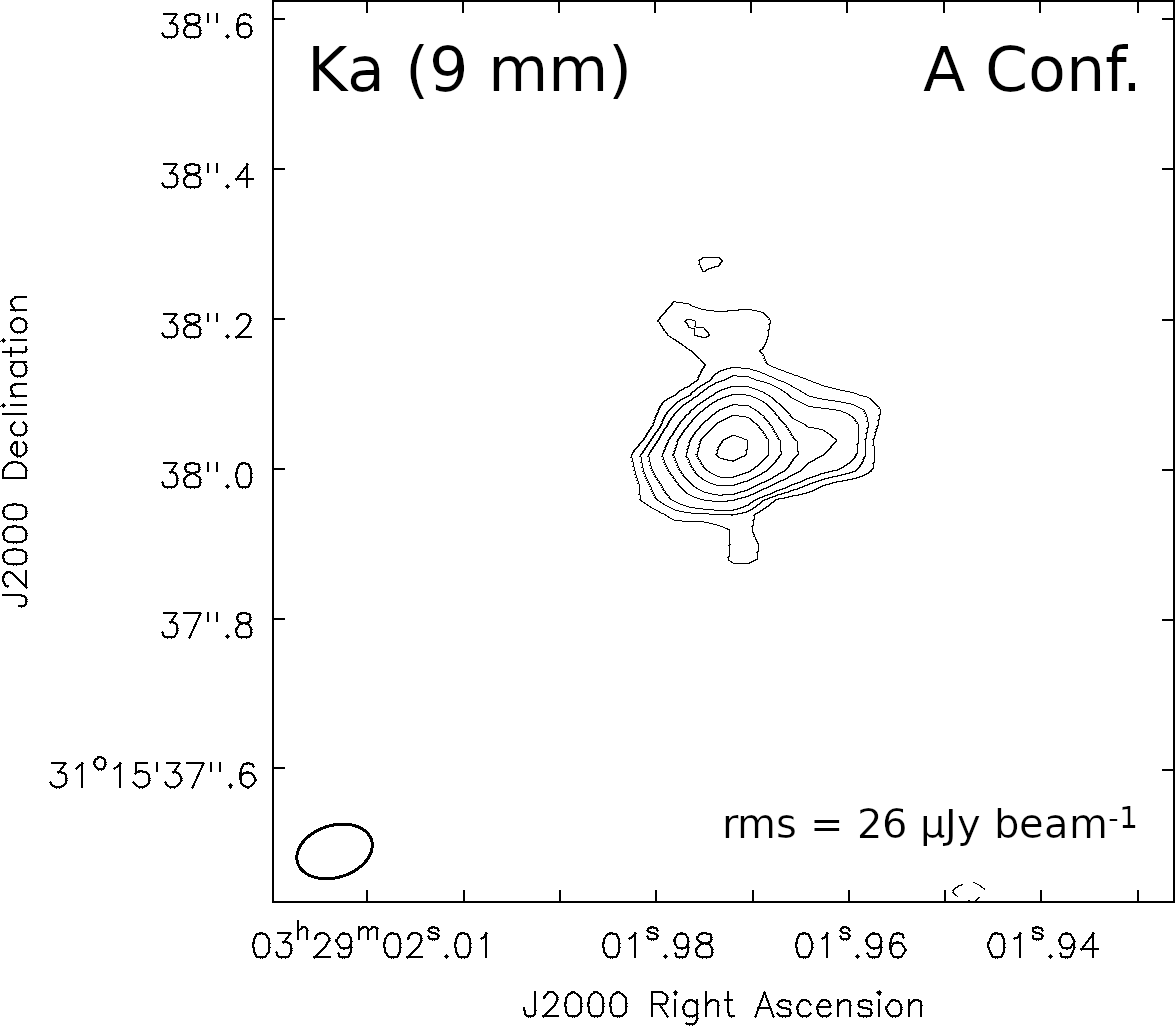}\hspace*{1.5em}\includegraphics[height=0.34\textheight]{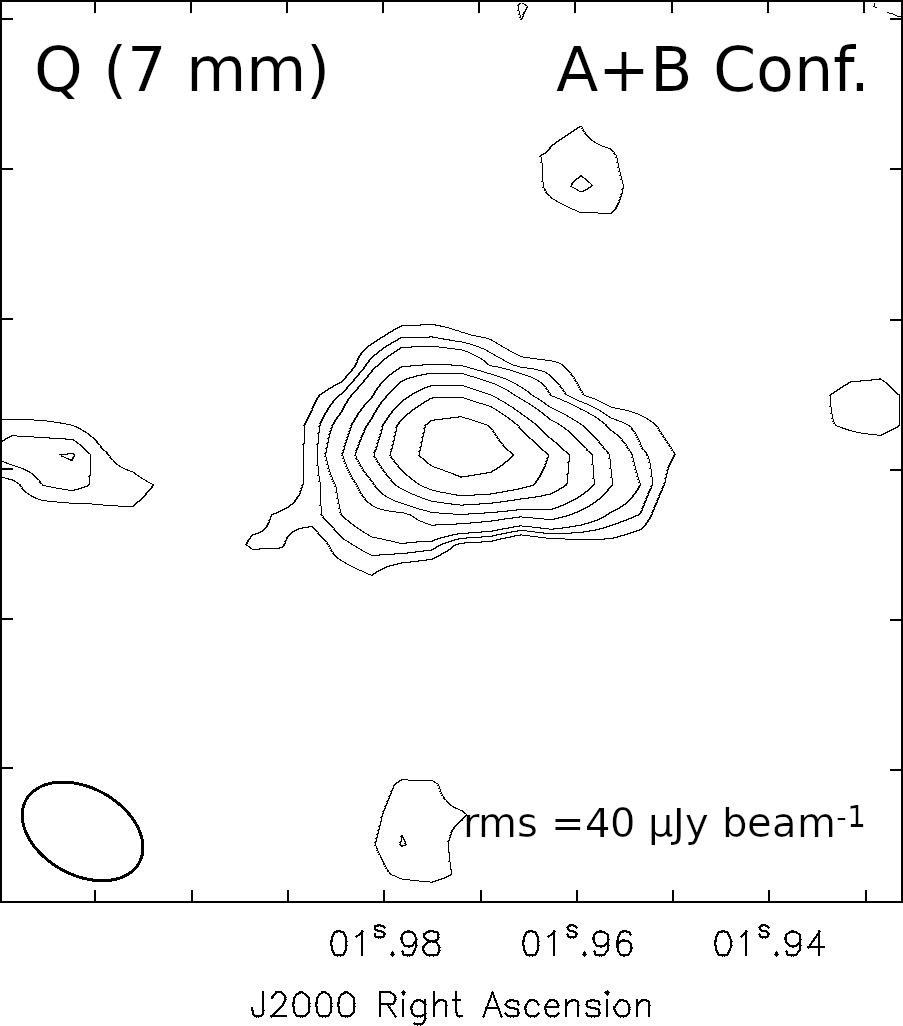}\hfill
\caption{Same as Fig.~\ref{fig:vla3} but for source VLA 2. Contour levels are $-$3, 3, 4, 5, 7, 9, 12, 15, 20, 25, 30, 40, 50, and 60 times the rms of each image. 
\label{fig:vla2}}
\end{center}
\end{figure}

\begin{figure}[h]
\begin{center}
\includegraphics[height=0.2\textheight]{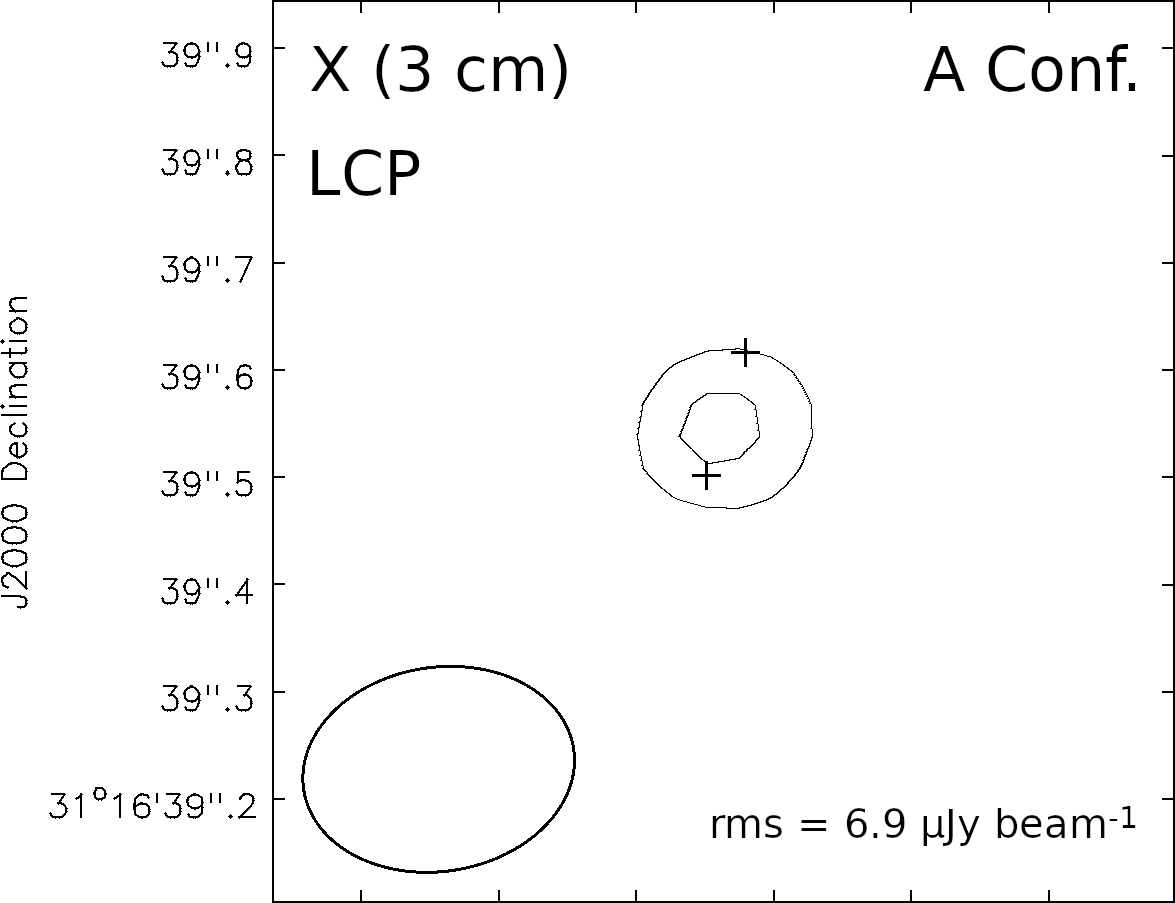}\hspace*{1.5em}\includegraphics[height=0.2\textheight]{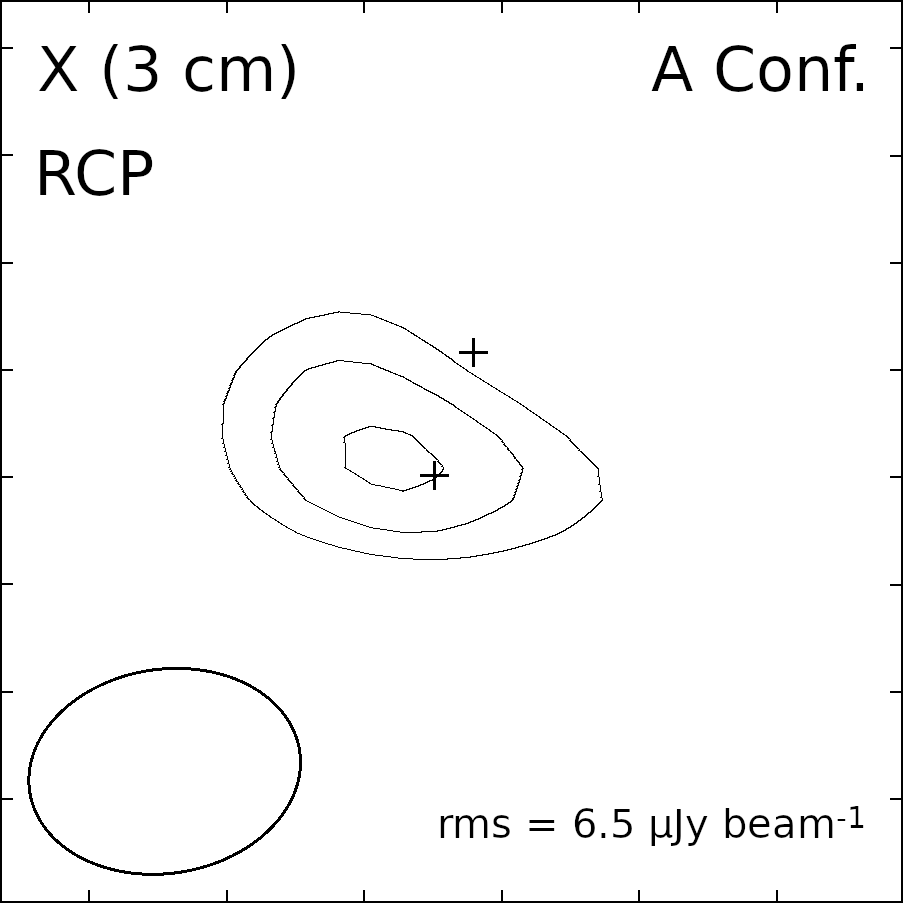}\hfill\\
\vspace*{1.0em}
\includegraphics[height=0.2\textheight]{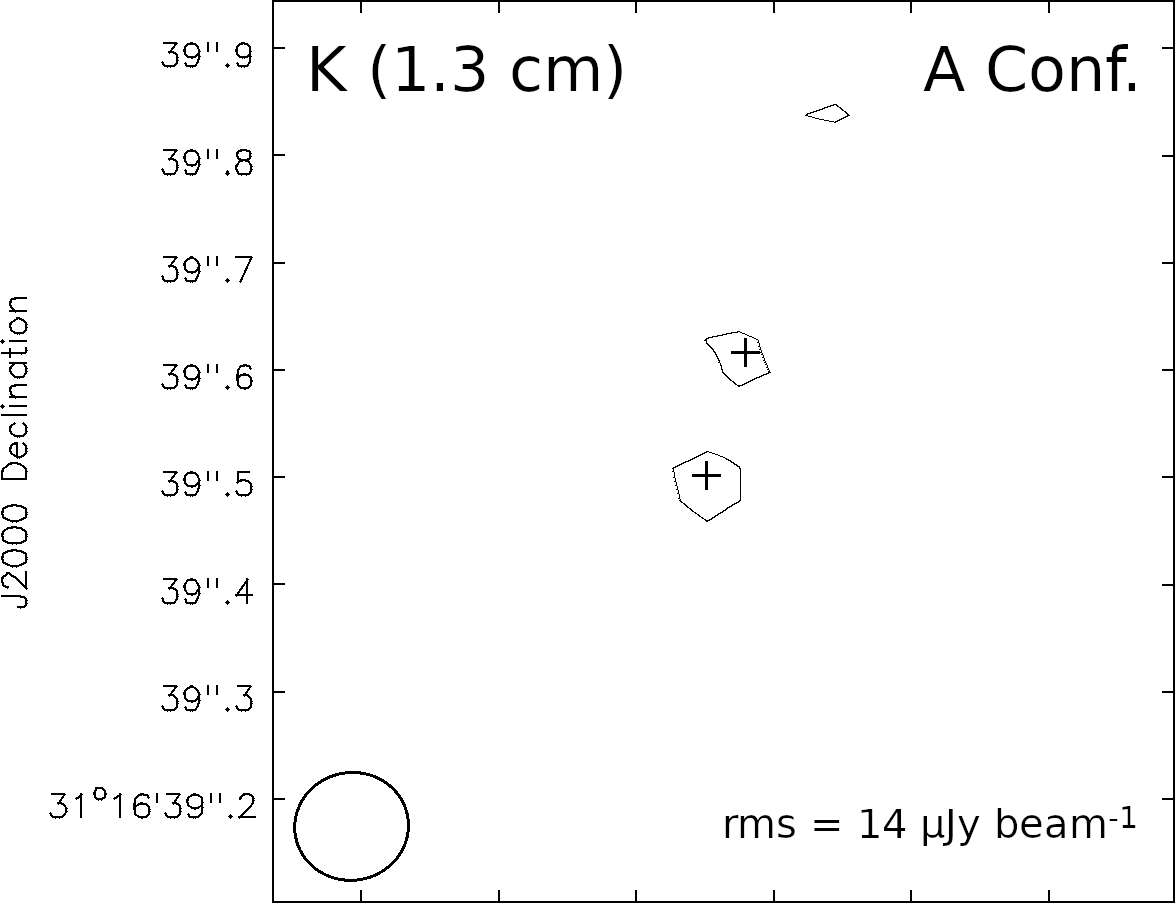}\hspace*{1.5em}\includegraphics[height=0.2\textheight]{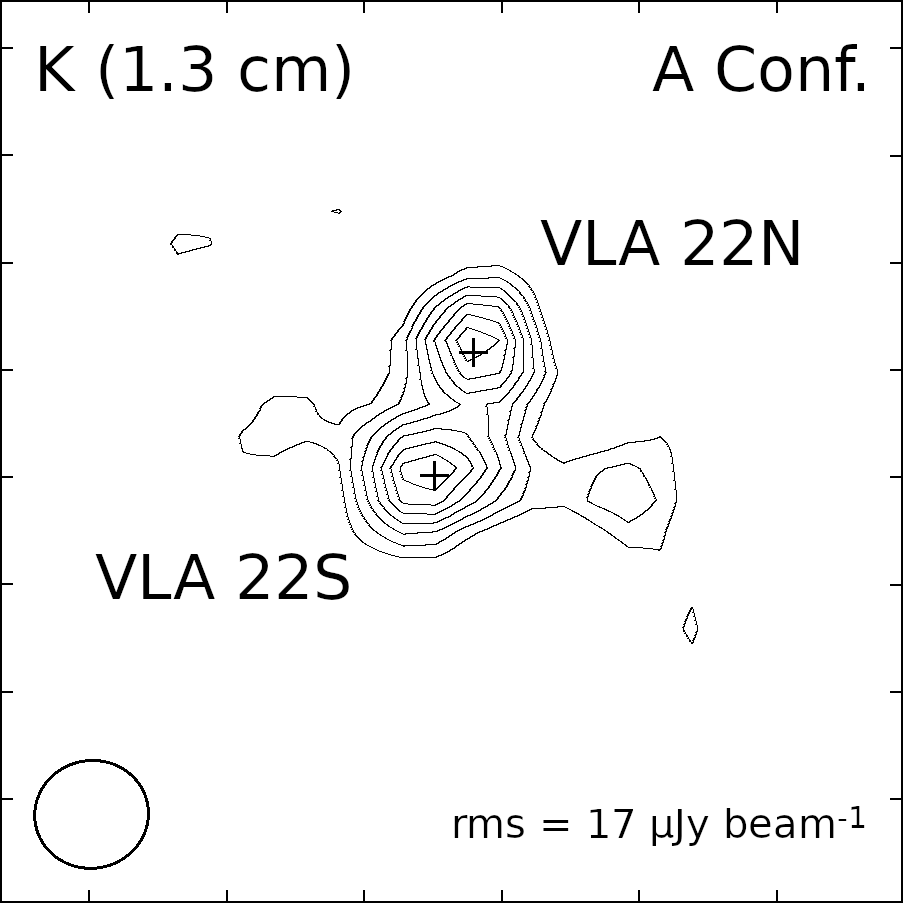}\hfill\\
\vspace*{1.0em}
\includegraphics[height=0.2\textheight]{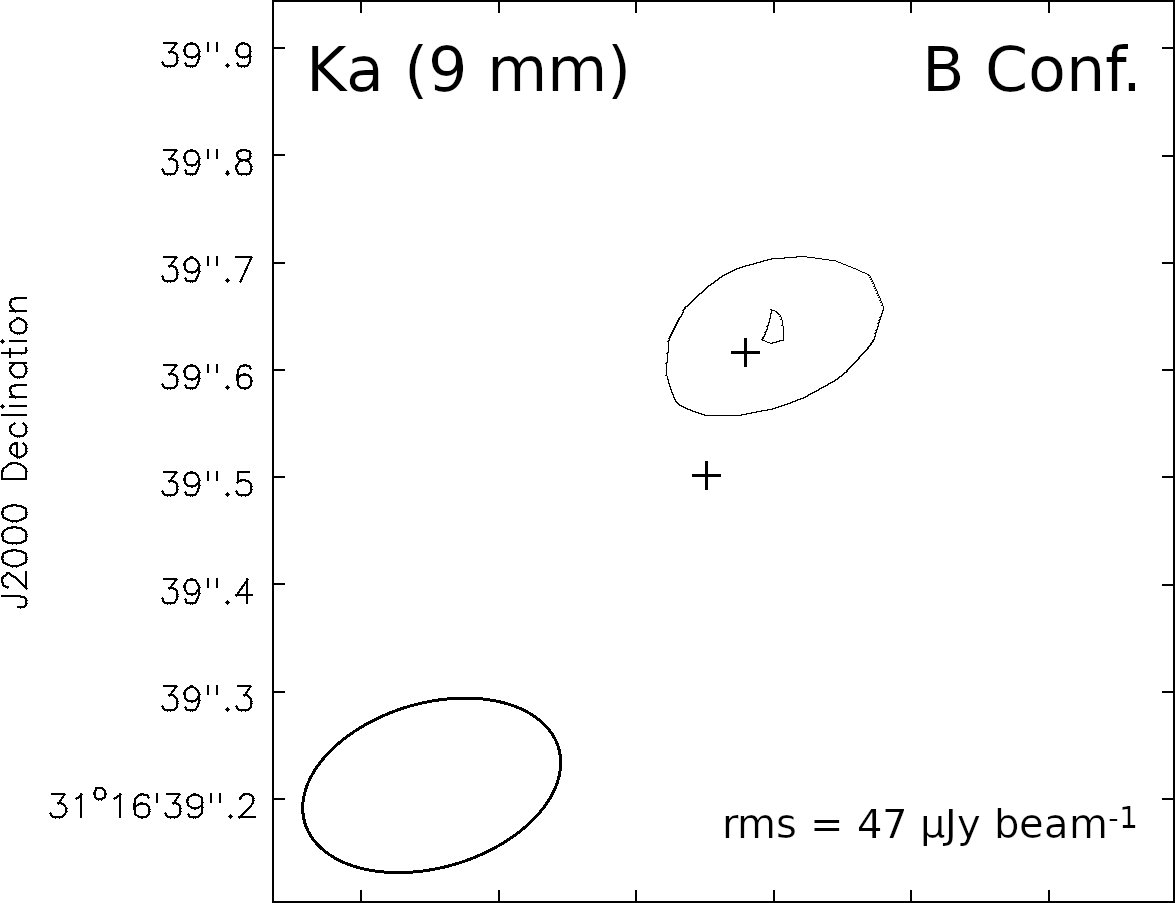}\hspace*{1.5em}\includegraphics[height=0.2\textheight]{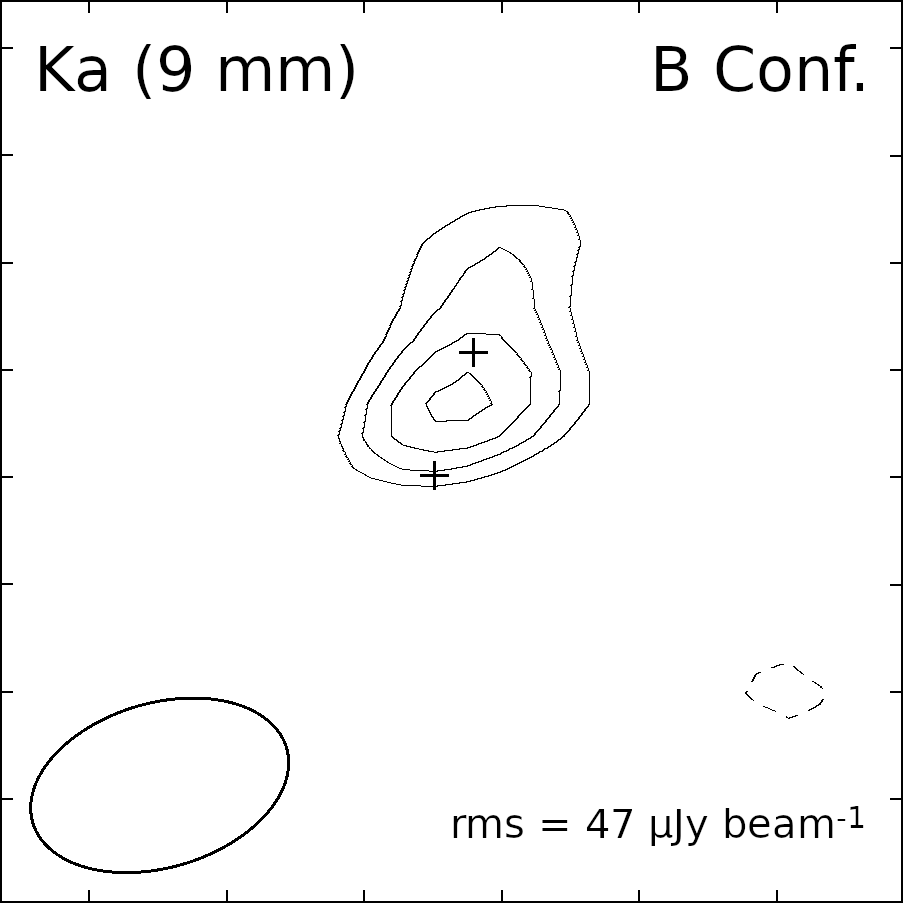}\hfill\\
\vspace*{1.0em}
\includegraphics[height=0.228\textheight]{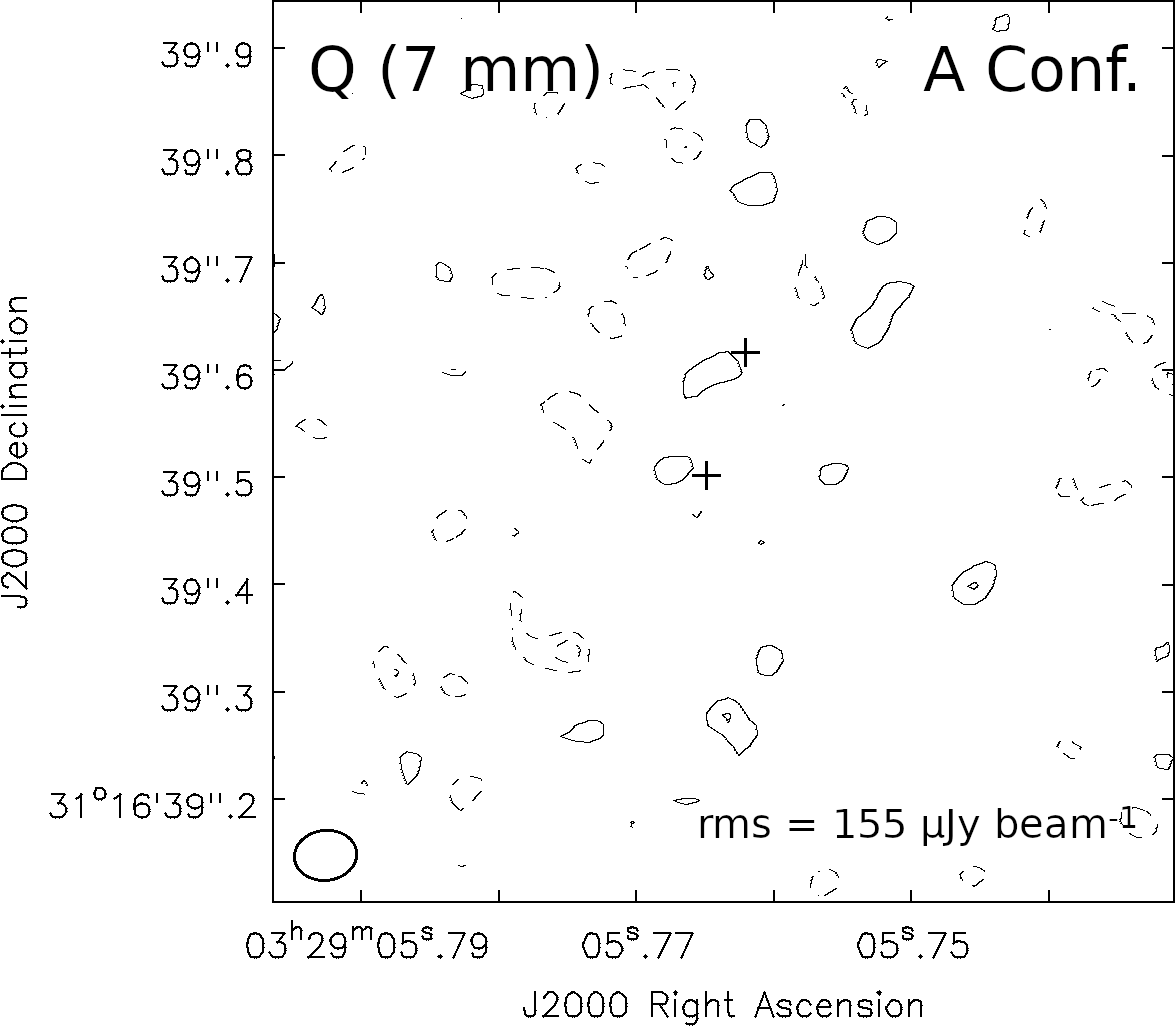}\hspace*{1.5em}\includegraphics[height=0.228\textheight]{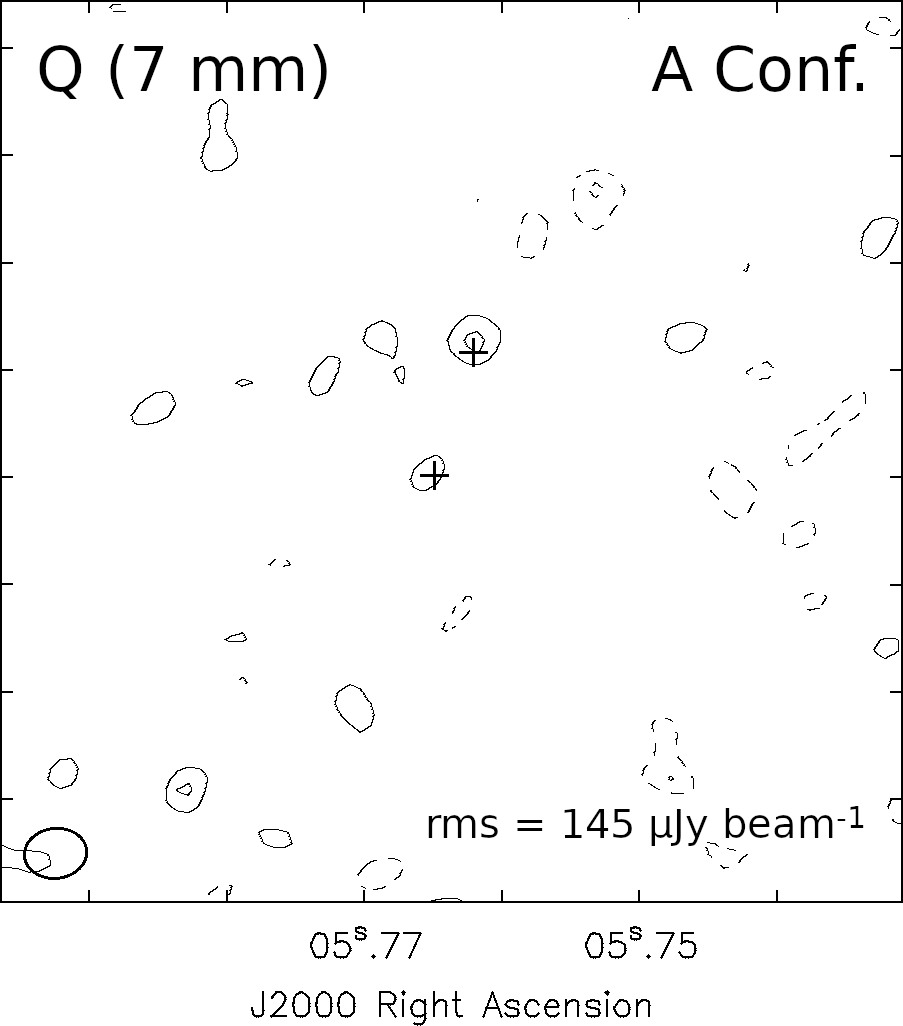}\hfill
\caption{VLA images of source VLA 22 in left (LCP; left column) and right circular polarization (RCP; right column). 
 The total intensity can be obtained as $I=(LCP+RCP)/2$. Contour levels are $-$3, 3, 4, 5, 6, 7, 8, and 9 times the average rms of the RCP and LCP images, 6.7 $\mu$Jy beam$^{-1}$ (X band), 15.5 $\mu$Jy beam$^{-1}$ (K band), and 47 $\mu$Jy beam$^{-1}$ (Ka band); and $-$3, $-$2, 2, and 3 times 150 $\mu$Jy beam$^{-1}$ for the Q band image. Plus signs indicate the positions of VLA 22N and VLA 22S, obtained from the K-band image (Table~\ref{tab:posSources}). We used Briggs weighting with the robust parameter set to 0.5 and set to 0 for the X- and Ka-band images, respectively, and natural weighting for the remaining images. The synthesized beams are $0\farcs20\times0\farcs14$, PA = $-81.3\degr$ (X band); $0\farcs11\times0\farcs10$, PA = $-80.7\degr$ (K band); $0\farcs25\times0\farcs15$, PA = $-73.6\degr$ (Ka band); and $0\farcs06\times0\farcs05$, PA = $-83.4\degr$ (Q band).
\label{fig:vla22}}
\end{center}
\end{figure}

Figures \ref{fig:vla3}, \ref{fig:vla20}, \ref{fig:vla17}, \ref{fig:vla2}, and \ref{fig:vla22} show, respectively, the VLA images of sources VLA 3, VLA 20, VLA 17, VLA 2, and VLA 22 (Table~\ref{tab:posSources}) in order of increasing (projected) distance to SVS 13.

\clearpage

\clearpage

\clearpage

\newpage

\section{Spectra and Identification of Molecular Transitions}\label{ap:spectra}

\restartappendixnumbering

\begin{figure*}[h]
\begin{center}
\includegraphics[height=0.25\textheight]{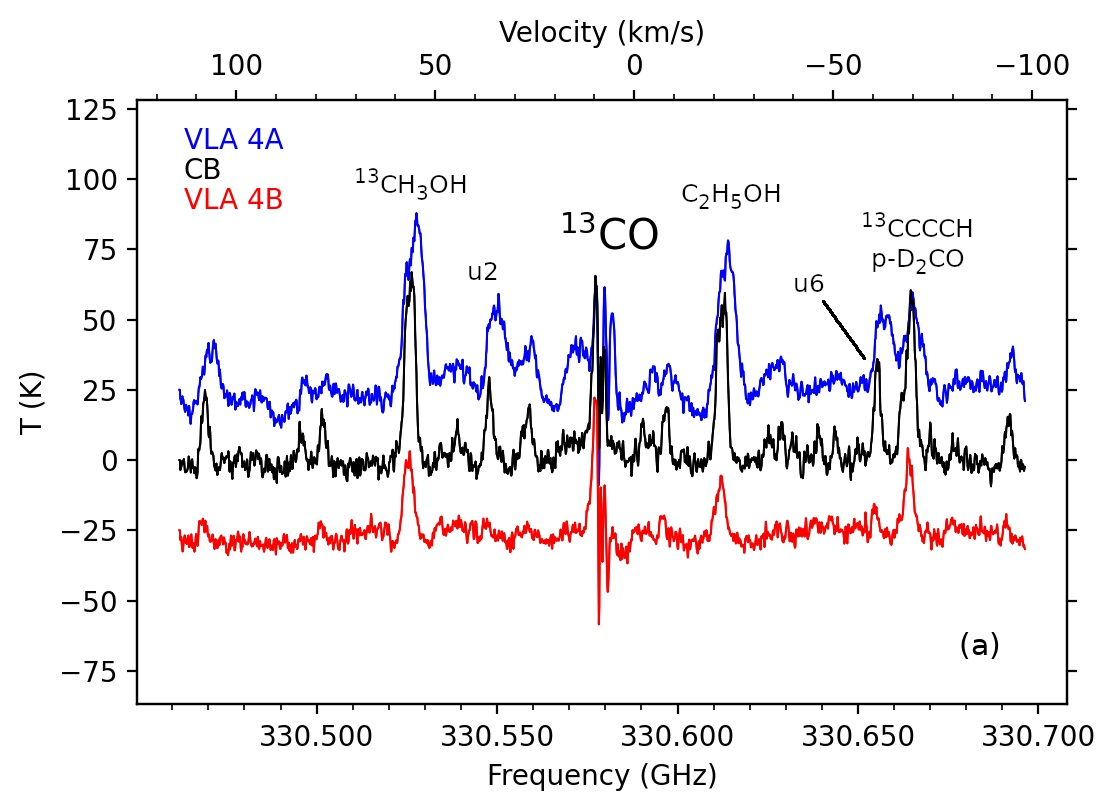}\hspace*{1.5em}\includegraphics[height=0.25\textheight]{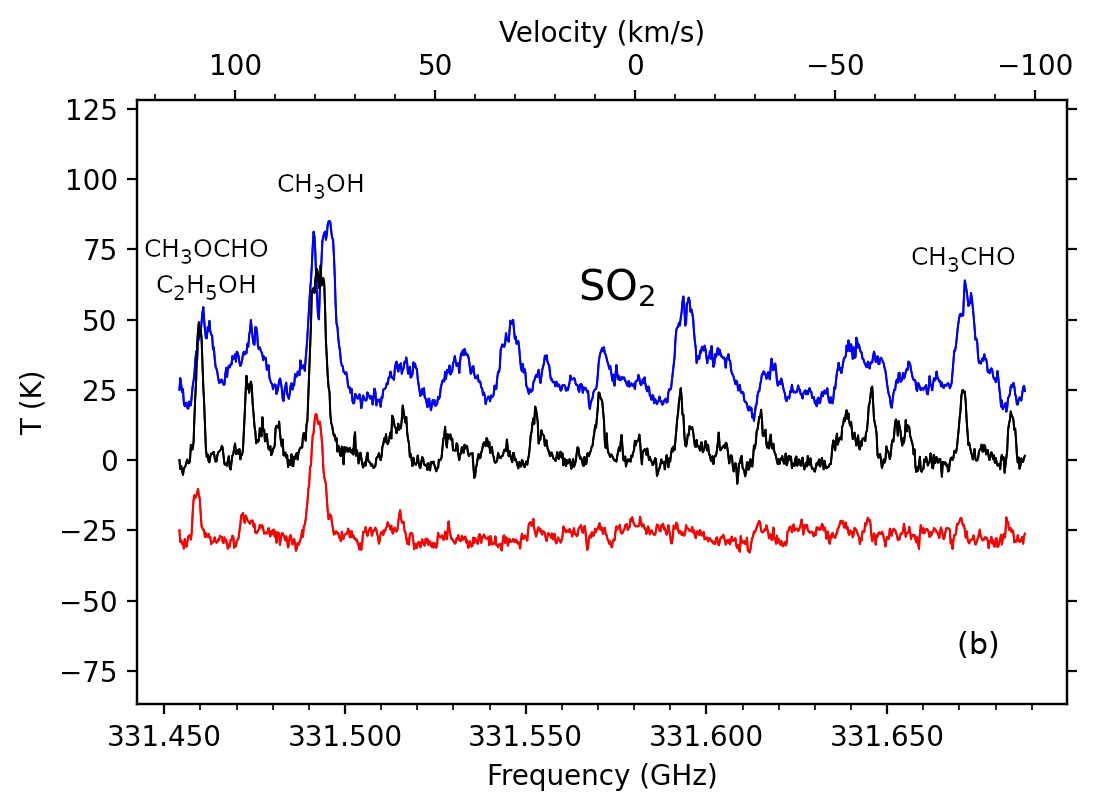}\hfill\\
\vspace*{1.0em}
\includegraphics[height=0.25\textheight]{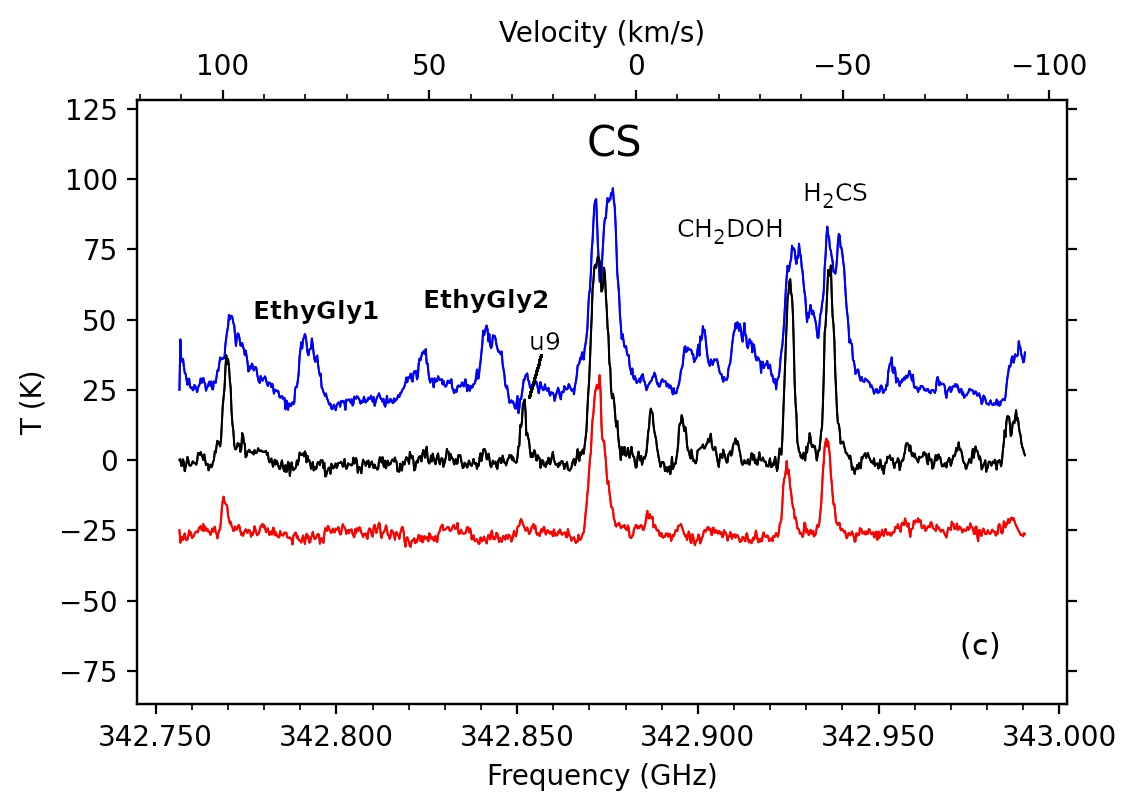}\hspace*{1.5em}\includegraphics[height=0.25\textheight]{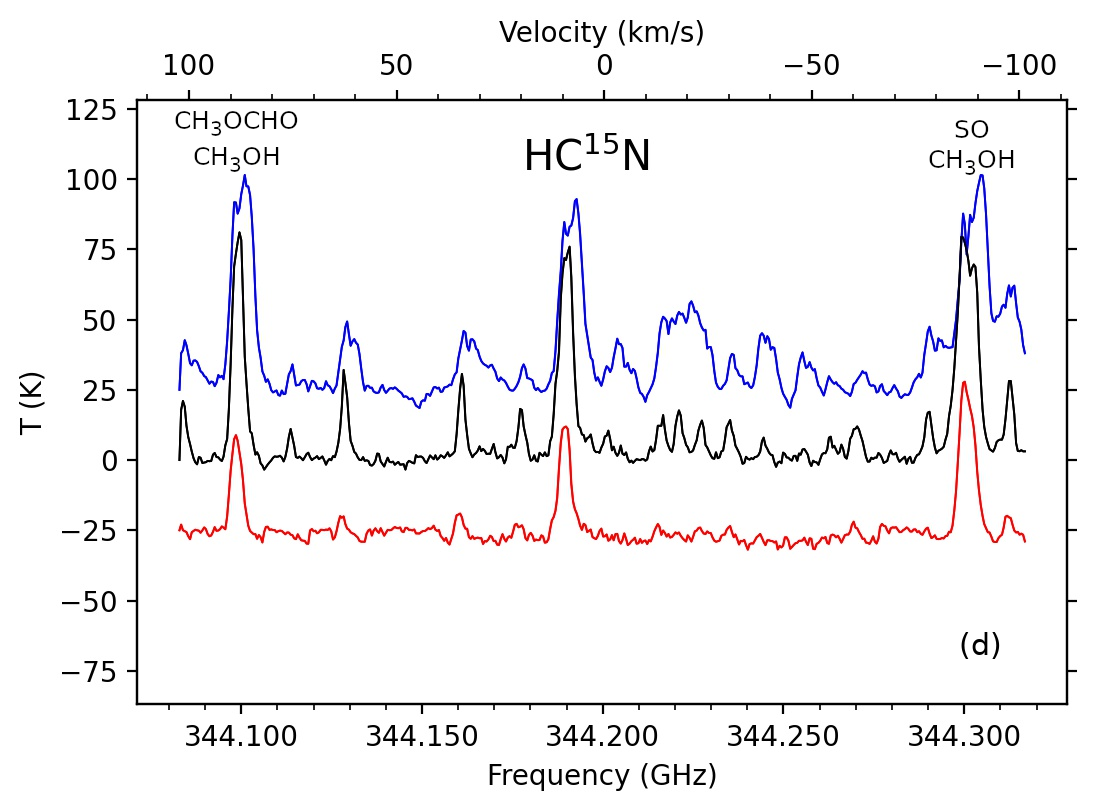}\hfill\\
\vspace*{1.0em}
\includegraphics[height=0.25\textheight]{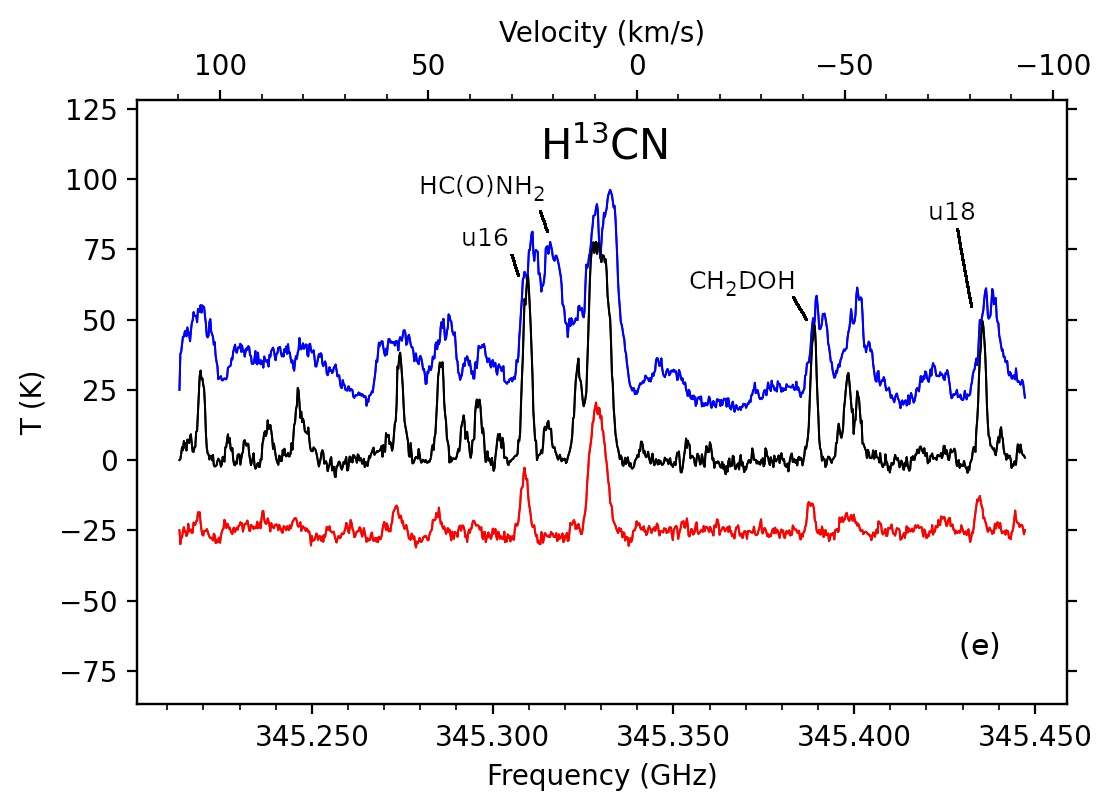}\hspace*{1.5em}\includegraphics[height=0.25\textheight]{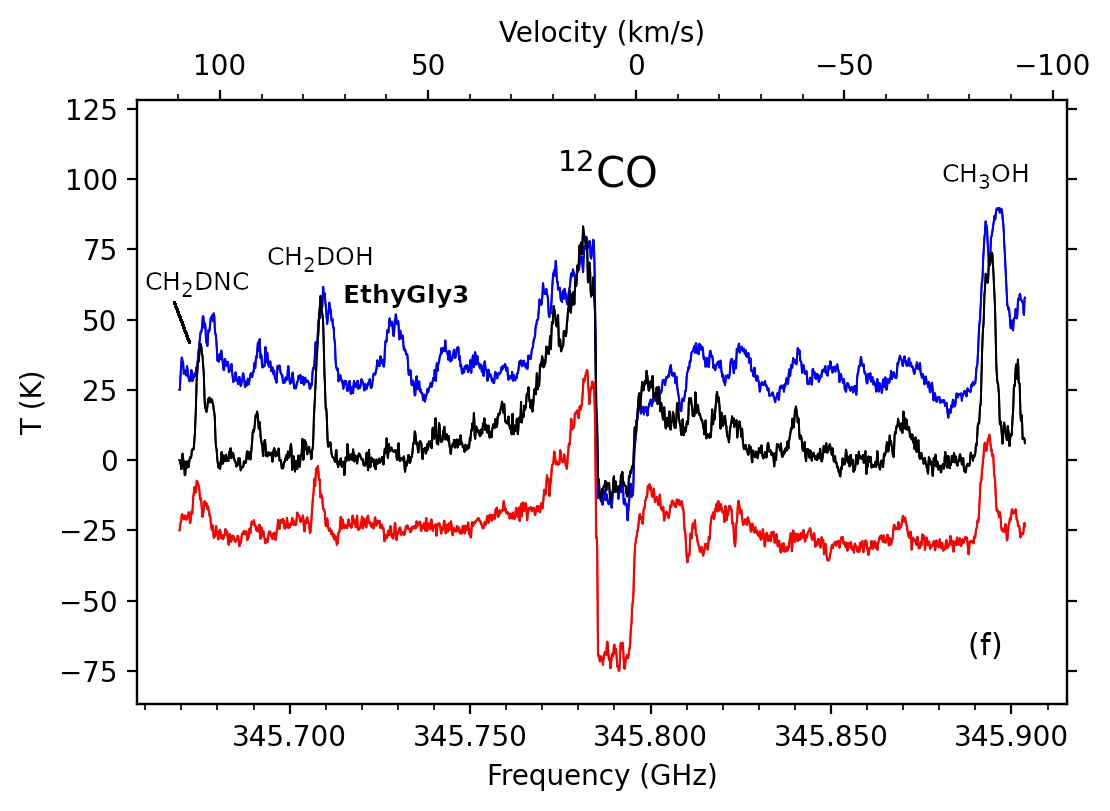}\hfill
\caption{Spectra taken in beam-size ($0\farcs16\times0\farcs08$, PA = 0$\degr$) regions toward VLA 4A (blue), VLA 4B (red), and the geometric center of the components (black). VLA 4A and VLA 4B spectra are offset by 25 K and $-$25 K, respectively, for illustrative purposes. The bottom x-axis shows the observed frequency, and the top x-axis shows the velocity referred to the frequency of $^{13}$CO (3-2) (a), SO$_2$ (11-12) (b), CS (7-6) (c), HC$^{15}$N (4-3) (d), H$^{13}$CN (4-3) (e), and $^{12}$CO (3-2) (f). The most prominent lines have been labeled; the ethylene glycol transitions associated only with VLA 4A (Fig.~\ref{fig:EGly}) are labeled following the Appendix~\ref{sec:keplerian} nomenclature.}
\label{fig:spectra}
\end{center}
\end{figure*}

Figure \ref{fig:spectra} shows the observed spectra toward the positions of VLA 4A, VLA 4B and the central position (CB). In Table \ref{tab:lines} we list the identified molecular transitions.

\newpage

\startlongtable
\centerwidetable
\begin{deluxetable*}{lcccccl}
\tabletypesize{\scriptsize}
\tablewidth{0.999\textwidth}
\tablecaption{Molecular Transitions Detected toward SVS 13\tablenotemark{a}\label{tab:lines}}
\tablehead{
\colhead{} 
& \colhead{} 
& \colhead{} 
& \colhead{Rest Frequency\tablenotemark{b}} 
& \colhead{E$_{u}/k$} 
& \colhead{A$_{ul}$} 
& \colhead{} \\
\colhead{Species} 
& \colhead{Formula} 
& \colhead{Quantum Numbers} 
& \colhead{(MHz)} 
& \colhead{(K)} 
& \colhead{(s$^{-1}$)} 
& \colhead{Notes\tablenotemark{c}}
 }
\startdata
Acetaldehyde        & CH$_3$CHO, vt=0 & 5 3 2 2 - 4 2 2 2     & 330566.705 & 34.26   & 1.15E-4 & nLTE \\
                    &                 & 17 1 16 3 - 16 1 15 3 & 330580.420 & 352.489 & 1.25E-3 &  \\ 
                    &                 & 17 1 16 2 - 16 1 15 2 & 331602.178 & 146.645 & 1.25E-3 & \\ 
                    &                 & 17 1 16 0 - 16 1 15 0 & 331680.380 & 146.6   & 1.25E-3 & \\ 
                    & CH$_3$CDO, vt=0 
                                      & 17 2 15 0 - 16 2 14 0 & 331490.785 & 149.161 & 1.28E-3 & CB \\\hline 
Acetone  & CH$_3$COCH$_3$, v=0        & 17 15 3 1 - 16 14 2 1 & 330506.310 & 138.132 & 1.28E-3 & \\ 
                &                     & 17 15 2 1 - 16 14 3 1 & 330512.291 & 138.132 & 1.28E-3 &  bl\\ 
                &                     & 17 15 3 1 - 16 14 3 2 & 330549.327 & 138.023 & 1.28E-3 & \\ 
                &                     & 15 8 7 0 - 14 7 8 1   & 331525.975 & 95.682  & 1.30E-4 &  bl\\ 
                &                     & 17 17 1 0 - 16 16 0 0 & 342780.035 & 147.049 & 1.85E-3 &  \\ 
                &                     & 17 17 0 0 - 16 16 1 0 & 342780.036 & 147.049 & 1.85E-3 &  \\ 
                &                     & 41 3 39 0 - 41 2 40 1 & 342836.464 & 447.674 & 1.53E-4 & bl, db \\ 
                &                     & 41 2 39 0 - 41 1 40 1 & 342836.464 & 447.674 & 1.53E-4 & bl, db \\ 
                &                     & 18 12 6 1 - 17 11 7 1 & 342836.912 & 141.302 & 4.35E-4 & bl, db \\ 
                &                     & 18 12 6 0 - 17 11 7 0 & 342864.593 & 141.23  & 4.41E-4 & $<$5$\sigma$ \\ 
                &                     & 18 12 6 0 - 17 11 7 1 & 342896.394 & 141.271 & 4.35E-4 & bl\\ 
                &                     & 18 15 4 0 - 17 14 3 0 & 345256.037 & 150.827 & 1.29E-3 & bl\\
                &                     & 18 15 3 0 - 17 14 4 0 & 345305.365 & 150.828 & 1.29E-3 & \\
                &                     & 24 12 12 1 - 23 13 11 1 & 345875.586 & 235.239 & 7.51E-4 & bl, $<$5$\sigma$ \\ 
                &                     & 24 12 12 1 - 23 13 11 2 & 345877.896 & 235.239 & 7.52E-4 & bl, $<$5$\sigma$ \\ \hline 
Butadiynyl      & $^{13}$CCCCH        & 36 36.5 37 36.5 - 35 35.5 36 35.5 & 330637.439 & 293.629 & 1.57E-4 & bl\\ 
                &                     & 36 36.5 37 37.5 - 35 35.5 36 36.5 & 330637.439 & 293.628 & 1.57E-4 & bl\\ 
                &                     & 36 36.5 36 36.5 - 35 35.5 35 35.5 & 330638.331 & 293.618 & 1.57E-4 & bl\\ 
                &                     & 36 36.5 36 35.5 - 35 35.5 35 34.5 & 330638.355 & 293.618 & 1.57E-4 & bl\\ 
                &                     & 36 35.5 36 35.5 - 35 34.5 35 34.5 & 330672.730 & 293.685 & 1.57E-4 & bl\\ 
                &                     & 36 35.5 36 36.5 - 35 34.5 35 35.5 & 330672.754 & 293.685 & 1.57E-4 & bl\\ 
                &                     & 36 35.5 35 34.5 - 35 34.5 34 33.5 & 330673.647 & 293.692 & 1.57E-4 & bl\\ 
                &                     & 36 35.5 35 35.5 - 35 34.5 34 34.5 & 330673.647 & 293.693 & 1.57E-4 & bl\\\hline 
Carbon Monoxide & CO, v=0             & 3 - 2                 & 345795.990 & 33.192  & 2.50E-6 & \\ 
                & $^{13}$CO, v=0      & 3 - 2                 & 330587.965 & 31.732  & 2.19E-6 & \\ \hline 
Carbon Monosulfide & CS, v=0          & 7 0 - 6 0             & 342882.850 & 65.828  & 8.40E-4 & \\ \hline 
Ethanol & C$_2$H$_5$OH, v=0 & 19 4 16 0 - 18 4 15 0 & 330478.629 & 234.672 & 3.23E-4 & \\ 
        &                   & 14 5 9 1 - 14 4 11 0  & 330599.812 & 179.62  & 1.04E-4 & bl\\ 
        &                   & 20 0 20 2 - 19 1 19 2 & 330621.123 & 169.18  & 3.82E-4 & \\ 
        &                   & 9 6 3 0 - 8 5 3 1     & 331469.617 & 138.161 & 1.61E-4 & \\ 
        &                   & 9 6 4 0 - 8 5 4 1     & 331469.664 & 138.161 & 1.61E-4 & bl\\ 
        &                   & 11 5 7 1 - 11 4 7 0   & 331648.876 & 147.183 & 1.02E-4 & bl\\ 
        &                   & 11 3 8 2 - 10 2 9 2   & 331655.211 & 66.945  & 1.26E-4 & \\ 
        &                   & 21 1 21 0 - 20 1 20 0 & 345229.285 & 241.552 & 3.71E-4 & bl\\ 
        &                   & 21 1 21 1 - 20 1 20 1 & 345295.355 & 246.22  & 3.67E-4 & \\ 
        &                   & 22 3 19 1 - 21 4 17 0 & 345311.658 & 286.007 & 7.58E-5 & \\ 
        &                   & 21 0 21 0 - 20 0 20 0 & 345333.442 & 241.539 & 3.72E-4 & \\ 
        &                   & 21 0 21 1 - 20 0 20 1 & 345408.165 & 246.207 & 3.67E-4 & bl \\ \hline 
Ethyl Cyanide & C$_2$H$_5$CN, v=0 & 37 12 25 - 36 12 24 & 331484.515 & 462.06 & 2.77E-3 & bl\\ 
              &                   & 37 12 26 - 36 12 25 & 331484.515 & 462.06 & 2.77E-3 & bl\\ 
              &                   & 37 11 26 - 36 11 25 & 331487.287 & 436.589 & 2.83E-3 & \\ 
              &                   & 37 11 27 - 36 11 26 & 331487.287 & 436.589 & 2.83E-3 & \\ 
              &                   & 37 10 28 - 36 10 27 & 331523.317 & 413.325 & 2.88E-3 & bl\\ 
              &                   & 37 10 27 - 36 10 26 & 331523.317 & 413.325 & 2.88E-3 & bl\\ 
              &                   & 37 14 23 - 36 14 22 & 331549.416 & 519.581 & 2.66E-3 & CB, $<$5$\sigma$ \\ 
              &                   & 37 14 24 - 36 14 23 & 331549.416 & 519.581 & 2.66E-3 & CB, $<$5$\sigma$ \\ 
              &                   & 37 9 29 - 36 9 28 & 331605.594 & 392.28 & 2.92E-3 & bl \\ 
              &                   & 37 9 28 - 36 9 27 & 331605.594 & 392.28 & 2.92E-3 & bl\\ 
              &                   & 37 15 22 - 36 15 21 & 331608.406 & 551.61 & 2.59E-3 & bl\\
              &                   & 37 15 23 - 36 15 22 & 331608.406 & 551.61 & 2.59E-3 & bl\\
              &                   & 37 2 35 - 36 2 34 & 331662.287 & 310.75 & 3.08E-3 & \\ 
              &                   & 10 6 5 - 9 5 4 & 344278.941 & 63.667 & 2.03E-4 & \\ 
              &                   & 10 6 4 - 9 5 5 & 344278.941 & 63.667 & 2.03E-4 & \\ \hline 
Ethylene Glycol & aGg'-(CH$_2$OH)$_2$ & 22 6 17 0 - 21 5 17 1   & 330510.922 & 142.638 & 3.93E-5 & bl\\
                        &             & 18 8 10 0 - 17 7 10 1   & 330539.249 & 115.544 & 2.45E-5 & bl\\ 
                        &             & 18 8 11 0 - 17 7 11 1   & 330544.758 & 115.544 & 2.44E-5 & bl\\ 
                        &             & 33 18 15 0 - 32 18 14 1 & 331479.210 & 434.667 & 6.27E-4 & 4A, bl\\ 
                        &             & 33 18 16 0 - 32 18 15 1 & 331479.210 & 434.667 & 6.27E-4 & 4A, bl\\
                        &             & 35 3 33 0 - 34 3 32 1   & 331553.542 & 304.273 & 8.86E-4 & 4A, bl \\
                        &             & 35 2 33 0 - 34 2 32 1   & 331554.797 & 304.273 & 8.86E-4 & 4A, bl \\
                        &             & 33 17 16 0 - 32 17 15 1 & 331608.371 & 417.526 & 6.56E-4 & bl\\
                        &             & 33 17 17 0 - 32 17 16 1 & 331608.371 & 417.526 & 6.56E-4 & bl\\
                        &             & 34 4 31 1 - 33 4 30 0   & 342800.519 & 297.127 & 9.61E-4 & 4A, nLTE\\ 
                        &             & 34 3 31 1 - 33 3 30 0   & 342850.907 & 297.124 & 9.81E-4 & 4A, nLTE\\ 
                        &             & 22 5 17 0 - 21 4 18 0   & 342899.671 & 138.24  & 4.26E-5 & bl\\
                        &             & 34 13 22 0 - 33 13 21 1 & 342920.095 & 375.646 & 8.43E-4 & 4A, bl\\ 
                        &             & 34 13 21 0 - 33 13 20 1 & 342920.100 & 375.646 & 8.43E-4 & 4A, bl \\ 
                        &             & 20 4 16 1 - 19 2 17 0   & 342941.447 & 113.509 & 8.79E-6 & db, $<$5$\sigma$ \\
                        &             & 34 5 30 1 - 33 4 29 1   & 344137.936 & 305.126 & 2.42E-4 & bl\\
                        &             & 33 18 15 1 - 32 18 14 0 & 344226.046 & 434.972 & 7.02E-4 & bl \\ 
                        &             & 33 18 16 1 - 32 18 15 0 & 344226.046 & 434.972 & 7.02E-4 & bl \\ 
                        &             & 20 17 3 1 - 20 16 5 0   & 344228.068 & 245.387 & 8.84E-6 & bl \\ 
                        &             & 20 17 4 1 - 20 16 4 0   & 344228.068 & 245.387 & 8.84E-6 & bl \\ 
                        &             & 33 19 14 1 - 32 19 13 0 & 344231.354 & 453.099 & 6.68E-4 & bl \\
                        &             & 33 19 15 1 - 32 19 14 0 & 344231.354 & 453.099 & 6.68E-4 & bl \\
                        &             & 33 17 16 1 - 32 17 15 0 & 344253.269 & 417.828 & 7.35E-4 & 4A \\
                        &             & 33 17 17 1 - 32 17 16 0 & 344253.269 & 417.828 & 7.35E-4 & 4A \\
                        &             & 33 20 13 1 - 32 20 12 0 & 344264.330 & 472.207 & 6.33E-4 & 4A \\
                        &             & 33 20 14 1 - 32 20 13 0 & 344264.330 & 472.207 & 6.33E-4 & 4A \\
                        &             & 33 16 17 1 - 32 16 16 0 & 344319.573 & 401.672 & 7.65E-4 & bl \\
                        &             & 33 16 18 1 - 32 16 17 0 & 344319.573 & 401.672 & 7.65E-4 & bl \\
                        &             & 33 21 12 1 - 32 21 11 0 & 344321.301 & 492.291 & 5.95E-4 & bl \\
                        &             & 33 21 13 1 - 32 21 12 0 & 344321.301 & 492.291 & 5.95E-4 & bl \\
                        &             & 33 12 22 1 - 32 12 21 0 & 345224.705 & 347.051 & 8.75E-4 & 4A, bl\\ 
                        &             & 33 12 21 1 - 32 12 20 0 & 345224.778 & 347.051 & 8.75E-4 & 4A, bl\\ 
                        &             & 21 7 15 0 - 20 6 14 0   & 345266.087 & 137.904 & 8.68E-5 & 4A, bl \\
                        &             & 35 4 31 0 - 34 4 30 1   & 345278.673 & 321.646 & 8.99E-4 & bl \\
                        &             & 33 11 23 1 - 32 11 22 0 & 345737.011 & 335.95  & 9.00E-4 & 4A, bl \\ 
                        &             & 33 11 22 1 - 32 11 21 0 & 345738.443 & 335.95  & 9.01E-4 & 4A, bl \\ 
                        &             & 34 9 26 0 - 33 9 25 1   & 345812.719 & 333.572 & 9.36E-4 & bl \\
                        &             & 34 8 27 0 - 33 8 26 1   & 345832.886 & 325.787 & 5.43E-4 &  \\ \hline 
Ethylene Oxide        & c-C$_2$H$_4$O & 9 5 5 - 8 4 4           & 345688.322 & 90.408  & 3.37E-4 & bl\\ \hline
Formaldehyde            & H$_2$C$^{18}$O  & 5 0 5 - 4 0 4 & 345881.039 & 49.871 & 1.19E-3 & bl \\
                        & p-D$_2$CO   & 6 1 6 - 5 1 5 & 330674.347 & 53.025 & 1.02E-3 & bl \\ \hline  
Formamide & HC(O)NH$_2$, v=0   & 16 3 14 - 16 2 15	&  331685.873 & 165.588 & 7.87E-5 & 4A, bl, db  \\
          &                    & 16 1 15 - 15 1 14	&  345325.391 & 145.155 & 3.02E-3 & \\ \hline 
Hydrogen Cyanide & HC$^{15}$N, v=0 & 4 - 3 & 344200.109 & 41.298 & 1.88E-3 & bl\\
                 & H$^{13}$CN, v=0 & 4 - 3 & 345339.769 & 41.435 & 1.90E-3 & bl\\\hline
Methanol & CH$_3$OH, vt=0         & 11 1 10 0 - 11 0 11 0 & 331502.319 & 169.008 & 3.93E-4 & nLTE\\
         &                        & 18 2 17 1 - 17 3 15 1 & 344109.039 & 419.401 & 6.80E-5 & bl \\
         &                        & 16 1 15 0 - 15 2 14 0 & 345903.916 & 332.651 & 1.04E-4 &  \\         
         & CH$_3$OH, vt=1 & 10 2 9 5 - 11 3 9 5   & 344312.267 & 491.913 & 1.77E-4 & bl \\
         & CH$_3$$^{18}$OH, vt=0 & 13 0 13 0 - 12 1 12 0 & 331482.670 & 202.281 & 1.99E-4 & bl\\
         & $^{13}$CH$_3$OH, vt=0 
                                   & 7 2 5 0 - 6 2 4 0     & 330535.222 & 85.799  & 1.44E-4 & bl\\ 
         &                         & 7 -2 6 0 - 6 -2 5 0   & 330535.890 & 89.448  & 1.46E-4 & bl\\ 
                       & CH$_2$DOH & 5 3 2 1 - 6 2 5 0     & 330606.339 & 77.042  & 7.70E-6 &  \\
                       &           & 9 0 9 1 - 8 2 6 0     & 330686.419 & 109.536 & 4.06E-7 & db \\
                       &           & 15 7 9 1 - 16 6 10 1  & 331647.951 & 458.992 & 9.48E-6 & bl, $<$5$\sigma$ \\ 
                       &           & 15 7 8 1 - 16 6 11 1  & 331647.959 & 458.992 & 9.48E-6 & bl, $<$5$\sigma$ \\ 
                       &           & 17 2 16 0 - 17 1 17 0 & 342935.645 & 342.828 & 1.67E-4 & \\
                       &           & 20 2 18 1 - 19 3 16 2 & 344138.882 & 478.684 & 7.97E-5 & bl\\
                       &           & 8 2 7 2 - 7 3 5 0     & 345284.157 & 112.579 & 1.88E-6 & bl\\
                       &           & 16 2 14 0 - 15 3 13 0 & 345398.904 & 310.213 & 5.83E-5 & \\
                       &           & 23 4 19 1 - 23 3 21 2 & 345701.510 & 662.813 & 1.34E-4 & \\ 
                       &           & 3 2 1 1 - 2 1 2 1     & 345718.718 & 39.435 & 4.23E-5  & bl, nLTE \\ 
                       &           & 19 1 19 1 - 18 2 17 2 & 345820.793 & 418.03 & 2.94E-5  & bl\\ 
                       &           & 22 4 19 1 - 22 3 19 2 & 345850.485 & 613.622 & 1.29E-4 &  \\ \hline 
Methyl Cyanide   & CH$_3$$^{13}$CN, v=0 & 18 6 - 17 6 & 330679.886 & 407.876 & 2.80E-3 & \\
                 &CH$_2$DCN             & 20 1 20 - 19 1 19 & 345685.380 & 179.63 & 3.60E-3 & bl \\ \hline 
Methyl Formate & CH$_3$OCHO v=0 & 27 13 15 4 - 26 13 14 4 & 330602.640 & 522.121 & 4.24E-4 & bl\\
                      &         & 27 11 16 5 - 26 11 15 5 & 330634.841 & 490.367 & 4.59E-4 & \\ 
                      &         & 28 3 25 2 - 27 4 24 1   & 330641.527 & 247.741 & 5.56E-5 & \\
                      &         & 28 3 25 0 - 27 4 24 0   & 330653.474 & 247.738 & 5.56E-5 & \\ 
                      &         & 27 12 16 3 - 26 12 15 3 & 330701.651 & 505.429 & 4.43E-4 & \\ 
                      &         & 27 12 15 3 - 26 12 14 3 & 330701.651 & 505.429 & 4.43E-4 & \\ 
                      &         & 28 3 25 0 - 27 3 24 0   & 331469.465 & 247.739 & 5.36E-4 & bl \\
                      &         & 27 10 17 5 - 26 10 16 5 & 331483.804 & 476.502 & 4.78E-4 & bl\\ 
                      &         & 27 22 5 2 - 26 22 4 2 & 331539.684 & 543.597 & 1.88E-4 & db, nLTE \\
                      &         & 27 21 6 0 - 26 21 5 0 & 331564.665 & 515.157 & 2.21E-4 &  \\ 
                      &         & 27 21 7 0 - 26 21 6 0 & 331564.665 & 515.157 & 2.21E-4 &  \\ 
                      &         & 27 21 6 2 - 26 21 5 2 & 331586.163 & 515.149 & 2.21E-4 & $<$5$\sigma$ \\ 
                      &         & 27 21 7 1 - 26 21 6 1 & 331589.894 & 515.146 & 2.21E-4 & $<$5$\sigma$ \\ 
                      &         & 27 20 7 0 - 26 20 6 0 & 331625.030 & 488.026 & 2.52E-4 &  \\
                      &         & 27 20 8 0 - 26 20 7 0 & 331625.030 & 488.026 & 2.52E-4 &  \\ 
                      &         & 27 20 8 1 - 26 20 7 1 & 331647.933 & 488.017 & 2.52E-4 & bl \\ 
                      &         & 27 20 7 2 - 26 20 6 2 & 331648.683 & 488.018 & 2.52E-4 & bl \\ 
                      &         & 28 10 18 5 - 27 10 17 5 & 344108.640 & 493.017 & 5.42E-4 & bl, nLTE \\ 
                      &         & 28 18 10 0 - 27 18 9 0  & 344170.930 & 454.246 & 3.66E-4 &  \\
                      &         & 28 18 11 0 - 27 18 10 0 & 344170.930 & 454.246 & 3.66E-4 &  \\ 
                      &         & 28 18 10 2 - 27 18 9 2  & 344187.246 & 454.239 & 3.66E-4 &  \\ 
                      &         & 28 18 11 1 - 27 18 10 1 & 344197.339 & 454.238 & 3.66E-4 & bl\\
                      &         & 23 13 10 2 - 23 12 11 2 & 344206.436 & 274.953 & 4.27E-5 & bl \\
                      &         & 23 13 11 1 - 23 12 12 1 & 344210.812 & 274.944 & 4.27E-5 & bl \\
                      &         & 23 13 10 0 - 23 12 11 0 & 344237.391 & 274.954 & 4.27E-5 &  \\ 
                      &         & 23 13 11 0 - 23 12 12 0 & 344237.414 & 274.954 & 4.27E-5 &  \\ 
                      &         & 28 11 18 4 - 27 11 17 4 & 344300.259 & 506.713 & 5.27E-4 & \\
                      &         & 28 17 11 0 - 27 17 10 0 & 344322.992 & 431.091 & 3.94E-4 & bl\\
                      &         & 28 17 12 0 - 27 17 11 0 & 344322.992 & 431.091 & 3.94E-4 & bl\\ 
                      &         & 18 13 5 2 - 18 12 6 2   & 345230.296 & 212.973 & 3.22E-5 & bl\\ 
                      &         & 18 13 6 1 - 18 12 7 1   & 345241.935 & 212.959 & 3.22E-5 & \\ 
                      &         & 28 10 19 4 - 27 10 18 4 & 345248.210 & 492.904 & 5.48E-4 & \\ 
                      &         & 18 13 5 0 - 18 12 6 0   & 345281.320 & 212.973 & 3.22E-5 & bl\\ 
                      &         & 18 13 6 0 - 18 12 7 0   & 345281.320 & 212.973 & 3.22E-5 & bl\\ 
                      &         & 17 13 4 2 - 17 12 5 2   & 345351.333 & 202.36 & 2.89E-5  &  \\ 
                      &         & 17 13 5 1 - 17 12 6 1   & 345364.235 & 202.345 & 2.89E-5 &  \\ 
                      &         & 17 13 4 0 - 17 12 5 0   & 345405.869 & 202.36 & 2.89E-5 & bl\\
                      &         & 17 13 5 0 - 17 12 6 0   & 345405.869 & 202.36 & 2.89E-5 & bl \\ 
                      &         & 16 13 3 2 - 16 12 4 2   & 345451.103 & 192.34 & 2.51E-5 & bl, $<$5$\sigma$ \\ 
                      &         & 13 13 0 0 - 13 12 1 0   & 345714.339 & 165.831 & 8.37E-6 & bl, $<$5$\sigma$\\ 
                      &         & 13 13 1 0 - 13 12 2 0   & 345714.339 & 165.831 & 8.37E-6 & bl, $<$5$\sigma$ \\
                      &         & 9 9 0 0 - 8 8 1 0       & 345718.662 & 80.32   & 9.86E-5 &  bl \\ 
                      &         & 9 9 1 0 - 8 8 0 0       & 345718.662 & 80.32   & 9.86E-5 &  bl \\ 
                      &         & 28 6 23 4 - 27 6 22 4   & 345828.557 & 451.764 & 5.99E-4 & bl \\
                      &         & 9 9 0 3 - 8 8 1 3       & 345908.676 & 268.174 & 1.00E-4 & bl \\
                      &         & 9 9 1 3 - 8 8 0 3       & 345908.676 & 268.174 & 1.00E-4 & bl\\\hline 
Nitrogen Sulfide      &  NS     & 8 1 1 8.5-7 1 -1 7.5 	& 345823.288 & 70.795  & 7.36E-4 & bl \\ 
                      &         & 8 1 1 7.5-7 1 -1 6.5 	& 345823.288 & 70.797  & 7.23E-4 & bl \\ 
                      &         & 8 1 1 6.5-7 1 -1 5.5 	& 345824.130 & 70.799  & 7.21E-4 & bl \\\hline 
Sulfur Dioxide & SO$_2$, v=0 & 11 6 6 - 12 5 7 & 331580.244 & 148.954 & 4.35E-5 &  \\ 
               &             & 13 2 12 - 12 1 11 & 345338.538 & 92.984 & 2.38E-4 & bl \\ 
               &$^{34}$SO$_2$& 21 2 20 - 21 1 21 & 330667.565 & 218.583 & 1.45E-4 & $<$5$\sigma$  \\ 
               &             & 10 4 6 - 10 3 7   & 344245.346 & 88.384  & 2.96E-4 &  \\ 
               &             & 9 4 6 - 9 3 7     & 345285.620 & 79.203  & 2.88E-4 & bl \\  \hline 
Sulfur Monoxide  & SO, v=0 & 8 8 - 7 7 & 344310.612 & 87.482 & 5.19E-4 & bl \\\hline 
Thioformaldehyde & H$_2$CS        & 10 0 10 - 9 0 9 & 342946.424 & 90.592  & 6.08E-4 & \\ 
                 & H$_2$$^{13}$CS & 10 2 8 - 9 2 7  & 330544.304 & 140.012 & 5.23E-4 & bl  \\ \hline 
Unknown-1 & \nodata & \nodata & 330505.15$\pm$0.11 & \nodata & \nodata & bl\\ 
Unknown-2 & \nodata & \nodata & 330557.55$\pm$0.04 &\nodata & \nodata & \\ 
Unknown-3 & \nodata & \nodata & 330568.61$\pm$0.05 &\nodata & \nodata & bl\\ 
Unknown-4 & \nodata & \nodata & 330622.86$\pm$0.03 &\nodata & \nodata & bl \\ 
Unknown-5 & \nodata & \nodata & 330648.68$\pm$0.06 &\nodata & \nodata &  \\ 
Unknown-6 & \nodata & \nodata & 330664.961$\pm$0.022 &\nodata & \nodata & bl \\
Unknown-7 & \nodata & \nodata & 331562.17$\pm$0.06 &\nodata & \nodata & bl \\
Unknown-8 & \nodata & \nodata & 331694.10$\pm$0.06 &\nodata & \nodata & CB \\ 
Unknown-9 & \nodata & \nodata & 342861.82$\pm$0.04 &\nodata & \nodata & bl \\
Unknown-10 & \nodata & \nodata & 342905.71$\pm$0.06 &\nodata & \nodata &   \\ 
Unknown-11 & \nodata & \nodata & 342995.72$\pm$0.06 &\nodata & \nodata & bl\\ 
Unknown-12 & \nodata & \nodata & 342998.21$\pm$0.05 &\nodata & \nodata & \\ 
Unknown-13 & \nodata & \nodata & 344123.71$\pm$0.04 &\nodata & \nodata & \\
Unknown-14 & \nodata & \nodata & 344280.85$\pm$0.07 &\nodata & \nodata & bl\\ 
Unknown-15 & \nodata & \nodata & 345302.06$\pm$0.13 &\nodata & \nodata &  \\ 
Unknown-16 & \nodata & \nodata & 345319.621$\pm$0.020 &\nodata & \nodata & \\
Unknown-17 & \nodata & \nodata & 345411.10$\pm$0.06 &\nodata & \nodata & bl\\
Unknown-18 & \nodata & \nodata & 345445.60$\pm$0.02 &\nodata & \nodata & \\
Unknown-19 & \nodata & \nodata & 345700.70$\pm$0.09 &\nodata & \nodata & bl\\ 
Unknown-20 & \nodata & \nodata & 345911.585$\pm$0.025 &\nodata & \nodata & \\ 
Unknown-21 & \nodata & \nodata & 342832.08$\pm$0.13 & \nodata & \nodata & 4A, bl \\ 
Unknown-22 & \nodata & \nodata & 342905.38$\pm$0.07 & \nodata & \nodata & 4A, bl\\ 
Unknown-23 & \nodata & \nodata & 342909.54$\pm$0.06 & \nodata & \nodata & 4A, bl\\
Unknown-24 & \nodata & \nodata & 342913.42$\pm$0.10 & \nodata & \nodata & 4A, bl\\ 
Unknown-25 & \nodata & \nodata & 342962.01$\pm$0.08 & \nodata & \nodata & 4A, bl\\ 
Unknown-26 & \nodata & \nodata & 344213.68$\pm$0.22 & \nodata & \nodata & 4A, bl\\ 
Unknown-27 & \nodata & \nodata & 345354.29$\pm$0.14 & \nodata & \nodata & 4A, bl\\
Unknown-28 & \nodata & \nodata & 345430.67$\pm$0.17 & \nodata & \nodata & 4A, bl, wide\\ 
\enddata
\tablenotetext{a}{From spectra obtained in a beam-size ($0\farcs16\times0\farcs08$, PA = 0$\degr$) region toward VLA 4A, VLA 4B, and toward the geometric center of the components (CB) that we adopt as representative of the circumbinary disk.} 
\tablenotetext{b}{For identified transitions, this is the laboratory frequency reported in the catalogs used (JPL, CDMS, VASTEL). For unidentified transitions, we list the frequency derived from a Gaussian fit to the line and the formal error of the fit. In general, we used for the fit the spectrum taken toward the geometric center of the two components, VLA 4A and VLA 4B, assuming $V_{\rm LSR}$ = 8.77 km s$^{-1}$. In the cases where the unidentified line is detected only toward VLA 4A, $V_{\rm LSR}$ = 7.36 km s$^{-1}$ is used. When the unidentified line is blended with known lines, a multiple Gaussian fit is performed assuming that the value of the FWHM is the same for all the lines. We only report intense ($>10\sigma$) unidentified lines.}
\tablenotetext{c}{4A: Detected only toward VLA 4A; CB: Detected only in the circumbinary material, toward the geometric center of the two components (CB); bl: Blended with another line; db: Doubtful identification; $<$5$\sigma$: Low ($<5\sigma$) signal-to-noise ratio; nLTE: Not well reproduced by the LTE model (differences $\gtrsim 50\%$); wide: Wide line with probable blending of several lines.}
\end{deluxetable*}

\onecolumngrid

\section{Error estimate of the inclination of a disk}\label{sec:errors}

\restartappendixnumbering

The inclination of the disks in SVS 13 is estimated from the ratio of the minor axis, $a_\mathrm{min}$, and major axis, $a_\mathrm{maj}$, of the projection of the disk on the sky, 
\begin{equation}
i= \arccos \frac{a_\mathrm{min}}{a_\mathrm{maj}}. 
\end{equation}

The inclination of the orbit is estimated from the ratio of the components of the orbital velocity along the line-of-sight (the difference in the LSR velocity of the VLA 4B and VLA 4A line emission), $v_\mathrm{los}$, and on the plane of the sky (the difference in the proper motion velocities of VLA 4B and VLA 4A), $v_\mathrm{sky}$, 
\begin{equation}
i= \arctan \frac{v_\mathrm{los}}{v_\mathrm{sky}}. 
\end{equation}
In both cases, the estimation of the inclination involves the ratio of two magnitudes known with a given uncertainty.
Let us call the two magnitudes $x$ and $y$, which we assume are independent, normally-distributed (Gaussian) random variables with expected values $\mu_x$ and $\mu_y$, and standard deviations $\sigma_x$ and $\sigma_y$. 
In the following we will calculate an approximation of the expected value and standard deviation of the ratio $z=x/y$, and after this, the expected value and standard deviation of the inclination, $\mu_i$ and $\sigma_i$.

\subsection{Error of the ratio}

The study of the error of the ratio of two normal random variables is not easy. 
For instance, the case of variables with zero mean is presented in \citet{Papoulis2002}. 
In the general case of non-zero mean, the probability density function of the ratio is obtained by \citet{Diaz2012}, but the expression is cumbersome and the integrals have to be evaluated numerically. 
Another approach, based on the expansion up to second or third order in $\sigma_x/\mu_x$ and $\sigma_y/\mu_y$, is more practical.

\subsubsection{Expected value of the ratio}

The probability density function of the Gaussian random variable $x$ is given by   
\begin{equation}
f(x)=\frac{1}{\sqrt{2\pi}\,\sigma_x} 
\exp\left[-\frac{1}{2}\left(\frac{x-\mu_x}{\sigma_x}\right)^2\right],
\end{equation}
and a similar expression for $y$.

The expected value of the ratio $z=x/y$ is defined as
\begin{equation}
E[z]= \int_{-\infty}^{+\infty}\!\!\!\int_{-\infty}^{+\infty} \frac{x}{y} 
\frac{1}{2\pi\sigma_x\sigma_y}
\exp\left[-\frac{1}{2}\left(\frac{x-\mu_x}{\sigma_x}\right)^2\right]
\exp\left[-\frac{1}{2}\left(\frac{y-\mu_y}{\sigma_y}\right)^2\right]
dx\,dy.
\end{equation}
The integral over $x$ gives the expected value of $x$, so 
\begin{equation}
E[z]= \mu_x \int_{-\infty}^{+\infty} \frac{1}{y} \frac{1}{\sqrt{2\pi}\,\sigma_y} 
\exp\left[-\frac{1}{2}\left(\frac{y-\mu_y}{\sigma_y}\right)^2\right]\,dy.
\end{equation}
Let us use the standard normally distributed variable
\begin{equation}
\epsilon=\frac{y-\mu_y}{\sigma_y},
\end{equation}
defined with zero expected value $\mu_\epsilon=0$ and unit standard deviation $\sigma_\epsilon=1$. The change of variable gives
\begin{equation}
E[z]= \frac{\mu_x}{\mu_y} \frac{1}{\sqrt{2\pi}} \int_{-\infty}^{+\infty} 
\frac{1}{1+(\sigma_y/\mu_y)\epsilon} \exp\left(-\frac{\epsilon^2}{2}\right) d\epsilon.
\end{equation}
Let us expand $1/[1+(\sigma_y/\mu_y)\epsilon]$ up to order 2,
\begin{equation}
\frac{1}{1+(\sigma_y/\mu_y)\epsilon} \simeq 
1 - \frac{\sigma_y}{\mu_y}\,\epsilon + \frac{\sigma_y^2}{\mu_y^2}\,\epsilon^2 +\ldots
\end{equation}
The odd terms of the expansion have zero integral (and thus the expansion is correct up to third order). The integral of the constant term is 1, and that of the quadratic term is $\sigma_y^2/\mu_y^2$ (because $\sigma_\epsilon=1$). The final result is
\begin{equation}
\mu_z\equiv E[z] \simeq \frac{\mu_x}{\mu_y} \left(1+\frac{\sigma_y^2}{\mu_y^2}\right).
\end{equation}
It is noticeable that there is a bias in the expected value of the ratio, so that $E[x/y]\ne E[x]/E[y]$.

\subsubsection{Standard deviation of the ratio}

The same procedure can be applied to estimate the variance of the ratio $z=x/y$. The expected value of $z^2$ is
\begin{equation}
E[z^2]= \int_{-\infty}^{+\infty}\!\!\!\int_{-\infty}^{+\infty} \frac{x^2}{y^2} 
\frac{1}{2\pi\sigma_x\sigma_y}
\exp\left[-\frac{1}{2}\left(\frac{x-\mu_x}{\sigma_x}\right)^2\right]
\exp\left[-\frac{1}{2}\left(\frac{y-\mu_y}{\sigma_y}\right)^2\right]
dx\,dy.
\end{equation}
The integral over $x$ gives the expected value of $x^2$, $E[x^2]= \mu_x^2+\sigma_x^2$, so 
\begin{equation}
E[z^2]= (\mu_x^2+\sigma_x^2) \int_{-\infty}^{+\infty} \frac{1}{y^2} \frac{1}{\sqrt{2\pi}\,\sigma_y} 
\exp\left[-\frac{1}{2}\left(\frac{y-\mu_y}{\sigma_y}\right)^2\right]\,dy.
\end{equation}
The same change of variables as before gives
\begin{equation}
E[z^2]= \frac{\mu_x^2+\sigma_x^2}{\mu_y^2} \frac{1}{\sqrt{2\pi}} \int_{-\infty}^{+\infty} 
\frac{1}{[1+(\sigma_y/\mu_y)\epsilon]^2} \exp\left(-\frac{\epsilon^2}{2}\right)\,d\epsilon.
\end{equation}
The expansion of $1/[1+(\sigma_y/\mu_y)\epsilon]^2$ is
\begin{equation}
\frac{1}{[1+(\sigma_y/\mu_y)\epsilon]^2} \simeq 
1 - 2\frac{\sigma_y}{\mu_y}\,\epsilon + 3\frac{\sigma_y^2}{\mu_y^2}\,\epsilon^2 +\ldots
\end{equation}
The integral of the constant term is 1, that of the quadratic term is $3\sigma_y^2/\mu_y^2$, and the result is
\begin{equation}
E[z^2] \simeq \frac{\mu_x^2+\sigma_x^2}{\mu_y^2} \left(1+3\frac{\sigma_y^2}{\mu_y^2}\right).
\end{equation}
The variance is $\sigma_z^2= E[z^2]-E^2[z]$. The square of $E[z]$, neglecting terms of fourth order in $\delta_x$ and $\delta_y$, is
\begin{equation}
E^2[z]\simeq \frac{\mu_x^2}{\mu_y^2} \left(1+2\frac{\sigma_y^2}{\mu_y^2}\right),
\end{equation}
resulting in a difference, neglecting again terms of fourth order,
\begin{equation}
\sigma_z^2 \simeq \frac{\mu_x^2}{\mu_y^2}\left(\frac{\sigma_x^2}{\mu_x^2}+\frac{\sigma_y^2}{\mu_y^2}\right),
\end{equation}
and the standard deviation is
\begin{equation}
\sigma_z\simeq \frac{\mu_x}{\mu_y}\sqrt{\frac{\sigma_x^2}{\mu_x^2}+\frac{\sigma_y^2}{\mu_y^2}}.
\end{equation}
In the limiting case $\sigma_y=0$ the expression reduces to the usual value $\sigma_z=\sigma_x/\mu_y$.

\subsection{Error of $i=\arccos z$}

Let us now calculate an approximation to the expected value and standard deviation of the inclination
\begin{equation}
i= \arccos z,
\end{equation}
given the expected value, $\mu_z$, and standard deviation, $\sigma_z$, of $z$.

\subsubsection{Expected value of $i=\arccos z$}

The expected value $E[i]$ is 
\begin{equation}
\mu_i\equiv E[i]= \int_{-\infty}^{+\infty}\!\!\! \arccos z \frac{1}{\sqrt{2\pi}\,\sigma_z} 
\exp\left[-\frac{1}{2}\left(\frac{z-\mu_z}{\sigma_z}\right)^2\right]\,dz.
\end{equation}
The expansion of $\arccos$ around the expected value $\mu_z$, using as before the  standard normally distributed variable
$\epsilon=(z-\mu_z)/\sigma_z$,
is 
\begin{equation}
\arccos z\simeq \arccos \mu_z
                -\frac{\sigma_z}{\sqrt{1-\mu_z^2}}\epsilon
                -\frac{1}{2!}\,\frac{\mu_z\sigma_z^2}{(1-\mu_z^2)^{3/2}}\epsilon^2
                +\ldots
\end{equation}
The integral is not null for the even terms of the expansion only, and thus the approximation is correct up to third order.
The result of the integration is
\begin{equation}
E[i]\simeq \arccos \mu_z -\frac{1}{2}\,\frac{\mu_z\sigma_z^2}{(1-\mu_z^2)^{3/2}}. 
\end{equation}
It is noticeable that there is a bias in the expected value of the inclination, because $E[\arccos z]\ne\arccos(E[z])$.

\subsubsection{Standard deviation of $i=\arccos z$}

In a similar way, let us calculate $E[i^2]$,
\begin{equation}
E[i^2]= \int_{-\infty}^{+\infty}\!\!\! \arccos^2 z \frac{1}{\sqrt{2\pi}\,\sigma_z} 
\exp\left[-\frac{1}{2}\left(\frac{z-\mu_z}{\sigma_z}\right)^2\right]\,dz.
\end{equation}
The expansion of $\arccos^2$ is
\begin{equation}
\arccos^2 z\simeq \arccos^2 \mu_z
                -2\frac{\sigma_z\arccos\mu_z}{\sqrt{1-\mu_z^2}}\epsilon
                -\frac{\sigma_z^2}{1-\mu_z^2}\left(\frac{\mu_z\arccos\mu_z}{\sqrt{1-\mu_z^2}}-1\right)\epsilon^2
                +\ldots
\end{equation}
The integration gives non-zero values for the even terms, resulting in
\begin{equation}
E[i^2]\simeq \arccos^2\mu_z 
            -\frac{\sigma_z^2}{1-\mu_z^2}\left(\frac{\mu_z\arccos\mu_z}{\sqrt{1-\mu_z^2}}-1\right). 
\end{equation}
The square of $E[i]$, discarding terms of fourth order, is
\begin{equation}
E^2[i]\simeq \arccos^2\mu_z -\frac{\sigma_z^2}{1-\mu_z^2}\,   
                               \frac{\mu_z\arccos\mu_z}{\sqrt{1-\mu_z^2}}.
\end{equation}
The difference gives us the variance $\sigma_i^2=E[i^2]-E^2[i]$, resulting in a standard deviation that is simply
\begin{equation}
\sigma_i \simeq  \frac{\sigma_z}{\sqrt{1-\mu_z^2}}.
\end{equation}

\subsection{Error of $i=\arctan z$}
Let us now calculate an approximation to the expected value and standard deviation of the inclination
\begin{equation}
i= \arctan z,
\end{equation}
given the expected value, $\mu_z$, and standard deviation, $\sigma_z$, of $z$. The procedure is the same as that used in the former section.

\subsubsection{Expected value of $i=\arctan z$}

The expansion of $\arctan$ around the expected value $\mu_z$, using the standard normally distributed variable
$\epsilon=(z-\mu_z)/\sigma_z$,
is 
\begin{equation}
\arctan z\simeq \arctan \mu_z
                +\frac{\sigma_z}{(1+\mu_z^2)^2}\epsilon
                -\frac{\mu_z\sigma_z^2}{(1+\mu_z^2)^2}\epsilon^2
                +\ldots
\end{equation}
The expected value of $i$, neglecting terms of order higher than three is
\begin{equation}
E[i]\simeq \arctan \mu_z -\frac{\mu_z\sigma_z^2}{(1+\mu_z^2)^2}. 
\end{equation}
Like in the former case, there is a bias in the expected value of the inclination, because $E[\arctan z]\ne\arctan(E[z])$.

\subsubsection{Standard deviation of $i=\arctan z$}

To calculate $E[i^2]$ we need the expansion
\begin{equation}
\arctan^2 z\simeq \arctan^2 \mu_z
                +2\frac{\sigma_z\arctan\mu_z}{1+\mu_z^2}\epsilon
                +\frac{\sigma_z^2}{(1+\mu_z^2)^2}(1-2\mu_z\arctan\mu_z)\epsilon^2
                +\ldots
\end{equation}
resulting in 
\begin{equation}
E[i^2]\simeq \arctan^2\mu_z 
            +\frac{\sigma_z^2}{(1+\mu_z^2)^2}(1-2\mu_z\arctan\mu_z).
\end{equation}
The difference, $\sigma_i^2=E[i^2]-E^2[i]$, neglecting terms of fourth order, gives
\begin{equation}
\sigma_i \simeq  \frac{\sigma_z}{1+\mu_z^2}.
\end{equation}

\subsection{Summary}

Given two independent Gaussian random variables  $x$ and $y$, with expected values $\mu_x$ and $\mu_y$, and standard deviations $\sigma_x$ and $\sigma_y$, the expected value and the standard deviation of the ratio $z=x/y$ are
\begin{eqnarray}
\mu_z    &\simeq& \displaystyle\frac{\mu_x}{\mu_y} \left(1+\frac{\sigma_y^2}{\mu_y^2}\right), \nonumber \\
\sigma_z &\simeq& \frac{\mu_x}{\mu_y}\sqrt{\frac{\sigma_x^2}{\mu_x^2}+\frac{\sigma_y^2}{\mu_y^2}}.
\end{eqnarray}

Given the ratio of the minor and major axes, $a_\mathrm{min}$ and $a_\mathrm{maj}$, of the projection of the disk on the plane  of the sky, $z=a_\mathrm{min}/a_\mathrm{maj}$, the expected value and the standard deviation of the disk inclination, $i=\arccos z$, are
\begin{eqnarray}
\mu_i    &\simeq& \arccos\mu_z -\displaystyle\frac{1}{2}\,\frac{\mu_z\sigma_z^2}{(1-\mu_z^2)^{3/2}},  \nonumber \\
\sigma_i &\simeq& \displaystyle\frac{\sigma_z}{\sqrt{1-\mu_z^2}}.
\end{eqnarray}

Given the ratio of the components of the orbital velocity along the line-of-sight and on the plane of the sky, $z= v_\mathrm{los}/v_\mathrm{sky}$, the expected value and the standard deviation of the orbit inclination, $i=\arctan z$, are
\begin{eqnarray}
\mu_i    &\simeq& \displaystyle\arctan \mu_z -\frac{\mu_z\sigma_z^2}{(1+\mu_z^2)^2},  \nonumber \\
\sigma_i &\simeq& \displaystyle\frac{\sigma_z}{1+\mu_z^2}.
\end{eqnarray}

\section{Fitting of a Keplerian Thin Disk Model}\label{sec:keplerian}

\restartappendixnumbering

We use the model of a thin disk with a Keplerian rotation developed by R. Estalella (see Appendix B in \citealt{Zapata2019} and references therein for details) to constrain the properties of the VLA 4A source. The model consists of a geometrically thin disk with an inner radius $R_{\rm inn}$ and an outer radius $R_{\rm out}$, where the intensity and velocity are described by power laws of the radius with index $-$1 and $-$0.5, respectively. The kinematics of the disk is computed by considering a cube composed of a grid of points in the plane of the disk, and for each point of the grid, a number of velocity channels. For each point of the grid, the projection of its rotation velocity onto the line-of-sight is calculated. A Gaussian line profile, of width $\Delta V$ (assumed constant for the whole disk), and proportional to the estimated intensity is added to the channels associated with the grid point. After this first cube is computed, its projection onto the plane of the sky is estimated. Finally, each channel of the plane-of-the-sky cube is convolved with a Gaussian with the size of the beam. The intensity depends on a scale factor, which is estimated by minimizing the sum of the squared differences between the data cube and the model cube. The parameter space is searched for the minimum of the rms fit residual for all the channel maps using a pseudo-random sequence as described in \citet{Estalella2012}, and \citet{Estalella2017}. Once a minimum of the rms fit residual is found, the uncertainty in the parameters fitted is found as described in \citet{Wall2003} and \citet{Estalella2017}. 

\subsection{VLA 4A Circumstellar Disk traced by Ethylene Glycol}

The ethylene glycol lines detected only toward VLA 4A (Table \ref{tab:gly}) were used to fit the rotation of a geometrically thin disk. The model was fitted to the spectral data, assuming as position of its center that of VLA 4A, and a reference velocity $V\s{sys}=7.36$ km s$^ {-1}$ (see Table \ref{tab:pos}).
The beam used to convolve the model intensity was that of the observation, that is, $0\farcs154\times0\farcs079$ ($\mathrm{PA}=-1.9^\circ$).
The free parameters of the model were the linewidth (without rotational broadening), $\Delta V$, the inclination of the disk, $i$, the inner and outer radii, $R\s{inn}$ and $R\s{out}$, and the rotation velocity at a reference radius of $1''$, times the sinus of the inclination, $V_0\sin i$.

\begin{table*}[hbtp]
\centering
\caption{Ethylene glycol (aGg'-(CH$_2$OH)$_2$) transitions.}
\label{tab:gly}
\begin{tabular}{lll}
\hline\hline
       &            & Frequency \\
Identifier & Transition & (GHz) \\
\hline
EthyGly1 & (34    04 31 1 -- 33    04 30 0) &  342.801     \\
EthyGly2 & (34    03 31 1 -- 33    03 30 0) &  342.851     \\
EthyGly3 & (33    11 23 1 -- 32    11 22 0) &  345.737$^a$ \\
         & (33    11 22 1 -- 33    11 21 0) &  345.738$^a$ \\ 
\hline
\end{tabular}

$^a$ The frequency difference between the two transitions is 1.432 MHz (1.242 km s$^{-1}$) and both appear blended.
\end{table*}

\begin{table*}[hbtp]
\centering
\caption{Positions and Reference Velocities.}
\label{tab:pos}
\begin{tabular}{llll}
\hline\hline
       & $\alpha(J2000)$ & $\delta(J2000)$ & $V\s{sys}$                 \\
Object &    (h:m:s)    &  (d:m:s)      & (km s$^{-1}$)         \\
\hline
VLA 4A     & 03:29:03.7460 & +31:16:03.769 & $7.36\pm0.21$   \\
VLA 4B     & 03:29:03.7691 & +31:16:03.761 & $9.33\pm0.18$       \\
Ambient Cloud     & \nodata & \nodata & 8.5$^a$       \\
\hline
\end{tabular}

\noindent $^a$ Adopted (Sect. \ref{sec:gas}). 
\end{table*}

The results from the fit are given in Table \ref{tab:4A_fit}, and the derived parameters, for a distance to SVS 13 of 300 pc, are given in Table \ref{tab:4A_deriv}.

\begin{table*}[htbp]
\centering
\tablewidth{0pt}
\caption{VLA 4A: Best Fit of a Geometrically-Thin Disk with Keplerian Rotation.}
\label{tab:4A_fit}
\begin{tabular}{llllll}
\hline\hline
         & $\Delta V$    & $i$          & $R\s{inn}$      & $R\s{out}$      & ${V_0\sin i}^{\,a}$      \\
Line     & (km s$^ {-1}$)& (deg)        & (arcsec)        & (arcsec)        & (km s$^{-1}$)    \\
\hline
EthyGly1 & $1.53\pm0.18$ & $22.6\pm5.6$ & $0.011\pm0.003$ & $0.109\pm0.004$ & $0.312\pm0.018$ \\
EthyGly2 & $1.87\pm0.22$ & $21.8\pm6.7$ & $0.006\pm0.003$ & $0.112\pm0.010$ & $0.327\pm0.019$ \\
EthyGly3 & $2.74\pm0.13$ & $22.3\pm3.2$ & $0.010\pm0.003$ & $0.110\pm0.004$ & $0.255\pm0.017$ \\
\hline
\end{tabular}

$^a$ At a radius of $1''$.
\end{table*}

\begin{table}[htbp]
\centering
\caption{VLA 4A: Parameters Derived from the Best Fit to the Ethylene Glycol Lines.}
\label{tab:4A_deriv}
\begin{tabular}{llllll}
\hline\hline
         & $R\s{inn}$    & $R\s{out}$   & ${V_0}^a$     & ${M_\ast}^b$  \\
Line     & (au)          & (au)         & (km s$^{-1}$) & ($M_\odot$)   \\
\hline
EthyGly1 & $3.3\pm0.9$ & $32.8\pm1.3$ & $0.81\pm0.20$ & $0.22\pm0.11$ \\
EthyGly2 & $1.9\pm0.8$ & $33.5\pm2.9$ & $0.88\pm0.26$ & $0.26\pm0.16$ \\
EthyGly3 & $2.9\pm0.9$ & $33.1\pm1.1$ & $0.67\pm0.10$ & $0.15\pm0.05$ \\
\hline
\end{tabular}

$^a$ At a radius of $1''$ (300 au). 
$^b$ Central mass.
\end{table}

\begin{figure}[htbp]
\centering
\includegraphics[width=0.325\textwidth]{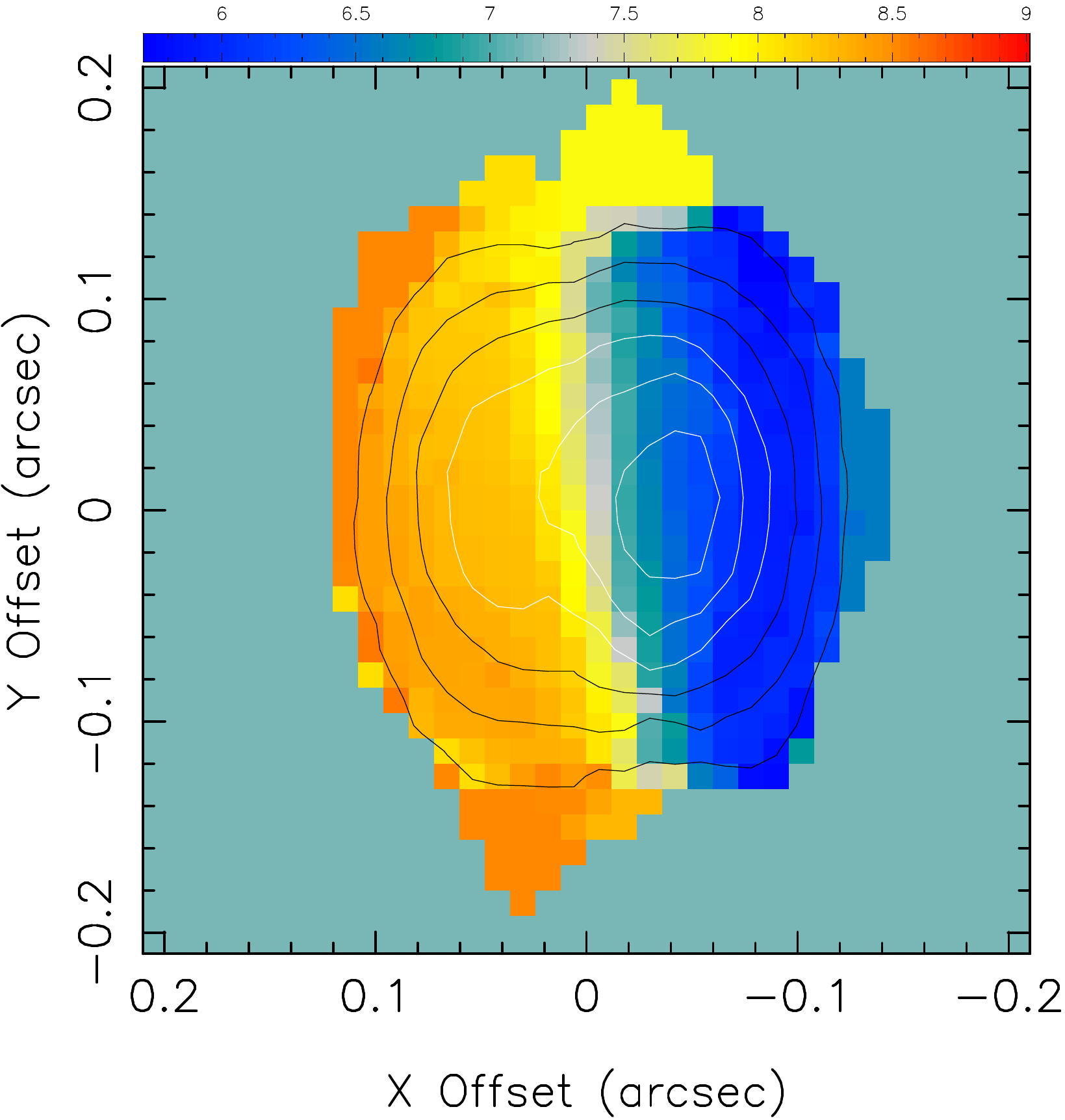}
\includegraphics[width=0.325\textwidth]{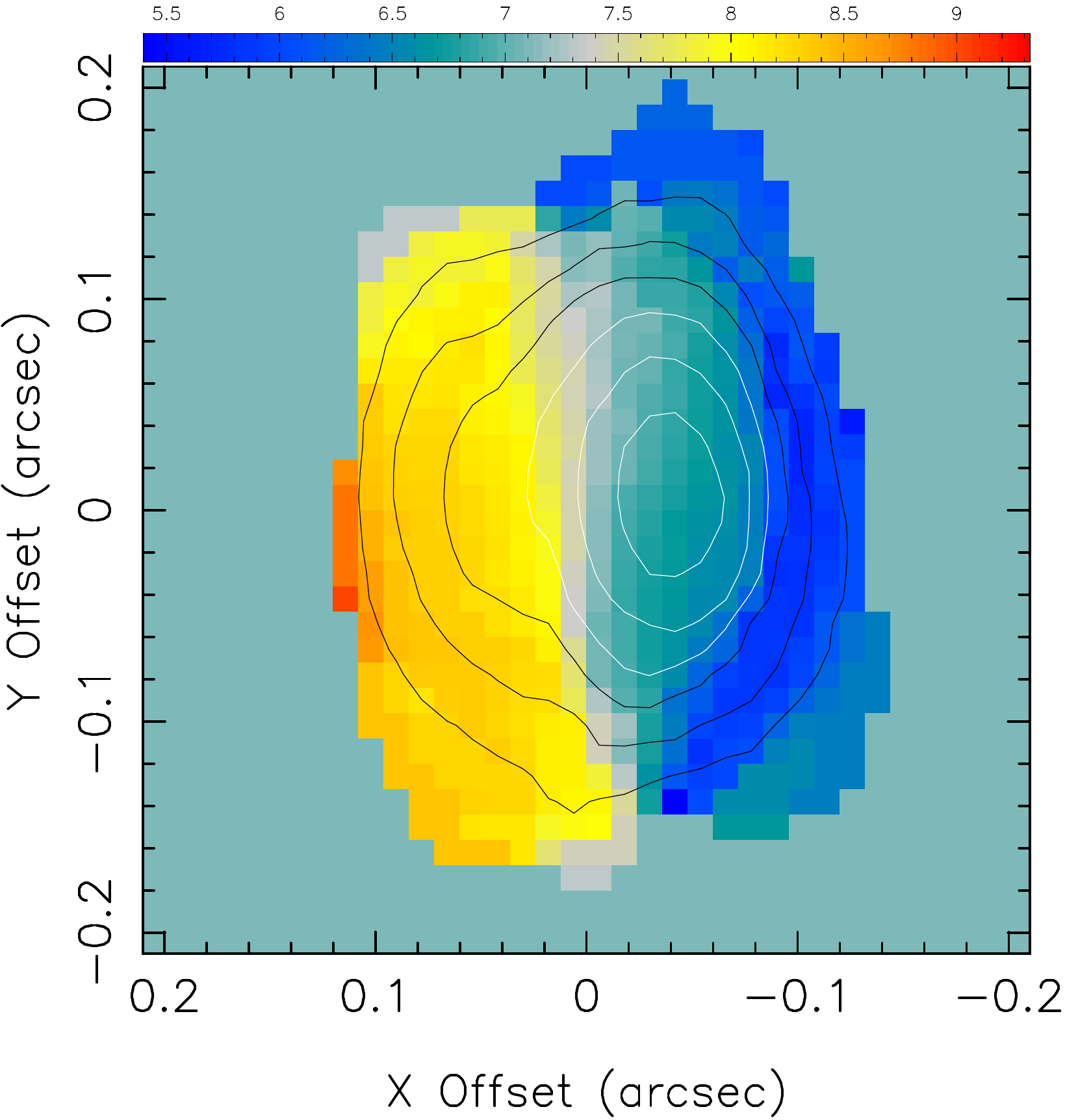}
\includegraphics[width=0.325\textwidth]{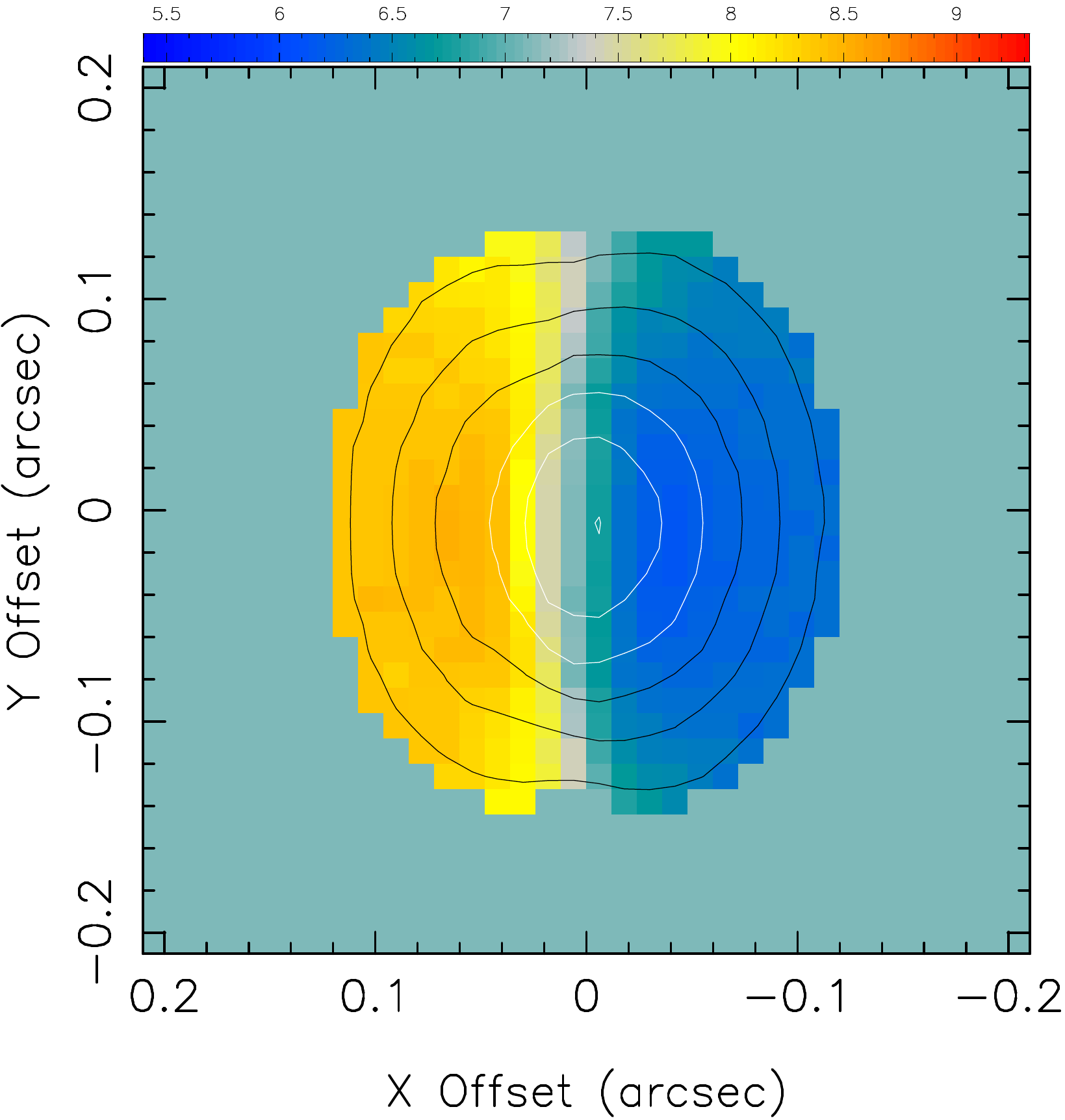}
\caption{
VLA 4A circumstellar disk: First-order moment (intensity weighted velocity in km~s$^{-1}$; color scale) overlaid on the zeroth-order moment (integrated intensity; contours) of EthyGly1 (\emph{left}), EthyGly2 (\emph{center}), and the thin-disk model of EthyGly2 (\emph{right}). For the three panels, contours range from 15\% to 90\% in steps of 15\% of the peak (0.120, 0.168 and 0.150 Jy beam$^{-1}$ km s$^{-1}$ for EthyGly1, EthyGly2 and the model of EthyGly2, respectively).  Offsets are relative to the position of VLA 4A (Table~\ref{tab:pos}), in right ascension (X axis) and declination (Y axis), positive toward east and north, respectively.
}
\label{fig:gly_mom1}
\end{figure}

\begin{figure*}[htb]
\centering
\includegraphics[width=0.39\textwidth]{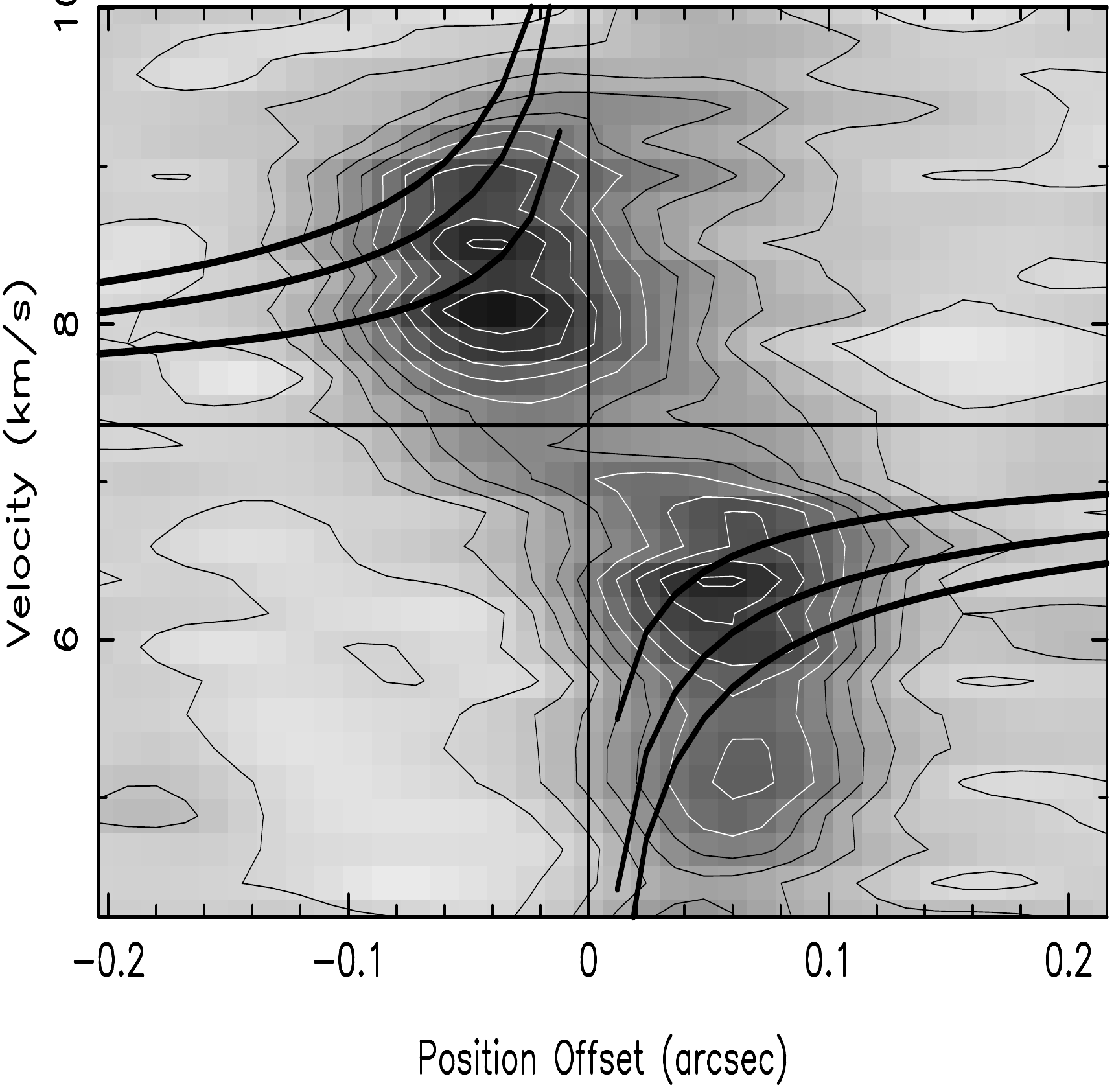}
\includegraphics[width=0.39\textwidth]{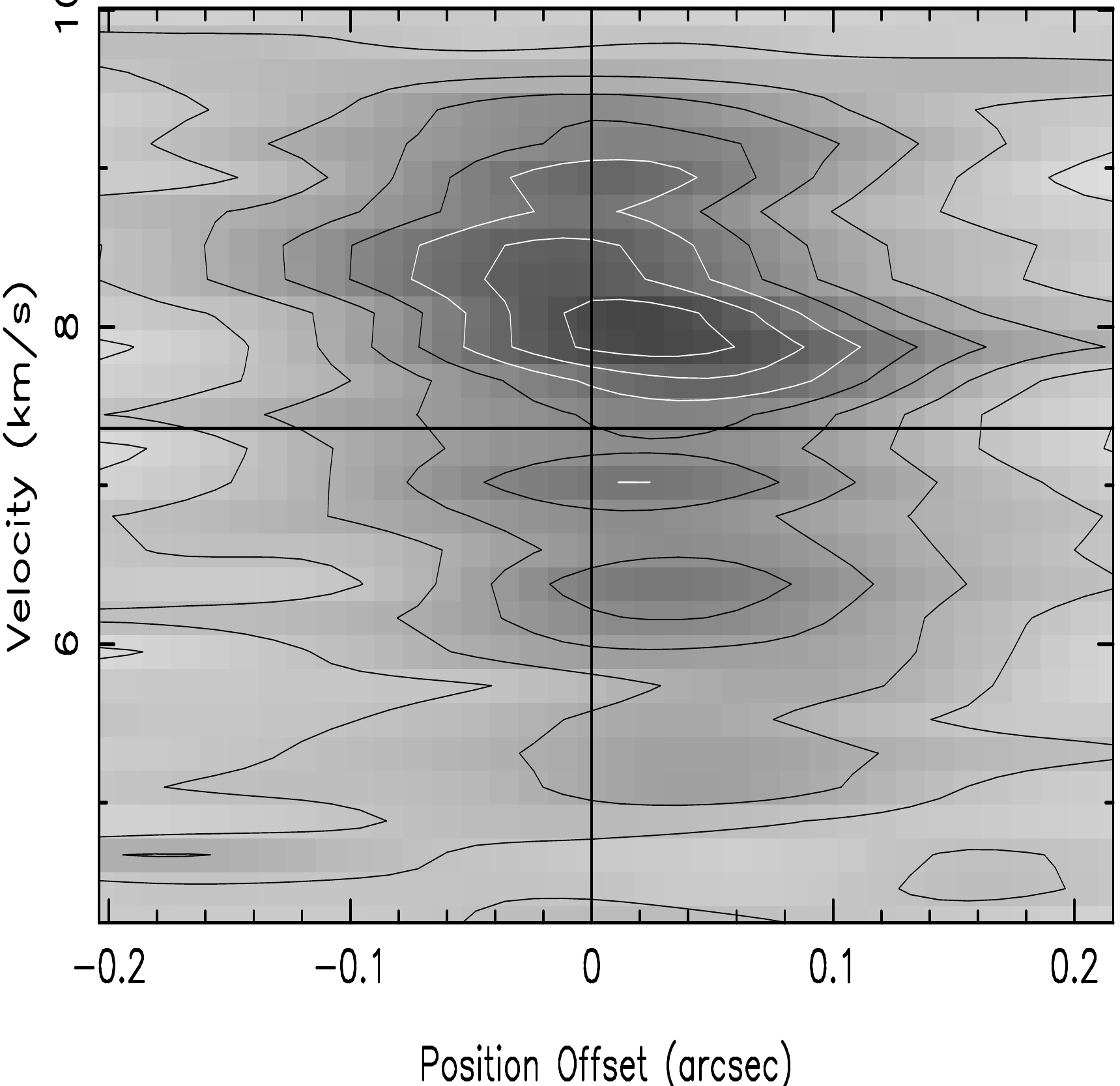}
\includegraphics[width=0.39\textwidth]{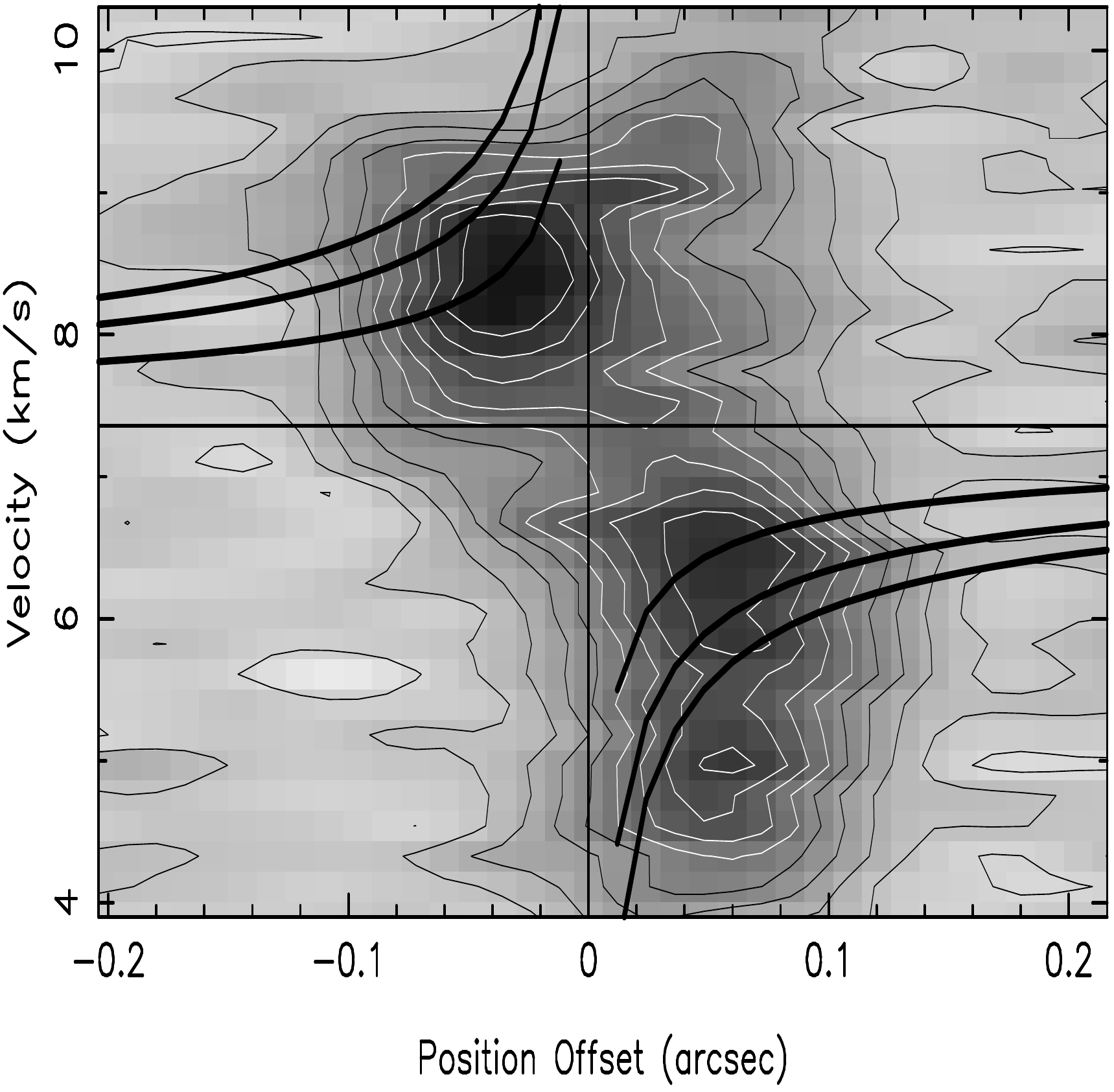}
\includegraphics[width=0.39\textwidth]{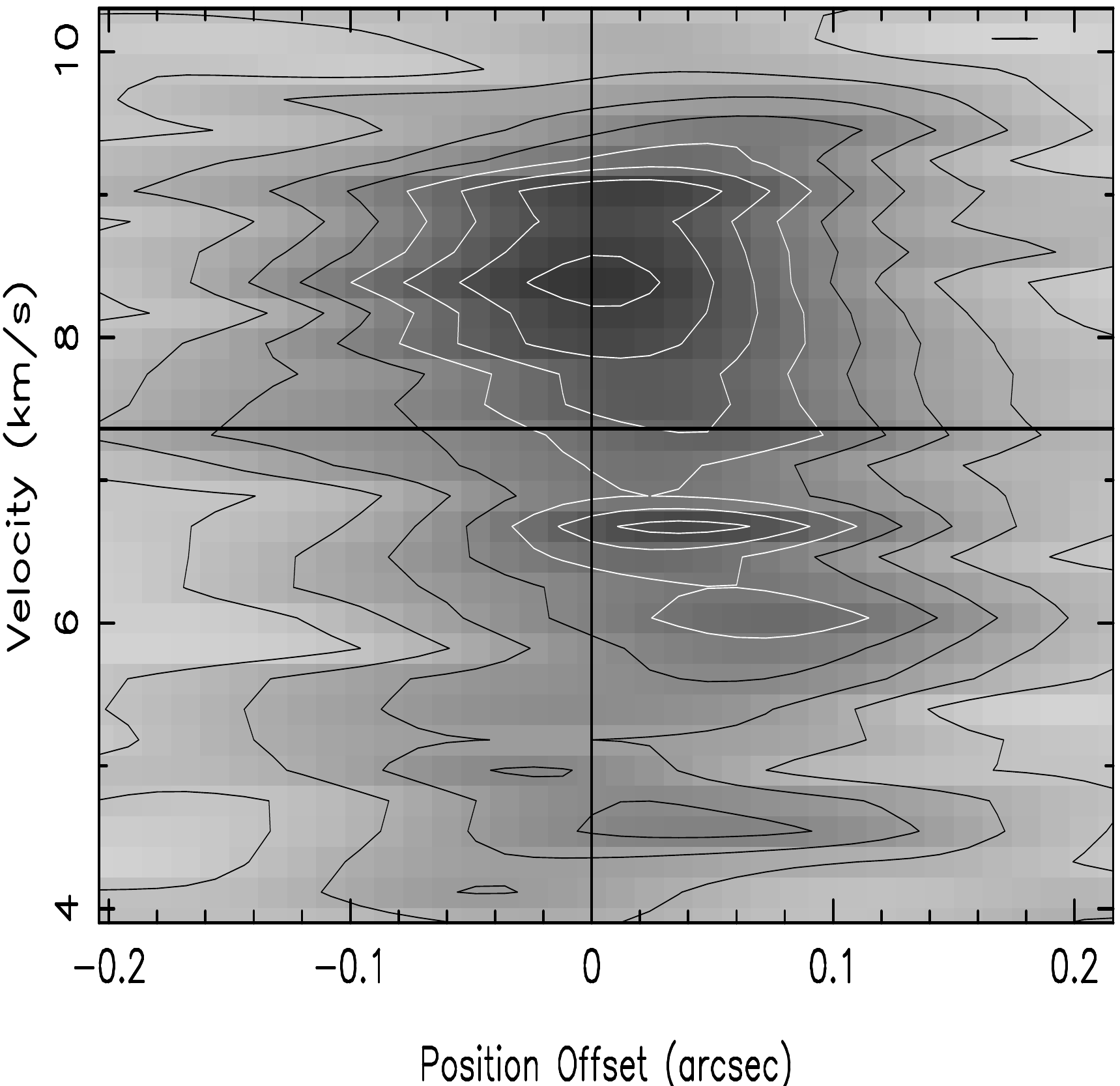}
\includegraphics[width=0.39\textwidth]{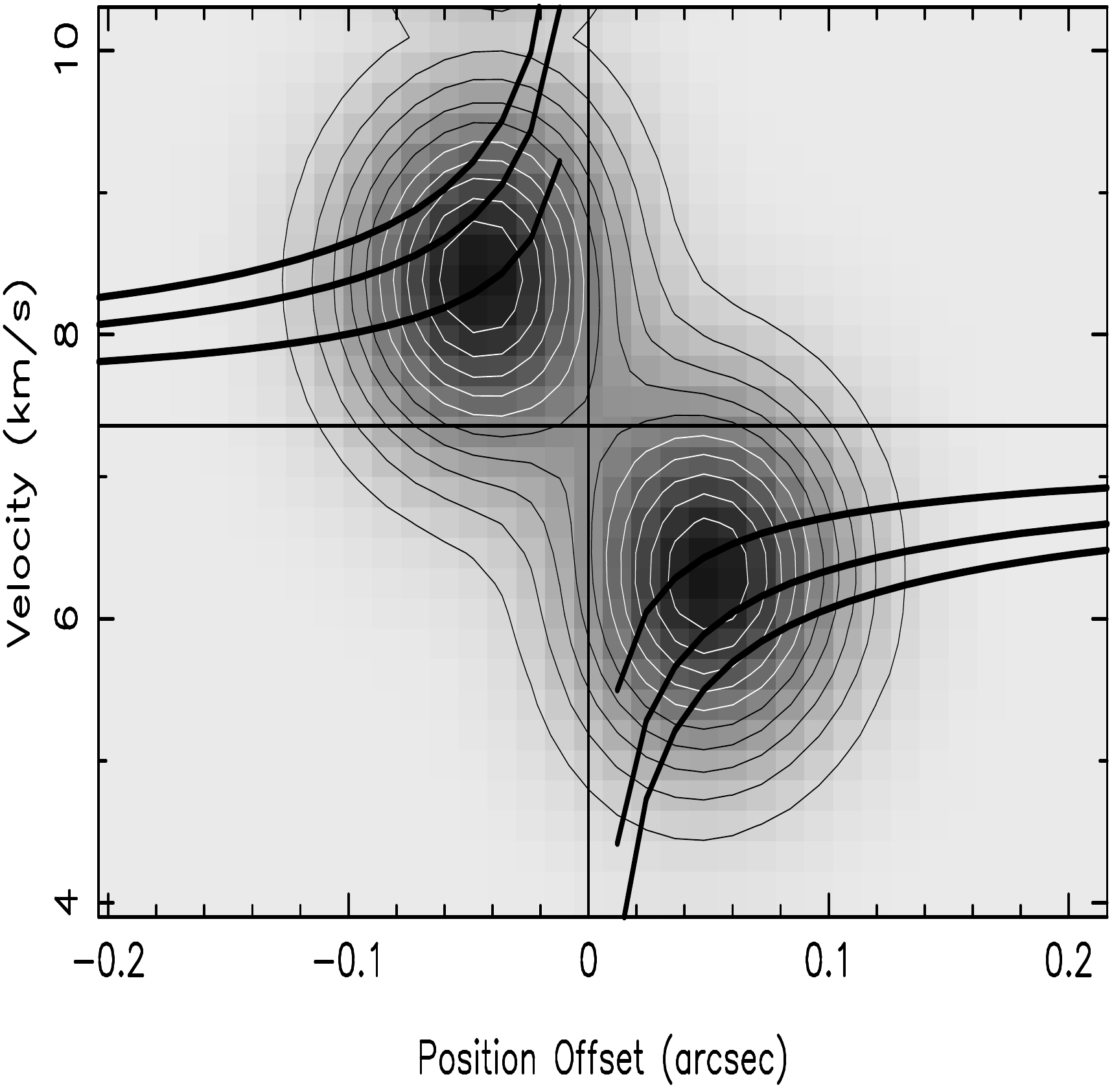}
\includegraphics[width=0.39\textwidth]{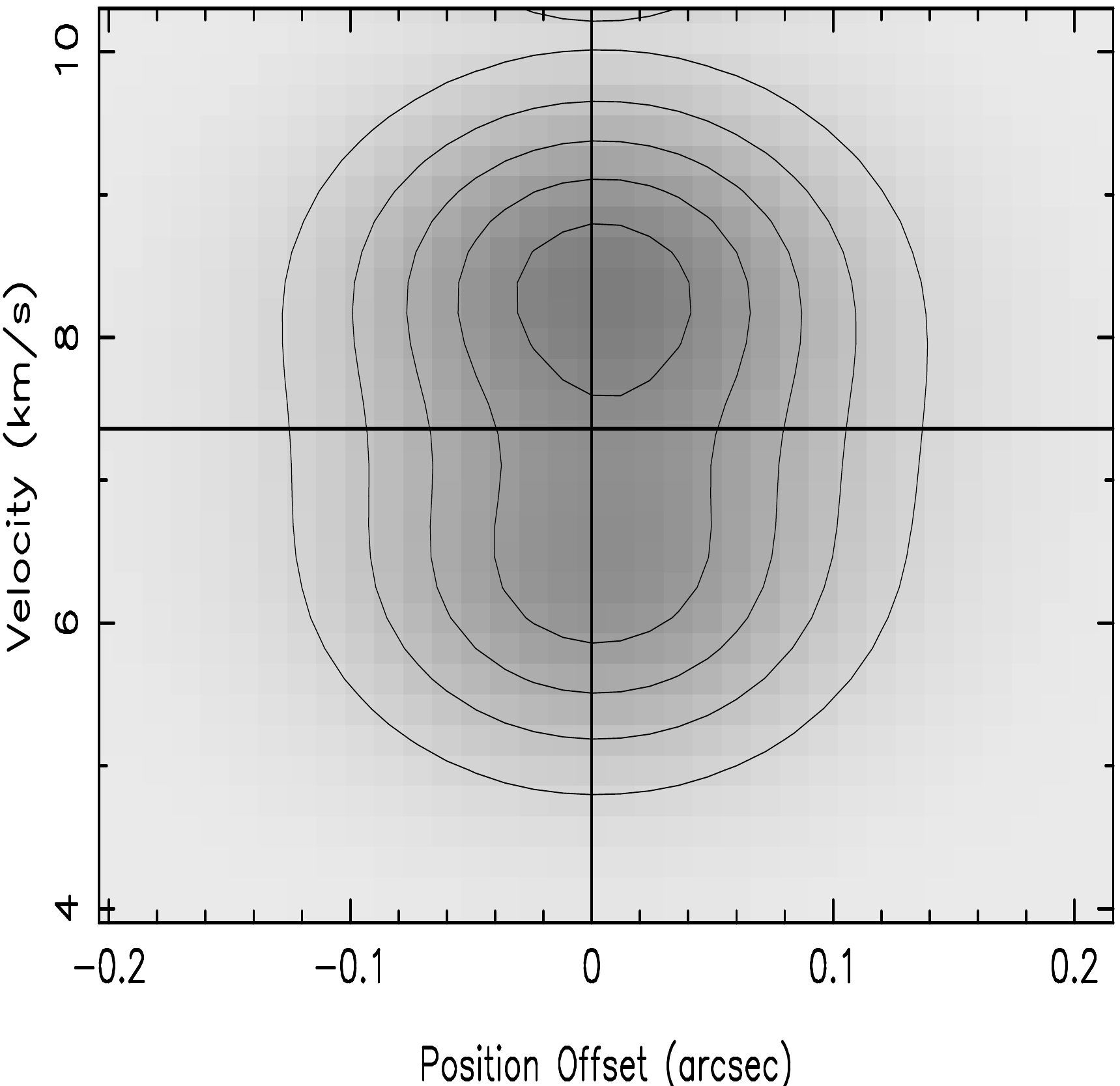}
\caption{
VLA 4A circumstellar disk: Position-velocity cuts for EthyGly1 (\emph{top panels}), EthyGly2 (\emph{middle panels}), and the thin-disk model (\emph{bottom panels}).
The left panels show the cuts along the major axis (east-west), while the right panels show the perpendicular direction. Offsets are positive westward and northward, respectively.
The model is for a central mass $M_\ast=0.25~M_\odot$, $R\s{inn}=1.8$ au, and $R\s{out}=33.5$ au, with an inclination $i=22^\circ$.
Contours start at 10\% of the peak of the EthyGly2 emission (peak = 0.047 Jy beam$^{-1}$), increasing in steps of 10\%. 
The curves in the major axis cuts are the Keplerian rotation curves for central masses $M_\ast=0.10$, 0.25, and 0.40 $M_\odot$, which are plotted as a reference.
}
\label{fig:gly_pvcuts}
\end{figure*}

\begin{figure}[htbp]
\centering
\includegraphics[width=0.47\textwidth]{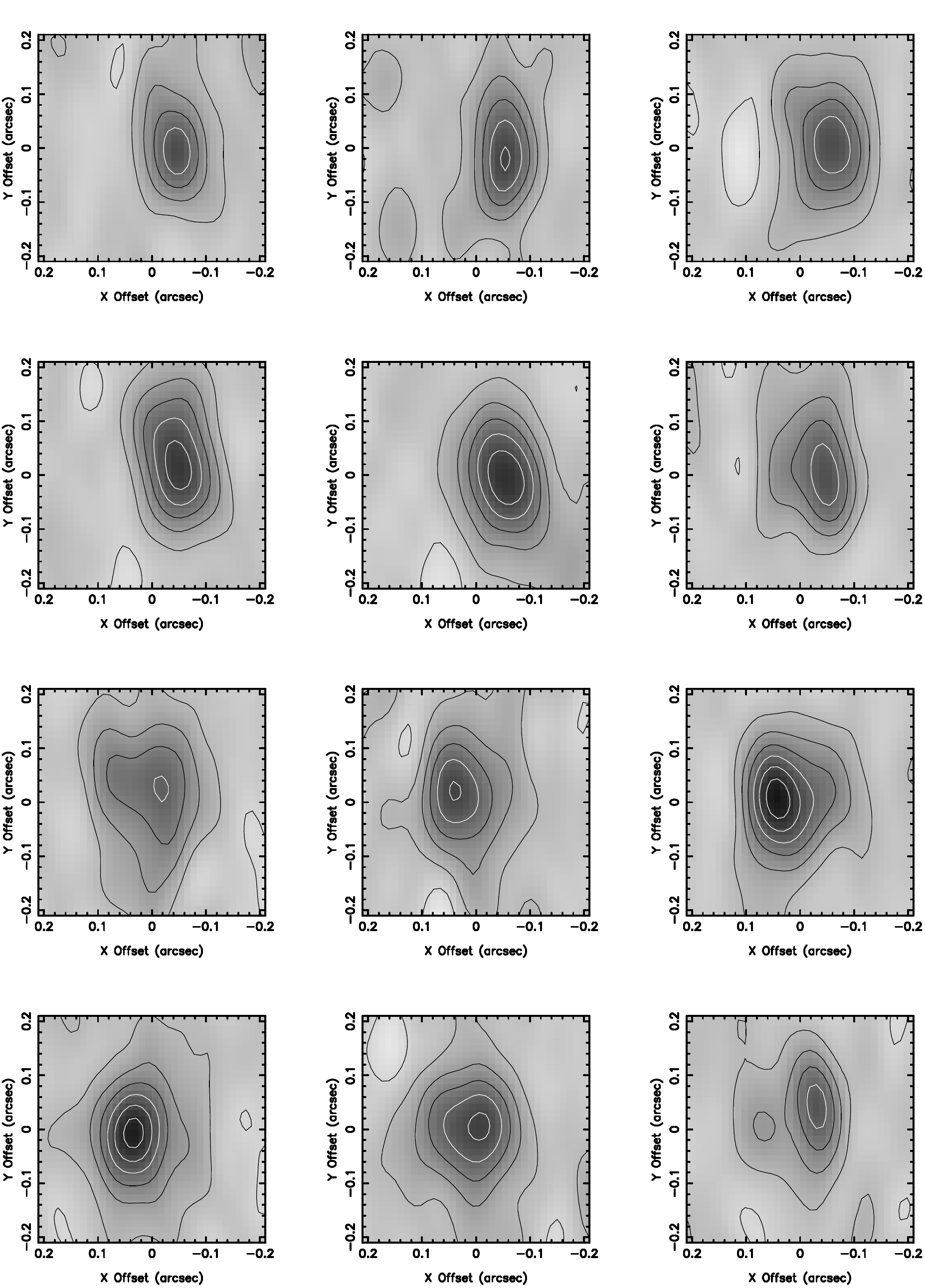}\hspace{\fill}
\includegraphics[width=0.47\textwidth]{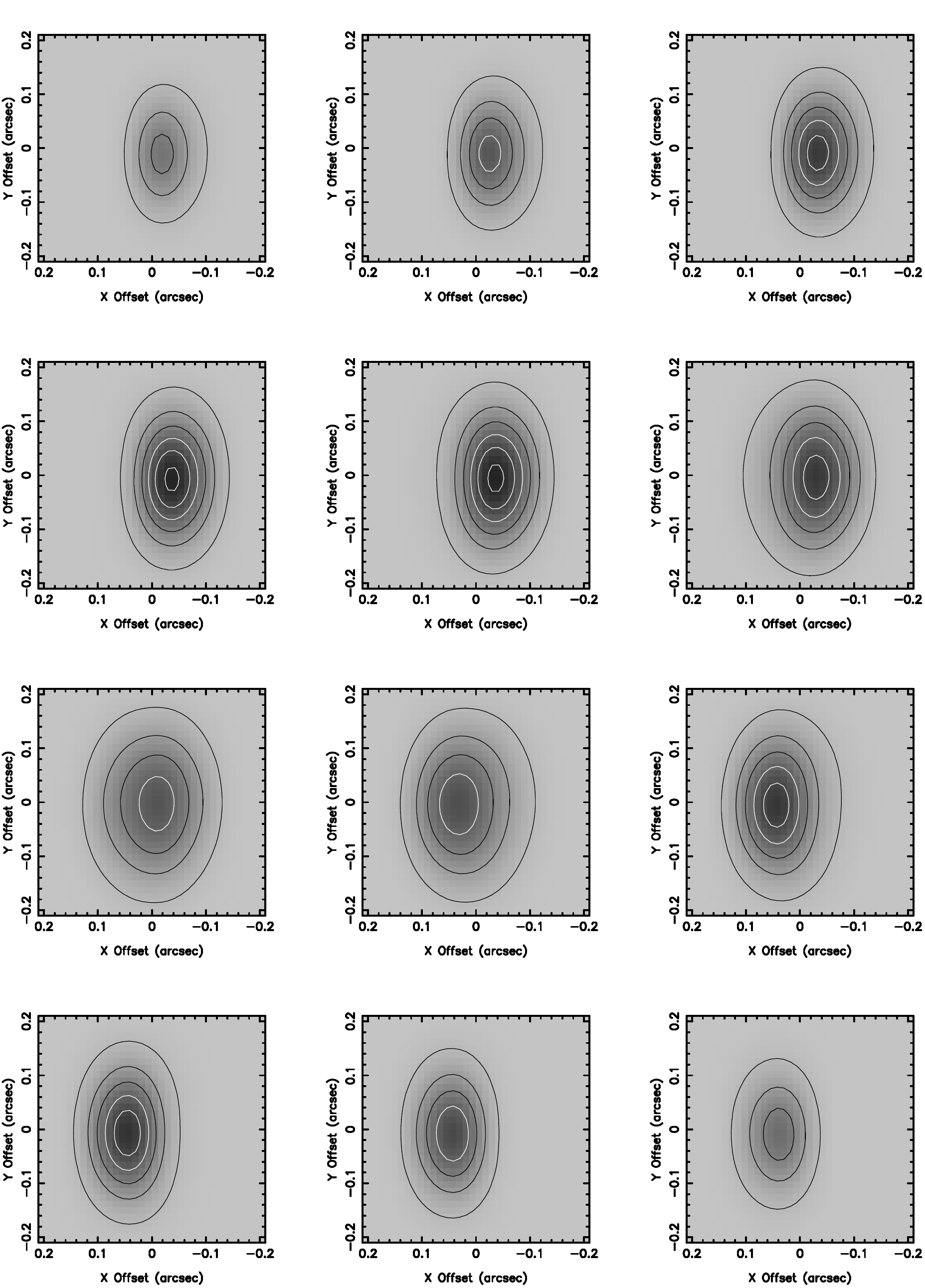}
\caption{
VLA 4A circumstellar disk: Channel maps around the reference velocity of the source, $V\s{sys}=7.36$ km s$^{-1}$, of EthyGly2 (\emph{three leftmost columns}) and the thin-disk model (\emph{three rightmost columns}). Velocities range from 4.76 km s$^{-1}$ (top-left corner) to 9.45 km s$^{-1}$ (bottom-right corner), in steps of 0.43 km s$^{-1}$. Contours start at 10\% of the peak (0.047 Jy beam$^{-1}$), increasing in steps of 13\%. 
Offsets are relative to the position of VLA 4A (Table~\ref{tab:pos}), in right ascension (X axis) and declination (Y axis), positive toward east and north respectively.
}
\label{fig:gly_chan}
\end{figure}

Figure \ref{fig:gly_mom1} shows a comparison of the zeroth-order and first-order moment maps of the EthyGly1 and EthyGly2 transitions with the thin-disk model. Figures~\ref{fig:gly_pvcuts} and \ref{fig:gly_chan} show a comparison of the observed and thin-disk model PV plots and channel maps, respectively.

The results obtained for the three lines are remarkably consistent. The linewidth of EthyGly3 is consistent with those of EthyGly1 and EthyGly2 if we take into account the separation of the two blended lines in EthyGly3 (see Table \ref{tab:gly}).
In summary, the ethylene glycol data indicate the presence of a rotating disk around VLA 4A, with an inner radius of $\sim2$ au, an outer radius of 33 au, an inclination of $22^\circ$ ($0^\circ$ for a face-on disk), a Keplerian rotation velocity with a value of 0.9 km s$^{-1}$ at a reference radius of 300 au, corresponding to  2.7 km s$^ {-1}$ at $r= 30$ au, or 9 km s$^{-1}$ at 3 au.
The Keplerian rotation implies a central mass of $\sim0.25$ $M_\odot$.

\section{Fitting of an Optically Thin Spherical Infalling Envelope Model}\label{sec:infall}

\restartappendixnumbering

\subsection{Infalling Spherical Envelope Centered on VLA 4A traced by CS (7--6)}

\begin{figure}[htbp]
\centering
\includegraphics[height=1.2\textwidth]{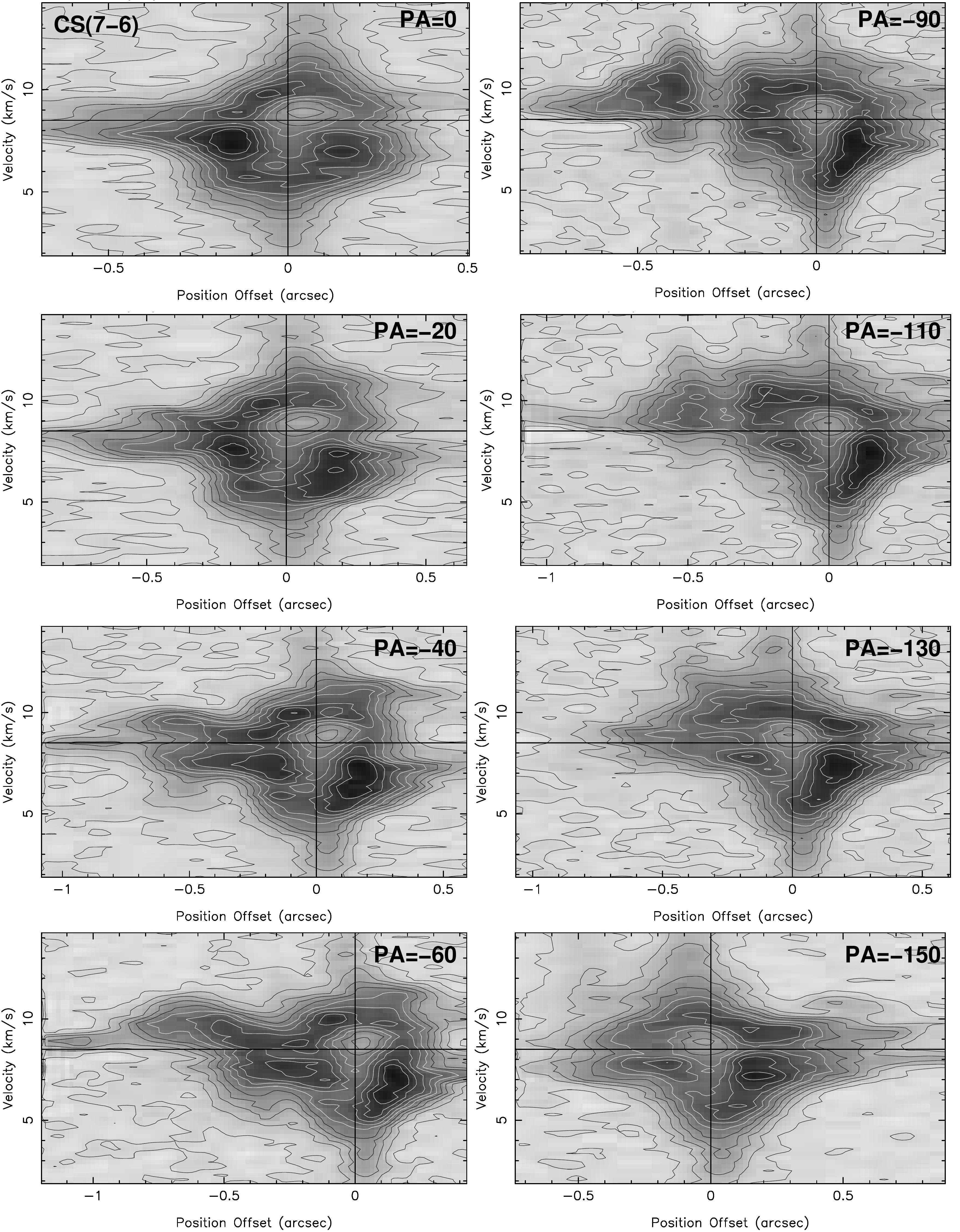}
\caption{
VLA 4A envelope: Position-velocity cuts of the CS (7--6) emission, centered on the position of VLA 4A, for different position angles. The position offsets are distances from VLA 4A along the position angle, which is indicated for each panel. Positive offsets correspond to points north of VLA 4A in the cut at $\mathrm{PA}= 0^\circ$ and to points west of VLA 4A in the cuts from $\mathrm{PA}= -20^\circ$ to $-150^\circ$. Contours start at 10\% of the peak (0.120 Jy beam$^{-1}$), increasing in steps of 10\%.}
\label{fig:cs_pvcuts}
\end{figure}

The integrated CS (7--6) emission is shown in Figure \ref{fig:CS} (left column panels). The PV cuts through the position of VLA 4A show high velocities near its center, suggesting gravitational acceleration (Fig.~\ref{fig:CSPV}). The morphology of the PV cuts is quite similar for all the position angles (see Figs.\ \ref{fig:CSPV} and \ref{fig:cs_pvcuts}), suggesting that assuming spherical symmetry is an acceptable approximation. The excess emission at offsets $\sim-0\farcs5$ in the cuts at PA = $-90^\circ$ and $-110^\circ$ is at the position of VLA 4B, and is associated with its own envelope. We assume that this emission does not belong to the envelope around VLA 4A.

Thus, the CS emission seems to be tracing an infalling envelope. We compared the observed emission from that of a spherically symmetric envelope in infall, assuming that the line emission is optically thin (perhaps only for the wings of the line), with infall velocity and temperature proportional to $r^{-1/2}$ (characteristic of the main accretion phase of infall), with no rotation, inner radius $R_\mathrm{inn}$, outer radius $R_\mathrm{out}$, and constant linewidth $\Delta V$. The volume density of the envelope, as a consequence of the continuity equation, must be $\rho\propto r^{-3/2}$. For such a model we computed the emission as a function of position and velocity, and calculated the 2-D convolution of the emission with a Gaussian beam. In particular, the intensity as a function of the projected radius was calculated at each line-of-sight velocity as proportional to the product of the temperature and density of the intersection of the isovelocity surface with the line-of-sight, and to the derivative of the line-of-sight distance with respect to the line-of-sight velocity (R. Estalella et al.\ in preparation).

\begin{figure}[htbp]
\centering
\includegraphics[height=8cm]{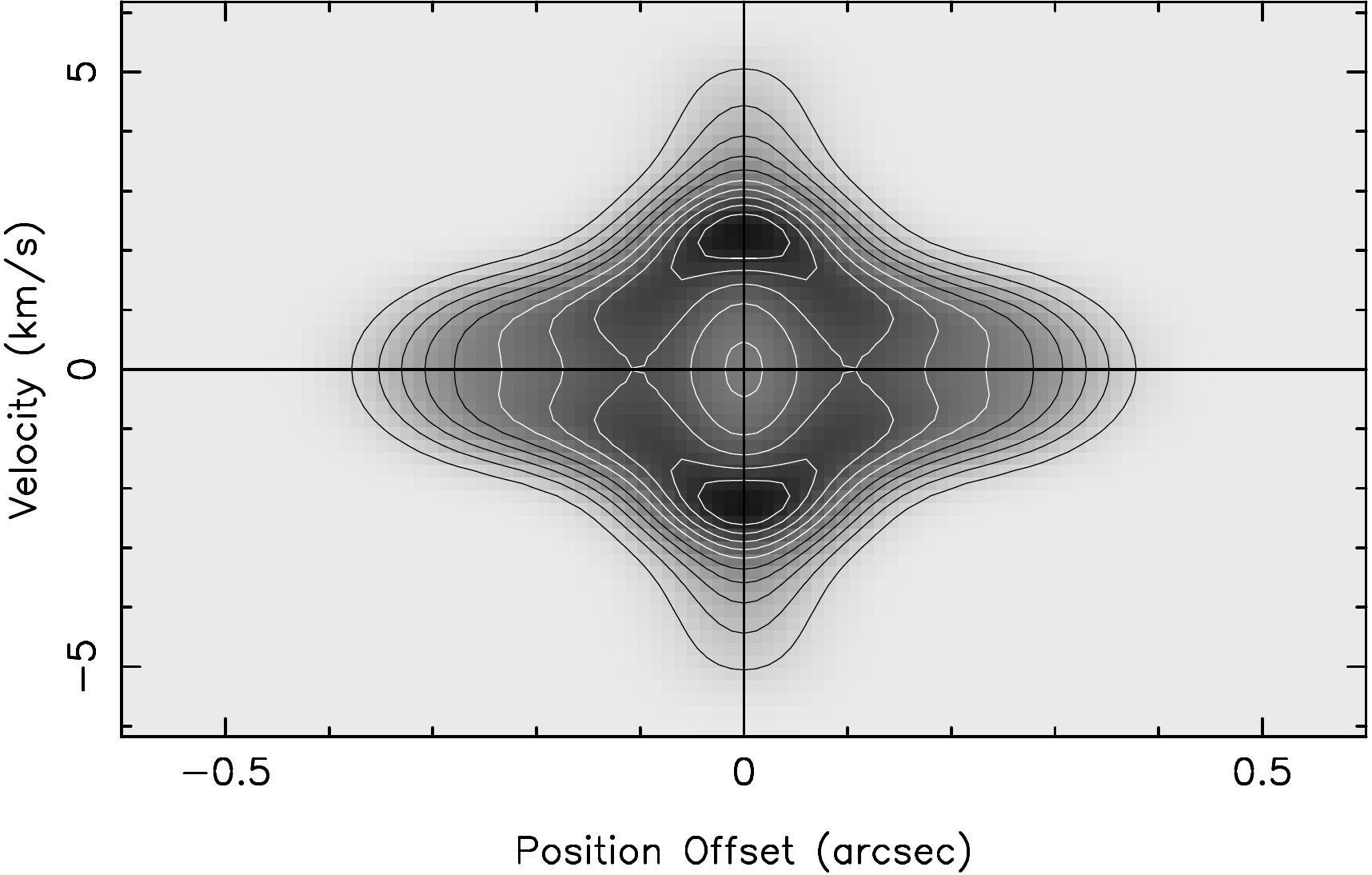}
\caption{
VLA 4A envelope: Position-velocity cut of a model of spherically symmetric envelope in free-fall, with 
$\Delta V= 1$ \kms{}, $R\s{inn}=  0\farcs05$ (15 au), $R\s{out}=  0\farcs35$ (105 au), $V_0= 1.2$ \kms{} (at a reference radius of 300 au), giving a central mass $M_\ast= 0.26$ $M_\odot$. Contours start at 10\% of the peak (arbitrary intensity), increasing in steps of 10\%. 
\label{fig:cs_pvcut_model}
}
\end{figure}

A satisfactory agreement was obtained qualitatively for the range of parameters shown in Table \ref{tab:env_fit}. The model PV cut is shown in Figure \ref{fig:cs_pvcut_model}. The inner radius of the envelope is $R\s{inn}=  10$--20 au, and the outer radius is $R\s{out}=  90$--120 au. The infall velocity is $V_0= 1.0$--1.5 \kms{} at a reference radius of $1''$ (300 au), corresponding to 1.7--2.6 \kms{} at 100 au. This range of infall velocities corresponds to the free-fall onto a central mass $M_\ast= 0.17$--0.38 $M_\odot$.

\begin{table}
\centering
\caption{VLA 4A: Infalling Spherical Envelope traced by the CS (7--6) line.}
\label{tab:env_fit}
\begin{tabular}{lcl}
\hline\hline
Parameter    & Value & Units     \\
\hline
Linewidth $\Delta V$          & 0.5--1.5    & \kms{}    \\ 
Inner radius $R\s{inn}$       & 0.04--0.07  & arcsec    \\ 
                              & 10--20      & au        \\ 
Outer radius $R\s{out}$       & 0.3--0.4    & arcsec    \\ 
                              & 90--120 & au        \\ 
Infall velocity ${V_0}^{\,a}$ & 1.0--1.5    & \kms{}    \\
Central mass $M_\ast$         & 0.17--0.38  & $M_\odot$ \\
\hline
\end{tabular}

$^a$ At a radius of $1''$ (300 au).
\end{table}

\subsection{Infalling Spherical Envelope Centered on VLA 4B traced by CS (7--6)}

\begin{figure}[htbp]
\centering
\includegraphics[width=\textwidth]{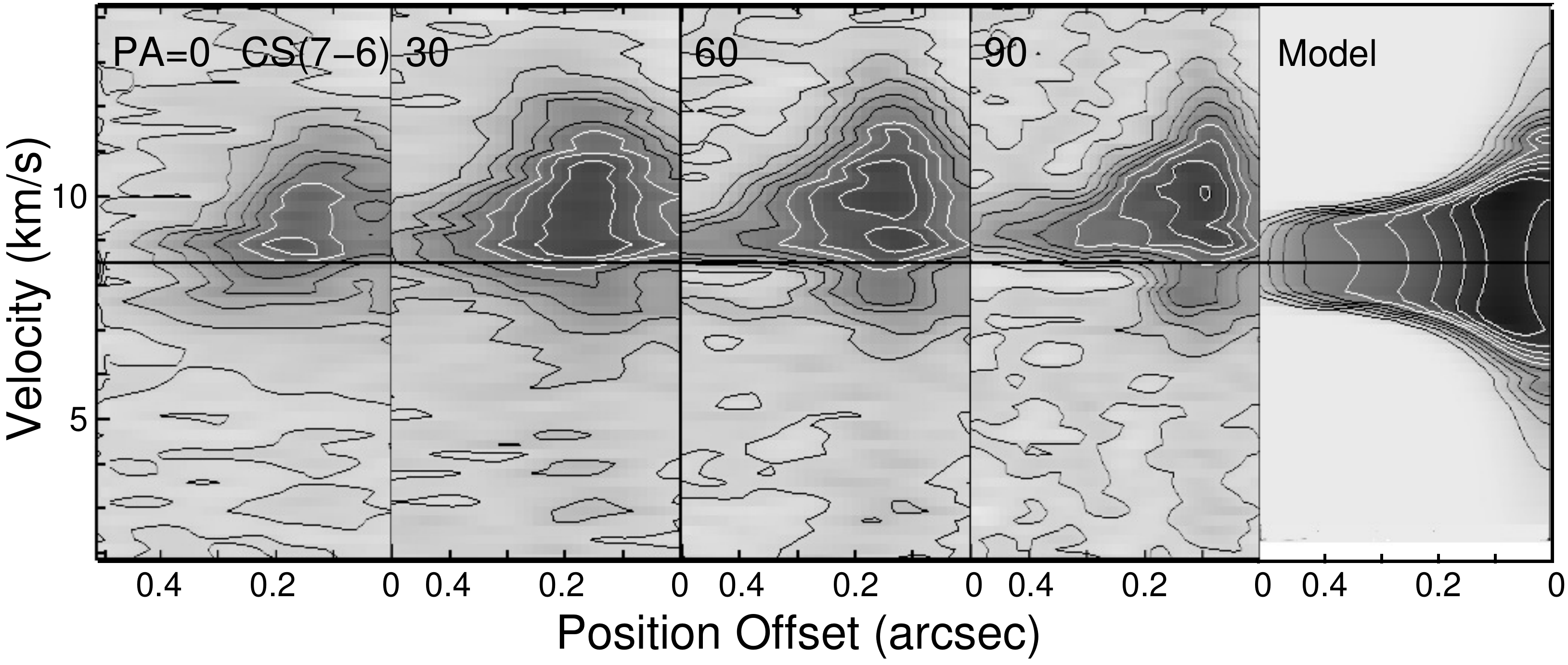}
\caption{
VLA 4B envelope: (\emph{Leftmost panels}) Position-velocity cuts of the CS (7--6) emission, through the position of VLA 4B, for different position angles, from $\mathrm{PA}=0^\circ$ to $90^\circ$. (\emph{Rightmost panel}) Position-velocity cut of a model of a spherically symmetric envelope in free-fall, with 
$\Delta V= 0.3$ \kms{}, 
$R\s{inn}=  0\farcs04$ (12 au), 
$R\s{out}=  0\farcs5$ (150 au), 
$V_0= 1.8$ \kms{} (at a reference radius of 300 au), 
giving a central mass $M_\ast= 0.6$ $M_\odot$.
Contours start at 10\% of the peak (0.100 Jy beam$^{-1}$), increasing in steps of 10\%. 
The position offsets are distances from VLA 4B along the position angle, which is indicated for each panel.
}
\label{fig:cs4b_pvcuts}
\end{figure}

Because of the presence of the large envelope around VLA 4A (Fig.\ \ref{fig:CS}), in the fitting of the VLA 4B data, we consider only positions east of VLA 4B, away from the position of VLA 4A. Also, there is a lack of emission to the south of VLA 4B, which could be due to the disruption of the southern half of the envelope by the blue-shifted jet (Fig.\ \ref{fig:VLA4B}). Therefore, we restrict our fitting to position angles in the range $0^\circ$ to 90$^\circ$. In Figure~\ref{fig:cs4b_pvcuts} we show position-velocity cuts centered on the position of VLA 4B, for position angles in this range. As can be seen in the figure, the red-shifted velocities, $V>9.3$ \kms{}, show a pattern similar to that expected for an infalling spherical envelope.
 
We fitted the model of an infalling envelope with optically thin emission, with infall velocity and temperature proportional to $r^{-1/2}$, with no rotation, inner radius $R_\mathrm{inn}$, outer radius $R_\mathrm{out}$, and constant linewidth $\Delta V$.  The range of values of the parameters that fit better the pattern observed in the PV cuts are given in Table \ref{tab:env4b_fit}. The model PV cut is shown in the rightmost panel of Figure \ref{fig:cs4b_pvcuts}. The inner radius of the envelope is $R\s{inn}\simeq 12$ au, and the outer radius is $R\s{out}\simeq 150$ au. The infall velocity is $V_0 =$ 1.5-2.2 \kms{} at a reference radius of $1''$ (300 au), corresponding to 2.6-3.8 \kms{} at 100 au. These infall velocities correspond to the free-fall onto a central mass $M_\ast \simeq 0.6$ $M_\odot$.

\begin{table}
\centering
\caption{VLA 4B. Infalling spherical envelope traced by the CS (7--6) line.}
\label{tab:env4b_fit}
\begin{tabular}{lcl}
\hline\hline
Parameter    & Value & Units     \\
\hline
Linewidth $\Delta V$          & $\sim0.3$   & \kms{} \\
Inner radius $R\s{inn}$       & $\sim0.04$  & arcsec    \\
                              & $\sim12$    & au        \\
Outer radius $R\s{out}$       & $\sim0.5$   & arcsec    \\
                              & $\sim150$   & au        \\
Infall velocity ${V_0}^{\,a}$ & 1.5--2.2    & \kms{}    \\
Central mass $M_\ast$         & 0.38--0.82  & $M_\odot$ \\
\hline
\end{tabular}

$^a$ At a radius of $1''$ (300 au).
\end{table}





\end{document}